\documentclass[acmsmall,dvipsnames,table,screen]{acmart}

\setcopyright{rightsretained}
\acmJournal{PACMPL}
\acmVolume{4}
\acmNumber{ICFP}
\acmArticle{109}
\acmYear{2020}
\acmMonth{8}
\acmDOI{10.1145/3408991}
\acmPrice{}
\copyrightyear{2020}
\acmSubmissionID{icfp20main-p104-p}
\startPage{1}

\usepackage{booktabs}   %% For formal tables:
\usepackage{subcaption} %% For complex figures with subfigures/subcaptions
\usepackage{mathpartir}

\usepackage{ulem}
\usepackage{enumitem}
\setlist
  { leftmargin = 5.5mm
  , itemsep = 3pt
  , topsep = 4pt
  }

\usepackage{fancyvrb}

\usepackage{setspace}

\usepackage{mathtools}

\usepackage[colorinlistoftodos]{todonotes}

\usepackage{listings} % Code highlighting

\usepackage{mdframed}

\newcommand{\mynote}[3]
  {\textcolor{#3}{{#2}}}

\def\parahead#1
  {\paragraph{\textbf{#1.}}}
\def\subparahead#1
  {\paragraph{#1.}}

\newcommand{\ie}{{i.e.}}
\newcommand{\eg}{{e.g.}}
\newcommand{\etc}{{etc.}}
\newcommand{\cf}{{cf.}}

\newcommand{\experimentTableSize}
  {\scriptsize}

\newcommand{\sns}
  {\ensuremath{\textsc{Sketch-n-Sketch}}}

\newcommand{\snsMyth}
  {\ensuremath{\textsc{Smyth}}}
\newcommand{\smyth}
  {\ensuremath{\snsMyth{}}}
\newcommand{\coreSnsMythPlain}
  {Core Smyth}
\newcommand{\coreSnsMyth}
  {\ensuremath{\textsc{Core Smyth}}}

\newcommand{\hazelnutLive}
  {\ensuremath{\textsc{Hazelnut Live}}}
\newcommand{\myth}
  {\ensuremath{\textsc{Myth}}}
\newcommand{\sketch}
  {\ensuremath{\textsc{Sketch}}}
\newcommand{\rosette}
  {\ensuremath{\textsc{Rosette}}}
\newcommand{\escher}
  {\ensuremath{\textsc{Escher}}}
\newcommand{\leon}
  {\ensuremath{\textsc{Leon}}}
\newcommand{\synquid}
  {\ensuremath{\textsc{Synquid}}}
\newcommand{\lambdaSquared}
  {\ensuremath{\textsc{$\lambda^2$}}}
\newcommand{\mythTwo}
  {\ensuremath{\textsc{Myth2}}}
\newcommand{\morpheus}
  {\ensuremath{\textsc{Morpheus}}}
\newcommand{\viser}
  {\ensuremath{\textsc{Viser}}}

\newif\ifarxiv

\arxivtrue

\ifarxiv

  \newcommand{\citepAppendix}[1]{(#1)}
  \newcommand{\citetAppendix}[1]{the appendix~(#1)}

  \newcommand{\refAppendixB}{\autoref{sec:appendix-random-graphs}}
  \newcommand{\refAppendixC}{\autoref{sec:appendix-polymorphism}}

  \newcommand{\refAppendixSyntax}{\autoref{sec:appendix-syntax}}
  \newcommand{\refAppendixTypeChecking}{\autoref{sec:appendix-type-checking}}
  \newcommand{\refAppendixTypeSoundness}{\autoref{sec:appendix-type-soundness}}
  \newcommand{\refAppendixResumption}{\autoref{sec:appendix-resumption}}
  \newcommand{\refAppendixConstraintMerge}{\autoref{sec:appendix-constraint-merge}}
  \newcommand{\refAppendixGuessing}{\autoref{sec:appendix-guessing}}

\else

  \newcommand{\citepAppendix}[1]{(#1)}
  \newcommand{\citetAppendix}[1]{the appendix~\mbox{(#1)}}

  \newcommand{\refAppendixB}{\S B}
  \newcommand{\refAppendixC}{\S C}

  \newcommand{\refAppendixSyntax}{\S A.1}
  \newcommand{\refAppendixTypeChecking}{\S A.2}
  \newcommand{\refAppendixTypeSoundness}{\S A.3}
  \newcommand{\refAppendixResumption}{\S A.4}
  \newcommand{\refAppendixConstraintMerge}{\S A.5}
  \newcommand{\refAppendixGuessing}{\S A.6}

\fi

\newcommand{\JudgementBox}[1]{{
  \setlength{\fboxsep}{3pt} % 3pt is the default
  \fbox{#1}
}}

\newcommand{\TightJudgementBox}[1]{{
  \setlength{\fboxsep}{2.5pt}
  \fbox{\hspace{0.5pt}#1\hspace{0.5pt}}
}}

\newcommand{\relDescription}[1]{\ensuremath{\textrm{\textbf{#1}}}}
\newcommand{\judgementHead}[2]
  {\ensuremath{\relDescription{#1}\hfill\JudgementBox{#2}}}
\newcommand{\tightJudgementHead}[2]
  {\ensuremath{\relDescription{#1}\hfill\TightJudgementBox{#2}}}
\newcommand{\tightJudgementHeadTwo}[3]
  {\ensuremath{\relDescription{#1}\ \textrm{#2}\hfill\TightJudgementBox{#3}}}

\newcommand{\judgementHeadNameOnly}[1]
  {\ensuremath{\relDescription{#1}\hfill}}
\newcommand{\judgementHeadNameOnlyTwo}[2]
  {\ensuremath{\relDescription{#1}\ \textrm{#2}\hfill}}

\newcommand{\ruleName}[1]{\mbox{\textsc{\begin{normalsize}#1\end{normalsize}}}}
\newcommand{\ruleNameFig}[1]{\textsc{\begin{scriptsize}[#1]\end{scriptsize}}}

\newcommand{\ruleNameCaption}[1]{\textsc{#1}}

\newcommand{\maybeUnderline}[1]
  {\underline{#1}}
\newcommand{\maybeUnderlineSyntaxFig}[1]
  {\underline{#1}}

\newcommand{\dimColor}[1]
  {\textcolor{gray}{#1}}

\newcommand{\SpacedBox}[2]{{
  \setlength{\fboxrule}{0.5pt}
  \setlength{\fboxsep}{3pt}
  \begin{tabular}{l}
    \textit{\ \ \ \color{gray}\scriptsize #1} \\
    \fcolorbox{gray}{white}{#2} \\[0.5em]
  \end{tabular}
}}

\newcommand{\highlightSubGoalCreation}[1]{
  \SpacedBox{New Goal}{#1}
}

\newcommand{\highlightSubGoalsCreationTwo}[1]{
  \SpacedBox{New Goals, $ \color{black} i = 1, 2 $}{#1}
}

\newcommand{\highlightSubGoalsCreation}[1]{
  \SpacedBox{New Goals, $ \color{black} i = 1, 2, \ldots, n $}{#1}
}

\definecolor{constraintLight}{HTML}{d5eced}
\definecolor{constraintDark}{HTML}{000000}

\definecolor{liveBiSatBackground}{HTML}{eaf9eb}
\definecolor{liveBiSatText}{HTML}{000000}

\definecolor{goalBackground}{HTML}{fff7e0}
\definecolor{goalText}{HTML}{000000}

\definecolor{assertionBackground}{HTML}{ffebe9}
\definecolor{assertionText}{HTML}{000000}

\definecolor{holeClosureBackground}{HTML}{EEEEEE}
\definecolor{indetBorder}{HTML}{888888}

\newcommand{\ColorBorderBox}[4]{
  \setlength{\fboxrule}{1.25pt}
  \setlength{\fboxsep}{1pt}
  \fcolorbox{#1}{#2}{\color{#3} #4}
}

\newcommand{\ColorBox}[3]{
  \setlength{\fboxrule}{0pt}
  \setlength{\fboxsep}{1pt}
  \fcolorbox{#1}{#1}{\color{#2} #3}
}

\newcommand{\highlightUnsolvedConstraint}[1]{
  \ColorBorderBox{constraintLight}{white}{constraintDark}{#1}
}

\newcommand{\highlightSolvedConstraint}[1]{
  \ColorBox{constraintLight}{constraintDark}{#1\vphantom{$_\varHoleName$}}
}

\newcommand{\highlightGoal}[1]{
  \ColorBox{goalBackground}{goalText}{#1}
}

\newcommand{\highlightAssertion}[1]{
  \ColorBox{assertionBackground}{assertionText}{\vphantom{$A_1$}#1}
}

\newcommand{\highlightHoleClosure}[1]{
  \ColorBox{holeClosureBackground}{black}{\vphantom{$E_1$}#1}
}

\newcommand{\highlightIndet}[1]{
  \ColorBox{holeClosureBackground}{black}{\vphantom{$E_1$}#1}
}

\newcommand{\beginVerbatim}
  {\begin{Verbatim}[commandchars=\\\{\},
                    codes={\catcode`\$=3\catcode`\^=7\catcode`\_=8},
                    xleftmargin=4mm %% TODO for Intro
                   ]}

\newcommand{\myVerb}
  {\Verb[commandchars=\\\{\}, codes={\catcode`\$=3\catcode`\^=7\catcode`\_=8}]}

\newcommand{\filling}[1]
  {\ensuremath{\hspace{0.02in}\fbox{$\mathtt{\vphantom{S}{#1}\vphantom{S}}$}\hspace{0.02in}}}

\newcommand{\fillingTightBoxName}[2]
  {\ensuremath{\fillingTightBox{\expHole{#1}\hspace{0.01in}=\hspace{0.01in}#2}}}

\newcommand{\fillingTightBox}[1]
  {\ensuremath{
     \setlength{\fboxsep}{1.5pt}
     \fcolorbox{white}{constraintLight}
       {\hspace{0.04in}$\mathtt{#1}$\hspace{0.04in}}}}

\newtheorem*{theorem*}{Theorem}

\newcommand{\breakAndIndent}
  {\mbox{}

   \hspace{0.00in}
  }

\newcommand{\justIndent}
  {\hspace{0.00in}
  }

\newcommand{\setComp}[2]
           {\ensuremath{\{\miniSepThree #1 \hspace{0.02in}
                            \mid\hspace{0.02in} #2 \miniSepThree\}}}

\newcommand{\grungyITE}[3]
  {\ensuremath{(\textrm{if\ } #1 \textrm{\ then\ } #2 \textrm{\ else\ } #3)}}

\newcommand{\sep}{\hspace{0.06in}}
\newcommand{\sepPremise}{\hspace{0.20in}}
\newcommand{\hsepRule}
  {\hspace{0.20in}}
\newcommand{\vsepRuleHeight}{0.12in}
\newcommand{\vsepRule}{\vspace{\vsepRuleHeight}}
\newcommand{\vsepRuleNoNeed}{}
\newcommand{\miniSepOne}{\hspace{0.01in}}
\newcommand{\miniSepTwo}{\hspace{0.02in}}
\newcommand{\miniSepThree}{\hspace{0.03in}}

\newcommand{\vsepBeforeCaption}{\vspace{0.05in}}
\newcommand{\vsepAfterCaptionBottomFig}{\vspace{0.04in}} %% this is maxed out...

\newcommand{\figSyntaxLineBreak}{\\[1pt]}
\newcommand{\figSyntaxSpaceNextCategory}{\\[1pt]}
\newcommand{\figSyntaxSpaceNextCategoryMoreSpace}
  {\\[8pt]}
\newcommand{\figSyntaxSpaceItem}{\sep\mid\sep}

\newcommand{\figSyntaxEnd}
  {\end{array}$}
\newcommand{\figSyntaxBegin}
  {$\begin{array}{rrcll}}
\newcommand{\figSyntaxRowLabel}[2]
  {{\textbf{#1}}\!&\!{#2}&{::=}&}
\newcommand{\figSyntaxRow}{&&\miniSepOne\mid\miniSepOne&}

\newcommand{\ttlcurly}{\ensuremath{\color{CadetBlue}\texttt{\char`\{}}\hspace{0.00in}\color{Black}}
\newcommand{\ttrcurly}{\ensuremath{\color{CadetBlue}\hspace{0.00in}\texttt{\char`\}}}\color{Black}}
\newcommand{\ttlparen}{\ensuremath{\texttt{(}}}
\newcommand{\ttrparen}{\ensuremath{\texttt{)}}}

\newcommand{\ttm}{\ensuremath{\texttt{m}}}
\newcommand{\ttn}{\ensuremath{\texttt{n}}}

\newcommand{\ttx}{\ensuremath{\texttt{x}}}

\newcommand{\twoThings}[2]{{#1}\miniSepTwo,\miniSepTwo{#2}}
\newcommand{\threeThings}[3]{{#1}\miniSepTwo,\miniSepTwo{#2}\miniSepTwo,\miniSepTwo{#3}}
\newcommand{\fourThings}[4]{{#1}\miniSepTwo,\miniSepTwo{#2}\miniSepTwo,\miniSepTwo{#3}\miniSepTwo,\miniSepTwo{#4}}

\newcommand{\parens}[1]{\ensuremath{\ttlparen{#1}\ttrparen}}

\newcommand{\pair}[2]{\color{CadetBlue}\parens{\twoThings{\color{Black}#1}{\color{Black}#2}}\color{Black}}

\newcommand{\varExp}{e}
\newcommand{\varResult}{r}
\newcommand{\varVar}{x}
\newcommand{\varVarF}{f}

\newcommand{\varEnv}{E}
\newcommand{\varEx}{ex}
\newcommand{\varSimpleVal}{v}

\newcommand{\world}[2]
  {(\HoleConstraint{#1}{#2})}
\newcommand{\varWorlds}{X} %% {W}
\newcommand{\varType}{T}
\newcommand{\varTypeEnv}{\Gamma}
\newcommand{\varDatatypeEnv}{\Sigma}
\newcommand{\varHoleEnv}{\Delta}
\newcommand{\varConstraints}{K}
\newcommand{\varAssertions}{A}
\newcommand{\varSolution}{F}
\newcommand{\varSolutionGuesses}{\varSolution'}
\newcommand{\varUnsolvedConstraints}
  {U}
\newcommand{\varDataCon}{C}
\newcommand{\varDataConChoice}[1]
  {\varDataCon_{{\alpha}_{#1}}}
\newcommand{\varTypeCon}{D}
\newcommand{\varFuel}{k}
\newcommand{\varHoleName}{h}
\newcommand{\varGoals}{G}
\newcommand{\varProgram}{p}

\newcommand\doubleplus{+\kern-1.3ex+\kern0.8ex}
\newcommand\mdoubleplus{\ensuremath{\mathbin{+\mkern-7mu+}}}

\newcommand{\emptyEnv}{-}
\newcommand{\envCat}[2]{\ensuremath{{#1},\hspace{0.02in}{#2}}}
\newcommand{\envCatThree}[3]{\ensuremath{\envCat{#1}{\envCat{#2}{#3}}}}
\newcommand{\envCatFour}[4]{\ensuremath{\envCat{#1}{\envCatThree{#2}{#3}{#4}}}}
\newcommand{\envAppend}[2]
  {\ensuremath{{#1}\mdoubleplus{#2}}}
\newcommand{\envBind}[2]{\ensuremath{{#1}\mapsto{#2}}}
\newcommand{\envBindType}[2]{\ensuremath{{#1}\!:\!{#2}}}
\newcommand{\envBindHole}[2]
  {\ensuremath{{#1}\mapsto{#2}}}

\newcommand{\holeFilling}[2]
  {\ensuremath{{#1}\mapsto{#2}}}

\newcommand{\holeFillingOverview}[2]
  {\ensuremath{{#1}\!\miniSepOne=\miniSepOne\!{#2}}}

\newcommand{\unsolvedConstraint}[2]
  {\ensuremath{#1}\mapsto{#2}}

\newcommand{\assertion}[2]
  {\ensuremath{{#1}\Rightarrow{#2}}}
\newcommand{\applySubst}[2]
  {\ensuremath{\llbracket{#1}\rrbracket}\miniSepTwo{#2}}

\newcommand{\dataConstructors}[3]
  {\ensuremath{{\color{CadetBlue}\{}{#1}\ {#2}{\color{CadetBlue}\}}^{\sequenceSyntax}}}

\newcommand{\envBindDatatype}[3]
  {\ensuremath{\text{\inlinecode{type}}\ {#1}\ \text{\inlinecode{=}}\ \dataConstructors{#2}{#3}{#1}}}

\newcommand{\lookupTypeConstructors}[3]
  {\varDatatypeEnv({#1})=\dataConstructors{#2}{#3}{#1}}

\newcommand{\lookupDataConArrowType}[3]
  {\varDatatypeEnv({#3})({#1})={#2}}

\newcommand{\emptySet}{\emptyEnv}

\newcommand{\setMinus}[2]{\ensuremath{{#1}\hspace{0.02in}\backslash\hspace{0.02in}{#2}}}
\newcommand{\setUnion}[2]{\envAppend{#1}{#2}}
\newcommand{\setUnionThree}[3]{\setUnion{#1}{\setUnion{#2}{#3}}}
\newcommand{\setUnionFour}[4]{\setUnion{#1}{\setUnionThree{#2}{#3}{#4}}}

\newcommand{\tUnit}{{\color{CadetBlue}\parens{}}}
\newcommand{\tArrow}[2]
  {\ensuremath{#1}\!\color{CadetBlue}\rightarrow\!\color{Black}{#2}}
\newcommand{\tPair}[2]{\ensuremath{\color{CadetBlue}\pair{{\color{Black}#1}}{{\color{Black}#2}}}\color{Black}}

\newcommand{\tHole}[2]
  {\ensuremath{({#1}\vdash{\expHoleDot{}}:{#2})}}

\newcommand{\hole}
  {\ensuremath{\texttt{\char`?\char`?}}}

\newcommand{\expUnit}{{\color{CadetBlue}\parens{}}}
\newcommand{\expFun}[2]{\ensuremath{{\color{Plum}\lambda}{#1}{\color{CadetBlue}.}\miniSepOne{#2}}}
\newcommand{\expFunAnnot}[3]
  {\ensuremath{\expFun{#1}{#3}}}
\newcommand{\expFixFun}[4]
  {\ensuremath{\valFixFun{#1}{#3}{#4}}}
\newcommand{\valFixFun}[3]
  {\ensuremath{\text{\inlinecode{fix}}\ {#1}\ 
  \color{CadetBlue}(\color{Plum}{\lambda}{\color{Black}#2}\color{CadetBlue}.\color{Black}\miniSepOne{#3}\color{CadetBlue})\color{Black}}}
\newcommand{\expApp}[2]{\ensuremath{{#1}\ \,{#2}}}
\newcommand{\expAppTwo}[3]
  {\ensuremath{\expApp{\expApp{#1}{#2}}{#3}}}
\newcommand{\expProj}[2]
  {\ensuremath{{\color{Plum}\mathtt{prj}}_{\pairIndex{#1}}\ {#2}}}
\newcommand{\expFst}[1]
  {\ensuremath{{\color{Plum}\mathtt{prj}_{{\color{Black}1}}}\ {#1}}}
\newcommand{\expSnd}[1]
  {\ensuremath{{\color{Plum}\mathtt{prj}_{{\color{Black}2}}}\ {#1}}}
\newcommand{\expUnwrap}[2]
  {\ensuremath{\expApp{{#1}^{\hspace{0.02in}-1}}{#2}}}
\newcommand{\expPbeConstraints}[2]
  {\ensuremath{\text{\inlinecode{assert}}\
  {\color{CadetBlue}\ttlparen\color{Black}{#1}}\hspace{0.02in}\text{\inlinecode{=}}\
  {#2}{\color{CadetBlue}\ttrparen\color{Black}}}}
\newcommand{\expHole}[1]
   {\ensuremath{\hole{}_{#1}}}
\newcommand{\expHoleDot}[1]
   {\ensuremath{\bullet_{#1}}}
\newcommand{\expMatchOneBranch}[4]
  {\ensuremath{\text{\inlinecode{case}}\ {#1}\ \text{\inlinecode{of}}\
  \color{CadetBlue}\{{\color{RoyalBlue}#2}\ {#3} \color{CadetBlue} \rightarrow
  \color{Black}{#4}\color{CadetBlue}\}}\color{Black}}
\newcommand{\expMatchTwoBranches}[7]
  {\ensuremath{\text{\inlinecode{case}}\ {#1}\ \text{\inlinecode{of}}\
  \color{CadetBlue}\{{\color{RoyalBlue}#2}\color{Black}\ {#3} \color{CadetBlue}
  \rightarrow \color{Black} {#4}{\color{CadetBlue}\spaceSemiSpace\,}
                                                   {\color{RoyalBlue}{#5}}\ {#6} \color{CadetBlue}
                                                   \rightarrow \color{Black}
                                                   {#7}
                                                   \color{CadetBlue}\}}\color{Black}}

\newcommand{\expMatchTwoBranchesOverview}[7]
  {\ensuremath{\text{\inlinecode{case}}\ {#1}\              \color{CadetBlue}\{
    \color{RoyalBlue} {#2}
  \color{CadetBlue}\rightarrow
  \color{Black} {#4} \color{CadetBlue}\spaceSemiSpace \color{RoyalBlue}
                                                   {#5}\ \color{Black} {#6}
                                                   \color{CadetBlue} \rightarrow
                                                   \color{Black} {#7}}
                                                 \color{CadetBlue}\}\color{Black}}

\newcommand{\expMatchOverview}[2]
  {\ensuremath{\text{\inlinecode{case}}\ {#1}\ {#2}}\color{Black}}

\newcommand{\expMatch}[4]
  {\ensuremath{\expMatchOneBranch{#1}{\color{Black}#2}{\color{Black}#3}{#4}^{\sequenceSyntax}}\color{Black}}
\newcommand{\expMatchShort}[4]
  {\ensuremath{\text{\inlinecode{case}}\ {#1}\ ...}\color{Black}}

\newcommand{\holeClosure}[2]
  {\ensuremath{\closure{#1}{\expHole{#2}}}}
\newcommand{\caseClosure}[5]
  {\ensuremath{\closure{#1}{\expMatch{#2}{#3}{#4}{#5}}}}

\newcommand{\expAppIndet}[2]
  {\ensuremath{\expApp{\highlightIndet{$#1$}}{#2}}}
\newcommand{\expProjIndet}[2]
  {\ensuremath{\expProj{{#1}}{\highlightIndet{$#2$}}}}
\newcommand{\expFstIndet}[1]
  {\ensuremath{\expFst{\highlightIndet{$#1$}}}}
\newcommand{\expSndIndet}[1]
  {\ensuremath{\expSnd{\highlightIndet{$#1$}}}}
\newcommand{\caseClosureIndet}[5]
  {\ensuremath{\caseClosure{#1}{\highlightIndet{$#2$}}{#3}{#4}{#5}}}

\newcommand{\spaceThree}[1]{\ensuremath{\miniSepThree{#1}\miniSepThree}}
\newcommand{\expLetLong}[3]
  {\ensuremath{\text{\inlinecode{let}}\miniSepThree{#1}\spaceThree{\text{\inlinecode{=}}}{#2}\spaceThree{\text{\inlinecode{in}}}{#3}}}

\newcommand{\expProgram}[3]
  {\ensuremath{\expLetLong{\texttt{main}}{#1}{\expPbeConstraints{#2}{#3}}}}
\newcommand{\programReducesTo}[3]
  {\ensuremath{{#1}\hspace{0.01in}\Rightarrow{#2}\spaceSemiSpace{#3}}}

\newcommand{\sequenceSyntax}
  {\miniSepTwo i\in[n]}
\newcommand{\generalSequenceSyntax}[2]
  {\miniSepTwo {#1}\in[{#2}]}
\newcommand{\pairIndex}[1]
  {\miniSepTwo i\in[2]}
\newcommand{\sequenceSyntaxTwo}[1]
  {\miniSepTwo \pairIndex{#1}}
\newcommand{\multiPremise}[2]
  {\{\hspace{0.02in}{#1}\hspace{0.02in}\}^{#2}}
\newcommand{\bigMultiPremise}[2]
  {{#1}^{#2}}

\newcommand{\closure}[2]
  {\ensuremath{[{#1]\miniSepTwo{#2}}}}

\newcommand{\exHole}
  {\top}

\newcommand{\ioExample}[2]
  {\ensuremath{\ttlcurly{#1}\color{CadetBlue}\rightarrow\color{Black}{#2}}\ttrcurly}

\newcommand{\ioExampleThree}[6]
  {\ensuremath{\ttlcurly\threeThings{{#1}\color{CadetBlue}\rightarrow\color{Black}{#2}}{{#3}\color{CadetBlue}\rightarrow\color{Black}{#4}}{{#5}\color{CadetBlue}\rightarrow\color{Black}{#6}}\ttrcurly}}
\newcommand{\ioExampleThreeDots}[6]
  {\ensuremath{\ttlcurly\fourThings{{#1}\color{CadetBlue}\rightarrow\color{Black}{#2}}{{#3}\color{CadetBlue}\rightarrow\color{Black}{#4}}{{#5}\color{CadetBlue}\rightarrow\color{Black}{#6}}{\ldots}\ttrcurly}}

\newcommand{\program}[2]
  {{#1}\vdash{#2}}

\newcommand{\spaceSemiSpace}
  {\hspace{0.02in};\hspace{0.01in}}
\newcommand{\typeCheck}[3]
  {\ensuremath{{\varDatatypeEnv}\spaceSemiSpace{\varHoleEnv}\spaceSemiSpace{#1}\vdash{#2}:{#3}}}
\newcommand{\typeCheckIn}[4]
  {\ensuremath{{\varDatatypeEnv}\spaceSemiSpace{#1}\spaceSemiSpace{#2}\vdash{#3}:{#4}}}
\newcommand{\typeCheckResult}[2]
  {\ensuremath{{\varDatatypeEnv}\spaceSemiSpace{\varHoleEnv}\vdash{#1}:{#2}}}
\newcommand{\typeCheckExample}[2]
  {\ensuremath{{\varDatatypeEnv}\spaceSemiSpace{\varHoleEnv}\vdash{#1}:{#2}}}
\newcommand{\typeCheckEnv}[2]
  {\ensuremath{{\varDatatypeEnv}\spaceSemiSpace{\varHoleEnv}\vdash{#2}:{#1}}}
\newcommand{\typeCheckWorlds}[4]
  {\ensuremath{{\varDatatypeEnv}\spaceSemiSpace{#1}\vdash{#2}:{#3}\spaceSemiSpace{#4}}}
\newcommand{\typeCheckUnsolved}[2]
  {\ensuremath{{\varDatatypeEnv}\vdash{#2}:{#1}}}
\newcommand{\typeCheckSolution}[2]
  {\ensuremath{{\varDatatypeEnv}\vdash{#2}:{#1}}}
\newcommand{\typeCheckProgram}[3]
  {\ensuremath{{\varDatatypeEnv}\spaceSemiSpace{\varHoleEnv}\vdash{#1}:{#2}\spaceSemiSpace{#3}}}
\newcommand{\typeCheckProgramWrapper}[2]
  {\ensuremath{{\varDatatypeEnv}\spaceSemiSpace{\varHoleEnv}\vdash{#1}:{#2}}}
\newcommand{\typeCheckAssertions}[2]
  {\ensuremath{{\varDatatypeEnv}\vdash{#2}:{#1}}}

\newcommand{\problemType}[2]
  {\ensuremath{\highlightGoal{$({#1}\vdash{\expHoleDot{}}:{#2})$}}}
\newcommand{\problemNameTypeWorlds}[4]
  {\ensuremath{\highlightGoal{$({#2}\vdash{\expHoleDot{#1}}:{#3}\models{#4})$}}}

\newcommand{\reducesTo}[3]
  {\ensuremath{\program{#1}{#2}\hspace{0.01in}\Rightarrow{#3}}}
\newcommand{\reducesToUK}[4]
  {\ensuremath{\program{#1}{#2}\hspace{0.01in}\Rightarrow{#3}\dashv{#4}}}
\newcommand{\resumesTo}[3]
  {\ensuremath{{#1}\vdash{#2}\Rightarrow{#3}}}
\newcommand{\resumesToMultiLine}[3]
  {\ensuremath{{#1}\vdash{#2}\Rightarrow\\{\displaystyle\quad\quad\quad\quad}{#3}}}
\newcommand{\resumesToUK}[4]
  {\ensuremath{{#1}\vdash{#2}\Rightarrow{#3}\dashv{#4}}}

\newcommand{\liveEval}[4]
  {\ensuremath{\dimColor{#1\spaceSemiSpace{#2}\vdash}\hspace{0.03in}{#3}\Rightarrow{#4}}}

\newcommand{\reducesToN}[4]
  {\ensuremath{\program{#1}{#2}\hspace{0.02in}\Rightarrow_{\hspace{0.01in}#4}{#3}}}
\newcommand{\reducesToUKN}[5]
  {\ensuremath{\program{#1}{#2}\hspace{0.02in}\Rightarrow_{\hspace{0.01in}#3}{#4}\dashv{#5}}}

\newcommand{\fails}[3]
  {\ensuremath{\program{#1}{#2}\hspace{0.01in}\Rightarrow_{#3 }\varnothing}}

\newcommand{\extraInputsTwo}[2]
  {\ensuremath{\dimColor{{#1}\spaceSemiSpace{#2}}}}
\newcommand{\extraInputsTwoSemi}[2]
  {\ensuremath{\extraInputsTwo{#1}{#2}\dimColor{\spaceSemiSpace{}}}}

\newcommand{\pairFillingExp}[2]
  {\ensuremath{\dimColor{#1\hspace{0.01in}\vdash}\hspace{0.05in}#2}}

\definecolor{CombinedRefineColor}{HTML}{DD0088}
\definecolor{CombinedBranchColor}{HTML}{4400AA}

\newcommand{\synthesisArrow}
  {\rightsquigarrow}
\newcommand{\synthesisArrowCaption}[1]
  {\ensuremath{\synthesisArrow_{\hspace{0.02in}\textrm{#1}}}}

\newcommand{\synthesisArrowGuess}
  {\ensuremath{\synthesisArrowCaption{guess}}}
\newcommand{\synthesisArrowGen}
  {\ensuremath{\synthesisArrowCaption{gen}}}
\newcommand{\synthesisArrowRefine}
  {\ensuremath{\synthesisArrowCaption{refine}}}
\newcommand{\synthesisArrowBranch}
  {\ensuremath{\synthesisArrowCaption{branch}}}
\newcommand{\synthesisArrowRefineOrBranch}
  {\ensuremath{\synthesisArrowCaption{
    \{{\color{CombinedRefineColor}\hspace{0.01in}refine\hspace{0.01in}},%
    {\color{CombinedBranchColor}\hspace{0.01in}branch\hspace{0.01in}}\}
  }}}

\newcommand{\synthesisArrowFill}
  {\ensuremath{\synthesisArrowCaption{fill}}}

\newcommand{\guess}[3]
  {\ensuremath{\problemType{#1}{#2}\synthesisArrowGuess{#3}}}
\newcommand{\guessFull}[4]
  {\ensuremath{\dimColor{#1\spaceSemiSpace{}}\guess{#2}{#3}{#4}}}

\newcommand{\fillHoleOmitOutputHoleEnv}[7]
  {\ensuremath{{#2}\spaceSemiSpace\problemNameTypeWorlds{#1}{#3}{#5}{#4}\synthesisArrowFill{#6}}}
\newcommand{\fillHole}[7]
  {\ensuremath{\fillHoleOmitOutputHoleEnv{#1}{#2}{#3}{#4}{#5}{#6}{#7}\dimColor{\spaceSemiSpace{#7}}}}
\newcommand{\fillHoleFull}[9]
  {\ensuremath{\extraInputsTwoSemi{#2}{#1}
               \fillHole{#3}{#4}{#5}{#6}{#7}{#8}{#9}}}

\newcommand{\refine}[5]
  {\ensuremath{\problemNameTypeWorlds{}{#1}{#3}{#2}\synthesisArrowRefine{#4}\dashv{#5}}}
\newcommand{\refineFull}[7]
  {\ensuremath{\extraInputsTwoSemi{#2}{#1}
               \refine{#3}{#4}{#5}{#6}{#7}}}

\newcommand{\branch}[7]
  {\ensuremath{#7\spaceSemiSpace\problemNameTypeWorlds{}{#1}{#3}{#2}\synthesisArrowBranch{#4}\dashv{#5}\spaceSemiSpace{#6}}}
\newcommand{\branchFull}[9]
  {\ensuremath{\extraInputsTwoSemi{#2}{#1}
               \branch{#3}{#4}{#5}{#6}{#7}{#8}{#9}}}

\newcommand{\refineOrBranchName}{%
  {\color{CombinedRefineColor} Refine},
  {\color{CombinedBranchColor} Branch}%
}

\newcommand{\refineOrBranch}[7]
  {\ensuremath{{\color{CombinedBranchColor} #1\spaceSemiSpace}
               \problemNameTypeWorlds{}{#2}{#3}{#4}\synthesisArrowRefineOrBranch{#5}\dashv{#6}
               {\color{CombinedBranchColor} \spaceSemiSpace{#7}}}}

\newcommand{\metaFunc}[1]
  {\ensuremath{\mathit{#1}}}

\newcommand{\dom}[1]
  {\metaFunc{dom}(#1)}

\newcommand{\resultConsistent}[3]
  {\ensuremath{{#1}\equiv_{#3}{#2}}}

\newcommand{\ungroup}[4] %% TODO fix this weird macro name / args
  {\ensuremath{{#4} = {#2}}}
\newcommand{\filterWorlds}[1]
  {\ensuremath{Filter({#1})}}

\newcommand{\solve}[1]
  {\ensuremath{\metaFunc{Solve}({#1})}}
\newcommand{\solveEquals}[2]
  {\ensuremath{\solve{#1}\synthesisArrow{#2}}}
\newcommand{\iterSolve}[1]
  {\ensuremath{\metaFunc{Solve}{#1}}}

\newcommand{\iterSolveEquals}[3]
  {\ensuremath{\iterSolve{#1}\synthesisArrow{#2}\dimColor{\spaceSemiSpace{#3}}}}
\newcommand{\iterSolveEqualsFull}[5]
  {\ensuremath{\extraInputsTwoSemi{#2}{#1}
               \iterSolveEquals{#3}{#4}{#5}}}
\newcommand{\simplifyArrow}
  {\rhd}

\newcommand{\simplify}[1]
  {\ensuremath{\metaFunc{Simplify}({#1})}}
\newcommand{\simplifyEquals}[2]
  {\ensuremath{\simplify{#1}\simplifyArrow{#2}}}

\newcommand{\simplifyConstraints}[3]
  {\ensuremath{\extraInputsTwoSemi{#2}{#1}
     \metaFunc{Merge}{#3}}
  }

\newcommand{\simplifyConstraintsEquals}[4]
  {\ensuremath{\simplifyConstraints{#1}{#2}{#3}\simplifyArrow{#4}}}

\newcommand{\resolve}[4]
  {\ensuremath{\extraInputsTwoSemi{#2}{#1}
     \metaFunc{Resolve}{\pairConstraints{#3}{#4}}}
  }

\newcommand{\resolveEquals}[5]
  {\ensuremath{\resolve{#1}{#2}{#3}{#4}\synthesisArrow{#5}}}

\newcommand{\stepSimplifyConstraints}[3]
  {\ensuremath{\extraInputsTwoSemi{#2}{#1}
     \metaFunc{Step}{#3}}
  }

\newcommand{\stepSimplifyConstraintsEquals}[4]
  {\ensuremath{\stepSimplifyConstraints{#1}{#2}{#3}\synthesisArrow{#4}}}

\newcommand{\uneval}[4]
  {\ensuremath{\pairFillingExp{#1}{#2}\Leftarrow{#3}\dashv{#4}}}
\newcommand{\unevalFull}[6]
  {\ensuremath{\extraInputsTwoSemi{#2}{#1}
               \uneval{#3}{#4}{#5}{#6}}}

\newcommand{\liveBiEval}[5]
  {\ensuremath{\worldConsistent{#1}{#3}{\world{#2}{#4}}{#5}}}
\newcommand{\mergeSolutions}[2]
  {\ensuremath{{#1}\oplus{#2}}}
\newcommand{\mergeSolutionsThree}[3]
  {\ensuremath{\mergeSolutions{#1}{\mergeSolutions{#2}{#3}}}}
\newcommand{\addSolutions}[2]
  {\ensuremath{\mergeSolutions{#1}{#2}}}

\newcommand{\pairConstraints}[2]
  {\ensuremath{(#2\spaceSemiSpace#1)}}

\newcommand{\mergeConstraints}[2]
  {\ensuremath{{#1}\oplus{#2}}}
\newcommand{\mergeConstraintsThree}[3]
  {\ensuremath{\mergeConstraints{#1}{\mergeConstraints{#2}{#3}}}}
\newcommand{\mergeConstraintsPurple}[2]
  {\ensuremath{{#1}{\color{CombinedBranchColor}\ \oplus\ {#2}}}}

\newcommand{\mergeConstraintsEquals}[3]
  {\ensuremath{{#1}\oplus{#2}={#3}}}

\newcommand{\assertionSat}[2]
  {\ensuremath{{#1}\models{#2}}}
\newcommand{\constraintSat}[2]
  {\ensuremath{{#1}\models{#2}}}
\newcommand{\exSat}[3]
  {\ensuremath{\pairFillingExp{#1}{#2}\models{#3}}}
\newcommand{\worldSat}[3]
  {\ensuremath{\exSat{#1}{#2}{#3}}}

\newcommand{\worldConsistent}[4]
  {\ensuremath{\pairFillingExp{#1}{#2}\rightleftharpoons{#3}\dashv{#4}}}
\newcommand{\worldConsistentFull}[6]
  {\ensuremath{\extraInputsTwoSemi{#2}{#1}
               \worldConsistent{#3}{#4}{#5}{#6}}}

\newcommand{\isFinal}[1]
  {\ensuremath{{#1}\ \texttt{final}}}
\newcommand{\isIndet}[1]
  {\ensuremath{{#1}\ \texttt{indet}}}
\newcommand{\isDet}[1]
  {\ensuremath{{#1}\ \texttt{det}}}
\newcommand{\coerceSyntax}[1]
  {\ensuremath{\lceil{#1}\rceil}}
\newcommand{\liftSyntax}[1]
  {\ensuremath{\lfloor{#1}\rfloor}}
\newcommand{\coerceUndet}[2]
  {\ensuremath{\coerceSyntax{#1}={#2}}}
\newcommand{\liftSimpleEx}[1]
  {\ensuremath{\liftSyntax{#1}}}
\newcommand{\liftSimpleExEquals}[2]
  {\ensuremath{\liftSimpleEx{#1}={#2}}}
\newcommand{\guessesForMatch}[3]
  {\metaFunc{Guesses}(#1,#2,#3)}
\newcommand{\collapseUnwrapWrap}[1]
  {\ensuremath{\llbracket}{#1}\rrbracket}

\newcommand{\varHoleConstraints}
  {\varConstraints}

\newcommand{\HoleConstraint}[2]
  {\ensuremath{\program{#1}{\expHoleDot{}}\hspace{0.00in}\models{#2}}}

\newcommand{\HoleConstraintOverview}[3]
  {\ensuremath{(\program{#2}{\expHoleDot{#1}}\hspace{0.00in}\models{#3})}}

\newcommand{\SimpleReducesTo}[2]
  {\ensuremath{#1}\hspace{0.01in}\Rightarrow{#2}}

\newcommand{\SubstReducesTo}[3]
  {\ensuremath{{#1}\hspace{0.01in}{#2}\hspace{0.01in}\Rightarrow{#3}}}

\newcommand{\LiveUneval}[3]
  {\ensuremath{{#1}\hspace{0.01in}\Leftarrow{#2}\dashv{#3}}}

\newcommand{\indzero}
  {}
\newcommand{\indone}
  {\indzero \indent}
\newcommand{\indtwo}
  {\indone \indent}

\newcommand{\HoleBox}[1]
  {\fboxsep=1.5pt \fcolorbox{white}{constraintLight}{{#1}}}

\lstdefinestyle{myElmStyle}
  { basewidth=0.5em
  , backgroundcolor=
  , basicstyle=\ttfamily\color{Black}
  , morecomment=[l]{--}
  , commentstyle=\color{Gray}
  , morekeywords={forall,fix,type,alias,case,of,let,in,raise,assert,spec2,spec,specifyFunction,specifyFunction2}
  , keywordstyle=\color{Plum}
  , aboveskip=0.5em
  , belowskip=0.5em
  , escapechar=`
  , emph={
      Left,Right,
      Nothing,Just,
			Nil,Cons,
			True,False,
			S,Z
      }
  , emphstyle=\color{RoyalBlue}
  , morestring=[b]"
  , stringstyle=\color{OliveGreen}
  , showstringspaces=false
  , otherkeywords={{\{},{\}},[,],+,-,.,=,|,(,),:,;,{,},<,>}
  , literate={
      {,}{{{\color{CadetBlue},}}}1
      {'}{{{'}}}1
      {LAMBDA}{{{\color{Plum}$\lambda$}}}1
      {LAND}{{{\color{CadetBlue}$\land$}}}1
      {LIMPLIES}{{{\color{CadetBlue}$\Rightarrow$}}}1
      {\_}{{{\color{CadetBlue}\_}}}1
			{0}{{{\color{Emerald}0}}}1
			{1}{{{\color{Emerald}1}}}1
			{2}{{{\color{Emerald}2}}}1
			{spec2}{{{\color{Plum}spec2}}}5
			{3}{{{\color{Emerald}3}}}1
			{4}{{{\color{Emerald}4}}}1
			{5}{{{\color{Emerald}5}}}1
			{6}{{{\color{Emerald}6}}}1
			{7}{{{\color{Emerald}7}}}1
			{8}{{{\color{Emerald}8}}}1
			{9}{{{\color{Emerald}9}}}1
    }
  , classoffset=1
  , morekeywords={{\{},{\}},[,],+,-,.,=,|,(,),:,;,{,},{\_},<,>}
  , keywordstyle=\color{CadetBlue}
  , emph={
      BooleanList,Tree,List,Nat,Maybe,String,Either,NatList,Bool,MaybeNat
      }
  , emphstyle=\color{VioletRed}
  }

\lstnewenvironment{blockcode}{
  \lstset{style=myElmStyle}
}{
}

\newcommand{\inlinecode}[1]
  {\lstinline[style=myElmStyle]{#1}}

\newcommand{\holeSolutionEqV}[1]
  {\HoleBox{\vphantom{Ap}#1}}

\newcommand{\holeSolutionOverview}[2]
  {\holeSolutionEqV{#2}}

\newcommand{\jtheorem}[2]{
  \vspace{0.5em}

  {\color{CadetBlue}\hrule height 0.5pt}

  \vspace{0.25em}

  \noindent \textbf{#1}

  \vspace{0.25em}

  {\color{CadetBlue}\hrule height 0.5pt}

  \vspace{0.5em}

  \noindent \textit{Proof.} #2 \qed

  \vspace{1em}
}

\newcommand{\jtref}[1]
  {\textsc{#1}}

\newmdenv[
  usetwoside=false,
  topline=false,
  bottomline=false,
  rightline=false,
  leftmargin=0.2in,
  linewidth=0.75pt,
  skipabove=\topsep,
  skipbelow=\topsep,
  nobreak=false
]{leftrule}

\newcommand{\allrule}[1]{
  \vspace{\topsep}

  \noindent\hspace{10pt}\fbox{#1}

  \vspace{\topsep}
}

\newcommand{\jgivengoal}[2]{
  \vspace{0.5em}

  {
    \setlength{\parskip}{0em}
    \noindent $\blacktriangleright$ \textbf{Given:}
    {
      \setlength{\parskip}{0.7em}
      \begin{adjustwidth}{0.1in}{0in}
        #1
      \end{adjustwidth}
      \vspace{0.7em}
    }

    \noindent $\blacktriangleright$ \textbf{Goal:}

    \vspace{0.35em}

    \begin{adjustwidth}{0.1in}{0in}
      \allrule{#2}
    \end{adjustwidth}

    \vspace{0.35em}
  }
  \noindent \ignorespaces
}

\newcommand{\jgivengoalTwo}[3]{
  \vspace{0.5em}

  {
    \setlength{\parskip}{0em}
    \noindent $\blacktriangleright$ \textbf{Given:}
    {
      \setlength{\parskip}{0.7em}
      \begin{adjustwidth}{0.1in}{0in}
        #1
      \end{adjustwidth}
      \vspace{0.7em}
    }

    \noindent $\blacktriangleright$ \textbf{Goal A:}

    \vspace{0.35em}

    \begin{adjustwidth}{0.1in}{0in}
      \allrule{#2}
    \end{adjustwidth}

    \vspace{0.35em}

    \noindent $\blacktriangleright$ \textbf{Goal B:}

    \vspace{0.35em}

    \begin{adjustwidth}{0.1in}{0in}
      \allrule{#3}
    \end{adjustwidth}

    \vspace{0.35em}
  }
  \noindent \ignorespaces
}

\newcommand{\jcase}[3]{
  \vspace{0.7em}

  \noindent $\blacktriangleright$ \textbf{Case #1:} \textit{#2.}

  {
    \setlength{\parskip}{0.7em}
    \begin{leftrule}
      \vspace{0.35em}
      #3
    \end{leftrule}
  }

  \noindent \ignorespaces
}

\newcommand{\jnocase}[1]{
  {
    \setlength{\parskip}{0.7em}
    #1
  }
}

\newcommand{\jsubcase}[3]{
  \noindent $\blacktriangleright$ \textbf{Subcase #1:} \textit{#2.}

  {
    \setlength{\parskip}{0.7em}
    \begin{adjustwidth}{0.2in}{0in}
      #3
    \end{adjustwidth}
  }

  \noindent \ignorespaces
}

\newcommand{\caseText}[1]
  {\noindent #1}

\newcommand{\caseFact}[1]
  {\noindent \hspace{10pt}(#1)\hspace{5pt}}

\newcommand{\caseFactPl}[1]
  {\caseFact{#1}}

\newcommand{\numBenchmarks}{38}
\newcommand{\numBenchmarksAll}{43}

\newcommand{\pctFewerExamplesTopOne}{61\%}
\newcommand{\pctFewerExamplesTopOneUpperBound}{66\%}

\newcommand{\numBenchmarksRecursive}{27}
\newcommand{\numBenchmarksBase}{25}

\newcommand{\pctFewerExamplesBaseCaseNoSketch}{57\%}
\newcommand{\pctFewerExamplesBaseCaseStrategy}{46\%}

\newcommand{\maxExamplesBase}{7}
\newcommand{\avgExamplesBase}{2.12}
\usepackage[belowskip=-13pt,aboveskip=6pt]{caption}

\usepackage{wrapfig}

\usepackage{amssymb}% http://ctan.org/pkg/amssymb
\usepackage{pifont}% http://ctan.org/pkg/pifont
\newcommand{\cmark}{\ding{51}}%
\newcommand{\xmark}{\ding{55}}%

\usepackage{changepage}

\usepackage[firstpage]{draftwatermark}
\SetWatermarkText{\hspace*{4.2in}\raisebox{8.8in}{%
  \includegraphics[scale=0.28]{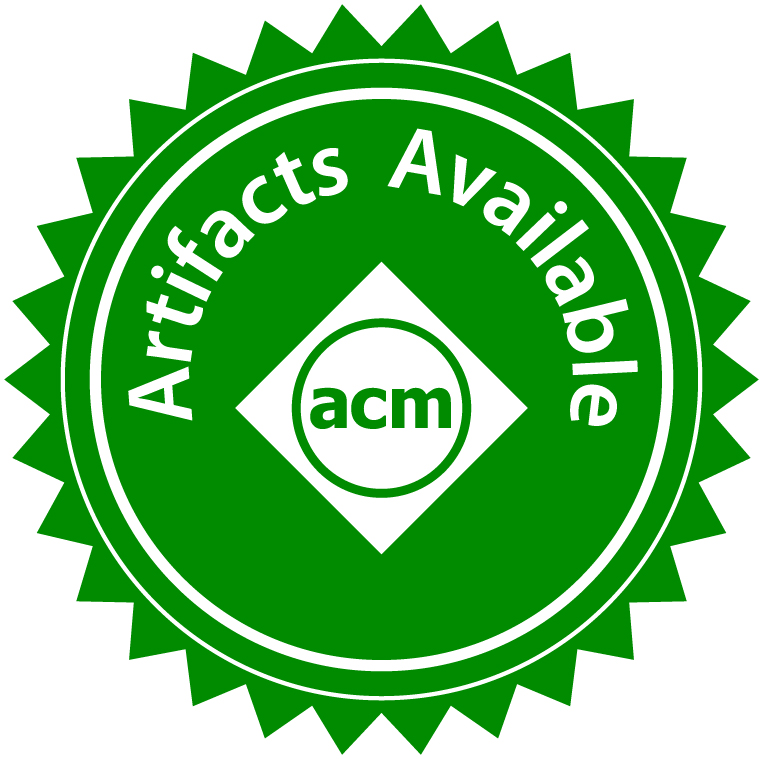}%
  \includegraphics[scale=0.28]{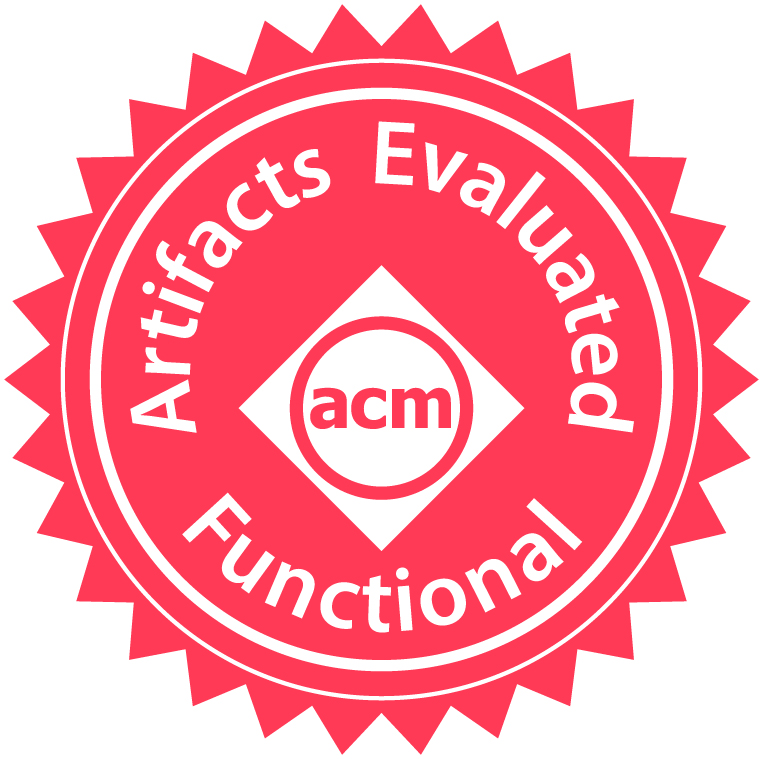}%
}}
\SetWatermarkAngle{0}

\begin{document}

% This needs to be inside \begin{document} to not screw up the columns.

%% \setlength{\columnsep}{7pt}
\setlength{\columnsep}{10pt}
\setlength{\intextsep}{0pt}

%% Title information
\title
  {Program Sketching with Live Bidirectional Evaluation}

%% Author information
%% Contents and number of authors suppressed with 'anonymous'.
%% Each author should be introduced by \author, followed by
%% \authornote (optional), \orcid (optional), \affiliation, and
%% \email.
%% An author may have multiple affiliations and/or emails; repeat the
%% appropriate command.
%% Many elements are not rendered, but should be provided for metadata
%% extraction tools.

%% Author with single affiliation.
\author{Justin Lubin}
%% \authornote{with author1 note}          %% \authornote is optional;
                                        %% can be repeated if necessary
%% \orcid{nnnn-nnnn-nnnn-nnnn}             %% \orcid is optional
\affiliation{
  %% \position{Position1}
  %% \department{Department1}              %% \department is recommended
  \institution{University of Chicago}            %% \institution is required
  %% \streetaddress{Street1 Address1}
  %% \city{City1}
  %% \state{State1}
  %% \postcode{Post-Code1}
  \country{USA}
}
\email{justinlubin@uchicago.edu}          %% \email is recommended

\author{Nick Collins}
\affiliation{\institution{University of Chicago} \country{USA}}
\email{nickmc@uchicago.edu}

\author{Cyrus Omar}
\affiliation{\institution{University of Michigan} \country{USA}}
\email{comar@umich.edu}

\author{Ravi Chugh}
\affiliation{\institution{University of Chicago} \country{USA}}
\email{rchugh@cs.uchicago.edu}

%% 2012 ACM Computing Classification System (CSS) concepts
%% Generate at 'http://dl.acm.org/ccs/ccs.cfm'.

%% End of generated code

%% Keywords
%% comma separated list
%% \keywords are mandatory in final camera-ready submission

%% Note: \begin{abstract}...\end{abstract} environment must come
%% before \maketitle command
\begin{abstract}

We present a system called \smyth{} for program sketching in a typed
functional language whereby the concrete evaluation of ordinary assertions
gives rise to input-output examples, which are then used to guide the search to
complete the holes.
The key innovation, called \emph{live bidirectional evaluation}, propagates
examples ``backward'' through partially evaluated sketches.
Live bidirectional evaluation enables \smyth{} to
(a)~synthesize recursive functions without trace-complete sets of examples and
(b)~specify and solve interdependent synthesis goals.
Eliminating the trace-completeness requirement resolves a significant limitation
faced by prior synthesis techniques when given partial specifications in the
form of input-output examples.

To assess the practical implications of our techniques, we ran several
experiments on benchmarks used to evaluate \myth{}, a state-of-the-art
example-based synthesis tool.
First, given expert examples (and no partial implementations), we find that
\smyth{} requires on average \pctFewerExamplesTopOneUpperBound{} of the number
of expert examples required by \myth{}.
Second, we find that \smyth{} is robust to randomly-generated examples,
synthesizing many tasks with relatively few more random examples than those
provided by an expert.
Third, we create a suite of small sketching tasks by systematically employing a
simple sketching strategy to the \myth{} benchmarks; we find that user-provided
sketches in \smyth{} often further reduce the total specification burden
(\ie{}~the combination of partial implementations and examples).
Lastly, we find that \leon{} and \synquid{}, two state-of-the-art logic-based
synthesis tools, fail to complete several tasks on which \smyth{} succeeds.

\end{abstract}

%% \maketitle
%% Note: \maketitle command must come after title commands, author
%% commands, abstract environment, Computing Classification System
%% environment and commands, and keywords command.
\maketitle

% !TEX root = ./main.tex
\section{Introduction}
\label{sec:intro}

Program synthesis is closer than ever to making its  
way into the working programmer's toolbox.
Synthesis techniques that operate on fine-grained logical specifications---%
such
as~\sketch{}~\citep{SketchingThesis},
\rosette{}~\citep{Rosette2013},
\leon{}~\cite{Kneuss2013}, and
\synquid{}~\cite{Polikarpova2016}---%
as well as techniques that operate on input-output examples---%
such as~\escher{}~\citep{Escher},
$\lambdaSquared$~\citep{Feser2015}
\myth{}~\citep{Myth}, and
``\mythTwo{}''~\citep{Frankle2016}---%
can synthesize a variety of challenging tasks, from subtle
bit-manipulating computations in imperative languages to recursive
functions over inductive datatypes in functional languages.

\clearpage %% arxiv-version

\begin{figure}[t]

%% TODO maybe use \expHoleVerb

%% TODO maybe ListNat to match MaybeNat

  %% case n of
    %% Z    -> ??$_1$    $\fillingTightBoxName{1}{[]}$
    %% S n' -> ??$_2$    $\fillingTightBoxName{2}{x \texttt:\texttt: replicate n\texttt' x}$

%% replicate  Z     x = ??$_1$
%% replicate (S n') x = ??$_2$
%%
%% stutter\_n n []       = []
%% stutter\_n n (x::xs') =
%%   replicate n x ++ stutter\_n n xs'

%% replicate n x =
%%   case n of \{Z -> ??$_1$; S n' -> ??$_2$\}

%% replicate : Nat -> Nat -> NatList
%% replicate n x =
%%   case n of
%%     Z    -> $\holeSolutionIntro{}{[]}$
%%     S n' -> $\holeSolutionIntro{}{x \texttt:\texttt: replicate n\texttt' x}$
%%
%% stutter\_n : Nat -> NatList -> NatList
%% stutter\_n n xs =
%%   case xs of
%%     []     -> []
%%     x::xs' -> replicate n x ++ stutter\_n n xs'

\newsavebox{\boxStutterI}
\begin{lrbox}{\boxStutterI}
  \inlinecode{[]}
\end{lrbox}

\newsavebox{\boxStutterII}
\begin{lrbox}{\boxStutterII}
  \inlinecode{x :: replicate n' x}
\end{lrbox}

{\small
\begin{minipage}[t]{0.58\textwidth}
\begin{blockcode}
  stutter_n : Nat -> NatList -> NatList
  stutter_n n xs =
    case xs of
      []     -> []
      x::xs' -> replicate n x ++ stutter_n n xs'

  assert (stutter_n 1 [1, 0] == [1, 0])
  assert (stutter_n 2 [3]    == [3, 3])
\end{blockcode}
\end{minipage}
\begin{minipage}[t]{0.4\textwidth}
\begin{blockcode}
replicate : Nat -> Nat -> NatList
replicate n x =
  case n of
    Z    -> `\color{CadetBlue}\texttt{??}`
    S n' -> `\color{CadetBlue}\texttt{??}`
\end{blockcode}
\end{minipage}
}

%%   $\fillingTightBoxName{1}{[]}$  $\fillingTightBoxName{2}{x \texttt:\texttt: replicate n\texttt' x}$

%%   [ (0, [1, 0], []    )
%%   , (1, [1, 0], [1, 0])
%%   , (2, [3],    [3, 3])
%%   ]

%%   [ (0, [1], [])
%%   , (1, [1], [1])
%%   , (2, [3], [3, 3])
%%   ]

%%   [ (1, [1], [1])
%%   , (2, [3], [3, 3])
%%   ]

\caption{A program sketch in \snsMyth{} to ``stutter'' each element of a list
\inlinecode{n} times. The desired solutions
for the holes in \inlinecode{replicate} are
\HoleBox{\usebox\boxStutterI}
for the \inlinecode{Z} branch and
\HoleBox{\usebox\boxStutterII}
for the \inlinecode{S} branch.
}
\label{fig:stutter-n}
\end{figure}

However, there remain commonplace program synthesis tasks that cannot be
completed by state-of-the-art techniques.
\autoref{fig:stutter-n} shows an incomplete program (a.k.a. ``program sketch''),
written in an ML-style functional language.
The implementation of the \inlinecode{stutter_n} function itself---which is intended to ``stutter'' each element of
a given list \inlinecode{n} times---is complete.
However, it depends on an incomplete
helper function \inlinecode{replicate} with \emph{holes} (written \inlinecode{??}) denoting missing
expressions that the programmer might hope to automatically synthesize.
The two \inlinecode{assert} statements provide simple test cases that constrain
the behavior of \inlinecode{stutter_n}. Because \inlinecode{stutter_n} applies
\inlinecode{replicate}, these assertions indirectly constrain the holes in
\inlinecode{replicate} as well.
Unfortunately, the aforementioned synthesis techniques are not able to
synthesize the desired hole completions shown in blue boxes in
\autoref{fig:stutter-n}.
In what ways do the prior techniques fall short for this task?

\parahead{Logic-Based Program Synthesis}

\leon{}~\citep{Kneuss2013} and \synquid{}~\citep{Polikarpova2016} support
sketching
for richly-typed, general-purpose
functional languages (as used in \autoref{fig:stutter-n}).
As pioneered in \sketch{}~\citep{SketchingThesis}, \leon{} and \synquid{} are
solver-based techniques that fill holes such that given specifications are satisfied.
Both systems synthesize many challenging benchmarks involving complex data
invariants, yet
neither can complete the task in \autoref{fig:stutter-n}.

The approach to synthesis and verification in \leon{} does not decompose the
\inlinecode{assert} constraints on \inlinecode{stutter_n} into constraints on
\inlinecode{replicate}, so the holes remain unspecified.
By using an approach based on \emph{liquid types}~\citep{LiquidTypes,Vazou2013},
\synquid{} is able to systematically decompose the given constraints into the following specification:

\begin{center}
\begin{minipage}{0.8\textwidth}
\begin{blockcode}
replicate :: (n : Nat) -> (x : Nat)
          -> { out : NatList | (n = 1 LAND x = 0 LIMPLIES out = [0])
                             LAND (n = 1 LAND x = 1 LIMPLIES out = [1])
                             LAND (n = 2 LAND x = 3 LIMPLIES out = [3, 3]) }
\end{blockcode}
\end{minipage}
\end{center}

\newcommand{\expAppReplicate}[2]
  {\expAppTwo{\texttt{replicate}}{\texttt{#1}}{\texttt{#2}}}

\noindent
However, because this specification is not inductive---it provides no information about
\expAppReplicate{0}{0}, \expAppReplicate{0}{1}, \expAppReplicate{1}{3}, or
\expAppReplicate{0}{3}---\synquid{} cannot type check the desired solution for
\inlinecode{replicate}, let alone synthesize it.

\parahead{Evaluator-Based Program Synthesis}

In contrast to logic-based techniques, another class of techniques operate on
input-output examples and rely on concrete evaluation to ``guess-and-check''
candidate terms.
We choose the term \emph{evaluator-based} to describe such techniques---rather
than \emph{example-based} or \emph{programming-by-example}---to distinguish how
the underlying algorithms work (using concrete evaluation) from the
specification mechanism they provide to users (examples).
Examples can also be encoded as partial logical specifications, as just
discussed.

Among evaluator-based techniques, \escher{}~\citep{Escher} and
\myth{}~\citep{Myth} can synthesize recursive functions, and \myth{} employs
several type-directed optimizations to navigate the search space.
(We discuss the remaining systems in \autoref{sec:related}.)
However, there are two fundamental reasons why these tools cannot complete the
task in \autoref{fig:stutter-n}.

\newcommand{\mythLimitation}[1]
  %% {\parahead{#1}}
  {\subparahead{#1}}

\mythLimitation{Limitation A: Trace-Complete Examples}
The user must provide input-output examples for recursive calls internal to the
eventual solution---this is the ``example analog'' to \synquid{}'s requirement
for inductive logical specifications.
\citet{Myth} acknowledge that providing \emph{trace-complete examples}
(\ie{}~serving as an \emph{oracle}~\cite{Escher}) ``proved to be difficult
initially'' even for experts, and ``discovering ways to get around this
restriction ... would greatly help in converting this type-directed synthesis
style into a usable tool.''
\citet{Hanoi} also observe the need to ``manage \myth{}'s requirement for trace
completeness.''

\mythLimitation{Limitation B: Independent, Top-Level Goals}
The user must factor all synthesis tasks into completely unimplemented top-level
functions, each of which must be equipped directly with (trace-complete)
example sets. The system attempts to synthesize each of these functions separately. 
Granular sketching, where holes appear in arbitrary positions
and are simultaneously solved, is not supported.

\parahead{Our Approach: Live Bidirectional Evaluation}
In this paper, we present a new evaluator-based synthesis technique that
addresses Limitations A and B.
Holes can appear in arbitrary expression positions and are constrained by
types and \inlinecode{assert} statements which give
rise to example constraints.
Given the sketch in \autoref{fig:stutter-n}, our implementation---called
\snsMyth{}---synthesizes the desired expressions to fill the holes.
(Our exposition employs certain syntactic conveniences not currently
implemented.
These are described in \autoref{sec:implementation}.)

In order to make evaluator-based synthesis techniques compatible with sketching,
we must formulate hole-aware notions
of (1)~\emph{concrete evaluation} and (2)~\emph{example satisfaction}---which
form the central term enumeration search strategy (\ie{}~\emph{guess-and-check})
for evaluator-based synthesis.
Our solution, called \emph{live bidirectional evaluation}, comprises two parts:

\begin{enumerate}

\item
A \emph{live evaluator} $\SimpleReducesTo{\varExp}{\varResult}$ that partially
evaluates a sketch $e$ by proceeding around holes, producing a result $\varResult$
which is either a value or a ``paused'' expression that, when the necessary
holes are filled, will ``resume'' evaluating; and

\item
A \emph{live unevaluator} $\LiveUneval{\varResult}{\varEx}{\varHoleConstraints}$
that, given a result $\varResult$ to be checked against example $\varEx$,
computes constraints $\varHoleConstraints$ (over possibly many holes in the
sketch) that, if satisfied, ensure the result will eventually produce a value
satisfying $\varEx$.

\end{enumerate}

\noindent
Live evaluation is adapted from \citet{HazelnutLive} to our setting and is not
a technical contribution of our work.
Live unevaluation is the key novel mechanism that---together with live
evaluation---enables us to ``combine sketching with \myth{}-style synthesis''
(hence the name \snsMyth{}).
Compared to the aforementioned logic-based and other symbolic evaluation
techniques~(\eg{}~\citep{Morpheus,Bornholt2018,Viser}), live bidirectional
evaluation employs \emph{concrete} evaluation to
collect example constraints ``globally'' across multiple holes in the sketch.

\parahead{Contributions}

This paper generalizes the theory of \myth{}~\citep{Myth}---the state-of-the-art
in type-directed, evaluator-based program synthesis---to support sketches and
live bidirectional evaluation.
Formally, we present a calculus of recursive functions, algebraic datatypes, and
holes---called \coreSnsMyth{}---which includes the following technical
contributions:

\begin{itemize}

\item
We present \emph{live unevaluation}, a novel technique that checks example
satisfaction of sketches.
The combination of \emph{live evaluation} to partially evaluate
sketches~\citep{HazelnutLive} and live unevaluation---which we call \emph{live
bidirectional evaluation}---forms a core guess-and-check strategy for programs
with holes.
Our formulation generalizes \myth{}, but the notion of live bidirectional
evaluation can also be developed for other evaluator-based synthesizers.
(\autoref{sec:live-uneval})

\item
We use live bidirectional evaluation to simplify program assertions into
input-output constraints and
generalize the \myth{} hole synthesis algorithm to employ live bidirectional
evaluation.
The resulting synthesis algorithm
(a)~alleviates the trace-completeness requirement and
(b)~globally solves the examples that arise from multiple interdependent tasks.
(\autoref{sec:synthesis})

\end{itemize}

\noindent
For simplicity, our formal system accounts only for top-level
\inlinecode{assert}s, but
we describe how subsequent work may extend our approach to allow assertions in
arbitrary program positions.

To empirically evaluate our approach, we implement \snsMyth{} and
perform several experiments:

\begin{itemize}

\item
We synthesize \numBenchmarks{} of \numBenchmarksAll{} tasks from the \myth{}
benchmark suite~\citep{Myth,OseraThesis} in \snsMyth{}.
Given expert examples (without sketches), \smyth{} requires
\pctFewerExamplesTopOneUpperBound{} of the number of expert examples required by
\myth{}.
Moreover, \smyth{} typically requires only a slightly larger set of examples
if they are generated randomly, rather than by an expert.
(\autoref{sec:experiment-two})

\item
To create a suite of sketching tasks, we identify a simple \emph{base case
sketching strategy} and apply it systematically to the \myth{} benchmarks.
As expected, base case sketches further reduce the number of examples that
\snsMyth{} requires to complete many tasks.
Furthermore, the total specification size with sketching (partial
implementation plus examples) is often smaller than without (just examples).
(\autoref{sec:experiment-three})

\item
We identify a handful of additional sketching tasks, similar in size and
    flavor to \inlinecode{stutter_n}, which \snsMyth{} can complete. 
(\autoref{sec:overview})

\item
To situate our experimental results in a broader context,
we run \leon{} and \synquid{} on our benchmarks.
We find several tasks for which \smyth{} succeeds but these tools do not.
(\autoref{sec:experiment-logic})

\end{itemize}

\noindent
The experimental results demonstrate
(i) that the theoretical advances in \snsMyth{} address Limitations A and B of
prior evaluator-based synthesizers, and
(ii) that even though examples can generally be encoded as logical
specifications, current logic-based synthesizers are not necessarily strictly
more powerful than evaluator-based ones.

Because our approach generalizes \myth{}, we provide comparison throughout the paper.
We further discuss related work in \autoref{sec:related}.
Additional definitions, proofs, and experimental data are available
in an extended technical report~\citep{smyth-full};
in the rest of the paper, we write \S A, \S B, and \S C{} to refer to
appendices in the technical report.

% !TEX root = ./main.tex
\section{Overview}
\label{sec:overview}

\newcommand{\varConstraintsOverview}
  {\varConstraints}

\newcommand{\fillHoleOverview}[3]
  {\ensuremath{{#1}\synthesisArrowCaption{}{#2}\dashv{#3}}}

\newcommand{\liveCheckOverview}[4]
  {\ensuremath{\SimpleReducesTo{#1}{#2}\Leftarrow{#3}\dashv{#4}}}

In this section, we work through several small programs to introduce how
\snsMyth{}:
(1) employs live bidirectional evaluation to check example satisfaction of guessed
expressions (which, in our formulation, may include
holes)~(\autoref{sec:overview-one});
(2) supports user-defined sketches~(\autoref{sec:overview-two}); and
(3) derives examples from \inlinecode{assert}s in the program~(\autoref{sec:overview-three}).

We write holes $\expHole{\varHoleName}$ below with explicit names
$\varHoleName$; our implementation automatically generates names for holes as in
\autoref{fig:stutter-n}.
Literals \inlinecode{0}, \inlinecode{1}, \inlinecode{2}, \etc{} are syntactic
sugar for the corresponding naturals of \inlinecode{type Nat = Z | S Nat}.
Some judgement forms below are simplified for expositional purposes.

\subsection{Synthesis without Trace-Completeness}
\label{sec:overview-one}

%% need \{ and \} if inline
\newsavebox{\boxPlusOneLine}
\begin{lrbox}{\boxPlusOneLine}
\inlinecode{fix plus LAMBDAm n -> case m of \{Z -> n; S m' -> S (plus m' n)\}}
\end{lrbox}

\setlength{\intextsep}{0pt}%
\setlength{\columnsep}{10pt}%

\begin{wrapfigure}{r}{0pt}
\small
\begin{blockcode}
plus : Nat -> Nat -> Nat
plus = ??`\vspace{0.10in}`
assert (plus 0 1 == 1)
assert (plus 2 0 == 2)
assert (plus 1 2 == 3)
\end{blockcode}
\end{wrapfigure}
\noindent
Consider the task to synthesize \inlinecode{plus} given the three test cases on the right.
Given this specification,
the resulting \emph{example constraint}
\mbox{
$ %% \begin{align*}
\varConstraintsOverview_0 =
\HoleConstraintOverview{0}{\emptyEnv}
{\ioExampleThree{0\ 1}{1}{2\ 0}{2}{1\ 2}{3}}
$ %% \end{align*}
}
requires that $\expHole{0}$ (hole name 0 generated for the definition of
  \inlinecode{plus}) be filled with a function expression that, in the empty
environment, $\emptyEnv$, conforms to the given input-output examples.
(We write \expHoleDot{\varHoleName} to distinguish the concrete syntax of
constraints from expression holes \expHole{\varHoleName}.)

Given a set of constraints $\varConstraintsOverview_\varHoleName$,
\snsMyth{} employs the \emph{hole synthesis} search procedure
$\mbox{\fillHoleOverview{\varConstraintsOverview_\varHoleName}{\varExp_\varHoleName}{\varConstraintsOverview'}}$
to fill the hole $\expHole{\varHoleName}$ with an expression
$\varExp_\varHoleName$ that is valid assuming new constraints $\varConstraintsOverview'$ over
other holes in the program. %% are satisfied.
Following \myth{}~\citep{Myth}, hole synthesis begins with a
guess-and-check approach that enumerates increasingly large terms
comprising variables and functions applied to variables.
This na\"ive search is limited to small terms, \ie{}, starting with AST size 1 in early ``stages'' of the search and increasing to size 13 in latter stages.
When enumerative search fails to find a solution in a particular stage,
hole synthesis performs \emph{example-directed refinement and branching}: introductory forms and case analyses are considered, and the examples are \emph{distributed} to create subgoals for new holes that arise.

We will describe the following search path---among many that \smyth{} will
consider---that yields the solution \HoleBox{\usebox\boxPlusOneLine}
for \inlinecode{plus}.
$$
\setlength{\arraycolsep}{3.5pt}
\begin{array}{rllcl}
\varConstraintsOverview_0 &
\synthesisArrowRefine &
\holeFillingOverview{\expHole{0}}{\valFixFun{\texttt{plus}}{\ttm\ \ttn}{\expHole{1}}} &
\dashv &
\varConstraintsOverview_1 \\
\varConstraintsOverview_1 &
\synthesisArrowBranch &
\holeFillingOverview{\expHole{1}}{\expMatchTwoBranchesOverview{\ttm}{\texttt{Z}}{BLAH}{\expHole{2}}{\texttt{S}}{\texttt{m\texttt'}}{\expHole{3}}} &
\dashv &
\envCat{\varConstraintsOverview_2}{\varConstraintsOverview_3} \\
\varConstraintsOverview_3 &
\synthesisArrowRefine &
  \holeFillingOverview{\expHole{3}}{\expApp{\text{\inlinecode{S}}}{\expHole{4}}} &
\dashv &
\varConstraintsOverview_4 \\
\varConstraintsOverview_4 &
\synthesisArrowGuess &
\holeFillingOverview{\expHole{4}}{\expAppTwo{\texttt{plus}}{\ttm\texttt'}{\ttn}} &
\dashv &
\varConstraintsOverview_2' \\
\envCat{\varConstraintsOverview_2}{\varConstraintsOverview_2'} &
\synthesisArrowGuess &
\holeFillingOverview{\expHole{2}}{\ttn} &
\dashv &
\emptySet
\end{array}
$$

\newcommand{\envPlus}[2]
  {\ensuremath{(\envCat
    {\envBind{\text{\inlinecode{m}}}{\text{\inlinecode{#1}}}}
    {\envBind{\text{\inlinecode{n}}}{\text{\inlinecode{#2}}}})}}

\newcommand{\envPlusS}[3]
  {\ensuremath{(\envCatThree
    {\envBind{\text{\inlinecode{m}}}{\text{\inlinecode{#1}}}}
    {\envBind{\text{\inlinecode{n}}}{\text{\inlinecode{#2}}}}
    {\envBind{\text{\inlinecode{m'}}}{\text{\inlinecode{#3}}}})}}

\newcommand{\envPlusRec}[2]
  {\ensuremath{(\envCatThree
    {\envBind{\text{\inlinecode{plus}}}{...}}
    {\envBind{\text{\inlinecode{m}}}{\text{\inlinecode{#1}}}}
    {\envBind{\text{\inlinecode{n}}}{\text{\inlinecode{#2}}}})}}

\newcommand{\envPlusRecPad}[2]
  {\ensuremath{(\envCatThree
    {\envBind{\text{\inlinecode{plus}}}{...}}
    {\envBind{\text{\inlinecode{m}}}{\text{\inlinecode{#1}}}}
    {\envBind{\text{\inlinecode{n}}}{\text{\inlinecode{#2}}}}}
    {\phantom{,\hspace{0.02in}\envBind{\text{\inlinecode{m'}}}{\text{\inlinecode{X}}}}})}

\newcommand{\envPlusSRec}[3]
  {\ensuremath{(\envCatFour
    {\envBind{\text{\inlinecode{plus}}}{...}}
    {\envBind{\text{\inlinecode{m}}}{\text{\inlinecode{#1}}}}
    {\envBind{\text{\inlinecode{n}}}{\text{\inlinecode{#2}}}}
    {\envBind{\text{\inlinecode{m'}}}{\text{\inlinecode{#3}}}})}}

First,
because the goal is a function type, \smyth{} synthesizes a recursive function literal, with subgoal
$\expHole{1}$ for the body.
The constraint set $\varConstraintsOverview_1$ (not shown) consists of three constraints created from
the three input-output examples in $\varConstraintsOverview_0$ by binding the
input values to \inlinecode{m} and
\inlinecode{n} in the environment and constraining the new subgoal with the corresponding output value.

Second, after guessing-and-checking fails to solve $\expHole{1}$, \smyth{} attempts to \emph{branch} by
guessing the scrutinee \inlinecode{m}. This scrutinee is evaluated in each
environment of the three constraints in $\varConstraintsOverview_1$.
One constraint from $\varConstraintsOverview_1$ is distributed to subgoal $\expHole{2}$
for the base case branch (this constraint $\varConstraintsOverview_{2.1}$ is shown below), and
the other two constraints from $\varConstraints_1$ are distributed to subgoal
$\expHole{3}$ for the recursive case (these constraints $\varConstraints_3$ are not shown).

Third, \smyth{} chooses to work on the recursive branch,
for which the two constraints in $\varConstraints_3$ involve output examples
\inlinecode{2} and \inlinecode{3}
(\ie{}~\expApp{\text{\inlinecode{S}}}{\color{CadetBlue}(\expApp{\text{\inlinecode{S}}}{\text{\inlinecode{Z}}})}
and
\expApp{\text{\inlinecode{S}}}{\color{CadetBlue}
(\expApp{\text{\inlinecode{S}}}{\color{CadetBlue}(\expApp{\text{\inlinecode{S}}}{\text{\inlinecode{Z}}})})
}).
\smyth{} \emph{refines} the task by synthesizing the literal
$\expApp{\text{\inlinecode{S}}}{\expHole{4}}$; the new subgoal is constrained by
two examples (in $\varConstraintsOverview_4$, shown below) obtained by removing
the shared constructor head \inlinecode{S} from the output examples in $\varConstraintsOverview_3$.
(\smyth{} synthesizes a literal of the form
$\expApp{\text{\inlinecode{S}}}{\color{CadetBlue}(\expApp{\text{\inlinecode{S}}}{\color{Black}\expHole{4}}\color{CadetBlue})}$
along other search paths, but those paths do not yield a solution as quickly as the one being described.)
\begin{align*}
\varConstraints_{2.1} &=
  \HoleConstraintOverview{2}{\envPlusRecPad{0}{1}}{\text{\inlinecode{1}}} \\
\varConstraints_{4.1} &=
  \HoleConstraintOverview{4}{\envPlusSRec{2}{0}{1}}{\text{\inlinecode{1}}} \\
\varConstraints_{4.2} &=
  \HoleConstraintOverview{4}{\envPlusSRec{1}{2}{0}}{\text{\inlinecode{2}}}
\end{align*}

\noindent
The remaining two subgoals, $\expHole{4}$ and $\expHole{2}$, are filled via
guess-and-check as discussed below.

\parahead{Live Bidirectional Example Checking}

To decide whether a guessed expression $\varExp$ conforms to a constraint
$\mbox{\HoleConstraintOverview{\varHoleName}{\varEnv}{\varEx}}$ in \snsMyth{},
the procedure $\SubstReducesTo{\varEnv}{\varExp}{\varResult}$ applies the
substitution (\ie{}~environment) $\varEnv$ to the expression and evaluates it to
a {result} $\varResult$, and
the \emph{live unevaluation} procedure
$\LiveUneval{\varResult}{\varEx}{\varConstraintsOverview}$ checks satisfaction
modulo new constraints $\varConstraintsOverview$.

Consider guesses to fill $\expHole{4}$.
Notice that \inlinecode{plus}---the function \snsMyth{} is working to synthesize---is
recursive and thus bound in the constraint environments above.
In addition to variables and calls to existing functions,
\snsMyth{} enumerates structurally-decreasing recursive calls
($\expAppTwo{\mathtt{plus}}{\ttm\texttt'}{\ttn}$,
$\expAppTwo{\texttt{plus}}{\ttm}{\ttn\texttt'}$, and
$\expAppTwo{\texttt{plus}}{\ttm\texttt'}{\ttn}$').

When considering $\expAppTwo{\texttt{plus}}{\ttm\texttt'}{\ttn}$, the name
\inlinecode{plus} binds the following value
comprising the first three fillings and the ``current'' guess:
$$
{\valFixFun{\texttt{plus}}{\ttm\ \ttn}{
{\expMatchTwoBranchesOverview{\ttm}{\texttt{Z}}{BLAH}{\expHole{2}}{\texttt{S}}{\texttt{m\texttt'}}{
  {\expApp{\text{\inlinecode{S}}}{
    \color{CadetBlue}(\color{Black}{\expAppTwo{\texttt{plus}}{\ttm\texttt'}{\ttn}}\color{CadetBlue})
}}}}}}
$$
\noindent
Given the environment in constraint $\varConstraintsOverview_{4.1}$, the guess
evaluates and unevaluates as follows:
$$
\setlength{\arraycolsep}{3.5pt}
\begin{array}{lclclcl}
\expAppTwo{\mathtt{plus}}{\ttm\texttt'}{\mathtt{\ttn}} &
\rightarrow^* &
  \expAppTwo{\mathtt{plus}}{\text{\inlinecode{1}}}{\text{\inlinecode{0}}} &
&
&
&
\\
&
\rightarrow^* &
  \expApp{\text{\inlinecode{S}}}{\color{CadetBlue}(\color{Black}\expAppTwo{\mathtt{plus}}{\text{\inlinecode{0}}}{\text{\inlinecode{0}}}\color{CadetBlue})} &
&
&
\\
&
\Rightarrow^{\phantom{*}} &
  \expApp{\text{\inlinecode{S}}}{\color{CadetBlue}(\color{Black}\holeClosure{\envPlusRec{0}{0}}{2}\color{CadetBlue})\color{Black}} &
\Leftarrow &
  \text{\inlinecode{1}} &
\dashv &
\varConstraintsOverview_{2.2}
\end{array}
$$
\noindent
(We write $\varExp\rightarrow^*\varExp'\Rightarrow\varResult$ to display
intermediate steps of the big-step evaluation, but $\varExp\rightarrow^*\varExp'$
does not appear in the formal system.)
Although the function is incomplete, \emph{live evaluation}~\citep{HazelnutLive}
resolves two recursive calls to \inlinecode{plus}, before the hole $\expHole{2}$ in
the base case reaches evaluation position;
the resulting \emph{hole closure}, of the form
$\holeClosure{\varEnv}{\varHoleName}$, captures the environment at that point.
Comparing the result to \inlinecode{1}
(\ie{}~\expApp{\text{\inlinecode{S}}}{\text{\inlinecode{Z}}}), unevaluation
removes an \inlinecode{S} from each side and
creates a new constraint $\varConstraintsOverview_{2.2}$ (shown below) for the
base case.

Similarly, the guess checks against constraint
$\varConstraintsOverview_{4.2}$, adding another new constraint
$\varConstraintsOverview_{2.3}$ (shown below) on the base case.
$$
\setlength{\arraycolsep}{3.5pt}
\begin{array}{lclclcl}
\expAppTwo{\mathtt{plus}}{\ttm\texttt'}{\mathtt{\ttn}} &
\rightarrow^* &
  \expAppTwo{\mathtt{plus}}{\text{\inlinecode{0}}}{\text{\inlinecode{2}}} &
&
&
&
\\
&
\Rightarrow^{\phantom{*}} &
\holeClosure{\envPlusRec{0}{2}}{2} &
\Leftarrow &
  \text{\inlinecode{2}} &
\dashv &
\varConstraintsOverview_{2.3}
\end{array}
$$
\noindent
Both checks succeed, so the fourth step of the search commits to the guess,
returning the two new constraints in $\varConstraints_2'$.
\begin{align*}
  \varConstraints_{2.2} &=
  \HoleConstraintOverview{2}{\envPlusRec{0}{0}}{\text{\inlinecode{0}}} \\
  \varConstraints_{2.3} &=
  \HoleConstraintOverview{2}{\envPlusRec{0}{2}}{\text{\inlinecode{2}}}
\end{align*}

The fifth and final step is to fill the base case $\expHole{2}$, subject to
constraints
$\varConstraintsOverview_{2.1}$, $\varConstraintsOverview_{2.2}$, and
$\varConstraintsOverview_{2.3}$.
The guess \inlinecode{n} evaluates to the required values
(\inlinecode{0}, \inlinecode{1}, and \inlinecode{2}, respectively), without assumption.
Together, the five filled holes comprise the final solution.

Notice that the test cases used to synthesize \inlinecode{plus} were \emph{not}
trace-complete: live bidirectional example checking recursively called
$\expAppTwo{\text{\inlinecode{plus}}}{\text{\inlinecode{1}}}{\text{\inlinecode{0}}}$,
$\expAppTwo{\text{\inlinecode{plus}}}{\text{\inlinecode{0}}}{\text{\inlinecode{0}}}$, and
$\expAppTwo{\text{\inlinecode{plus}}}{\text{\inlinecode{0}}}{\text{\inlinecode{2}}}$,
none of which were included in the examples. %% input-output examples.
Instead, \snsMyth{} \emph{generated} additional constraints that the user would
be required to provide in prior systems (\ie{}~\escher{}, \myth{}, \mythTwo{},
and \synquid{}).

\subsection{User-Defined Sketches}
\label{sec:overview-two}

\smyth{} is the first evaluator-based synthesis technique to support sketching, thus allowing users to split domain knowledge naturally across a partial implementation and examples.
For instance, if the user sketches the zero cases for \texttt{max}, as shown in
\autoref{fig:overview-examples}, just a few examples are sufficient for \smyth{}
to complete the recursive case.
(The library function \inlinecode{spec2} asserts input-output examples for a binary
function, as was written out fully for \inlinecode{plus} above.)

Sketches from the user are handled in the same way as the sketches, described above, created internally by the \smyth{} algorithm.
\myth{} and several other evaluator-based techniques~(\cf{}~\autoref{sec:related}) can also be described as creating sketches internally, but \smyth{} uniquely supports \emph{concrete evaluation} of sketches---with holes in arbitrary positions---as a way to generate new example constraints.

%% \begin{figure}[b]
\begin{figure}[t]
%\begin{minipage}{3.70in}
\small

\begin{center}

\newcommand{\sketchesLeftCol}{0.45\textwidth}
\newcommand{\sketchesRightCol}{0.5\textwidth}

%% \vspace{0.20in} %% TODO full col figure? or squeeze?

%% specifyFunction2 plus
%%   [ (0, 0, 0), (0, 1, 1), (2, 0, 2), (1, 2, 3) ]

%% plus : Nat -> Nat -> Nat
%% max : Nat -> Nat -> Nat
%% minus : Nat -> Nat -> Nat
%% mult : Nat -> Nat -> Nat

%% plus m n = $\expHole{0}$

%%  $\fillingTightBoxName{0}{\texttt{$\lambda$m n -> case m of Z -> n ; S m' -> S (plus m' n)}}$

%%  $\fillingTightBoxName{5}{S (max m\textrm' n\textrm')}$

%%  $\fillingTightBoxName{6}{XXX XXX}$

%%  $\fillingTightBoxName{7}{a\texttt'}$  $\fillingTightBoxName{8}{b}$  $\fillingTightBoxName{9}{a}$

%%   $\fillingTightBoxName{10}{q}$  $\fillingTightBoxName{11}{p\texttt'}$  $\fillingTightBoxName{12}{q}$

\newsavebox{\boxMax}
\begin{lrbox}{\boxMax}
  \inlinecode{S (max m' n')}
\end{lrbox}

\newsavebox{\boxOdd}
\begin{lrbox}{\boxOdd}
  \inlinecode{Just 1}
\end{lrbox}

\newsavebox{\boxMinusI}
\begin{lrbox}{\boxMinusI}
  \inlinecode{a'}
\end{lrbox}

\newsavebox{\boxMinusII}
\begin{lrbox}{\boxMinusII}
  \inlinecode{b'}
\end{lrbox}

\newsavebox{\boxMinusIII}
\begin{lrbox}{\boxMinusIII}
  \inlinecode{a}
\end{lrbox}

\newsavebox{\boxMultI}
\begin{lrbox}{\boxMultI}
  \inlinecode{q}
\end{lrbox}

\newsavebox{\boxMultII}
\begin{lrbox}{\boxMultII}
  \inlinecode{p'}
\end{lrbox}

\newsavebox{\boxMultIII}
\begin{lrbox}{\boxMultIII}
  \inlinecode{q}
\end{lrbox}

{ \arrayrulecolor{CadetBlue}
\begin{tabular}{p{\sketchesLeftCol{}}|p{\sketchesRightCol{}}}
\vspace{-1em}
\begin{blockcode}
  max  m      Z     = m
  max  Z      n     = n
  max (S m') (S n') = `\holeSolutionOverview{5}{\usebox\boxMax}`

  spec2 max
    [(1, 1, 1), (1, 2, 2), (3, 1, 3)]
\end{blockcode}
&
\vspace{-1em}
\begin{blockcode}
  odd n =                 unJust mx =
    case n of               case mx of
      Z       -> False        Nothing -> 0
      S Z     -> True         Just x  -> x
      S S n'' -> odd n''

  assert (odd (unJust `\holeSolutionOverview{6}{\usebox\boxOdd}\miniSepThree`) == True)`\vspace{-1.5em}`
\end{blockcode}
\\ \hline
\vspace{-0.6em}
\begin{blockcode}
  minus (S a') (S b') = minus `\holeSolutionOverview{7}{\usebox\boxMinusI}` `\holeSolutionOverview{8}{\usebox\boxMinusII}`
  minus  a      b     = `\holeSolutionOverview{9}{\usebox\boxMinusIII}`

  spec2 minus
    [(2, 0, 2), (3, 2, 1), (3, 1, 2)]
\end{blockcode}
&
\vspace{-0.6em}
\begin{blockcode}
  mult p q =
    case p of
      Z    -> Z
      S p' -> plus `$\holeSolutionOverview{10}{\usebox\boxMultI}$` (mult `$\holeSolutionOverview{11}{\usebox\boxMultII}$` `$\holeSolutionOverview{12}{\usebox\boxMultIII}\miniSepThree$`)

  spec2 mult
    [(2, 1, 2), (3, 2, 6)]`\vspace{-1.4em}`
\end{blockcode}
\end{tabular}
}

\end{center}

\caption{\smyth{} fills the holes \texttt{??} (not shown) with the code shown in blue boxes.}
\label{fig:overview-examples}
\end{figure}

\subsection{Deriving Examples from Assertions}
\label{sec:overview-three}

\newcommand{\nameBranches}[1]
  {\mathit{#1}}

% https://tex.stackexchange.com/a/7045
% https://tex.stackexchange.com/a/48540
\newcommand*\circled[1]{\tikz[baseline=(char.base)]{
            \node[shape=circle,draw,inner sep=0pt,minimum size=11pt] (char) {#1};}}

\newcommand{\circleCase}[1]
  {\circled{\footnotesize #1}}

\newcommand{\circleCaseItem}[1]
  {\noindent \circleCase{#1}}

For the \inlinecode{plus} and \inlinecode{max} programs so far, evaluating assertions
provided examples ``directly'' on holes.
In general, however, an assertion may involve more complicated results.

For instance, consider the definitions of
\inlinecode{odd : Nat -> Bool} and
\inlinecode{unJust : MaybeNat -> Nat}
in \autoref{fig:overview-examples},
and the evaluation of the expression
$\expAppTwo{\mathtt{odd}}{\color{CadetBlue}(\color{Black}\mathtt{unJust}}{\expHole{5}\color{CadetBlue})}$:
$$
\setlength{\arraycolsep}{3.5pt}
\begin{array}{lcl}
  \expApp{\mathtt{odd}}{\color{CadetBlue}(\color{Black}\expApp{\mathtt{unJust}}{\expHole{5}}\color{CadetBlue})} &
\rightarrow^* &
  \expApp{\mathtt{odd}}{\color{CadetBlue}(\color{Black}\expApp{\mathtt{unJust}}{(\holeClosure{\emptyEnv}{5}\color{CadetBlue})})}
\\
&
\rightarrow^* &
\expApp
  {\mathtt{odd}}
  {\color{CadetBlue}(\expMatchOverview{(\color{Black}\holeClosure{\emptyEnv}{5}\color{CadetBlue})}{\color{Black}\nameBranches{unJust}}\color{CadetBlue})}
\\
&
\Rightarrow^{\phantom{*}} &
\expMatchOverview
  {\color{CadetBlue}(\expMatchOverview{(\color{Black}\holeClosure{\emptyEnv}{5}\color{CadetBlue})}{\color{Black}\nameBranches{unJust}})}
  {\nameBranches{odd}}
\end{array}
$$
\noindent
(For clarity, we omit the recursive environment bindings for \texttt{odd} and
\texttt{unJust}.)
First, evaluation produces the hole closure
$\holeClosure{\emptyEnv}{5}$, which is passed to \inlinecode{unJust}.
Then, the \inlinecode{case} expression in \inlinecode{unJust}---we write
$\nameBranches{unJust}$ to refer to its two branches---scrutinizes the hole
closure.
The form of the constructor application has not yet been determined,
so evaluation ``pauses'' by returning the \emph{indeterminate}~\citep{HazelnutLive}
result
$\expMatchOverview{(\holeClosure{\emptyEnv}{5})}{\nameBranches{unJust}}$,
which records the fact that, when the scrutinee resumes to a constructor head
\inlinecode{Nothing} or \inlinecode{Just}, evaluation of the \inlinecode{case} will proceed down
the appropriate branch.
This indeterminate case result is passed to the \inlinecode{odd} function.
Finally, the \inlinecode{case} inside \inlinecode{odd}---we write $\nameBranches{odd}$
to refer to its three branches---scrutinizes it, building up a nested
indeterminate result.

How can we ``indirectly'' constrain the expression $\expHole{5}$ to ensure that
the partially evaluated expression
$\expMatchOverview
  {\color{CadetBlue}(\expMatchOverview{(\color{Black}\holeClosure{\emptyEnv}{5}\color{CadetBlue})}{\color{Black}\nameBranches{unJust}}\color{CadetBlue})}
  {\nameBranches{odd}}$
evaluates to \inlinecode{True} as asserted?

\parahead{Unevaluating Case Expressions}

Unevaluation will run each of the three branches of $\nameBranches{odd}$ ``in
reverse,'' attempting to reconcile each with the required example, \inlinecode{True};
we write \circleCaseItem{1}, \circleCaseItem{2}, \circleCaseItem{3}, \etc{} to
help discuss different branches of the search considered by \smyth{}:
$$
\setlength{\arraycolsep}{3.5pt}
\begin{array}{lclcl}
\expMatchOverview
  {\color{CadetBlue}(\expMatchOverview{(\color{Black}\holeClosure{\emptyEnv}{5}\color{CadetBlue})}{\color{Black}\nameBranches{unJust}}\color{CadetBlue})}
  {\nameBranches{odd}} &
\Leftarrow &
\text{\inlinecode{True}} &
\dashv &
\circleCase{1} \, \circleCase{2} \, \circleCase{3}
\end{array} $$

\begin{itemize}
\item[\circleCaseItem{1}] 
The first branch expression, \inlinecode{False}, is inconsistent with
\inlinecode{True}
(\ie{}~${\text{\inlinecode{False}}}\hspace{0.01in}\Leftarrow{\text{\inlinecode{True}}}\not\dashv$).

\item[\circleCaseItem{2}]
The second branch expression, \inlinecode{True}, is equal to the example.
However, to take this branch, unevaluation must ensure that the scrutinee---an
indeterminate case result itself---will match the pattern
$\expApp{\text{\inlinecode{S}}}{\text{\inlinecode{Z}}}$ (\ie{}~\inlinecode{1});
that is,
$\!\!
\setlength{\arraycolsep}{3.5pt}
\begin{array}{lclcl}
  \expMatchOverview{\color{CadetBlue}(\color{Black}\holeClosure{\emptyEnv}{5}\color{CadetBlue})}{\color{Black}\nameBranches{unJust}} &
\Leftarrow &
  \text{\inlinecode{1}} &
\dashv &
\circleCase{\textrm{2a}} \, \circleCase{\textrm{2b}}.
\end{array}
$
\begin{itemize}
\item[\circleCaseItem{2a}]
The first branch expression, \inlinecode{0}, is inconsistent with \inlinecode{1}.

\item[\circleCaseItem{2b}]
Reasoning about the second branch expression is more involved:
the variable \inlinecode{x} must bind the argument of \inlinecode{Just}, but we have not yet
ensured that this branch will be taken!
To bridge the gap, we bind \inlinecode{x} to
the symbolic, and indeterminate, \emph{inverse constructor application}
$\expUnwrap{{\color{RoyalBlue}\mathtt{Just}}}{\color{CadetBlue}(\color{Black}\holeClosure{\emptyEnv}{5}\color{CadetBlue})}$
when evaluating the branch expression;
unevaluation ``transfers'' the resulting example from the symbolic result to the
scrutinee:
$$
\setlength{\arraycolsep}{3.5pt}
\begin{array}{lclclcl}
\mathtt{x} &
\Rightarrow^{\phantom{*}} &
  \expUnwrap{\text{\inlinecode{Just}}}{\color{CadetBlue}(\color{Black}\holeClosure{\emptyEnv}{5}\color{CadetBlue})} &
\Leftarrow &
  \text{\inlinecode{1}} &
\dashv &
  \HoleConstraintOverview{5}{\emptyEnv}{\expApp{\text{\inlinecode{Just}}}{\text{\inlinecode{1}}}}
\end{array}
$$
\noindent
This constraint ensures that the \inlinecode{case} in \inlinecode{unJust} will resolve to
the second branch ($\expApp{\text{\inlinecode{Just}}}{\ttx}$) and that its expression will
produce \inlinecode{1}, and thus that
the \inlinecode{case} in \inlinecode{odd} will resolve to the second branch
($\expApp{\text{\inlinecode{S}}}{\text{\inlinecode{Z}}}$) and produce \inlinecode{True}, as asserted.
\end{itemize}
\vspace{-0.5em}
\item[\circleCaseItem{3}]
By recursively unevaluating the third branch, \inlinecode{odd n''}, case unevaluation
can derive additional solutions:
$\expApp{\text{\inlinecode{Just}}}{\text{\inlinecode{3}}}$,
$\expApp{\text{\inlinecode{Just}}}{\text{\inlinecode{5}}}$, \etc{}
Na\"ively unevaluating all branches, however, would introduce a significant
degree of non-determinism---even non-termination.
Therefore, our formulation and implementation impose simple
restrictions---described in \autoref{sec:live-bi-eval} and
\autoref{sec:implementation}---on case unevaluation to trade expressiveness for
performance.
\end{itemize}

Altogether, live bidirectional evaluation untangles the interplay between
indeterminate branching and assertions so that \snsMyth{} can, for instance,
fill the holes in \inlinecode{minus} and \inlinecode{mult} in
\autoref{fig:overview-examples}.

% !TEX root = ./main.tex

%%%%%%%%%%%%%%%%%%%%%%%%%%%%%%%%%%%%%%%%%%%%%%%%%%%%%%%%%%%%%%%%%%%%%%%%%%%%%%%%

\section{Live Bidirectional Evaluation}
\label{sec:live-bi-eval}

In this section, we formally define
\emph{live evaluation}
$\liveEval{\varEnv}{\varSolution}{\varExp}{\varResult}$
and \emph{live unevaluation}
$\uneval{\varSolution}{\varResult}{\varEx}{\varConstraints}$
for a calculus called \coreSnsMyth{}.
We choose a natural semantics (big-step, environment-style)
presentation~\citep{Kahn:1987}, though our techniques can be re-formulated for a
small-step, substitution-style model.
Compared to our earlier notation, here we refer to environments $\varEnv$ and
$\varSolution$---often typeset in light gray, because environments would ``fade
away'' in a substitution-style presentation.

Our formulation proceeds as follows.
First, in \autoref{sec:syntax} and \autoref{sec:types}, we define the syntax and
type checking judgements of \coreSnsMyth{}.
Next, in \autoref{sec:live-eval}, we present live evaluation, which
adapts the \emph{live programming with holes} technique~\citep{HazelnutLive}
to our setting;
minor differences are described in \autoref{sec:related-work-techniques}.
Lastly,
we define example satisfaction in
\autoref{sec:ex-sat} and live unevaluation in \autoref{sec:live-uneval}.
In \autoref{sec:synthesis}, we build a synthesis pipeline around the combination
of live evaluation and unevaluation.

%%%%%%%%%%%%%%%%%%%%%%%%%%%%%%%%%%%%%%%%%%%%%%%%%%%%%%%%%%%%%%%%%%%%%%%%%%%%%%%%

\subsection{Syntax}
\label{sec:syntax}

%% this is a one-col version of fig-syntax-0
%%
\begin{figure}[b]

\newcommand{\textEnvironments}
  {Environments}
  %% {Env.}

\newcommand{\textContexts}
  {Contexts}
  %% {Ctx.}

\newcommand{\textConstraints}
  {Constraints}
  %% {Con.}

\centering

%% \judgementHeadNameOnly{Syntax of Programs}

%% \vsepRule

%% \newcommand{\metaVarsSpaceOne}
%%   {\hspace{0.05in}}
%% \newcommand{\metaVarsSpaceBetween}
%%   {\hspace{0.20in}}
%%   %% {\hspace{0.35in}}
%% $$
%% \textbf{\maybeUnderlineSyntaxFig{D}atatypes} \metaVarsSpaceOne \varTypeCon
%% \metaVarsSpaceBetween
%% \textbf{Variables} \metaVarsSpaceOne \varVarF,\varVar
%% \metaVarsSpaceBetween
%% \textbf{\maybeUnderlineSyntaxFig{C}onstructors} \metaVarsSpaceOne \varDataCon
%% \metaVarsSpaceBetween
%% \textbf{\maybeUnderlineSyntaxFig{H}ole Names} \metaVarsSpaceOne \varHoleName
%% $$
%% 
%% \vspace{0.10in} %% TODO

\newcommand{\metaVarDef}[2]
  {\hspace{0.50in}\textbf{#1}\ \ #2}

\figSyntaxBegin
\figSyntaxRowLabel{\maybeUnderlineSyntaxFig{T}ypes}{\varType}
%% \figSyntaxRowLabel{\maybeUnderlineSyntaxFig{T}yp.}{\varType}
  \tArrow{\varType_1}{\varType_2}
  \figSyntaxSpaceItem
  %% \tTriple{\varType_1}{\ldots}{\varType_n}
  \tUnit
  \figSyntaxSpaceItem
  \tPair{\varType_1}{\varType_2}
  \figSyntaxSpaceItem
  \varTypeCon
  &
  \metaVarDef{\maybeUnderlineSyntaxFig{D}atatypes}{\varTypeCon}
\figSyntaxSpaceNextCategoryMoreSpace
%
%% \figSyntaxRowLabel{Expressions (Sketches)}{\varExp}
\figSyntaxRowLabel{\maybeUnderlineSyntaxFig{E}xpressions}{\varExp}
%% \figSyntaxRowLabel{\maybeUnderlineSyntaxFig{E}xp.}{\varExp}
  %% \dimColor{\expFunAnnot{\varVar}{\varType}{\varExp}}
  %% \dimColor{\figSyntaxSpaceItem}
  \expFixFun{\varVarF}{\tArrow{\varType_1}{\varType_2}}{\varVar}{\varExp}
  \figSyntaxSpaceItem
  \expApp{\varExp_1}{\varExp_2}
  \figSyntaxSpaceItem
  \varVar
  &
  \metaVarDef{Variables}{\varVarF,\varVar}
\figSyntaxLineBreak
\figSyntaxRow
%%   \figSyntaxSpaceItem
  %% \triple{\varExp_1}{\ldots}{\varExp_n}
  \expUnit
  \figSyntaxSpaceItem
  \pair{\varExp_1}{\varExp_2}
  \figSyntaxSpaceItem
  \expProj{i}{\varExp}
  %% \expFst{\varExp}
  %% \figSyntaxSpaceItem
  %% \expSnd{\varExp}
\figSyntaxLineBreak
\figSyntaxRow
%%   \figSyntaxSpaceItem
  \expApp{\varDataCon}{\varExp}
  \figSyntaxSpaceItem
  \expMatch{\varExp}{\varDataCon_i}{\varVar_i}{\varExp_i}
  &
  \metaVarDef{\maybeUnderlineSyntaxFig{C}onstructors}{\varDataCon}
\figSyntaxLineBreak
\figSyntaxRow
%%   \figSyntaxSpaceItem
  %% \expHole{i}{\varType}
  \expHole{\varHoleName}
  &
  \metaVarDef{\maybeUnderlineSyntaxFig{H}ole Names}{\varHoleName}
\figSyntaxSpaceNextCategoryMoreSpace
%
%% \figSyntaxRowLabel{Values}{\varVal}
%%   \closure{\varEnv}{\expFun{\varVar}{\varExp}}
%%   \figSyntaxSpaceItem
%%   \closure{\varEnv}{\valFixFun{\varVarF}{\varVar}{\varExp}}
%%   \figSyntaxSpaceItem
%%   \triple{\varVal_1}{\ldots}{\varVal_n}
%%   \figSyntaxSpaceItem
%%   \expApp{\varDataCon}{\varVal}
%% %
%% \figSyntaxSpaceNextCategory
%
%% \figSyntaxRowLabel{(Unfinished) Results}{\varResult}
%% \figSyntaxRowLabel{(Unfinished) Results}{\varVal,\varResult}
%% \figSyntaxRowLabel{Results}{\varResult}
%% \figSyntaxRowLabel{Results}{\varVal,\varResult,\varResult}
%% \figSyntaxRowLabel{\maybeUnderlineSyntaxFig{R}esults}{\varResult}
%% \figSyntaxRowLabel{{\normalfont ``Destructible''} \maybeUnderlineSyntaxFig{R}esults}{\varResult}
\figSyntaxRowLabel{\maybeUnderlineSyntaxFig{R}esults}{\varResult}
%% \figSyntaxRowLabel{\maybeUnderlineSyntaxFig{R}es.}{\varResult}
  %% \dimColor{\closure{\varEnv}{\expFun{\varVar}{\varExp}}}
  %% \dimColor{\figSyntaxSpaceItem}
  %% \overbrace{
  \multicolumn{2}{l}{ %% b/c of Metavariable column
  \closure{\varEnv}{\valFixFun{\varVarF}{\varVar}{\varExp}}
  \figSyntaxSpaceItem
  %% \triple{\varResult_1}{\ldots}{\varResult_n}
  \expUnit
  \figSyntaxSpaceItem
  \pair{\varResult_1}{\varResult_2}
  \figSyntaxSpaceItem
  \expApp{\varDataCon}{\varResult}
  }
  %% }^{\textrm{``destructible'' results}}
\figSyntaxLineBreak
%% \figSyntaxRow
%% \figSyntaxRowLabelPipe{\normalfont ``Indestructible'' Results}{}
\figSyntaxRow %% LabelPipe{}{}
  %% \underbrace{
  \multicolumn{2}{l}{ %% b/c of Metavariable column
  \closure{\varEnv}{\expHole{\varHoleName}}
  %% \closure{\varEnv}{\expHole{i}{\varType}}
  \figSyntaxSpaceItem
  \expApp{\varResult_1}{\varResult_2}
  \figSyntaxSpaceItem
  \expProj{i}{\varResult}
  %% \expFst{\varResult}
  %% \figSyntaxSpaceItem
  %% \expSnd{\varResult}
  %% \figSyntaxSpaceItem
  %% &
  %% \textbf{Undet. }\ \varResult\ (\varResult \textrm{ s.t. } \isIndet{\varResult})
%% \figSyntaxLineBreak
%% \figSyntaxRow
%% \figSyntaxLineBreak
%% \figSyntaxRow %% LabelPipe{}{}
  \figSyntaxSpaceItem
  \closure{\varEnv}{\expMatch{\varResult}{\varDataCon_i}{\varVar_i}{\varExp_i}}
  }
\figSyntaxLineBreak
\figSyntaxRow
  \expUnwrap{\varDataCon}{\varResult}
%
%% \figSyntaxLineBreak
%% \figSyntaxRow
%%   \closure{\varEnv}{\expHole{i}{\varType}}
%% \figSyntaxLineBreak
%% \figSyntaxRow
%%   \varEx \textrm{ (only for user-manipulated results $\varResultEx$)}
%
%% \figSyntaxSpaceNextCategoryMoreSpace
\figSyntaxEnd

\vspace{0.10in} %% TODO

\figSyntaxBegin
%
%% \figSyntaxRowLabel{\textEnvironments}{\varEnv}
%% \figSyntaxRowLabel{\maybeUnderlineSyntaxFig{E}valuation \textEnvironments}{\varEnv}
\figSyntaxRowLabel{\maybeUnderlineSyntaxFig{E}nvironments}{\varEnv}
%% \figSyntaxRowLabel{\maybeUnderlineSyntaxFig{E}nv.}{\varEnv}
  \emptyEnv
  \figSyntaxSpaceItem
  \envCat{\varEnv}{\envBind{\varVar}{\varResult}}
  %% \multiPremise
  %%   {\envBind{\varVar_i}{\varResult_i}}
  %%   {\sequenceSyntax}
  %% \envCat{\varEnv}{\envBindBoth{\varVar}{\varResult}{\varType}}
%
\figSyntaxSpaceNextCategory
\figSyntaxRowLabel{Hole \maybeUnderlineSyntaxFig{F}illings}{\varSolution}
  \emptyEnv
  \figSyntaxSpaceItem
  \envCat{\varSolution}{\highlightSolvedConstraint{\holeFilling{\varHoleName}{\varExp}}}
  %% \multiPremise
  %%   {\highlightSolvedConstraint{\holeFilling{\varHoleName}{\varExp}}}
  %%   {\sequenceSyntax}
%
\figSyntaxSpaceNextCategoryMoreSpace
%% \figSyntaxSpaceNextCategory
%
%% \figSyntaxRowLabel{Type \textEnvironments}{\varTypeEnv}
\figSyntaxRowLabel{Type \textContexts}{\varTypeEnv}
  \emptyEnv
  \figSyntaxSpaceItem
  \envCat{\varTypeEnv}{\envBindType{\varVar}{\varType}}
  %% \multiPremise
  %%   {\envBindType{\varVar_i}{\varType_i}}
  %%   {\sequenceSyntax}
%
\figSyntaxSpaceNextCategory
%
%% \figSyntaxRowLabel{Data Type \textEnvironments}{\varDatatypeEnv}
\figSyntaxRowLabel{Datatype \textContexts}{\varDatatypeEnv}
  \emptyEnv
  \figSyntaxSpaceItem
%% \figSyntaxLineBreak
%% \figSyntaxRow
  \envCat
    {\varDatatypeEnv}
    %% {\{ \envBind{\varDataCon_i}{(\tArrow{\varType_i}{\varTypeCon})} \}}
    {\envBindDatatype{\varTypeCon}{\varDataCon_i}{\varType_i}}
    %% {\threeThings{(\envBindType{\varDataCon_1}{\tArrow{\varType_1}{\varTypeCon})}}{\ldots}
    %%              {(\envBindType{\varDataCon_n}{\tArrow{\varType_n}{\varTypeCon})}}}
  %% \multiPremise
  %%   {\envBindDatatype{\varTypeCon_ij}{\varDataCon_ij}{\varType_ij}}
  %%   {\generalSequenceSyntax{j}{m}}
%
\figSyntaxSpaceNextCategory
\figSyntaxRowLabel{Hole Type \textContexts}{\varHoleEnv}
%% \figSyntaxRowLabel{Hole \textContexts}{\varHoleEnv}
  \emptyEnv
  \figSyntaxSpaceItem
  \envCat{\varHoleEnv}{\envBindHole{\varHoleName}{\tHole{\varTypeEnv}{\varType}}}
  %% \multiPremise
  %%   {\envBindHole{\varHoleName_i}{\tHole{\varTypeEnv_i}{\varType_i}}}
  %%   {\sequenceSyntax}
%
%% \figSyntaxSpaceNextCategory
%% %
%% \figSyntaxRowLabel{\rkc{Hole Example \textContexts}}{\varUnsolvedConstraints}
%%   \emptyEnv
%%   \figSyntaxSpaceItem
%%   \envCat{\varUnsolvedConstraints}
%%          {\highlightUnsolvedConstraint{$\unsolvedConstraint{\varHoleName}{\varWorlds}$}}
%
%% \figSyntaxSpaceNextCategory
%% %
%% \figSyntaxRowLabel{Solutions (Hole-Fillings)}{\varSolution}
%% \figSyntaxRowLabel{\maybeUnderlineSyntaxFig{H}ole-Fillings}{\varSolution}
%% \figSyntaxRowLabel{Hole \maybeUnderlineSyntaxFig{F}illings}{\varSolution}
%%   \emptyEnv
%%   \figSyntaxSpaceItem
%%   \envCat{\varSolution}{\holeFilling{\varHoleName}{\varExp}}
%
\figSyntaxSpaceNextCategoryMoreSpace
\figSyntaxRowLabel{Synthesis \maybeUnderlineSyntaxFig{G}oals}{\varGoals}
%% \figSyntaxRowLabel{Synth. \maybeUnderlineSyntaxFig{G}oals}{\varGoals}
%
  \emptyEnv
  \figSyntaxSpaceItem
  \envCat{\varGoals}
         {\problemNameTypeWorlds{\varHoleName}{\varTypeEnv}{\varType}{\varWorlds}}
  %% \multiPremise{
  %% \problemNameTypeWorlds{\varHoleName_i}{\varTypeEnv_i}{\varType_i}{\varWorlds_i}
  %% }{\sequenceSyntax}
%
\figSyntaxSpaceNextCategoryMoreSpace
%
%% \figSyntaxRowLabel{\maybeUnderlineSyntaxFig{W}orlds}{\varWorlds}
%% \figSyntaxRowLabel{E\maybeUnderlineSyntaxFig{x}ample Worlds}{\varWorlds}
%% \figSyntaxRowLabel{E\maybeUnderlineSyntaxFig{x}ample Constraints}{\varWorlds}
\figSyntaxRowLabel{E\maybeUnderlineSyntaxFig{x}ample \textConstraints}{\varWorlds}
%% \figSyntaxRowLabel{E\maybeUnderlineSyntaxFig{x}. \textConstraints}{\varWorlds}
  \emptyEnv
  \figSyntaxSpaceItem
  \envCat{\varWorlds}{\world{\varEnv}{\varEx}}
  %% \multiPremise{
  %% \world{\varEnv_i}{\varEx_i}
  %% }{\sequenceSyntax}
%
\figSyntaxSpaceNextCategory
\figSyntaxRowLabel{Simple \maybeUnderlineSyntaxFig{V}alues}{\varSimpleVal}
%% \figSyntaxRowLabel{Simple \maybeUnderlineSyntaxFig{V}al.}{\varSimpleVal}
%
  \expUnit
  \figSyntaxSpaceItem
  \pair{\varSimpleVal_1}{\varSimpleVal_2}
  \figSyntaxSpaceItem
  \expApp{\varDataCon}{\varSimpleVal}
\figSyntaxSpaceNextCategory
%% \figSyntaxSpaceNextCategoryMoreSpace
%
%% \figSyntaxRowLabel{\maybeUnderlineSyntaxFig{Ex}amples}{\varEx}
%% \figSyntaxRowLabel{\maybeUnderlineSyntaxFig{Ex}ample Values}{\varEx}
\figSyntaxRowLabel{\maybeUnderlineSyntaxFig{Ex}amples}{\varEx}
  \expUnit
  \figSyntaxSpaceItem
  \pair{\varEx_1}{\varEx_2}
  \figSyntaxSpaceItem
  \expApp{\varDataCon}{\varEx}
  \figSyntaxSpaceItem
%% \figSyntaxLineBreak
%% \figSyntaxRow %% LabelPipe{}{}
  %% \ioExample{\varResult}{\varExOrIO}
  \ioExample{\varSimpleVal}{\varEx}
  \figSyntaxSpaceItem
  \exHole
\figSyntaxSpaceNextCategoryMoreSpace
%
%% \figSyntaxRowLabel{Uneval \maybeUnderlineSyntaxFig{C}onstraints}{\varConstraints}
%% \figSyntaxRowLabel{Uneval. \textConstraints}{\varConstraints}
\figSyntaxRowLabel{Unevaluation \textConstraints}{\varConstraints}
  \pairConstraints{\varSolution}{\varUnsolvedConstraints}
%
%% \figSyntaxSpaceNextCategory
%% %
%% %% \figSyntaxRowLabel{Solved Constraints (Hole \maybeUnderlineSyntaxFig{F}illings)}{\varSolution}
%% \figSyntaxRowLabel{Hole \maybeUnderlineSyntaxFig{F}illings}{\varSolution}
%%   \emptyEnv
%%   \figSyntaxSpaceItem
%%   \envCat{\varSolution}
%%          {\highlightSolvedConstraint{\holeFilling{\varHoleName}{\varExp}}}
%
\figSyntaxSpaceNextCategory
%
%% \figSyntaxRowLabel{\maybeUnderlineSyntaxFig{U}nsolved Constraints}{\varUnsolvedConstraints}
%% \figSyntaxRowLabel{Hole Ex. \textContexts}{\varUnsolvedConstraints}
%% \figSyntaxRowLabel{\maybeUnderlineSyntaxFig{U}nfilled Hole \textContexts}{\varUnsolvedConstraints}
\figSyntaxRowLabel{\maybeUnderlineSyntaxFig{U}nfilled Holes}{\varUnsolvedConstraints}
%% \figSyntaxRowLabel{\maybeUnderlineSyntaxFig{U}nfilled}{\varUnsolvedConstraints}
  \emptyEnv
  \figSyntaxSpaceItem
  \envCat{\varUnsolvedConstraints}
         {\highlightUnsolvedConstraint{$\unsolvedConstraint{\varHoleName}{\varWorlds}$}}
  %% \multiPremise
  %%   {\highlightUnsolvedConstraint{$\unsolvedConstraint{\varHoleName_i}{\varWorlds_i}$}}
  %%   {\sequenceSyntax}
%
\figSyntaxEnd

\vsepBeforeCaption

\caption{Syntax of \coreSnsMyth{}.}

\vsepAfterCaptionBottomFig

\label{fig:syntax}
\end{figure}

\autoref{fig:syntax} defines the syntax of \coreSnsMyth{}, a calculus of
recursive functions, unit, pairs, and (named, recursive) algebraic datatypes.
We say ``products'' to mean unit and pairs.

\parahead{Datatypes}

We assume a fixed datatype context $\varDatatypeEnv$. A datatype
$\varTypeCon$ has some number $n$ of constructors $\varDataCon_i$, each of which
carries a single argument of type $\varType_i$---the type of $\varDataCon_i$ is
$\tArrow{\varType_i}{\varTypeCon}$.

\parahead{Expressions and Holes}

The expression forms on the first three lines are standard function, product,
and constructor forms, respectively.
The expressions $\expFst{\varExp}$ and $\expSnd{\varExp}$ project the first and
second components of a pair.
Each \inlinecode{case} expression has one branch for each of the $n$
constructors $\varDataCon_i$ corresponding to the type of the scrutinee
$\varExp$;
for simplicity, nested patterns are not supported.

Holes $\expHole{\varHoleName}$ can appear anywhere in expressions
(\ie{}~expressions are sketches).
We assume each hole in a sketch has a unique name $\varHoleName$, but
we sometimes write $\expHole{}$ when the name is not referred to.
Hole contexts $\varHoleEnv$ define a \emph{contextual type}
$\tHole{\varTypeEnv}{\varType}$ to describe the type and the type context that is available to expressions that can ``fill'' a
given hole~\citep{CMTT,HazelnutLive}.

\parahead{Results}

We define a separate grammar of \emph{results} $\varResult$---with evaluation
environments $\varEnv$ that map variables to results---to support the definition
of big-step, environment-style evaluation
$\reducesTo{\varEnv}{\varExp}{\varResult}$ below.
Because of holes, results are not conventional values.
Terminating evaluations produce two kinds of \emph{final} results; neither kind
of result is stuck (\ie{}~erroneous).

The four result forms on the first line of the result grammar would---on their
own---correspond to values in a conventional natural semantics (without holes).
In \coreSnsMyth{}, these \emph{determinate} results can be eliminated
in a type-appropriate position; \citetAppendix{\refAppendixSyntax} defines a simple
predicate $\isDet{\varResult}$ to identify such results, and type checking is
discussed below.
Note that a recursive function closure
$\closure{\varEnv}{\valFixFun{\varVarF}{\varVar}{\varExp}}$ stores an
environment $\varEnv$ that binds the free variables of the function body~$\varExp$,
except the name $\varVarF$ of the function itself.
We sometimes write $\expFun{\varVar}{\varExp}$ for non-recursive functions.

The four \emph{indeterminate} result forms on the second line of the grammar are unique to the
presence of holes.
Rather than aborting evaluation with an error when a hole reaches elimination
position (\eg{}, \inlinecode{raise "Hole"}), an indeterminate result $
\varResult $ (defined by the predicate
$ \isIndet{\varResult} $~\citepAppendix{\refAppendixSyntax})
serves as a placeholder for where to continue
evaluation if and when the hole is later filled (either by the programmer or
synthesis engine) with a well-typed expression.
The primordial indeterminate result is a \emph{hole closure}
$\closure{\varEnv}{\expHole{\varHoleName}}$---the environment binds the free variables that
a hole-filling expression may refer to.
An indeterminate application $\expApp{\varResult_1}{\varResult_2}$ appears when
the function has not yet evaluated to a function closure
(\ie{}~$\isIndet{\varResult_1}$); we require that $\varResult_2$ be final in
accordance with our eager evaluation semantics, discussed below.
An indeterminate projection $\expProj{i}{\varResult}$ appears when the argument
has not yet evaluated to a pair (\ie{}~$\isIndet{\varResult}$).
An indeterminate case closure
$\closure{\varEnv}{\expMatch{\varResult}{\varDataCon_i}{\varVar_i}{\varExp_i}}$
appears when the scrutinee has not yet evaluated to a constructor application
(\ie{}~$\isIndet{\varResult}$)---like with function and hole closures, the
environment $\varEnv$ is used when evaluation resumes with the appropriate
branch.
Because they record how ``paused'' expressions should ``resume,'' we sometimes refer to indeterminate results as ``partially evaluated expressions.''

The \emph{inverse constructor application} form
$\expUnwrap{\varDataCon}{\varResult}$ on the third line of the result grammar
is internal to live unevaluation and is discussed in \autoref{sec:live-uneval}.

\parahead{Examples}

A \emph{synthesis goal}
$\problemNameTypeWorlds{\varHoleName}{\varTypeEnv}{\varType}{\varWorlds}$
describes a hole $\expHole{\varHoleName}$ to be filled according to the
contextual type $\tHole{\varTypeEnv}{\varType}$ and \emph{example constraints}
$\varWorlds$.
Each example constraint $\world{\varEnv}{\varEx}$ requires that an expression
to fill the hole must, in the environment $\varEnv$, satisfy example $\varEx$.

Examples include \emph{simple values} $\varSimpleVal$, which are first-order
product values or constructor applications;
\emph{input-output} examples $\ioExample{\varSimpleVal}{\varEx}$, which
constrain function-typed holes;
and \emph{top} $\exHole$, which imposes no constraints.
We sometimes refer to example constraints simply as ``examples'' when the
meaning is clear from context.
The coercion $\liftSimpleEx{\varSimpleVal}$ ``upcasts'' a simple value to a
result.
The coercion $\coerceUndet{\varResult}{\varSimpleVal}$ ``downcasts'' a result to
a simple value, if possible. %%\footnote{\
%% %
%% Examples and example constraints are essentially the same as described by \citet{Myth}; we
%% say example constraints to refer to ``worlds.''

Examples are essentially the same as described by \citet{Myth}.
\smyth{} additionally includes top examples.
For simplicity \coreSnsMyth{} includes only
first-order function examples, though our implementation
(\autoref{sec:implementation}) supports higher-order function examples
like \myth{}.

%%%%%%%%%%%%%%%%%%%%%%%%%%%%%%%%%%%%%%%%%%%%%%%%%%%%%%%%%%%%%%%%%%%%%%%%%%%%%%%%

\subsection{Type Checking}
\label{sec:types}

Type checking $\typeCheck{\varTypeEnv}{\varExp}{\varType}$ (\autoref{fig:eval})
takes a hole type context $\varHoleEnv$ as input, used by the \ruleName{T-Hole}
rule to decide valid typings for a hole $\expHole{\varHoleName}$.
The remaining rules are standard~\citepAppendix{\refAppendixTypeChecking}.

%%%%%%%%%%%%%%%%%%%%%%%%%%%%%%%%%%%%%%%%%%%%%%%%%%%%%%%%%%%%%%%%%%%%%%%%%%%%%%%%

\subsection{Live Evaluation}
\label{sec:live-eval}

%% \begin{figure*}[t]
\begin{figure}[b]

%% \judgementHeadNameOnlyTwo
%%   {Type Checking}
%%   {(excerpt from \autoref{sec:appendix-type-checking})}
%%   {\JudgementBox{\typeCheck{\varTypeEnv}{\varExp}{\varType}}}

\judgementHeadNameOnlyTwo
  {Type Checking}
  {(excerpt from {\refAppendixTypeChecking}) \textbf{and Live Eval.}} %% uation}}
  {$\JudgementBox{\typeCheck{\varTypeEnv}{\varExp}{\varType}}$
   \hspace{0.00in}
   $\JudgementBox{\liveEval{\varEnv}{\varSolution}{\varExp}{\varResult}}$}

\vsepRule

$
\inferrule*[lab=\ruleNameFig{T-Hole}]
  {
   \varHoleEnv(\expHole{\varHoleName}) = \tHole{\varTypeEnv}{\varType}
  }
  {\typeCheck
    {\varTypeEnv}
    {\expHole{\varHoleName}}
    {\varType}}
$
\hsepRule
\hsepRule
\hsepRule
$
\inferrule* %% [lab=\ruleNameFig{Eval}]
  {
   \reducesTo{\varEnv}{\varExp}{\varResult}
   \sepPremise
   \reducesTo{\varSolution}{\varResult}{\varResult'}
  }
  {\liveEval{\varEnv}{\varSolution}{\varExp}{\varResult'}}
$

\begin{comment}

\vsepRule

\judgementHead
  {Live Evaluation}
  %% {\reducesTo{\varEnv}{\varExp}{\varResult}}
  {\liveEval{\varEnv}{\varSolution}{\varExp}{\varResult}}

\vsepRuleNoNeed

$
\inferrule* %% [lab=\ruleNameFig{Eval}]
  {
   \reducesTo{\varEnv}{\varExp}{\varResult}
   \sepPremise
   \reducesTo{\varSolution}{\varResult}{\varResult'}
  }
  {\liveEval{\varEnv}{\varSolution}{\varExp}{\varResult'}}
$

\end{comment}

\vsepRule

%% \judgementHead
%%   %% {\maybeUnderline{E}valuation}
%%   {Expression \maybeUnderline{E}valuation}
%%   {\reducesTo{\varEnv}{\varExp}{\varResult}}
%%   %% {\reducesToN{\varEnv}{\varExp}{\varResult}{\dimColor{\varFuel}}}

\judgementHeadNameOnlyTwo
  {Expression \maybeUnderline{E}valuation}
  %% {(excerpt from \autoref{sec:appendix-dynamics})}
  {(excerpt from {\refAppendixTypeSoundness})}
  %% {(excerpt)}
  {\JudgementBox{\reducesTo{\varEnv}{\varExp}{\varResult}}}

\vsepRule

%% TODO MAYBE ADD BACK IN

\begin{comment}
$
\inferrule*[lab=\ruleNameFig{E-Unit}]
  {
  }
  {\reducesTo
    {\varEnv}
    {\expUnit}
    {\tUnit}}
$
%
%% \hsepRule
\hfill %% TODO
%
$
%% \inferrule*[lab=\ruleNameFig{E-Tuple}]
\inferrule*[lab=\ruleNameFig{E-Pair}]
  {
   \multiPremise{\reducesTo{\varEnv}{\varExp_i}{\varResult_i}}{\sequenceSyntaxTwo{i}}
  }
  {\reducesTo
    {\varEnv}
    {\pair{\varExp_1}{\varExp_2}}
    {\pair{\varResult_1}{\varResult_2}}}
    %% {\triple{\varExp_1}{\ldots}{\varExp_n}}
    %% {\triple{\varResult_1}{\ldots}{\varResult_n}}}
$
%
%% \hsepRule
\hfill %% TODO
%
$
\inferrule*[lab=\ruleNameFig{E-Ctor}]
  {
   \reducesTo{\varEnv}{\varExp}{\varResult}
  }
  {\reducesTo{\varEnv}{\expApp{\varDataCon}{\varExp}}{\expApp{\varDataCon}{\varResult}}}
$
%
%% \hsepRule

\vsepRule

%
$
\inferrule*[lab=\ruleNameFig{E-Fix}]
  {\varExp = \valFixFun{\varVarF}{\varVar}{\varExp}
  }
  {\reducesTo
    {\varEnv}
    {\varExp}
    {\closure{\varEnv}{\varExp}}}
    %% {\expFixFun{\varVarF}{\tArrow{\varType_1}{\varType_2}}{\varVar}{\varExp}}
    %% {\closure{\varEnv}{\valFixFun{\varVarF}{\varVar}{\varExp}}}}
$
%
\hsepRule
%
$
\inferrule*[lab=\ruleNameFig{E-Var}]
  {
   \envBind{\varVar}{\varResult} \in \varEnv
   %% \envBindBoth{\varVar}{\varResult}{\varType} \in \varEnv
  }
  {\reducesTo{\varEnv}{\varVar}{\varResult}}
$
%
\hsepRule
%
\end{comment}
$
\inferrule*[lab=\ruleNameFig{E-Hole}]
  {
  }
  {\reducesTo
    {\varEnv}
    {\expHole{\varHoleName}}
    {\highlightHoleClosure{\holeClosure{\varEnv}{\varHoleName}}}}
$
%% TODO weird location for space...
%
\hsepRule
$
\inferrule*[lab=\ruleNameFig{E-App}]
  {
   %% \reducesTo{\varEnv}{\varExp_1}{\closure{\varEnv_f}{\valFixFun{\varVarF}{\varVar}{\varExp_f}}}
   %% \reducesTo{\varEnv}{\varExp_1}{\varVal_1}
   \reducesTo{\varEnv}{\varExp_1}{\varResult_1}
   %% \reducesToN{\varEnv}{\varExp_1}{\varResult_1}{\varFuel}
   \sepPremise
   \reducesTo{\varEnv}{\varExp_2}{\varResult_2}
   %% \varVal_1 = \closure{\varEnv_f}{\valFixFun{\varVarF}{\varVar}{\varExp_f}}
   \\\\
   %% \reducesToN{\varEnv}{\varExp_2}{\varResult_2}{\varFuel}
   \varResult_1 = \closure{\varEnv_f}{\valFixFun{\varVarF}{\varVar}{\varExp_f}}
   %% \sepPremise
   \\\\
   \reducesTo %% N
             {\envCatThree{\varEnv_f}
                          %% {\envBind{\varVarF}{\closure{\varEnv_f}{\valFixFun{\varVarF}{\varVar}{\varExp_f}}}}
                          %% {\envBind{\varVarF}{\varVal_1}}
                          {\envBind{\varVarF}{\varResult_1}}
                          {\envBind{\varVar}{\varResult_2}}}
             {\varExp_f}{\varResult}
             %% {\varFuel - 1}
  }
  %% {\reducesToN{\varEnv}{\expApp{\varExp_1}{\varExp_2}}{\varResult}{\varFuel}}
  {\reducesTo{\varEnv}{\expApp{\varExp_1}{\varExp_2}}{\varResult}}
$
\hsepRule
$
\inferrule*[lab=\ruleNameFig{E-App-Indet}]
  {
   \reducesTo{\varEnv}{\varExp_1}{\highlightIndet{$\varResult_1$}}
   \sepPremise
   \reducesTo{\varEnv}{\varExp_2}{\varResult_2}
   %% \sepPremise
   \\\\
   \varResult_1 \neq \closure{\varEnv_f}{\valFixFun{\varVarF}{\varVar}{\varExp_f}}
   %% \dimColor{\varResult_1 \neq \closure{\varEnv_f}{\expFun{\varVar}{\varExp_f}}}
  }
  {\reducesTo
    {\varEnv}
    {\expApp{\varExp_1}{\varExp_2}}
    {\expAppIndet{\varResult_1}{\varResult_2}}}
$

\vsepRule

%% \judgementHeadThree
%%   {\maybeUnderline{R}esumption}
%%   {(\autoref{fig:resumption} of \autoref{sec:appendix-dynamics})}
%%   {\resumesTo{\varSolution}{\varResult}{\varResult'}}

\judgementHeadNameOnlyTwo
  {\maybeUnderline{R}esumption}
  %% {(excerpt from \autoref{fig:resumption} of \autoref{sec:appendix-dynamics})}
  %% {\JudgementBox{\resumesTo{\varSolution}{\varResult}{\varResult'}} \
  %%  \JudgementBox{\resumesTo{\varSolution}{\varEnv}{\varEnv'}}}
  %% {(excerpt from \autoref{sec:appendix-dynamics})}
  {(excerpt from {\refAppendixResumption})}
  %% {(excerpt)}
  {\JudgementBox{\resumesTo{\varSolution}{\varResult}{\varResult'}}}

\vsepRule

$
\inferrule*[lab=\ruleNameFig{R-Hole-Resume}]
  {
   \varSolution(\varHoleName) = \varExp_{\varHoleName}
   %% \sepPremise
   %% \rkc{\resumesTo{\varSolution}{\varEnv}{\varEnv'} before eval?}
   \sepPremise
   \reducesTo{\varEnv}{\varExp_{\varHoleName}}{\varResult}
   \sepPremise
   \resumesTo{\varSolution}{\varResult}{\varResult'}
  }
  {\resumesTo{\varSolution}{\closure{\varEnv}{\expHole{\varHoleName}}}{\varResult'}}
$
\hsepRule
$
\inferrule*[lab=\ruleNameFig{R-Hole-Indet}]
  {
   \varHoleName \notin dom(\varSolution)
   \sepPremise
   \resumesTo{\varSolution}{\varEnv}{\varEnv'}
  }
  {\resumesTo
    {\varSolution}
    {\closure{\varEnv}{\expHole{\varHoleName}}}
    {\closure{\varEnv'}{\expHole{\varHoleName}}}}
$

\vsepBeforeCaption

%% \vsepRule
%% \caption{Big-step, environment-style evaluation.}
%% \caption{Evaluation and Resumption (\autoref{sec:appendix}).
%% \caption{Evaluation and Resumption.
\caption{Type Checking, Evaluation, and Resumption.
%% (Resumption in \autoref{fig:resumption} of \autoref{sec:appendix-dynamics}.)
}
\label{fig:eval}
%% \end{figure*}

\vsepAfterCaptionBottomFig

\end{figure}

\autoref{fig:eval} defines \emph{live evaluation}
$\liveEval{\varEnv}{\varSolution}{\varExp}{\varResult}$, which
%
%% for \coreSnsMyth{}.
%
first uses \emph{expression evaluation}
$\reducesTo{\varEnv}{\varExp}{\varResult}$ to produce a final result
$\varResult$, and
then \emph{resumes} evaluation
$\resumesTo{\varSolution}{\varResult}{\varResult'}$ of the result $\varResult$
in positions that were paused because of holes now filled by $\varSolution$.

\parahead{Expression Evaluation}

Compared to a conventional natural semantics, there are four new
rules---\ruleName{E-Hole}, \ruleName{E-App-Indet}, \ruleName{E-Prj-Indet}, and
\ruleName{E-Case-Indet}---one for each indeterminate result form.
The \ruleName{E-Hole} rule creates a hole closure
$\closure{\varEnv}{\expHole{\varHoleName}}$ that captures the evaluation
environment.

The other three rules, suffixed ``\ruleName{-Indet},'' are counterparts to rules
\ruleName{E-App}, \ruleName{E-Prj}, and \ruleName{E-Case} for determinate forms.
For example, when a function evaluates to a result $\varResult_1$ that is not a
function closure, the
\ruleName{E-App-Indet} rule creates the indeterminate application result
$\expApp{\varResult_1}{\varResult_2}$.
The remaining rules are similar~\citepAppendix{\refAppendixTypeSoundness}.
Evaluation is deterministic and produces final results;
\citetAppendix{\refAppendixTypeSoundness} formally establishes these propositions, as well as
a suitable notion of type safety.

\parahead{Resumption}

Result resumption resembles expression evaluation.
For closures $\holeClosure{\varEnv}{\varHoleName}$ over holes that
$\varSolution$ fill with an expression $\varExp_\varHoleName$,
\ruleName{R-Hole-Resume} evaluates $\varExp_\varHoleName$ in the closure
environment, producing a result $\varResult$.
Because $\varExp_\varHoleName$ may refer to other holes now filled by
$\varSolution$, $\varResult$ is recursively resumed to $\varResult'$.

%%%%%%%%%%%%%%%%%%%%%%%%%%%%%%%%%%%%%%%%%%%%%%%%%%%%%%%%%%%%%%%%%%%%%%%%%%%%%%%%

\vspace{-0.03in} %% HACK for arxiv PDF layout

\subsection{Example Satisfaction}
\label{sec:ex-sat}

%% \begin{figure*}[t]
\begin{figure}[b]

%% \input{fig-syntax-1}

%% \vsepRule

\begin{comment} %% defined in the text of Syntax section

\judgementHeadNameOnly
  %% {Simple Example Coercion}
  %% {Simple Value Coercion}
  %% {Result to Simple Value Coercion}
  {Result-Value Coercions}
  {
   $\JudgementBox{\coerceUndet{\varResult}{\varSimpleVal}}$
   \hspace{0.00in}
   $\JudgementBox{\liftSimpleExEquals{\varSimpleVal}{\varResult}}$
  }

\vsepRule

$
\inferrule*
  {
  }
  {\coerceUndet
    {\expUnit}
    {\expUnit}}
$
%
\hsepRule
%
$
\inferrule*
  %% {\coerceUndet{\varResult_i}{\varSimpleVal_i}
  {\coerceUndet{\varResult_1}{\varSimpleVal_1}
   \sepPremise
   \coerceUndet{\varResult_2}{\varSimpleVal_2}
  }
  {\coerceUndet
    {\pair{\varResult_1}{\varResult_2}}
    {\pair{\varSimpleVal_1}{\varSimpleVal_2}}}
$
%
\hsepRule
%
$
\inferrule*
  {\coerceUndet{\varResult}{\varSimpleVal}
  }
  {\coerceUndet
    {\expApp{\varDataCon}{\varResult}}
    {\expApp{\varDataCon}{\varSimpleVal}}}
$
%
\hsepRule
\hsepRule
\hsepRule
%
$
\inferrule*
  {
  }
  {\liftSimpleExEquals{\varSimpleVal}{\varSimpleVal}}
$

%% \vsepRule
%% 
%% \input{fig-ex-typing}

\vsepRule

\end{comment}

\newcommand{\hsepXSat}
  {\hsepRule}
  %% {\hspace{0.10in}} %% TODO

%% \judgementHead
%%   {E\maybeUnderline{x}ample \maybeUnderline{S}atisfaction}
%%   {\exSat{\varSolution}{\varResult}{\varEx}}

%% \judgementHeadNameOnly
%%   %% {Constraint, Example World, and
%%   {Example World and
%%    E\maybeUnderline{x}ample \maybeUnderline{S}atisfaction}
%%   {
%%   %% $\JudgementBox{\unsolvedConstraintSat{\varSolution}{\varUnsolvedConstraints}}$ \hspace{0.00in}
%%   $\JudgementBox{\worldSat{\varSolution}{\varExp}{\varWorlds}}$ \hspace{0.00in}
%%   $\JudgementBox{\exSat{\varSolution}{\varResult}{\varEx}}$
%%   }

\judgementHeadNameOnly
  %% {Example \maybeUnderline{Sat}isfaction (of Expressions)}
  %% {Example (Constraint) \maybeUnderline{Sat}isfaction}
  {Example Constraint \maybeUnderline{Sat}isfaction}
  {
  $\JudgementBox{\worldSat{\varSolution}{\varExp}{\varWorlds}}$
  }

\vsepRule %% NoNeed

%% $
%% \inferrule* %% [lab=\ruleNameFig{XS-Constraints}]
%%   {
%%    \multiPremise{
%%    \worldSat{\varSolution}{\expHole{\varHoleName_i}}{\varWorlds_i}
%%    }{\sequenceSyntax}
%%   }
%%   {\hSat{\varSolution}
%%         {(\envCatThree{\unsolvedConstraint{\varHoleName_1}{\varWorlds_1}}
%%                       {\ldots}
%%                       {\unsolvedConstraint{\varHoleName_n}{\varWorlds_n}})}}
%% $
%
\hsepRule
$
%% \inferrule* %% [lab=\ruleNameFig{XS-Worlds}]
\inferrule*[lab=\ruleNameFig{Sat}]
  {
   \multiPremise{
   \liveEval{\varEnv_i}{\varSolution}{\varExp}{\varResult_i}
   %% \reducesTo{\varEnv_i}{\varExp}{\varResult_i}
   %% \sepPremise
   %% \resumesTo{\varSolution}{\varResult_i}{\varResult_i'}
   \sepPremise
   \exSat{\varSolution}{\varResult_i}{\varEx_i}
   %% \exSat{\varSolution}{\varResult_i'}{\varEx_i}
   }{\sequenceSyntax}
  }
  {\worldSat{\varSolution}
            {\varExp}
            {\multiPremise{\world{\varEnv_i}{\varEx_i}}{\sequenceSyntax}}}
            %% {(\envCatThree{\world{\varEnv_1}{\varEx_1}}
            %%               {\ldots}
            %%               {\world{\varEnv_n}{\varEx_n}})}}
$

\vsepRule

\judgementHeadNameOnly
  %% {E\maybeUnderline{x}ample \maybeUnderline{S}atisfaction (of Results)}
  {E\maybeUnderline{x}ample \maybeUnderline{S}atisfaction}
  {
  $\JudgementBox{\exSat{\varSolution}{\varResult}{\varEx}}$
  }

\vsepRuleNoNeed

$
\inferrule*[lab=\ruleNameFig{XS-Top}]
  {
  }
  {\exSat{\varSolution}{\varResult}{\exHole}}
$
\hsepXSat
$
\inferrule*[lab=\ruleNameFig{XS-Unit}]
  {
  }
  {\exSat{\varSolution}{\expUnit}{\expUnit}}
$
\hsepXSat
$
\inferrule*[lab=\ruleNameFig{XS-Pair}]
  {\multiPremise{\exSat{\varSolution}{\varResult_i}{\varEx_i}}{\sequenceSyntaxTwo{i}}}
  {\exSat{\varSolution}{\pair{\varResult_1}{\varResult_2}}{\pair{\varEx_1}{\varEx_2}}}
$

%% %
%% \hsepXSat
%% %

\vsepRule

$
\inferrule*[lab=\ruleNameFig{XS-Ctor}]
  {\exSat{\varSolution}{\varResult}{\varEx}}
  {\exSat{\varSolution}{\expApp{\varDataCon}{\varResult}}{\expApp{\varDataCon}{\varEx}}}
$
\hsepXSat
$
\inferrule*[lab=\ruleNameFig{XS-Input-Output}]
  {
   %% \varResult_2 = \liftSimpleEx{\varSimpleVal_2}
   %% \\\\
   \resumesTo{\varSolution}
             %% {\expApp{\varResult_1}{\varResult_2}}
             {\expApp{\varResult_1}{\liftSimpleEx{\varSimpleVal_2}}}
             {\varResult}
   \sepPremise
   \exSat{\varSolution}{\varResult}{\varEx}
   %% \varWorlds = \world{\emptyEnv}{\varEx}
   %% \\\\
   %% \worldSat{\varSolution}
   %%          {\expApp{\varResult_1}{\liftSimpleEx{\varSimpleVal_2}}}
   %%          {\varWorlds}
  }
  {\exSat{\varSolution}{\varResult_1}{\ioExample{\varSimpleVal_2}{\varEx}}}
$

\vsepRule

\judgementHead
  %% {Constraint Satisfaction}
  %% {(Unevaluation) Constraint Satisfaction}
  {Unevaluation Constraint Satisfaction}
  {\constraintSat{\varSolution}{\varConstraints}}

\vsepRule

$
\inferrule*
  {
   \varSolution \supseteq \varSolution_0
   \sepPremise
   \multiPremise{
   \worldSat{\varSolution}{\expHole{\varHoleName_i}}{\varWorlds_i}
   }{\sequenceSyntax}
  }
  {\constraintSat
    {\varSolution}
    {\pairConstraints
      {\varSolution_0}
      {(\envCatThree{\unsolvedConstraint{\varHoleName_1}{\varWorlds_1}}
                    {\ldots}
                    {\unsolvedConstraint{\varHoleName_n}{\varWorlds_n}})}}}
$

\vsepBeforeCaption

%% \caption{Example, Constraint, and World Satisfaction.}
%% \caption{Example Typing and Satisfaction.}
%% \caption{Example Syntax, Typing, and Satisfaction.}
%% \caption{Example Satisfaction.}
\caption{Example and Constraint Satisfaction.}
\label{fig:ex-satisfaction}
%% \end{figure*}

\vsepAfterCaptionBottomFig

\end{figure}

Live evaluation partially evaluates a sketch to a result, and
\autoref{fig:ex-satisfaction} defines what it means for a result to satisfy an
example.
To decide whether expression $\varExp$ satisfies example constraint
$\world{\varEnv}{\varEx}$,
the \ruleName{Sat} rule evaluates the expression to a result $\varResult$ and
then checks whether $\varResult$ satisfies $\varEx$.
The \ruleName{XS-Top} rule accepts all results.
The remaining rules break down input-output examples
(\ruleName{XS-Input-Output}) into equality checks for products and constructors
(\ruleName{XS-Unit}, \ruleName{XS-Pair}, and \ruleName{XS-Ctor}).

Hole closures may appear in a satisfying result, but they may \emph{not} be
directly checked against product, constructor, or input-output examples.
The purpose of \emph{live unevaluation} is to provide a
notion of example \emph{consistency} to accompany this ``ground-truth'' notion
of example satisfaction.

%%%%%%%%%%%%%%%%%%%%%%%%%%%%%%%%%%%%%%%%%%%%%%%%%%%%%%%%%%%%%%%%%%%%%%%%%%%%%%%%

\vspace{-0.03in} %% HACK for arxiv PDF layout

\subsection{Live Unevaluation}
\label{sec:live-uneval}

\begin{figure*}[b]

\judgementHeadNameOnlyTwo
  %% {Constraint Merging}
  %% {(Unevaluation) Constraint Merging}
  {Unevaluation Constraint Merging}
  %% {(\autoref{sec:appendix-dynamics})}
  {(in {\refAppendixConstraintMerge})}
  %% {(\autoref{fig:constraint-merge} of \autoref{sec:appendix-dynamics})}
  {
   %% $\JudgementBox{\mergeConstraintsEquals{\varSolution_1}{\varSolution_2}{\varSolution}}$
   %% \hspace{0.00in}
   %% $\JudgementBox{\mergeConstraintsEquals{\varUnsolvedConstraints_1}{\varUnsolvedConstraints_2}{\varUnsolvedConstraints}}$
   %% \hspace{0.00in}
   $\JudgementBox{\mergeConstraintsEquals{\varConstraints_1}{\varConstraints_2}{\varConstraints}}$
   \hspace{0.00in}
   $\JudgementBox{
     \simplifyConstraintsEquals
       {\varHoleEnv}
       {\varDatatypeEnv}
       {(\varConstraints)}
       {\varConstraints'}}$
  }

\vsepRule

\tightJudgementHead
  %% {Live Bidirectional Example \maybeUnderline{Check}ing}
  {\maybeUnderline{Live} Bidirectional Example \maybeUnderline{Check}ing}
  %% {Live Bidirectional Example Satisfaction}
  %% {Live Bidirectional World Satisfaction}
   %% \dimColor{Live \maybeUnderline{W}orld \maybeUnderline{C}onsistency}}
  {\worldConsistentFull
    {\varHoleEnv}{\varDatatypeEnv}
    {\varSolution}{\varExp}{\varWorlds}{\varConstraints} %% {\varSolution'}
  }

\vsepRule

%% $
%% \inferrule*[lab=\ruleNameFig{WC-Empty}]
%%   {
%%   }
%%   {\iterConsistentFill{\varSolution}{\varExp}{\emptySet}{\emptySet}{\varSolution}}
%% $
%% %
%% \hsepRule
%% %
%% $
%% \inferrule*[lab=\ruleNameFig{WC-World}]
%%   {
%%    \iterConsistentFill{\varSolution}{\varExp}{\varWorlds}{\varConstraints_1}{\varSolution_1}
%%    \\\\
%%    %% \sepPremise
%%    \reducesToUK{\varEnv}{\varExp}{\varResult}{\emptySet}
%%    \sepPremise
%%    \resumesToUK{\varSolution}{\varResult}{\varResult'}{\emptySet}
%%    \sepPremise
%%    \xBackpropFill{\varSolution}{\varResult'}{\varEx}{\varConstraints_2}{\varSolution_2}
%%   }
%%   {\iterConsistentFill
%%     {\varSolution}
%%     {\setAdd{\varWorlds}{\world{\varEnv}{\varEx}}}
%%     {\varExp}
%%     {\setUnion{\varConstraints_1}{\varConstraints_2}}
%%     {\mergeSolutions{\varSolution_1}{\varSolution_2}}
%%   }
%% $

$
%% \inferrule* %% [lab=\ruleNameFig{WC-World}]
%% \inferrule*[lab=\ruleNameFig{Check}]
\inferrule*[lab=\ruleNameFig{Live-Check}]
  {
   \multiPremise{
   \liveEval{\varEnv_i}{\varSolution}{\varExp}{\varResult_i}
   %% \reducesTo{\varEnv_i}{\varExp}{\varResult_i}
   %% \sepPremise
   %% \resumesTo{\varSolution}{\varResult_i}{\varResult_i'}
   \sepPremise
   \uneval{\varSolution}{\varResult_i}{\varEx_i}{\varConstraints_i} %% {\varSolution_i}
   %% \uneval{\varSolution}{\varResult_i'}{\varEx_i}{\varConstraints_i} %% {\varSolution_i}
   }{\sequenceSyntax}
  }
  {\worldConsistent
    {\varSolution}
    {\varExp}
    {\envCatThree{\world{\varEnv_1}{\varEx_1}}
                 {\ldots}
                 {\world{\varEnv_n}{\varEx_n}}}
    {\mergeConstraintsThree{\varConstraints_1}{\cdots}{\varConstraints_n}}
    %% {\mergeSolutionsThree{\varSolution_1}{\cdots}{\varSolution_n}}
  }
$

\vsepRule

\judgementHead
  {Live \maybeUnderline{U}nevaluation}
  %% {\unevalExtraIn{\varSolution}{\varResult}{\varEx}{\varConstraints}{\varSolution'}}
  %% {\unevalExtraIn{(\varSolution}{\varResult)}{\varEx}{\varConstraints)}{(\varSolution'}}
  {\unevalFull
     {\varHoleEnv}{\varDatatypeEnv}
     {\varSolution}{\varResult}{\varEx}{\varConstraints}
  }

\vsepRuleNoNeed

$
\inferrule*[lab=\ruleNameFig{U-Top}]
  {
  }
  {\uneval{\varSolution}{\varResult}{\exHole}{\emptySet}} %% {\varSolution}}
$
\hsepRule
$
\inferrule*[lab=\ruleNameFig{U-Unit}]
  {
  }
  {\uneval
    {\varSolution}
    {\expUnit}
    {\expUnit}
    {\emptySet}
    %% {\varSolution}
  }
$
%
%% \hsepRule
%

\vsepRule

$
\inferrule*[lab=\ruleNameFig{U-Pair}]
  {
   \uneval{\varSolution}{\varResult_1}{\varEx_1}{\varConstraints_1} %% {\varSolution_1}
   \sepPremise
   %% \\\\
   \uneval{\varSolution}{\varResult_2}{\varEx_2}{\varConstraints_2} %% {\varSolution_2}
   %% \\\\
   %% \varConstraints =
   %%   {\setUnion{\varConstraints_1}{\varConstraints_2}}
   %% \sepPremise
   %% \varSolution' =
   %%   {\mergeSolutions{\varSolution_1}{\varSolution_2}}
  }
  {\uneval
    {\varSolution}
    {\pair{\varResult_1}{\varResult_2}}
    {\pair{\varEx_1}{\varEx_2}}
    {\mergeConstraints{\varConstraints_1}{\varConstraints_2}}
    %% {\mergeSolutions{\varSolution_1}{\varSolution_2}}
    %% {\varConstraints}
    %% {\varSolution'}
  }
$
\hsepRule
$
\inferrule*[lab=\ruleNameFig{U-Ctor}]
  {
   \uneval{\varSolution}{\varResult}{\varEx}{\varConstraints} %% {\varSolution'}
  }
  {\uneval
    {\varSolution}
    {\expApp{\varDataCon}{\varResult}}
    {\expApp{\varDataCon}{\varEx}}
    {\varConstraints}
    %% {\varSolution'}
  }
$

\vsepRule

$
\inferrule*[lab=\ruleNameFig{U-Fix}]
  {
   %% \overbrace{
   %% \varEnv' =
   %%   {\envCatThree
   %%     {\varEnv}
   %%     {\envBind{\varVarF}{\closure{\varEnv}{\valFixFun{\varVarF}{\varVar}{\varExp}}}}
   %%     {\envBind{\varVar}{\liftSimpleEx{\varSimpleVal}}}}
   %% \sepPremise
   \liveBiEval
     {\varSolution}
     %% {\varEnv'}
     {\envCatThree
       {\varEnv}
       {\envBind{\varVarF}{\closure{\varEnv}{\valFixFun{\varVarF}{\varVar}{\varExp}}}}
       {\envBind{\varVar}{\liftSimpleEx{\varSimpleVal}}}}
     {\varExp}
     {\varEx}
     {\varConstraints}
   %% }^{
   %% \highlightLiveBiEval{
   %% \resumesToUK
   %%   {\varSolution}
   %%   {\expApp{({\closure{\varEnv}{\valFixFun{\varVarF}{\varVar}{\varExp}}})}
   %%           {\liftSimpleEx{\varSimpleVal}}}
   %%   {\varResult}
   %%   {\emptySet}
   %% \sepPremise
   %% \uneval{\varSolution}{\varResult}{\varEx}{\varConstraints} %% {\varSolution'}
   %% }
   %% }
  }
  {\uneval
    {\varSolution}
    {\closure{\varEnv}{\valFixFun{\varVarF}{\varVar}{\varExp}}}
    {\ioExample{\varSimpleVal}{\varEx}}
    {\varConstraints}
    %% {\varSolution'}
  }
$
\hsepRule
$
\inferrule*[lab=\ruleNameFig{U-Hole}]
  {
   %% \varConstraints =
   \varUnsolvedConstraints =
     %% {\holeConstraint{\varHoleName}{\varEnv}{\varEx}}
     \highlightUnsolvedConstraint{
     $\unsolvedConstraint{\varHoleName}{\world{\varEnv}{\varEx}}$%
     }
  }
  {\uneval
    {\varSolution}
    {\highlightHoleClosure{\holeClosure{\varEnv}{\varHoleName}}}
    {\varEx}
    {\pairConstraints{\emptyEnv}{\varUnsolvedConstraints}}
    %% {\varConstraints}
    %% {\varSolution}
  }
$

\vsepRule

$
\inferrule*[lab=\ruleNameFig{U-App}]
  {
   \coerceUndet{\varResult_2}{\varSimpleVal_2}
   \sepPremise
   \uneval
    {\varSolution}
    {\varResult_1}
    {\ioExample{\varSimpleVal_2}{\varEx}}
    {\varConstraints}
    %% {\varSolution'}
  }
  {\uneval
    {\varSolution}
    {\expAppIndet{\varResult_1}{\varResult_2}}
    {\varEx}
    {\varConstraints}
    %% {\varSolution'}
  }
$
\hsepRule
%% \hspace{0.04in} %% TODO
%
$
\inferrule*[lab=\ruleNameFig{U-Prj-1}]
  {
   \uneval{\varSolution}{\varResult}{\pair{\varEx}{\exHole}}{\varConstraints} %% {\varSolution'}
  }
  {\uneval{\varSolution}{\expFstIndet{\varResult}}{\varEx}{\varConstraints}} %% {\varSolution'}}
$
\hsepRule
%% \hspace{0.04in} %% TODO
%
$
\inferrule*[lab=\ruleNameFig{U-Prj-2}]
  {
   \uneval{\varSolution}{\varResult}{\pair{\exHole}{\varEx}}{\varConstraints} %% {\varSolution'}
  }
  {\uneval{\varSolution}{\expSndIndet{\varResult}}{\varEx}{\varConstraints}} %% {\varSolution'}}
$

\vsepRule

$
%% \inferrule*[lab=\ruleNameFig{XB-Match}]
\inferrule*[lab=\ruleNameFig{U-Case}]
  {
   %% \exists j \in [1,n]
   j \in [1,n]
   \sepPremise
   \uneval{\varSolution}{\varResult}{\expApp{\varDataCon_j}{\exHole}}{\varConstraints_1}
   %% \sepPremise
   \\\\
   \liveBiEval
     {\varSolution}
     {\envCat{\varEnv}{\envBind{\varVar_j}{\expUnwrap{\varDataCon_j}{\varResult}}}}
     {\varExp_j}
     {\varEx}
     {\varConstraints_2}
   %% \resumesTo
   %%   {\emptyEnv}
   %%     {\expApp{(\closure{\varEnv}{\expFun{\varVar_j}{\varExp_j}})}
   %%                       {(\expUnwrap{\varDataCon_j}{\varResult})}}
   %%   {\varResult_j}
   %% \sepPremise
   %% \uneval{\varSolution}{\varResult_j}{\varEx}{\varConstraints_2}
  }
  {\uneval
    {\varSolution}
    {\closure{\varEnv}{\expMatch{\varResult}{\varDataCon_i}{\varVar_i}{\varExp_i}}}
    {\varEx}
    {\mergeConstraints{\varConstraints_1}{\varConstraints_2}}}
$
\hsepRule
$
%% \inferrule*[lab=\ruleNameFig{U-Unwrap-Ctor}]
\inferrule*[lab=\ruleNameFig{U-Inverse-Ctor}]
  {\uneval
    {\varSolution}
    {\varResult}
    {\expApp{\varDataCon}{\varEx}}
    {\varConstraints}
  }
  {\uneval
    {\varSolution}
    {\expUnwrap{\varDataCon}{\varResult}}
    {\varEx}
    {\varConstraints}}
$

\vsepRule

$
\inferrule*[lab=\ruleNameFig{U-Case-Guess}]
  {
   j \in [1,n]
   \sepPremise
   \varSolutionGuesses =
     \highlightSolvedConstraint{$
     \guessesForMatch{\varHoleEnv}{\varDatatypeEnv}{\varResult}
     $}
   \sepPremise
   %% \varSolution' =
   %%    \addSolutions
   %%      {\varSolution}
   %%      {\varSolution_{\textit{guesses}}}
   %% \sepPremise
   \resumesTo
     %% {\varSolution'}
     {\addSolutions
        {\varSolution}
        {\varSolutionGuesses}}
     {\varResult}
     {\expApp{\varDataCon_j}{\varResult'}}
   \\\\
   %% \sepPremise
   %% \varConstraints_1 =
   %%   \highlightSolvedConstraint{\liftSyntax{\varSolution_{\textit{guesses}}}}
   %% \varConstraints_1 =
   %%   \pairConstraints{\varSolution_{\textit{guesses}}}{\emptyEnv}
   %% \sepPremise
   %% \highlightLiveBiEval{
   %% \resumesToUK
   %%   {\varSolution'}
   %%   {\expApp{(\closure{\varEnv}{\expFun{\varVar_j}{\varExp_j}})}{\varResult'}}
   %%   {\varResult_j}
   %%   {\emptySet}
   %% \sepPremise
   %% \uneval
   %%   {\varSolution'}
   %%   {\varResult_j}
   %%   {\varEx}
   %%   {\varConstraints_2}
   %%   %% {\varSolution''}
   %% }
   %% \varEnv' =
   %%   {\envCat
   %%     {\varEnv}
   %%     {\envBind{\varVar_j}{\varResult'}}}
   %% \sepPremise
   \liveBiEval
     %% {\varSolution'}
     {\addSolutions
        {\varSolution}
        {\varSolutionGuesses}}
     %% {\varEnv'}
     {\envCat
       {\varEnv}
       {\envBind{\varVar_j}{\varResult'}}}
     {\varExp_j}
     {\varEx}
     {\varConstraints}
     %% {\varConstraints_2}
  }
  {\uneval
    {\varSolution}
    {\caseClosureIndet{\varEnv}{\varResult}{\varDataCon_i}{\varVar_i}{\varExp_i}}
    {\varEx}
    {\mergeConstraints
      %% {\varConstraints_1}
      {\pairConstraints{\varSolutionGuesses}{\emptyEnv}}
      {\varConstraints}
      %% {\varConstraints_2}
    }
    %% {\varConstraints}
    %% {\varSolution''}
  }
$

\vsepBeforeCaption

%% \caption{Live Bidirectional Evaluation.}

\caption{Live Bidirectional Example Checking via Live Unevaluation.}
\label{fig:ex-backprop}

\vsepAfterCaptionBottomFig

\end{figure*}

\autoref{fig:ex-backprop} defines \emph{live unevaluation}
$\uneval{\varSolution}{\varResult}{\varEx}{\varConstraints}$,
which produces constraints $\varConstraints$ over holes
that are sufficient to ensure example satisfaction
$\exSat{\varSolution}{\varResult}{\varEx}$.
The \emph{live bidirectional example checking} judgement
$\worldConsistent{\varSolution}{\varExp}{\varWorlds}{\varConstraints}$
%
%% (\autoref{fig:ex-backprop})
%
lifts this notion to example constraints:
\ruleName{Live-Check} appeals to evaluation followed by unevaluation
to check each constraint in $\varWorlds$.

\renewcommand{\dimColor}[1]{\textcolor{black}{#1}} %% HACK

%% \vspace{-0.03in} %% HACK for arxiv PDF layout

\begin{theorem*}[Soundness of Live Unevaluation]

\breakAndIndent
If $\uneval{\varSolution}{\varResult}{\varEx}{\varConstraints}$
and $\constraintSat{\addSolutions{\varSolution}{\varSolution'}}{\varConstraints}$
and $\resumesTo{\addSolutions{\varSolution}{\varSolution'}}{\varResult}{\varResult'}$,
then $\exSat{\addSolutions{\varSolution}{\varSolution'}}{\varResult'}{\varEx}$.

\end{theorem*}

\vspace{-0.03in} %% HACK for arxiv PDF layout

\begin{theorem*}[Soundness of Live Bidirectional Example Checking]

\breakAndIndent
If $\worldConsistent{\varSolution}{\varExp}{\varWorlds}{\varConstraints}$
and $\constraintSat{\addSolutions{\varSolution}{\varSolution'}}{\varConstraints}$,
then $\worldSat{\addSolutions{\varSolution}{\varSolution'}}{\varExp}{\varWorlds}$.

\end{theorem*}

\renewcommand{\dimColor}[1]{\textcolor{gray}{#1}} %% undo HACK

\parahead{Unevaluation Constraints}

Two kinds of constraints $\varConstraints$ are generated by unevaluation
(\cf{}~\autoref{fig:syntax}).
The first is a context $\varUnsolvedConstraints$ of bindings
$\highlightUnsolvedConstraint{$\unsolvedConstraint{\varHoleName}{\varWorlds}$}$
that maps unfilled holes $\expHole{\varHoleName}$ to sets $\varWorlds$ of
example constraints $\world{\varEnv}{\varEx}$.
The second is a hole-filling $\varSolution$ which, as discussed below, is used
to optimize unevaluation of \inlinecode{case} expressions.
The former are ``hole example contexts,'' analogous to hole type contexts
$\varHoleEnv$; the metavariable $\varUnsolvedConstraints$ serves as a mnemonic
for holes left \underline{u}nfilled by a hole-filling $\varSolution$.
(In the simpler presentation of \autoref{sec:overview}, only example constraints
were generated, and each was annotated with a hole name.)

To define what it means for a filling $\varSolution$ to constitute a
valid solution for a set of constraints
$\varConstraints=\pairConstraints{\varSolution_0}{\varUnsolvedConstraints}$,
\autoref{fig:ex-satisfaction} defines constraint satisfaction
$\constraintSat{\varSolution}{\varConstraints}$ by checking that
(i) $\varSolution$ subsumes any fillings $\varSolution_0$ in $\varConstraints$
and
(ii) $\varSolution$ satisfies the examples $\varWorlds_i$ for each hole
$\expHole{\varHoleName_i}$ constrained by $\varConstraints$.

When analyzing multiple subexpressions, several unevaluation
rules---discussed below---generate multiple sets of constraints that
must be combined.
\autoref{fig:ex-backprop} shows the signature of two constraint merge
operators.
The ``syntactic'' merge operation
$\mergeConstraints{\varConstraints_1}{\varConstraints_2}$ pairwise combines
example contexts $\varUnsolvedConstraints$ and
fillings $\varSolution$
in a straightforward way.
Syntactically merged constraints may describe holes $\expHole{\varHoleName}$
both with
example constraints $\varWorlds$ in $\varUnsolvedConstraints$ and
fillings in $\varSolution$;
the ``semantic'' operation $\metaFunc{Merge}(\varConstraints)$ uses live
bidirectional example checking to check consistency in such situations.
The full definitions can be found in \citetAppendix{\refAppendixConstraintMerge}.

\parahead{Simple Unevaluation Rules}

Analogous to the five example satisfaction rules (prefixed ``\ruleName{XS-}'' in
\autoref{fig:ex-satisfaction}) are
the \ruleName{U-Top} rule to unevaluate any result with $\exHole$ and
the \ruleName{U-Unit}, \ruleName{U-Pair}, \ruleName{U-Ctor}, and
\ruleName{U-Fix} rules to unevaluate determinate results.
The base case in which unevaluation generates example constraints is for
hole closures $\closure{\varEnv}{\expHole{\varHoleName}}$---the \ruleName{U-Hole}
rule generates the (named) example constraint
$\highlightUnsolvedConstraint{$\unsolvedConstraint{\varHoleName}{\world{\varEnv}{\varEx}}$}$.

The \ruleName{U-Fix} rule refers to bidirectional example
checking---evaluation followed by unevaluation---to ``test'' that a function
is consistent with an input-output example.
For instance, to unevaluate the function
closure $\closure{\envBind{\text{\inlinecode{zero}}}{\text{\inlinecode{0}}}}{\expFun{\varVar}{\expHole{\varHoleName}}}$
with $\ioExample{\text{\inlinecode{1}}}{\text{\inlinecode{2}}}$,
first, the function application is evaluated:
the closure environment is extended to bind the input example
$\envBind{\varVar}{\liftSimpleEx{\text{\inlinecode{1}}}}$, and
the function body is evaluated to result
$\closure{\envCat{\envBind{\text{\inlinecode{zero}}}{\text{\inlinecode{0}}}}{\envBind{\varVar}{\liftSimpleEx{\text{\inlinecode{1}}}}}}{\expHole{\varHoleName}}$.
Second, the output example $\text{\inlinecode{2}}$ is unevaluated to this result, for which \ruleName{U-Hole} generates the constraint
$\highlightUnsolvedConstraint{$\unsolvedConstraint{\varHoleName}{\world{\envCat{\envBind{\text{\inlinecode{zero}}}{\text{\inlinecode{0}}}}{\envBind{\varVar}{\liftSimpleEx{\text{\inlinecode{1}}}}}}{\text{\inlinecode{2}}}}$}$.
(Valid fillings for $\expHole{\varHoleName}$ include
\expApp{\text{\inlinecode{S}}}{\color{CadetBlue}(\color{Black}\expApp{\text{\inlinecode{S}}}{\text{\inlinecode{Z}}}\color{CadetBlue})\color{Black}},
\expApp{\text{\inlinecode{S}}}{\text{\inlinecode{x}}}, and
\expApp{\text{\inlinecode{S}}}{\color{CadetBlue}(\color{Black}\expApp{\text{\inlinecode{S}}}{\text{\inlinecode{zero}}}\color{CadetBlue})\color{Black}}.)

The remaining rules, discussed below, transform ``indirect'' unevaluation goals
for more complex indeterminate results into ``direct'' examples on holes.

\parahead{Indeterminate Function Applications}

Consider an indeterminate function application
$\expApp{\varResult_1}{\varResult_2}$, with the goal to satisfy $\varEx$.
For results $\varResult_2$ that are simple (first-order) values $\varSimpleVal_2$,
the \ruleName{U-App} rule unevaluates
the indeterminate function $\varResult_1$ with the input-output example
$\ioExample{\varSimpleVal_2}{\varEx}$.

In general, the argument $\varResult_2$ may include holes that would later
appear in elimination position when $\varResult_1$ is filled and the application
resumes.
For results $\varResult_2$ that are not simple values,
it is not possible to generate sufficient constraints locally to ensure that
\expApp{\varResult_1}{\varResult_2} satisfies $\varEx$.
For instance, if $\varResult_2$ is of the form $\closure{\varEnv}{\expHole{\varHoleName}}$,
the hypothetical constraint ``$\ioExample{(\closure{\varEnv}{\expHole{\varHoleName}})}{\varEx}$''
would not provide any information about which input values the function $\varResult_1$ must
map to results that satisfy $\varEx$.
As such, there is no unevaluation rule for arbitrary indeterminate application forms.

\parahead{Indeterminate Projections}

The \ruleName{U-Prj-1} and \ruleName{U-Prj-2} rules use $\exHole$ for the
component to be left unconstrained.
For example, unevaluating $\expFst{\closure{\varEnv}{\expHole{\varHoleName}}}$
with \inlinecode{1} generates
$\highlightUnsolvedConstraint{$\unsolvedConstraint{\varHoleName}{\world{\varEnv}{\pair{\text{\inlinecode{1}}}{\exHole}}}$}$.

\parahead{Indeterminate Case Expressions}

Recall from \autoref{sec:overview-three} the goal to unevaluate an indeterminate case expression with the number
\text{\inlinecode{1}}:
$
\expMatchTwoBranches{\closure{\emptyEnv}{\expHole{\varHoleName}}}{\text{\inlinecode{Nothing}}}{\_}{\text{\inlinecode{0}}}{\text{\inlinecode{Just}}}{\varVar}{\varVar}
\ \Leftarrow\ \text{\inlinecode{1}}.
$
Intuitively, this should require
$\highlightUnsolvedConstraint{$\unsolvedConstraint{\varHoleName}{\world{\emptyEnv}{\text{\inlinecode{Just 1}}}}$}$.

To compute this constraint, the \ruleName{U-Case} rule considers each branch
$j$.
The first premise unevaluates the scrutinee $\varResult$ with
$\expApp{\varDataCon_j}{\exHole}$ to the scrutinee $\varResult$, generating
constraints $\varConstraints_1$ required for $\varResult$ to produce an
application of constructor $\varDataCon_j$.
If successful, the next step is to evaluate the corresponding branch expression
$\varExp_j$ and check that it is consistent with the goal $\varEx$.
However, the argument to the constructor will only be available after all
constraints are solved and evaluation resumes.

We introduce the \emph{inverse constructor application}
$\expUnwrap{\varDataCon_j}{\varResult}$ (\autoref{fig:syntax}) to
bridge this gap between constraint generation and constraint solving.
To proceed down the branch expression, we bind the pattern variable $\varVar_j$
to $\expUnwrap{\varDataCon_j}{\varResult}$.
Locally, this allows the third premise of \ruleName{U-Case} to check whether
the branch expression $\varExp_j$ satisfies $\varEx$.
For the example above, the result of evaluating the second branch expression,
$x$, is
$\expUnwrap{\text{\inlinecode{Just}}}{\color{CadetBlue}(\color{Black}\holeClosure{\emptyEnv}{\varHoleName}\color{CadetBlue})\color{Black}}$.
Unevaluating
$\expUnwrap{\text{\inlinecode{Just}}}{\color{CadetBlue}(\color{Black}\holeClosure{\emptyEnv}{\varHoleName}\color{CadetBlue})\color{Black}}$
with \inlinecode{1} generates the constraint
$\unsolvedConstraint{\varHoleName}{\world{\emptyEnv}{\expUnwrap{\text{\inlinecode{Just}}}{\text{\inlinecode{1}}}}}$.
Finally, the \ruleName{U-Inverse-Ctor} rule transfers the example from the
inverse constructor application to a constructor application,
producing
$\highlightUnsolvedConstraint{$\unsolvedConstraint{\varHoleName}{\world{\emptyEnv}{\expApp{\text{\inlinecode{Just}}}{\text{\inlinecode{1}}}}}$}$.

\parahead{Indeterminate Case Expressions: Guessing Scrutinees}

The interplay between \ruleName{U-Case} and \ruleName{U-Inverse-Ctor} allows
unevaluation to resolve branching decisions by generating constraints
without the obligation to synthesize expressions that satisfy them.
A downside of this ``lazy'' approach is the significant degree of
non-determinism;
indeed, many of the generated sets of constraints may be unsatisfiable.

As a more efficient approach in situations where the full expressiveness of
\ruleName{U-Case} is not needed,
the \ruleName{U-Case-Guess} rule ``eagerly'' resolves the direction of the
branch by
guessing a hole-filling $\varSolution'$ via a non-deterministic uninterpreted
function $\guessesForMatch{\varHoleEnv}{\varDatatypeEnv}{\varResult}$, and
checking whether this filling resumes the scrutinee $\varResult$ to
an application of a constructor $\varDataCon_j$, where $\varDataCon_j$ is one 
of the $n$ data constructors for the datatype $\varTypeCon$ of the scrutinee.
If so, the direction of the branch has been determined, so the last
step is to unevaluate the $j$th branch expression $\varExp_j$ with
the goal example $\varEx$, in an appropriately extended environment.

For instance, consider again the goal
$
\expMatchTwoBranches{\closure{\varEnv}{\expHole{\varHoleName}}}{\text{\inlinecode{Nothing}}}{\_}{\text{\inlinecode{0}}}{\text{\inlinecode{Just}}}{\varVar}{\varVar}
\ \Leftarrow\ \text{\inlinecode{1}}
$
but here with the environment
$
\varEnv =
\envCatThree
{\envBind{\text{\inlinecode{nothing}}}{\text{\inlinecode{Nothing}}}}
{\envBind{\text{\texttt{just0}}}{\expApp{\text{\inlinecode{Just}}}{\text{\inlinecode{0}}}}}
{\envBind{\text{\texttt{just1}}}{\expApp{\text{\inlinecode{Just}}}{\text{\inlinecode{1}}}}}.
$
The $\metaFunc{Guesses}$ function
might choose the filling
$\varSolution'=\highlightSolvedConstraint{$\holeFilling{\varHoleName}{\text{\texttt{just1}}}$}$,
which resumes the scrutinee $\closure{\varEnv}{\expHole{\varHoleName}}$ to
$\expApp{\text{\inlinecode{Just}}}{\text{\inlinecode{1}}}$.
In the environment extended with $\envBind{\varVar}{\text{\inlinecode{1}}}$,
the corresponding branch expression $\varVar$ evaluates to the result
\inlinecode{1}.
Unevaluating this result with the example \inlinecode{1} succeeds via \ruleName{U-Ctor}
and \ruleName{U-Unit} without generating additional constraints.
(If guessing fills $\expHole{\varHoleName}$ with \text{\inlinecode{nothing}} or
\text{\texttt{just0}},
the result, \inlinecode{0}, of the branch expression would fail to unevaluate to
\inlinecode{1}.)

Whereas the \ruleName{U-Hole} rule is the source of example constraints
$\varUnsolvedConstraints$ produced by unevaluation,
the \ruleName{U-Case-Guess} rule is the source of hole-filling constraints
$\varSolution$.
We describe our concrete implementation of $\metaFunc{Guesses}$ in
\autoref{sec:implementation}.

% !TEX root = ./main.tex
\section{Synthesis Pipeline}
\label{sec:synthesis}

Live bidirectional evaluation addresses the challenge of checking example
satisfaction for programs with holes.
In this section, we define a synthesis pipeline that uses live bidirectional evaluation to
(1)~derive example constraints from \inlinecode{assert}s and
(2)~solve the resulting constraints.
\newcommand{\phaseCaption}[3]
  {\ensuremath{\substack{
    %% \textrm{(#1) #2}
    \textrm{#2 (#1)}
    }}
  }
\newcommand{\sepPhase}
  {\hsepRule}
  %% {\hsepRule\hsepRule}
  %% {\hspace{0.10in}}
%
$$
\overbrace{
\makebox[1.85in]{
%
%% \reducesToUK{\emptyEnv}{\varExp}{\varResult}{\varAssertions}
\programReducesTo{\varProgram}{\varResult}{\varAssertions}
\sepPhase
\simplifyEquals{\varAssertions}{\varConstraints}
}}^{\phaseCaption{\autoref{sec:constraint-collection}}{Constraint Collection}{from Live Evaluation}}
\sepPhase
%
%% \varConstraints=\pairConstraints{\varSolution_0}{\varUnsolvedConstraints}
%% %
%% \sepPhase
%% %
\overbrace{
\makebox[0.95in]{
\solveEquals{\varConstraints}{\varSolution}
}}^{\phaseCaption{\autoref{sec:constraint-solving}}{Constraint Solving}{to Synthesize Hole-Fillings}}
$$

\vspace{-0.25in} %% TODO HACK

%%%%%%%%%%%%%%%%%%%%%%%%%%%%%%%%%%%%%%%%%%%%%%%%%%%%%%%%%%%%%%%%%%%%%%%%%%%%%%%%

\parahead{Overview Program: Plus}

Before describing each of these components formally, we summarize how they will
fit together to synthesize the \inlinecode{plus} function in \autoref{sec:overview-one}:

\begin{center}
\begin{tabular}{c}
\begin{blockcode}
let plus = `\expHole{0}` in assert ([plus 0 1, plus 2 0, plus 1 2] == [1, 2, 3])
\end{blockcode}
\end{tabular}
\end{center}

\noindent
First, when evaluating the program, the left-hand side of the \inlinecode{assert}
produces three nested, indeterminate function calls:
\myVerb+[{\expAppIndet{\color{CadetBlue}(\expAppIndet{\closure{\emptyEnv}{\expHole{0}}}{\color{Emerald}\texttt{0}}\color{CadetBlue})}{\color{Emerald}\!\!\texttt{1}}}, {\expAppIndet{\color{CadetBlue}(\expAppIndet{\closure{\emptyEnv}{\expHole{0}}}{\color{Emerald}\texttt{2}}\color{CadetBlue})}{\color{Emerald}\!\!\texttt{0}}}, {\expAppIndet{\color{CadetBlue}(\expAppIndet{\closure{\emptyEnv}{\expHole{0}}}{\color{Emerald}\texttt{1}}\color{CadetBlue})}{\color{Emerald}\!\!\texttt{2}}}]+.
Structurally comparing this list of indeterminate results with the list of values
\inlinecode{[1, 2, 3]} yields three \emph{assertion} predicates $\varAssertions$ as a
side-effect
(via rules \ruleName{Eval-and-Assert}, \ruleName{RC-Ctor}, and \ruleName{RC-Assert-1},
discussed below):
$$
\varAssertions =
\envCatThree
   {\assertion{(\expAppIndet{\color{CadetBlue}(\expAppIndet{\closure{\emptyEnv}{\expHole{0}}}{\color{Emerald}\texttt{0}}\color{CadetBlue})}{\color{Emerald}\texttt{1}})}{\color{Emerald}\texttt{1}}}
   {\assertion{(\expAppIndet{\color{CadetBlue}(\expAppIndet{\closure{\emptyEnv}{\expHole{0}}}{\color{Emerald}\texttt{2}}\color{CadetBlue})}{\color{Emerald}\texttt{0}})}{\color{Emerald}\texttt{2}}}
   {\assertion{(\expAppIndet{\color{CadetBlue}(\expAppIndet{\closure{\emptyEnv}{\expHole{0}}}{\color{Emerald}\texttt{1}}\color{CadetBlue})}{\color{Emerald}\texttt{2}})}{\color{Emerald}\texttt{3}}}
$$

Second, we use live bidirectional example checking (\ruleName{Live-Check}) to
convert---\ie{}~\metaFunc{Simplify}---the assertions $\varAssertions$ into example
constraints $\varUnsolvedConstraints$ (via \ruleName{U-App} and \ruleName{U-Hole}):
$$
\varUnsolvedConstraints =
\highlightUnsolvedConstraint{$\unsolvedConstraint{0}{(
  \envCatThree
    {\world{\emptyEnv}{\ioExample{\text{\inlinecode{0}}}{\ioExample{\text{\inlinecode{1}}}{\text{\inlinecode{1}}}}}}
    {\world{\emptyEnv}{\ioExample{\text{\inlinecode{2}}}{\ioExample{\text{\inlinecode{0}}}{\text{\inlinecode{2}}}}}}
    {\world{\emptyEnv}{\ioExample{\text{\inlinecode{1}}}{\ioExample{\text{\inlinecode{2}}}{\text{\inlinecode{3}}}}}}
)}$}
$$

\noindent
The simplified constraints
$\varConstraints=\pairConstraints{\emptyEnv}{\varUnsolvedConstraints}$
contain an empty hole-filling because \ruleName{U-Case-Guess} is not invoked to
resolve any indeterminate case expressions.

Finally, the holes in $\varUnsolvedConstraints$ are solved one at a time;
here there is only $\expHole{0}$.
Solving one hole may generate new subgoals (\ruleName{Refine} and \ruleName{Branch})
or new constraints on existing goals (\ruleName{Guess-and-Check}).
The search path sketched in \autoref{sec:overview-one} produces the
solution $\varSolution$ below that solves the constraints
$\varConstraints=\pairConstraints{\emptyEnv}{\varUnsolvedConstraints}$.
Each step is annotated with the rules used to conclude the subderivation.

\def\solvedboxit#1{%
  \smash{\fboxsep=0pt\llap{\rlap{\fcolorbox{constraintLight}{constraintLight}{\strut\makebox[#1]{}}}~}}\ignorespaces
}

\newcommand{\explainFillStepTable}[3]
  {\solvedboxit{2.275in}{#1}&$\mapsto$&{#2}&\textrm{#3}}

\vspace{0.10in}

\begin{tabular}{ccll}
\explainFillStepTable
  {0}
  {$\expFixFun{\varVarF_1}{}{\ttm}{\expFixFun{\varVarF_2}{}{\ttn}{\expHole{{1}}}}$}
  {
   %% Twice:
   \ruleName{Solve-One},
   \ruleName{Refine},
   \ruleName{Refine-Fix}
   (twice)
  }
\\
\explainFillStepTable
  {1}
  %% copied from overview
  {$\expMatchTwoBranchesOverview{\ttm}{\texttt{Z}}{BLAH}{\expHole{2}}{\texttt{S}}{\texttt{m\texttt'}}{\expHole{3}}$}
  {
   \ruleName{Solve-One},
   \ruleName{Branch},
   \ruleName{Branch-Case}
  }
\\
\explainFillStepTable
  {3}
  {$\expApp{\text{\inlinecode{S}}}{\expHole{4}}$}
  {
   \ruleName{Solve-One},
   \ruleName{Refine},
   \ruleName{Refine-Ctor}
  }
\\
\explainFillStepTable
  {4}
  {$\expAppTwo{\texttt{plus}}{\ttm\textrm'}{\ttn}$}
  {
   \ruleName{Solve-One},
   \ruleName{Guess-and-Check},
   \ruleName{Live-Check}
  }
\\
\explainFillStepTable
  {2}
  {$\ttn$}
  {
   \ruleName{Solve-One},
   \ruleName{Guess-and-Check},
   \ruleName{Live-Check}
  }
\end{tabular}

%%%%%%%%%%%%%%%%%%%%%%%%%%%%%%%%%%%%%%%%%%%%%%%%%%%%%%%%%%%%%%%%%%%%%%%%%%%%%%%%

\subsection{Constraint Collection}
\label{sec:constraint-collection}

%% \begin{figure*}[t]
\begin{figure}[b]

%% \judgementHeadNameOnlyTwo
%%   {Expression and Assertion Syntax}
%%   {(extends \autoref{fig:syntax})}

%% \judgementHeadNameOnly
%%   {Program and Assertion Syntax}
%% 
%% \vsepRule

%% \rkc{for now, limiting assert to (single) top-level.}
%% \rkc{so ignore the changes to typing and evaluation below.}
%% \rkc{if/when allowing arbitrary assertions in expressions, then make red changes to Sat and Live Bi Sat.}

%% if this figure is at the bottom,
%% then put this header above p and A syntax definitions
\judgementHeadNameOnly
  {Program Evaluation}
  {$\JudgementBox{\programReducesTo{\varProgram}{\varResult}{\varAssertions}}$}

\vspace{0.06in} %% HACK

\figSyntaxBegin
%
%% \figSyntaxRowLabel{\maybeUnderlineSyntaxFig{E}xpressions}{\varExp}
%%   \cdots
%%   \figSyntaxSpaceItem
%%   %% \varPf
%%   %% \figSyntaxSpaceItem
%%   \expPbeConstraints{\varExp_1}{\varExp_2}
%% %
%% \figSyntaxSpaceNextCategory
%
\figSyntaxRowLabel{\maybeUnderlineSyntaxFig{P}rograms}{\varProgram}
  \expProgram{\varExp}{\varExp_1}{\varExp_2}
\figSyntaxSpaceNextCategory
%
%% \figSyntaxRowLabel{Resumption \maybeUnderlineSyntaxFig{A}ssertions}{\varAssertions}
\figSyntaxRowLabel{\maybeUnderlineSyntaxFig{A}ssertions}{\varAssertions}
  \multiPremise
    {\highlightAssertion{\assertion{\varResult_i}{\varSimpleVal_i}}}
    {\sequenceSyntax}
  %% \emptyEnv
  %% \figSyntaxSpaceItem
  %% \envCat{\varAssertions}
  %%        {\highlightAssertion{\assertion{\varResult}{\varSimpleVal}}}
%
\figSyntaxEnd

\vsepRule

$
\inferrule*[lab=\ruleNameFig{Eval-and-Assert}]
  {
   \reducesTo{\emptyEnv}{\varExp}{\varResult}
   \sepPremise
   \multiPremise{
   \reducesTo{\envBind{\texttt{main}}{\varResult}}
             {\varExp_i}
             {\varResult_i}
   }{\pairIndex{i}}
   \sepPremise
   \highlightAssertion{%
   \resultConsistent{\varResult_1}{\varResult_2}{\varAssertions}%
   }
  }
  {\programReducesTo{\expProgram{\varExp}{\varExp_1}{\varExp_2}}
                    {\varResult}{\varAssertions}
  }
$

%\vsepRule

\begin{comment}

\judgementHeadNameOnlyTwo
  %% {Expression \maybeUnderline{T}yping and \maybeUnderline{E}valuation}
  %% {(extended)}
  {\maybeUnderline{T}yping and \maybeUnderline{E}valuation}
  {(extends \autoref{fig:typing} and \autoref{fig:eval})}
  {
   $\JudgementBox{\typeCheck{\varTypeEnv}{\varExp}{\varType}}$
   \hspace{0.00in}
   $\JudgementBox{\reducesToUK{\varEnv}{\varExp}{\varResult}{\varAssertions}}$
  }

\vsepRule

$
\inferrule*[lab=\ruleNameFig{T-Assert}]
  {
   \typeCheck{\varTypeEnv}{\varExp_1}{\varType}
   \sepPremise
   \typeCheck{\varTypeEnv}{\varExp_2}{\varType}
  }
  {\typeCheck
    {\varTypeEnv}
    {\expPbeConstraints{\varExp_1}{\varExp_2}}
    {\tUnit}}
$
%
\hsepRule
%
$
\inferrule*[lab=\ruleNameFig{E-Assert}]
  {\reducesToUK{\varEnv}{\varExp_1}{\varResult_1}{\varAssertions_1}
   \sepPremise
   \reducesToUK{\varEnv}{\varExp_2}{\varResult_2}{\varAssertions_2}
   \sepPremise
   \highlightAssertion{%
   \resultConsistent{\varResult_1}{\varResult_2}{\varAssertions_3}%
   }
  }
  {\reducesToUK
    {\varEnv}
    {\expPbeConstraints{\varExp_1}{\varExp_2}}
    {\expUnit}
    {\setUnionThree{\varAssertions_1}{\varAssertions_2}{\varAssertions_3}}}
$

\vsepRule

{\small
Changes to evaluation rules in \autoref{fig:eval}: %% are straightforward:
constraints are propagated from premises to conclusions. \\
See \autoref{fig:eval-n} in \autoref{sec:appendix-dynamics} for more details.
}

\end{comment}

\vsepRule

\judgementHead
  %% {E\maybeUnderline{x}ample \maybeUnderline{C}ollection}
  {\maybeUnderline{R}esult \maybeUnderline{C}onsistency}
  {\resultConsistent{\varResult}{\varResult'}{\varAssertions}}

%% \vsepRule

%% $
%% \inferrule*[lab=\ruleNameFig{XC-Example}]
%%   {\coerceUndet{\varResult}{\varEx}
%%   }
%%   {\resultConsistent
%%     {\closure{\varEnv}{\expHole{i}}}
%%     %% {\varEx}
%%     {\varResult}
%%     {\holeConstraint{i}{\varEnv}{\varEx}}}
%% $
%% %
%% \hsepRule
%% %
%% $
%% \inferrule*[lab=\ruleNameFig{XC-Example-Symm}]
%%   {\coerceUndet{\varResult}{\varEx}
%%   }
%%   {\resultConsistent
%%     %% {\varEx}
%%     {\varResult}
%%     {\closure{\varEnv}{\expHole{i}}}
%%     {\holeConstraint{i}{\varEnv}{\varEx}}}
%% $
%% 
%% \vsepRule

%% $
%% \inferrule*[lab=\ruleNameFig{...}]
%%   {
%%   }
%%   {\resultConsistent
%%     {\closure{\varEnv}{\expHole{i}}}
%%     {\closure{\varEnv}{\expHole{i}}}
%%     {\emptySet}}
%% $
%% %
%% \hsepRule
%% %
%% $
%% \inferrule*[lab=\ruleNameFig{...}]
%%   {
%%   }
%%   {\resultConsistent
%%     {\closure{\varEnv}{\expFun{\varVar}{\varExp}}}
%%     {\closure{\varEnv}{\expFun{\varVar}{\varExp}}}
%%     {\emptySet}}
%% $

\vsepRule %% NoNeed

%% don't need XC-Unit, b/c XC-Refl

$
%% \inferrule*[lab=\ruleNameFig{XC-Refl}]
\inferrule*[lab=\ruleNameFig{RC-Refl}]
  {
  }
  {\resultConsistent
    {\varResult}
    {\varResult}
    {\emptySet}}
$
\hsepRule
$
%% \inferrule*[lab=\ruleNameFig{XC-Tuple}]
%% \inferrule*[lab=\ruleNameFig{XC-Pair}]
\inferrule*[lab=\ruleNameFig{RC-Pair}]
  {
   %% \resultConsistent{\varResult_i}{\varResult'_i}{\varAssertions_i}
   \resultConsistent{\varResult_1}{\varResult'_1}{\varAssertions_1}
   \sepPremise
   \resultConsistent{\varResult_2}{\varResult'_2}{\varAssertions_2}
  }
  {\resultConsistent
    {\pair{\varResult_1}{\varResult_2}}
    {\pair{\varResult'_1}{\varResult'_2}}
    {\setUnion{\varAssertions_1}{\varAssertions_2}}}
    %% {\triple{\varResult_1}{\ldots}{\varResult_n}}
    %% {\triple{\varResult'_1}{\ldots}{\varResult'_n}}
    %% {\setUnion{\varConstraints_1}{\setUnion{\ldots}{\varConstraints_n}}}}
$
\hsepRule
$
%% \inferrule*[lab=\ruleNameFig{XC-Ctor}]
\inferrule*[lab=\ruleNameFig{RC-Ctor}]
  {
   \resultConsistent{\varResult}{\varResult'}{\varAssertions}
  }
  {\resultConsistent
    {\expApp{\varDataCon}{\varResult}}
    {\expApp{\varDataCon}{\varResult'}}
    {\varAssertions}}
$
\hsepRule
$
%% \inferrule*[lab=\ruleNameFig{XC-Backprop-1}]
%% \inferrule*[lab=\ruleNameFig{XC-Collect-1}]
%% \inferrule*[lab=\ruleNameFig{RC-Constraint-1}]
%% \inferrule*[lab=\ruleNameFig{RC-Val-2}]
\inferrule*[lab=\ruleNameFig{RC-Assert-1}]
  %% {\xBackprop{\varResult}{\varEx}{\varConstraints}}
  %% {\resultConsistent{\varResult}{\varEx}{\varConstraints}}
  {\coerceUndet{\varResult_2}{\varSimpleVal_2}
   \\\\ %% \sepPremise
   %% \sepPremise
   %% \xBackprop{\varResult_1}{\varEx}{\varConstraints}
   \varAssertions =
     \highlightAssertion{\assertion{\varResult_1}{\varSimpleVal_2}}
  }
  {\resultConsistent{\highlightIndet{$\varResult_1$}}{\varResult_2}{\varAssertions}}
$
\hsepRule
$
%% \inferrule*[lab=\ruleNameFig{XC-Backprop-2}]
%% \inferrule*[lab=\ruleNameFig{XC-Collect-2}]
%% \inferrule*[lab=\ruleNameFig{RC-Constraint-2}]
%% \inferrule*[lab=\ruleNameFig{RC-Val-1}]
\inferrule*[lab=\ruleNameFig{RC-Assert-2}]
  %% {\xBackprop{\varResult}{\varEx}{\varConstraints}}
  %% {\resultConsistent{\varEx}{\varResult}{\varConstraints}}
  {\coerceUndet{\varResult_1}{\varSimpleVal_1}
   \\\\ %% \sepPremise
   %% \sepPremise
   %% \xBackprop{\varResult_2}{\varEx}{\varConstraints}
   \varAssertions =
     \highlightAssertion{\assertion{\varResult_2}{\varSimpleVal_1}}
  }
  {\resultConsistent{\varResult_1}{\highlightIndet{$\varResult_2$}}{\varAssertions}}
$

%% \vsepRule
%% 
%% $
%% \inferrule*[lab=\ruleNameFig{XC-App}]
%%   {
%%    \resultConsistent{\varResult_1}{\varResult'_1}{\varConstraints_1}
%%    \sepPremise
%%    \resultConsistent{\varResult_2}{\varResult'_2}{\varConstraints_2}
%%   }
%%   {\resultConsistent
%%     {\expApp{\varResult_1}{\varResult_2}}
%%     {\expApp{\varResult'_1}{\varResult'_2}}
%%     {\setUnion{\varConstraints_1}{\varConstraints_2}}}
%% $
%% %
%% \hsepRule
%% %
%% $
%% \inferrule*[lab=\ruleNameFig{XC-Get}]
%%   {
%%    \resultConsistent{\varResult}{\varResult'}{\varConstraints}
%%   }
%%   {\resultConsistent
%%     {\expProj{i}{\varResult}}
%%     {\expProj{i}{\varResult'}}
%%     {\varConstraints}}
%% $
%% 
%% \vsepRule
%% 
%% $
%% \inferrule*[lab=\ruleNameFig{XC-Match}]
%%   {
%%    \resultConsistent{\varResult}{\varResult'}{\varConstraints}
%%   }
%%   {\resultConsistent
%%     {\closure{\varEnv}{\expMatch{\varResult}{\varDataCon_i}{\varVar_i}{\varExp_i}}}
%%     {\closure{\varEnv}{\expMatch{\varResult'}{\varDataCon_i}{\varVar_i}{\varExp_i}}}
%%     {\varConstraints}}
%% $

%% "step 2"

\vsepRule

\judgementHeadNameOnly
  %% {Result Constraint Satisfaction and Simplification}
  {Assertion Satisfaction and Simplification}
  {
   $\JudgementBox{\assertionSat{\varSolution}{\varAssertions}}$
   \hspace{0.00in}
   $\JudgementBox{\simplifyEquals{\varAssertions}{\varConstraints}}$
  }

%% \judgementHeadNameOnly
%%   {Assertion Satisfaction}
%%   {$\JudgementBox{\assertionSat{\varSolution}{\varAssertions}}$}

\vsepRule

$
\inferrule* %% [lab=\ruleNameFig{Blah}]
  {
   \multiPremise{
   \resumesTo{\varSolution}{\varResult_i}{\varResult'_i}
   %% \resumesToUK{\varSolution}{\varResult_i}{\varResult'_i}{\varAssertions_i}
   %% \sepPremise
   %% {\assertionSat{\varSolution}{\varAssertions_i}}
   \sepPremise
   \coerceUndet{\varResult'_i}{\varSimpleVal_i}
   }{\sequenceSyntax}
  }
  {\assertionSat{\varSolution}
        {\multiPremise{\assertion{\varResult_i}{\varSimpleVal_i}}{\sequenceSyntax}}}
        %% {(\envCatThree{\assertion{\varResult_1}{\varSimpleVal_1}}
        %%               {\ldots}
        %%               {\assertion{\varResult_n}{\varSimpleVal_n}})}}
$
\hsepRule
$
\inferrule* %% [lab=\ruleNameFig{Blah}]
  {
   \multiPremise{
   \isFinal{\varResult_i}
   %% }{\sequenceSyntax}
   \sepPremise
   %% \multiPremise{
   \uneval
     {\emptyEnv}
     {\varResult_i}{\liftSimpleEx{\varSimpleVal_i}}
     {\varConstraints_i}
     %% {\varSolution_i}
   }{\sequenceSyntax}
  }
  {\simplifyEquals
    {\multiPremise{\assertion{\varResult_i}{\varSimpleVal_i}}{\sequenceSyntax}}
    %% {\envCatThree{\assertion{\varResult_1}{\varSimpleVal_1}}
    %%              {\ldots}
    %%              {\assertion{\varResult_n}{\varSimpleVal_n}}}
    {\mergeConstraintsThree{\varConstraints_1}{\cdots}{\varConstraints_n}}
    %% {\mergeSolutionsThree{\varSolution_1}{\cdots}{\varSolution_n}}
  }
$

\vsepBeforeCaption

\caption{Constraint Collection.}
\label{fig:ex-collection}
%% \end{figure*}

\vsepAfterCaptionBottomFig

\end{figure}

\autoref{fig:ex-collection} defines a \emph{program} to be an expression
followed by an $\expPbeConstraints{\varExp_1}{\varExp_2}$ statement.
Changes to allow \inlinecode{assert}s in arbitrary expressions are discussed in
\autoref{sec:related}.

\parahead{Assertions via Result Consistency}

A typical semantics for \inlinecode{assert} would require the expression results
$\varResult_1$ and $\varResult_2$ to be equal, otherwise raising an exception.
Instead, rather than equality, the \ruleName{Eval-and-Assert} rule in
\autoref{fig:ex-collection} checks \emph{result consistency},
$\resultConsistent{\varResult_1}{\varResult_2}{\varAssertions}$, a notion of
equality modulo assumptions $\varAssertions$ about indeterminate results.
Determinate results are consistent if structurally equal, as checked by the
\ruleName{RC-Refl}, \ruleName{RC-Pair}, and \ruleName{RC-Ctor} rules.
Indeterminate results $\varResult$ are consistent with simple
values $\varSimpleVal$---the \ruleName{RC-Assert-1} and \ruleName{RC-Assert-2}
rules generate \emph{assertion} predicates
$\assertion{\varResult}{\varSimpleVal}$ in such cases.
\autoref{fig:ex-collection} also defines assertion satisfaction
$\assertionSat{\varSolution}{\varAssertions}$:
for each assertion \mbox{$\assertion{\varResult_i}{\varSimpleVal_i}$} in
$\varAssertions$, the indeterminate result $\varResult_i$ should resume under
filling $\varSolution$ and produce the value $\varSimpleVal_i$.

\parahead{Assertion Simplification}

For each assertion $\assertion{\varResult_i}{\varSimpleVal_i}$,
the $\metaFunc{Simplify}$ procedure in \autoref{fig:ex-collection} converts the
simple value into an example $\liftSimpleEx{\varSimpleVal_i}$ and unevaluates it
to $\varResult_i$ to generate example constraints.

\begin{theorem*}[Soundness of Assertion Simplification]

\breakAndIndent
If $\simplifyEquals{\varAssertions}{\varConstraints}$
and $\constraintSat{\varSolution}{\varConstraints}$,
then $\assertionSat{\varSolution}{\varAssertions}$.

\end{theorem*}

%%%%%%%%%%%%%%%%%%%%%%%%%%%%%%%%%%%%%%%%%%%%%%%%%%%%%%%%%%%%%%%%%%%%%%%%%%%%%%%%

\subsection{Constraint Solving}
\label{sec:constraint-solving}

\begin{figure*}[b]

\judgementHeadNameOnly
  {Constraint Solving}
  %% $\JudgementBox
  %%  {\solveEqualsFull
  %%    {\varHoleEnv}{\varDatatypeEnv}{\varConstraints}{\varSolution}}$
  %% \hspace{0.00in}
  $\JudgementBox
   {\iterSolveEqualsFull
     {\varHoleEnv}
     {\varDatatypeEnv}
     %% {\pairConstraints{\varSolution}{\varUnsolvedConstraints}}
     {(\varConstraints)}
     {\varSolution}
     {\varHoleEnv'}}$

\vsepRuleNoNeed

%% $
%% \inferrule* %% [lab=\ruleNameFig{Iter-Solve}]
%%   {
%%    %% \splitEquals{\varConstraints}
%%    %%             {\varSolution_0}
%%    %%             {\varUnsolvedConstraints}
%%    %% \sepPremise
%%    \iterSolveEquals
%%      {\pairConstraints{\varSolution_0}{\varUnsolvedConstraints}}
%%      {\varSolution}
%%      {\varHoleEnv'}
%%   }
%%   {\solveEquals
%%     %% {\varConstraints}
%%     {\pairConstraints{\varSolution_0}{\varUnsolvedConstraints}}
%%     {\varSolution}}
%% $
%% \hsepRule
$
\inferrule*[lab=\ruleNameFig{Solve-Done}]
  {
  }
  {\iterSolveEqualsFull
     {\varHoleEnv}{\varDatatypeEnv}
     {\pairConstraints{\varSolution}{\emptySet}}
     {\varSolution}{\varHoleEnv}
  }
$
%% \hsepRule
\vsepRule

$
\inferrule*[lab=\ruleNameFig{Solve-One}]
  {
   \varHoleName \in \dom{\varUnsolvedConstraints}
   \sepPremise
   %% \varHoleEnv(\expHole{\varHoleName}) = \tHole{\varTypeEnv}{\varType}
   \varHoleEnv({\varHoleName}) = \tHole{\varTypeEnv}{\varType}
   \sepPremise
   %% \varUnsolvedConstraints(\expHole{\varHoleName}) = \varWorlds
   \varUnsolvedConstraints({\varHoleName}) = \varWorlds
   %% \\\\
   \sepPremise
   \fillHole
     {\varHoleName}
     {\varSolution}
     {\varTypeEnv}
     {\varWorlds}
     {\varType}
     {\varConstraints}
     {\varHoleEnv'}
   \\\\
   \simplifyConstraintsEquals
     {\setUnion{\varHoleEnv}{\varHoleEnv'}}
     {\varDatatypeEnv}
     {(\mergeConstraints
       %% {\pairConstraints{\varSolution}{\varUnsolvedConstraints}}
       %% {\pairConstraints{\varSolution}{\setMinus{\varUnsolvedConstraints}{\expHole{\varHoleName}}}}
       {\pairConstraints{\varSolution}{\setMinus{\varUnsolvedConstraints}{\varHoleName}}}
       {\varConstraints})}
     {\varConstraints'}
   %% \\\\
   \sepPremise
   \iterSolveEqualsFull
     {\setUnion{\varHoleEnv}{\varHoleEnv'}}
     {\varDatatypeEnv}
     {(\varConstraints')}
     {\varSolution'}
     {\varHoleEnv''}
  }
  {\iterSolveEqualsFull
     {\varHoleEnv}
     {\varDatatypeEnv}
     {\pairConstraints
       {\varSolution}
       %% {\twoThings{\unsolvedConstraint{\varHoleName}{\varWorlds}}
       %%            {\varUnsolvedConstraints}}}
       {\varUnsolvedConstraints}}
     {\varSolution'}
     {\varHoleEnv''}
  }
$

\vsepRule

\input{fig-synthesis-fill}

\vsepRule

%% TODO
%% \input{fig-synthesis-irefine}

\vsepBeforeCaption

%% \caption{Constraint Solving.}
\caption{Constraint Solving with Guessing, Refinement, and Branching.
%
% We abuse notation to at once define
% \ruleNameCaption{Refine} and \ruleNameCaption{Branch}.
%% because they are very similar.
%
\textit{We at once define \ruleNameCaption{Refine} and \ruleNameCaption{Branch}
by differentiating the two by color; the signature of the branching judgement
extends that of the refinement judgement with an additional input
$ \varSolution $ and an additional output $ \varConstraints $.}
% The signature of the branching judgement extends that of the refinement
% judgement with an additional input $\varSolution$ and an additional output
% $\varConstraints$.
% %
% The combined rule ``uses'' these syntactic elements---rendered in purple---to
% define \ruleNameCaption{Branch} and ``ignores'' them to define
% \ruleNameCaption{Refine}.
%
}
\label{fig:synthesis-solve}
\label{fig:synthesis-fill}
%% \label{fig:synthesis-refine}

\vsepAfterCaptionBottomFig

\end{figure*}

\begin{figure*}[b]
\input{fig-synthesis-irefine}
\caption{Guessing, Refinement, and Branching.}
\label{fig:synthesis-refine}

\vsepAfterCaptionBottomFig

\end{figure*}

The constraints $\varConstraints$, of the form
$\pairConstraints{\varSolution_0}{\varUnsolvedConstraints}$, include
filled holes $\varSolution_0$ from constraint simplification
(\cf{}~\ruleName{U-Case-Guess}) and
a set $\varUnsolvedConstraints$ of unfilled holes constrained by examples.
\autoref{fig:synthesis-solve} and \autoref{fig:synthesis-refine} define an
algorithm to synthesize expressions for unfilled holes, generalizing \myth{}
to use live bidirectional evaluation and to fill interdependent holes.

The $\iterSolve{\pairConstraints{\varSolution}{\varUnsolvedConstraints}}$
procedure in \autoref{fig:synthesis-solve} is the entry point for filling the
holes in $\varUnsolvedConstraints$.
The \ruleName{Solve-Done} rule handles the terminal case, when no unfilled holes
remain.
Otherwise, the \ruleName{Solve-One} rule chooses an unfilled hole
$\expHole{\varHoleName}$ and forms the synthesis goal
$\problemNameTypeWorlds{\varHoleName}{\varTypeEnv}{\varType}{\varWorlds}$ from
the hole type and example contexts $\varHoleEnv$ and $\varUnsolvedConstraints$.
The \emph{hole synthesis} procedure---discussed next---completes the task,
which, in \smyth{}, may assume constraints $\varConstraints$ over other holes.
Any such constraints $\varConstraints$ are combined with the existing ones using
the semantic $\metaFunc{Merge}$ operation (\cf{}~\autoref{sec:live-uneval}), and
the resulting constraints $\varConstraints'$ are recursively solved.

\parahead{Hole Synthesis}

For each unfilled hole, the {hole synthesis} procedure
${\fillHole{\varHoleName}{\varSolution}{\varTypeEnv}{\varWorlds}{\varType}{\varConstraints}{\varHoleEnv'}}$
augments guessing-and-checking (\ruleName{Guess-and-Check}) with
example-directed refinement (\ruleName{Refine}) and branching
(\ruleName{Branch});
these rules are discussed in turn below.

The structure of hole synthesis in \coreSnsMyth{} closely follows
\myth{}~\citep{Myth}, which presents a novel approach to synthesis by analogy to
proof search for \emph{bidirectional type checking}~\citep{PierceTurner}.
We refer the reader to their paper for a comprehensive account of their ideas;
we limit our discussion to the most important technical differences.

Besides modifications to notation and organization, the primary differences of
our formulation are that hole synthesis:
(i)~refers to the filling $\varSolution$ from previous synthesis tasks completed
by $\metaFunc{Solve}$;
(ii)~may generate example constraints over other holes in the program;
(iii)~may fill other holes in the program besides the goal
$\expHole{\varHoleName}$; and
(iv)~includes a rule, \ruleName{Defer}, to ``fill'' the hole with
$\expHole{\varHoleName}$ when all examples are top---these constraints are not
imposed directly from program assertions, but are created internally by
unevaluation.

\parahead{Guessing-and-Checking}

The \ruleName{Guess-and-Check} rule uses the procedure
$\guess{\varTypeEnv}{\varType}{\varExp}$ in \autoref{fig:synthesis-refine}
to \emph{guess} a well-typed expression without holes.
Guessing amounts to straightforward inversion of expression type checking rules;
\citetAppendix{\refAppendixGuessing} provides the full definition.

The candidate expression $\varExp$ is checked for consistency against the
examples $\varWorlds$ using live bidirectional example checking
(\cf{}~\autoref{sec:live-uneval} and \autoref{fig:ex-backprop}).
Whereas example checking in \myth{} produces a Boolean outcome, example checking
in \coreSnsMyth{} may assume constraints $\varConstraints$ over other holes.
The constraints that arise from (live bidirectional) example checking are the
source of the aforementioned differences (i), (ii), and (iii) compared to the
\myth{} hole synthesis procedure.

\parahead{Refinement}

The \ruleName{Refine} rule refers to the \emph{refinement} procedure
$\refine{\varTypeEnv}{\varWorlds}{\varType}{\varExp}{\varGoals}$
in \autoref{fig:synthesis-refine} to quickly synthesize a partial solution
$\varExp$ which refers to freshly created holes $\expHole{\varHoleName_1}$
through $\expHole{\varHoleName_n}$ described by subgoals $\varGoals$.
Using these results, \ruleName{Refine} generates output constraints comprising
the partial solution
$\highlightSolvedConstraint{\holeFilling{\varHoleName}{\varExp}}$
and the new unfilled holes
$\highlightUnsolvedConstraint{$\unsolvedConstraint{\varHoleName_1}{\varWorlds_1}$}$
through
$\highlightUnsolvedConstraint{$\unsolvedConstraint{\varHoleName_n}{\varWorlds_n}$}$.
For the purposes of metatheory, the typings for fresh holes are recorded in
the hole type context $\varHoleEnv'$.

Each refinement rule first uses $\filterWorlds{\varWorlds}$ to remove
top examples and then inspects the structure of the remaining examples.
For unit-type goals, \ruleName{Refine-Unit} simply synthesizes the unit
expression~$\expUnit{}$.
For pair-type goals, \ruleName{Refine-Pair} synthesizes the partial solution
$\pair{\expHole{\varHoleName_1}}{\expHole{\varHoleName_2}}$, creating two
subgoals from the type and examples of each component.
The \ruleName{Refine-Ctor} rule for datatype goals $\varTypeCon$ works
similarly when all of the examples share the same constructor
$\varDataCon$.

The refinement rules described so far are essentially the same as proposed by
\citet{Myth}.
But rather than explicitly naming subgoals $\varGoals$ and ``sending'' them to a
top-level \metaFunc{Solve} procedure, the refinement rules in \myth{}
recursively call hole synthesis to solve subgoals immediately.
In \coreSnsMyth{}, we separate the creation of subgoals from solving in order to
facilitate the ``global'' reasoning necessary to synthesize recursive function
literals without trace-complete examples, discussed next.

For function-type goals, the \ruleName{Refine-Fix} rule synthesizes the function
sketch
$\expFixFun{\varVarF}{}{\varVar}{\expHole{\varHoleName_1}}$.
The environments inside example constraints $\varWorlds_1$ for the function body
$\expHole{\varHoleName_1}$ bind $\varVarF$ to this function sketch (closed by
the appropriate environments $\varEnv_i$).
As a result, any recursive calls to $\varVarF$ will evaluate to closures of
$\expHole{\varHoleName_1}$ (to be constrained by live bidirectional example
checking), thus avoiding the need for trace-complete examples.\footnote{\hspace{-0.011in} %% SPACING HACK
For the constraint environments in
$\varConstraintsOverview_{4.1}$ and $\varConstraintsOverview_{4.2}$
in~\autoref{sec:overview-one},
the refinement rule for recursive functions in \myth{} would
bind \texttt{plus} to trace-complete examples
{\ioExampleThreeDots{\text{\inlinecode{0
1}}}{\text{\inlinecode{1}}}{\text{\inlinecode{2
0}}}{\text{\inlinecode{2}}}{\text{\inlinecode{1 2}}}{\text{\inlinecode{3}}}}.
In addition to usability obstacles of trace-completeness, their theory %% technical formulation
is complicated by a non-standard \emph{value compatibility}
notion~\citep[\S\! 3.3]{Myth} to approximate value equality because
input-output examples serve as a ``lookup table'' to resolve recursive calls.
}

\parahead{Branching}

Lastly, the \ruleName{Branch} rule refers to the procedure
$\branch{\varTypeEnv}{\varWorlds}{\varType}{\varExp}{\varGoals}{\varConstraints}{\varSolution}$
in \autoref{fig:synthesis-refine} to guess an expression on which to
\emph{branch}.
(As mentioned, the signature of the branching procedure extends refinement with
the additional input $\varSolution$ and additional output $\varConstraints$.)

The single rule, \ruleName{Branch-Case}, chooses an arbitrary expression
$\varExp$ (of arbitrary datatype $\varTypeCon$) to scrutinize, synthesizing the
sketch 
$\expMatch{\varExp}{\varDataCon_i}{\varVar_i}{\expHole{\varHoleName_i}}$
with subgoals $\varHoleName_i$ for each of the the constructors $\varDataCon_1$
through $\varDataCon_n$ for the datatype $\varTypeCon$.
The main task is to \emph{distribute} the examples $\varWorlds$ onto appropriate
subgoals.
To determine which subgoal should be responsible for the $j$th example, the
guessed scrutinee $\varExp$ is evaluated under the example constraint
environment $\varEnv_j$ to a result $\varResult_j$.

Consider the particular scenario in which
$\varResult_j$ has determinate form
$\expApp{\varDataCon_i}{\varResult_j'}$, for some constructor $\varDataCon_i$.
The $\varDataCon_i$ branch will surely be taken under environment
$\varEnv_j$, so
the constraint
$\world{\envCat{\varEnv_j}{\envBind{\varVar_i}{\varResult_j'}}}{\varEx_j}$
is added to the examples $\varWorlds_i$ for the subgoal of that branch.
If $\varResult_j$ is \emph{indeterminate}, however, we cannot be sure ``which
way'' the scrutinee will evaluate and thus which branch to ``assign'' the
subgoal.

Therefore, in general, $\ruleName{Branch-Case}$ non-deterministically chooses a
branch $\alpha_j \in [n]$ for each example $j$
and relies on unevaluation to determine whether $\varResult_j$ can satisfy
$\expApp{\varDataCon_{\alpha_j}}{\exHole}$ (assuming some constraints
$\varConstraints_j$).
The \emph{constructor simplification} operation 
$\collapseUnwrapWrap{\varResult} =$
$
{\grungyITE
  {\varResult = \expUnwrap{\varDataCon_i}{(\expApp{\varDataCon_i}{\varResult'})}}
  {\varResult'}
  {\varResult}}
$
helps streamline the determinate and indeterminate scenarios in the definition
of \ruleName{Branch-Case}.
This flexibility---analogous to \ruleName{U-Case}
(\cf{}~\autoref{fig:ex-backprop})---is needed to synthesize
several \emph{inside-out} recursive functions (without trace-complete examples),
as described in the next section.

\renewcommand{\dimColor}[1]{\textcolor{black}{#1}} %% HACK

\begin{theorem*}[Soundness of Synthesis]
\label{prop:sound-e2e}

\breakAndIndent
If $\typeCheckProgramWrapper{\varProgram}{\varType}$
and $\programReducesTo{\varProgram}{\varResult}{\varAssertions}$
and $\simplifyEquals{\varAssertions}{\varConstraints}$
and
  $\iterSolveEqualsFull
    {\varHoleEnv}
    {\varDatatypeEnv}
    {(\varConstraints)}
    {\varSolution}
    {\varHoleEnv'}$,

\justIndent
then $\typeCheckSolution{\varHoleEnv'}{\varSolution}$
and $\assertionSat{\varSolution}{\varAssertions}$.

\end{theorem*}

\renewcommand{\dimColor}[1]{\textcolor{gray}{#1}} %% undo HACK

% !TEX root = ./main.tex
\section{Implementation}
\label{sec:implementation}

We implemented \smyth{} (\url{https://github.com/UChicago-PL/smyth})
in approximately 6,500 lines of OCaml code, not including
the front-end to \smyth{} nor the experimental setup.
Compared to the core language in \autoref{fig:syntax}, our implementation
supports Haskell/Elm-like syntax, $n$-ary tuples, \inlinecode{let}-bindings,
\inlinecode{let}-bound recursive function definitions, and user-defined datatypes.
Our implementation also supports higher-order function examples (used in the
experiments below) and polymorphism (not used below, but described in
{\refAppendixC}) following
\citet{Myth} and \citet{OseraThesis}, respectively; these
features are orthogonal to our contributions.

Our prototype lacks many of the syntactic conveniences used in code listings in
\autoref{sec:intro} and \autoref{sec:overview} such as nested pattern matching,
infix list operators \verb+(::)+ and \verb|(++)|, and type inference for holes.
Following \myth{}, we synthesize only structurally decreasing recursive
functions, and we further require that the first
argument to a recursive call be structurally decreasing.
These are not fundamental challenges, but they result in slightly different code
than shown in the paper.

\parahead{Optimizations}

We adopt two primary optimizations from \myth{}. %% ~\citep{Myth,OseraThesis}.
The first is to guess and cache only \emph{proof
relevant}~\citep{ProofRelevance} elimination forms---variables $\varVar$ or
calls $\expApp{\expAppTwo{\varVarF}{\varExp_1}{\cdots}}{\varExp_n}$ to
variable-bound functions.
The second is a \emph{staging} approach to incrementally increase the maximum
branching depth, the size of terms to guess as scrutinees, and the size of terms
to guess in other goal positions.
We generally adopt the same parameters used by \citet{OseraThesis}, but with
additional intermediate stages to favor small solutions.
Furthermore, our parameters are ``sketch-sensitive'': \inlinecode{case}
expressions in the sketch, if any, count against the branching depth budget.

To rein in the non-determinism of case unevaluation, our implementation is configured, first,
to guess only variables and projections for the
$\guessesForMatch{\varHoleEnv}{\varDatatypeEnv}{\varResult}$ procedure
in the ``eager'' \ruleName{U-Case-Guess} rule and, second,
to bound the number of nested uses of the ``lazy'' \ruleName{U-Case} rule.

\section{Experiments}
\label{sec:experiments}

We consider several questions regarding how our techniques---which address
Limitations A and B of prior evaluator-based synthesis
(\autoref{sec:intro})---translate into practical gains for users of
synthesis tools.

\newcommand{\questionA}
  {Compared to prior evaluator-based synthesizers, does \smyth{} reduce the number of
   examples required to synthesize top-level, single-hole tasks?}
\newcommand{\questionB}
  {Unlike prior evaluator-based synthesizers, does \smyth{} support sketching tasks?
   Is the total specification burden less than when using examples alone?}
\newcommand{\questionC}
  {Can state-of-the-art logic-based synthesizers complete all tasks that \smyth{} can?}

\begin{itemize}

\item
\questionA

\item
\questionB

\item
\questionC
\end{itemize}
To shed light on these questions, we designed four experiments based on the
benchmarks used to evaluate \myth{}.
Expert examples are the de facto method for evaluating the \textit{raw
expressiveness} of synthesis
techniques~(\eg{}~\citep{Escher,Myth,Frankle2016,Feser2015}).
A notable exception is how \citet{Feser2015} evaluate the \textit{robustness} of
$\lambdaSquared{}$ using randomly-generated examples as a ```lower bound' on a
human user ... who has no prior exposure to program synthesis tools.''
Inspired by these approaches, our experiments consider both ``expert'' and
``random'' users to investigate \smyth{}'s expressiveness and robustness.

We ran each of the \smyth{} experiments on a Mid 2012 MacBook Pro with a 2.5
GHz Intel Core i5 CPU and 16 GB of RAM.
We describe each experimental setup and summarize the results
(\autoref{fig:experiments}) in turn, followed by a discussion including
limitations.

\subsection{Experiment 1: No Sketches + Trace-Complete Examples}
\label{sec:experiment-one}

%% NOTE: Don't change macros in this file.
%% Make the changes in the smyth repo and copy over via make ours.
\newcommand{\benchmarkName}[1]
  {#1}
\newcommand{\benchmarkNameBool}[1]
  {\benchmarkName{bool\_#1}}
\newcommand{\benchmarkNameList}[1]
  {\benchmarkName{list\_#1}}
\newcommand{\benchmarkNameNat}[1]
  {\benchmarkName{nat\_#1}}
\newcommand{\benchmarkNameTree}[1]
  {\benchmarkName{tree\_#1}}

\definecolor{skippedColor}{HTML}{c0c0c0}

\newcommand{\labelColorSkipped}[1]
  {\textcolor{skippedColor}{#1}}
\newcommand{\labelColorFailed}[1]
  {\textcolor{orange}{#1}}
\newcommand{\labelTimeout}
  {\labelColorFailed{timeout}}
\newcommand{\labelOverspec}
  {\labelColorFailed{overspec}}
\newcommand{\labelRandomFailed}
  %% {\labelColorFailed{failed}}
  {\labelColorFailed{(---,---)}}
\newcommand{\labelIncorrect}
  {\labelColorFailed{incorrect}}
\newcommand{\labelBlank}
  %% {---}
  {$\bullet$}
\newcommand{\labelBlankOneFailed}
  {\labelColorSkipped{\labelBlank$^1$}}
\newcommand{\labelBlankHigherOrder}
  {\labelColorSkipped{\labelBlank$^2$}}
\newcommand{\labelBlankNonRec}
  {\labelColorSkipped{\labelBlank$^3$}}
\newcommand{\labelBlankSameExpertExamples}
  {\labelColorSkipped{\labelBlank$^4$}}
\newcommand{\labelRandomTime}[1]
  %% {\textcolor{orange}{#1}}
  %% {#1}
  {t=#1}

\newcommand{\benchmarkExperimentOne}[3]
  {&{#1}&{#3}}
\newcommand{\benchmarkExperimentOneFailedTimeOut}[1]
  {&{#1}&{\scriptsize{timeout}}}
\newcommand{\benchmarkExperimentOneFailedOverSpecialized}[1]
  {&{#1}&{\scriptsize{overspec}}}
\newcommand{\benchmarkExperimentOneFailedNoSolutions}[1]
  {&{#1}&{\scriptsize{none}}}
\newcommand{\benchmarkExperimentThreeFailedTimeOut}
  {&{\scriptsize{timeout}}}
\newcommand{\benchmarkExperimentThreeFailedOverSpecialized}
  {&{\scriptsize{overspec}}}
\newcommand{\benchmarkExperimentTwo}[9]
  {&{#1}
  }
%% TODO rename: used for experiment three also
\newcommand{\benchmarkExperimentTwoRand}[3]
  {&(#1, #2)$^{#3}$}
\newcommand{\benchmarkExperimentTwoRandFailedHigherOrder}
  {&\blankEntry}
\newcommand{\benchmarkExperimentTwoRandFailedTimeout}
  {&\scriptsize{timeout}}
%% TODO rename
\newcommand{\benchmarkExperimentTwoRandFailedNoNinety}
  {&\scriptsize{failed}}
\newcommand{\benchmarkExperimentThree}[9]
  {&{#1}}

\newcommand{\insideOut}
  %% {*}
  {}
\newcommand{\upperBound}
  {*}
\newcommand{\blankEntry}
  {---}
\newcommand{\benchmarkExperimentTwoBlank}
  {&\blankEntry}
\newcommand{\benchmarkExperimentThreeBlank}
  {&\blankEntry}
\newcommand{\displayPct}[1]
  {\phantom{0 (}{#1}\phantom{)}}
\newcommand{\displayPctUpperBound}[1]
  {\phantom{0 (}{#1}\upperBound}

\newcommand{\leonquidBlank}
  {\textcolor{gray}{---}}
\newcommand{\leonquidCorrectNoPhantom}
  {\textcolor{gray}{\cmark}}
\newcommand{\leonquidCorrect}
  {\textcolor{gray}{\cmark\phantom{$^1$}}}
\newcommand{\leonquidIncorrect}
  {\textcolor{orange}{\xmark$^1$}}
\newcommand{\leonquidError}
  {\textcolor{orange}{\xmark$^0$}}
\newcommand{\leonquidHigherOrderFunc}
  {\textcolor{orange}{\xmark$^2$}}
\newcommand{\synquidDatatypeAxiomsNoPhantom}
  {\textcolor{orange}{\textbf{?}}}
\newcommand{\synquidDatatypeAxioms}
  {\textcolor{orange}{\textbf{?\ }}}

\begin{figure}

\experimentTableSize

\begin{tabular}{l|cc|cc|cc||cc|cc}
& \multicolumn{6}{c||}{\textbf{\snsMyth{}}}
& \multicolumn{2}{c}{\textbf{\leon{}}}
& \multicolumn{2}{|c}{\textbf{\synquid{}}}
\\\hline
\multicolumn{1}{r|}{\textbf{Experiment}} &
\multicolumn{2}{c|}{\textbf{1}} &
\textbf{2a} & \textbf{2b} & \textbf{3a} & \textbf{3b}
& \multicolumn{2}{c|}{\textbf{4}} & \multicolumn{2}{c}{\textbf{4}}
\\\hline
\multicolumn{1}{r|}{{Sketch / Objective}} &
\multicolumn{2}{c|}{\textit{None / Top-1}} &
\multicolumn{2}{c|}{\textit{None / Top-1}} &
\multicolumn{2}{c||}{\textit{Base Case / Top-1-R}}
%% \multicolumn{1}{r|}{{Sketch}} &
%% \multicolumn{4}{c|}{\textit{None}} &
%% \multicolumn{2}{c||}{\textit{Base Case}}
& \multicolumn{2}{c|}{}
\\\hline
%% \multicolumn{1}{r|}{{\#Benchmarks}} &
%% \multicolumn{4}{c|}{\textit{\numBenchmarks{}/\numBenchmarksAll{} \myth{} benchmarks}} &
%% \multicolumn{2}{c||}{\textit{\numBenchmarksBase{}/\numBenchmarksRecursive{} rec. benchmarks}}
%% \multicolumn{4}{c|}{\textit{\numBenchmarksAll{} \myth{} benchmarks}} &
%% \multicolumn{2}{c||}{\textit{\numBenchmarksRecursiveAll{} recursive benchmarks}}
%% & \multicolumn{2}{c|}{}
%% \\\hline
%% \multicolumn{1}{r|}{{Objective}} &
%% \multicolumn{2}{c|}{\makebox[0.38in]{\textit{Top-1}}} &
%% \multicolumn{2}{c|}{\makebox[0.38in]{\textit{Top-1}}} &
%% \multicolumn{2}{c||} {\makebox[0.38in]{\textit{Top-1-R}}}
%% & \multicolumn{2}{c|}{}
%% \\\hline
\textbf{Name} &
\textbf{Expert} & \textbf{Time} &
\textbf{Expert} & \textbf{Random} &
\textbf{Expert} & \textbf{Random} &
\textbf{1} & \textbf{2a} &
\textbf{1} & \textbf{2a}
\\
&
& &
& {(50\%, 90\%)} &
& {(50\%, 90\%)} &
& &
\\
\input{figure-10-data}
\hline
\textbf{Averages} &
&
&
\displayPctUpperBound{\pctFewerExamplesTopOne} &
&
\displayPct{\phantom{1+}\pctFewerExamplesBaseCaseStrategy} &
&&
\end{tabular}

\vsepBeforeCaption
  \captionsetup{justification=centering}
  \caption{
    Experiments.
    \\
      \textbf{Top-1(-R)}:
      1st (recursive) solution valid.
    \textbf{Time}:
       Average of 10 runs, in seconds.
    \\
    \textbf{2a Average}:
      \pctFewerExamplesTopOne{} for \numBenchmarks{} non-blank rows.
      (*Upper bound: \pctFewerExamplesTopOneUpperBound{} for all
     \numBenchmarksAll{} rows.)
    \\
    \textbf{3a Average}:
      \pctFewerExamplesBaseCaseStrategy{} for
      \numBenchmarksBase{} non-blank, non-error rows.
  }
\label{fig:experiments}
\end{figure}

As a baseline experiment, we first run \snsMyth{} on each \myth{} benchmark---a
top-level, single-hole task specified with the ``full'' set of trace-complete
expert examples reported by \citet{OseraThesis}.
\autoref{fig:experiments} (column 1) indicates that \snsMyth{} passes
\numBenchmarks{} of the same 43 benchmarks (without sketches) in a similar
amount of time (\cf{}~\citep{OseraThesis}).

Of the five \myth{} benchmarks that failed in Experiment 1,
\snsMyth{} produced an over-specialized solution for one
(\texttt{list\_even\_parity}) and did not terminate within 120 seconds for the
remaining four
(\texttt{list\_compress}, \texttt{tree\_binsert},
\texttt{tree\_nodes\_at\_level}, and \texttt{tree\_postorder}).
The overspecialized term \snsMyth{} synthesized for
\texttt{list\_even\_parity} was smaller
(AST size 14) than the desired term (size 16), which was correctly
synthesized by \myth{}.
(\snsMyth{} synthesizes and ranks the desired term second.)
It is unclear why \myth{} did not find and return the smaller solution, which is
consistent with the examples provided;
nevertheless, we classify this task as a failure.
The four benchmarks for which \snsMyth{} did not terminate are discussed further in \autoref{sec:limitations}.

Our validation process---which checks synthesized terms against a random set of
examples from a reference implementation---revealed that the solution for
\texttt{list\_filter} reported by
\citet[p.171]{OseraThesis} is incorrect.
As a workaround, we added one more (trace-complete) example to the reported set of 8 examples and observed that \smyth{} synthesized a correct solution.
We treat these 9 examples (marked with an asterisk in \autoref{fig:experiments}) as the set of \myth{} expert examples for this task.

\subsection{Experiment 2: No Sketches + Non-Trace-Complete Examples}
\label{sec:experiment-two}

Second, we measured how many examples---both expert and random---\snsMyth{}
requires to synthesize the \myth{} tasks when not limited to the
trace-complete examples from Experiment 1.

\parahead{Experiment 2a: No Sketches + Expert Examples}

To construct expert examples for \snsMyth{} on each of the \numBenchmarks{}
benchmarks it can synthesize, we manually removed sets of examples from the full
test suite until \snsMyth{} no longer synthesized a correct solution, \ie{}~a
solution that conforms to a reference implementation of the desired solution.
As such, there are no corresponding tasks for the five benchmarks that failed Experiment 1, as indicated by ``\labelBlankOneFailed{}'' in \autoref{fig:experiments}.

Of the \numBenchmarks{} benchmarks, \autoref{fig:experiments} (column 2a) shows
that \snsMyth{} required fewer examples to synthesize all but four benchmarks
(\verb+bool_neg+, \texttt{bool\_xor}, \verb+list_length+, and \verb+nat_max+), requiring on average
\pctFewerExamplesTopOne{} of the number of expert examples required by
\myth{}, with similar running times as in the baseline configuration (timing
data not shown).
To account for the 5 missing benchmarks, if we were to assume that \snsMyth{}
were extended with the \myth{}-style trace-complete approach to synthesizing
recursive functions as a backup synthesis procedure and that the remaining
benchmarks would require all of the expert examples, then \snsMyth{} would
require on average \pctFewerExamplesTopOneUpperBound{} of the number of examples
for the entire benchmark suite.

\parahead{Experiment 2b: No Sketches + Random Examples}

To evaluate the robustness of \myth{}, we implemented a random example
generator.
For simplicity, our random generator does not support function types; therefore,
we did not consider the 4 higher-order function benchmarks (\verb+list_filter+,
\verb+list_fold+, \verb+list_map+, and \verb+tree_map+; these are
marked ``\labelBlankHigherOrder{}'' in \autoref{fig:experiments}).
We also did not consider the four benchmarks that timed out in Experiment 1.

For each of the remaining 35 tasks,
we generated $N\!=\!50$ sets of $k$ random input examples (where $k$ ranges from 1
to a reasonable upper bound depending on the benchmark) and used a task
reference implementation to compute the corresponding outputs, thus producing
$N$ sets of input-output example sets of size $k$ for each $k$. 
We fixed relatively small upper bounds on the AST sizes of the input examples generated to ensure
the examples could reasonably be provided by a human, and, rather than sampling
inputs uniformly at random---in which case, \eg{}, a list of length 3 would be
twice as likely as a list of length 2---we first sampled different \emph{shapes}
for the data structures (\inlinecode{List}s and \inlinecode{Tree}s) uniformly at random,
then filled in base values at the AST leaves uniformly at random.
Furthermore, we required that each set of examples (regardless of size) contains
the unique ``minimal input'' to the function, that is, the input that consists
of the minimal value for each type of each argument of the function, where, for
\inlinecode{Nat}s, the minimal value is \inlinecode{0}, for \inlinecode{List}s, it is the empty
list, and for \inlinecode{Tree}s, it is a leaf.

Entries in \autoref{fig:experiments} (column 2b) show two values: the minimum $k$ for which \smyth{}
synthesized the desired solution within a $t\!=\!1$ second timeout for 50\% of the
$N$ sets of examples, and the minimum such $k$ to achieve 90\% success;
\citetAppendix{\refAppendixB} includes graphs for each benchmark.
Several entries require explanation.
Two benchmarks are marked with a superscript \mbox{``$t\!=\!3$''} (\verb+tree_collect_leaves+
and \verb+tree_preorder+) and
one benchmark is marked with a superscript \mbox{``$t\!=\!10$''} (\verb+tree_count_nodes+)
to indicate they they required a longer timeout.
For \verb+tree_count_nodes+, we do not report the minimum $k$ value for 90\% (marked
``$\downarrow$''), because the percentage dips below for subsequent values of
$k$.
One benchmark is marked ``(---,---)'' (\verb+list_even_parity+) and did not achieve
a 50\% success for reasonably-small values of $k$.
For this benchmark, we hypothesize that our simply-typed approach cannot
glean enough information from its input type, \inlinecode{BooleanList}.

\setlength{\intextsep}{0pt}%
\setlength{\columnsep}{-10pt}%

\begin{wrapfigure}{r}{0pt}
% Data copied from fig-k-prime-3.tex

{
\newcommand{\kfMedian}{0}
\newcommand{\kfMax}{2}
\newcommand{\kfAboveOne}{2 (7\%)}
\newcommand{\kfAboveTwo}{0 (0\%)}
\newcommand{\knMedian}{1}
\newcommand{\knMax}{9}
\newcommand{\knAboveOne}{8 (27\%)}
\newcommand{\knAboveTwo}{5 (17\%)}

  \footnotesize
  \begin{tabular}{c|cc}
    & median $k'_p$
    & max $k'_p$
    \\
    \hline
    $ p = 50\% $ & \kfMedian{} & \kfMax{} \\
    $ p = 90\% $ & \knMedian{} & \knMax{}
  \end{tabular}
}

\end{wrapfigure}
To analyze these $ k $-values, we consider the difference \mbox{$ k'_p := k_p -
k_\text{expert} $} for each benchmark that was successfully synthesized in this
experiment, where $ p $ is the required success rate (either 50\% or 90\%) and
$ k_\text{expert} $ is the number of \snsMyth{} expert examples for Experiment
2a.
The value $ k'_p $ thus represents how many more examples are needed, compared
to the expert set, to achieve success $ p\% $ of the time.
The adjacent table
summarizes the distribution of $ k'_p $ for Experiment~2b;
additional statistics and corresponding histograms can be found in
\citetAppendix{\refAppendixB}.

\subsection{Experiment 3: Base Case Sketching Strategy}
\label{sec:experiment-three}

Experiments 1 and 2 considered tasks without sketches from the user.
As a third experiment, we systematically converted the \myth{} benchmarks into a
suite of small sketching tasks by employing a simple \emph{base case sketch
strategy}---performing case analysis on the correct argument of the function,
filling in the base case properly, and leaving a hole in the recursive branch.
Of the \numBenchmarks{} tasks, \numBenchmarksRecursive{} are recursive and thus
subject to this strategy.
The remaining, non-recursive tasks are marked ``\labelBlankNonRec{}''.

In \autoref{fig:experiments} and the following, we write $1+n$ to denote a
specification with $n$ examples in addition to the base case sketch; our
accounting treats the specification burden of the base case sketching strategy
as equivalent to 1 example.
(We could report AST sizes of sketches and examples, but even these would be
just a rough proxy for the ``complexity'' of a specification.)

\parahead{Experiment 3a: Base Case Sketches + Expert Examples}

Analogous to Experiment 2a, we manually removed sets of examples from the full
trace-complete expert examples until \snsMyth{} no longer successfully completed
the task.
For this experiment, however, because the base case strategy pertains to
recursive functions, we considered a task successful if the smallest
\emph{recursive} solution was correct, rather than simply the smallest
solution overall.
\autoref{fig:experiments} (column 3a) shows the results of this experiment.

For \numBenchmarksBase{} of these \numBenchmarksRecursive{} tasks that succeeded, \snsMyth{}
on average required smaller total specifications with base case sketches than
with no sketches.
On average, specifications were \pctFewerExamplesBaseCaseStrategy{}
the size of the full trace-complete examples---compared to \pctFewerExamplesBaseCaseNoSketch{}
without a sketch
(average, not shown, of \numBenchmarksBase{} rows in the Experiment 2a column).
Given the sketches, the average number of examples required was
\avgExamplesBase{}; \texttt{list\_sorted\_insert} required \maxExamplesBase{},
while the rest required between 1 and 4.

Three tasks that succeeded
(\texttt{list\_filter}, \texttt{list\_pairwise\_swap}, and \texttt{list\_sorted\_insert})
required sketch-sensitive staging parameters~(\autoref{sec:implementation}).
This is because \smyth{}'s staging parameters increase branching depth
before scrutinee size, and a relatively large scrutinee is needed for the
desired solution;
compared to when no sketch is provided, sketch-insensitive staging parameters
effectively ``penalize'' the sketch for having introduced a case.
Before we accounted for branching depth in the user-provided sketch, \snsMyth{}
synthesized overspecialized solutions for these three tasks even with the full
set of \myth{} expert examples.

Two of the \numBenchmarksRecursive{} tasks failed this experiment.
For \texttt{list\_even\_parity}, \snsMyth{} synthesized an over-specialized solution (even with sketch-sensitive staging parameters).
For \texttt{list\_concat}, \snsMyth{} actually synthesized
``\texttt{list\_rev\_concat},'' which appends together a list of lists in
\textit{reverse} order.
The \myth{} expert examples are not sufficient to distinguish these two
functions; \snsMyth{} returns both, but they have the same AST size and the
desired solution is arbitrarily ranked second.

\begin{wrapfigure}{r}{0pt}
% Data copied from fig-k-prime-2.tex.

{
\newcommand{\kfMedian}{2}
\newcommand{\kfMax}{6}
\newcommand{\kfAboveOne}{4 (22\%)}
\newcommand{\kfAboveTwo}{3 (17\%)}
\newcommand{\knMedian}{4}
\newcommand{\knMax}{14}
\newcommand{\knAboveOne}{4 (22\%)}
\newcommand{\knAboveTwo}{4 (22\%)}

  \footnotesize
  \begin{tabular}{c|cc}
    & median $k'_p$
    & max $k'_p$
    \\
    \hline
    $ p = 50\% $ & \kfMedian{} & \kfMax{} \\
    $ p = 90\% $ & \knMedian{} & \knMax{}
  \end{tabular}
}

\end{wrapfigure}
\parahead{Experiment 3b: Base Case Sketches + Random Examples}

\mbox{} %% LAYOUT HACK to make newline after "."

\noindent %% LAYOUT HACK
Analogous to Experiment 2b, we generated random input-output examples for the
benchmarks, this time in addition to providing the base case sketches.
We again consider the difference $ k'_p := k_p - k_\text{expert} $ for each
benchmark that was successfully synthesized in this experiment, where $
k_\text{expert} $ is now the number of \snsMyth{} expert examples for Experiment
3a rather than for Experiment 2a.
The adjacent table
summarizes the distribution of $ k'_p $ for Experiment 3b;
additional data can be found in \citetAppendix{\refAppendixB}.

\subsection{Experiment 4: Programming-by-Example in \leon{} and \synquid{}}
\label{sec:experiment-logic}

The previous experiments evaluate the improvements in \smyth{} compared to prior
evaluator-based techniques.
In our final experiment, we run several of our ``programming-by-example'' tasks
on \leon{} and \synquid{}.
The goal is to understand whether---from the perspective of a user who wishes to
specify tasks through examples---\leon{} or \synquid{} are strictly more powerful
than \smyth{}.
That is, can \leon{} or \synquid{} solve every task that \smyth{} can?

We systematically generated Scala and Haskell versions of our
benchmarks to test \leon{} and \synquid{}, respectively.
Because this experiment is designed to answer a very simple question,
we did not develop a thorough experimental environment with random
examples or multiple trials.
Instead, we used web interfaces to \leon{} and \synquid{} to test
benchmarks.\footnote{\
\url{https://leon.epfl.ch/} and
\url{http://comcom.csail.mit.edu/comcom/\#Synquid}.
Accessed February 2020 and May 2020.
%
%% \rkc{maybe e=True and e=False flags.}
%
}

First, we tested the small sketching tasks from \autoref{sec:intro}
and \autoref{sec:overview}.
As described in \autoref{sec:intro}, both tools fail to complete the
\verb+stutter_n+ task.
We also found that \synquid{} fails to complete the four sketching tasks from
\autoref{fig:overview-examples} and that \leon{} successfully completes \verb+max+
and \verb+odd+ but fails on \verb+minus+ and \verb+mult+.

We then tested the tools for the top-level, single-hole tasks used in
Experiments 1 and 2a with trace-complete and non-trace-complete expert
examples, respectively.
Besides the function to synthesize, we used simple types (without
examples or precise logical predicates) for all functions in the context.
Four benchmarks had the same number of expert examples in Experiment~2a as
they did in Experiment 1 and thus do not have corresponding tasks in
Experiment~4 (marked ``\labelBlankSameExpertExamples{}'').

\autoref{fig:experiments} (columns 4) show the results.
\leon{} and \synquid{} successfully completed many tasks (marked
\leonquidCorrectNoPhantom{}), but failed several tasks for a variety
of reasons:
terminating without producing solutions or not terminating within a timeout (\leonquidError{});
returning over-specialized solutions (\leonquidIncorrect{}); and
not being able to directly express higher-order function examples
(\leonquidHigherOrderFunc{}).
As expected, \synquid{} failed to synthesize recursive functions
without inductive (\ie{}~trace-complete) specifications
(column 4, 2a).\footnote{\hspace{-0.011in} %% SPACING HACK
Earlier results from this experiment revealed an implementation issue
in \synquid{} involving the axiomatization of recursive datatypes in the
underlying logic.
This issue---which prevented the desired solutions for many benchmarks from typechecking,
even when given trace-complete examples---has since been fixed~\citep{Nadia}.
}

These results are not entirely surprising, as the underlying
techniques are not necessarily tailored to the structure of examples
encoded as conjunctions-of-implications.
This suggests opportunities for further
improvements to both evaluator- and logic-based techniques, for
instance, by integrating live bidirectional evaluation into more
fine-grained logic-based techniques.

As a final note, this experiment was \emph{not} intended to evaluate
whether \smyth{} is ``better'' than the logic-based tools.
Indeed, many tasks involving complex invariants are beyond the reach
of evaluator-based techniques, \smyth{} included.
\mbox{\citet[\S\! 4.3]{Polikarpova2016}} provide some empirical comparison
between example-based and logic-based specifications on several common
benchmarks.

\subsection{Limitations and Discussion}
\label{sec:limitations}

\parahead{Failing Benchmarks}

One major optimization in \myth{} that we have not implemented is to cache
solutions $\varSolution$---which correspond to \myth{}'s ``refinement trees''---across
branches of the search.
This optimization does not directly carry over to our setting because, unlike
in \myth{}, synthesized terms in \snsMyth{} may introduce different, conflicting assumptions across
different branches of search.
Thus, our first hypothesis is that suitably extending caching to our
setting could help synthesize the remaining tasks (although the difficulty of
this task is unclear).

Of the five benchmarks not successfully synthesized in our implementation,
\myth{} finds four solutions with \emph{inside-out
recursion}~\citep{OseraThesis}, which pattern match on a recursive call to the
function being synthesized.
Inside-out solutions are smaller than more ``natural'' ones, and sometimes they are
the only solutions to tasks in \myth{} and \smyth{} because only elimination
forms are enumerated and \inlinecode{let}-bindings are not synthesized~\citep{OseraThesis}.
Although \smyth{} does synthesize an inside-out solution for one benchmark
(\verb+list_pairwise_swap+), inside-out recursion
relies heavily on the non-determinism of \ruleName{Branch-Case} and
\ruleName{U-Case}.
Accordingly, our second hypothesis is that additional tuning for these
sources of non-determinism could help synthesize the necessary inside-out
recursion.

\parahead{Scalability}

Each benchmark in our experiments included the minimal context---as defined in the \myth{} benchmarks---required to
synthesize the desired solution.
In addition to minimal contexts, the \myth{} paper also reported results in the
presence of a slightly larger context and ran into scalability issues on
some benchmarks.
Though we did not run these versions of the benchmarks, we inherit any
scalability issues of the prior techniques.

Moreover, our approach introduces new sources of non-determinism.
To scale to much larger programs with complex control flow, static reasoning
(interleaved with concrete evaluation) could be used to prune unsatisfiable or
heuristically ``difficult'' sets of example constraints.
Orthogonal techniques for scaling to large contexts with additional
components~\citep{Gvero:2013,Feng:2017,TYGAR} might also be incorporated into
our approach in future work.

\parahead{Assertions}

Our formulation and thus our benchmarks support only top-level \verb+assert+s.
To allow \verb+assert+s in arbitrary expressions (as needed for larger and more realistic
sketching tasks), evaluation and resumption could be extended to generate
assertions $\varAssertions$ as a side-effect, to be translated by
\metaFunc{Simplify} into constraints for synthesis.
We expect the algorithmic changes to be straightforward, but the extended
definition of assertion satisfaction along with the corresponding correctness
properties and proofs are more delicate; we leave this task for future
work.

\parahead{Polymorphism}

Of the \numBenchmarks{} tasks that \smyth{} successfully synthesized in Experiment~1, 23 can be
specified with a polymorphic type signature rather than a monomorphic one. We
re-ran Experiments~2~and~3 with polymorphic type signatures, which are supported in our
implementation but are not included in our formal development.
As described in \citetAppendix{\refAppendixC}, polymorphic
type signatures lead to a modest reduction in the number of examples needed
for synthesis.

% !TEX root = ./main.tex

\section{Related Work}
\label{sec:related}

Our work generalizes the theory of evaluator-based synthesis techniques to
(a)~eliminate the need for trace-complete examples and
(b)~to support sketching---addressing Limitations A and B from
\autoref{sec:intro}.
We build directly on the work of \citet{Myth}, so we discussed \myth{}
throughout the paper.
To conclude, we discuss several additional directions of related work.

\subsection{Live Evaluation and Bidirectional Evaluation}
\label{sec:related-work-techniques}

The key technical mechanism underlying our approach is live bidirectional
evaluation, the combination of live evaluation and live unevaluation.
We choose the term ``live'' to describe partial evaluation of sketches,
following terminology of \citet{HazelnutLive}.
Future work must address important usability and scalability questions to
further develop and deploy our techniques in interactive,
\emph{live programming} environments~\citep{Tanimoto:2013,Kubelka:2018}.

\parahead{Live Evaluation (\hazelnutLive{})}

We adapt the technique for partially evaluating sketches from \hazelnutLive{}
\citep{HazelnutLive}.
In contrast to solver-based and symbolic execution techniques for partially
evaluating programs with holes~(\eg{}~\citep{Morpheus,Bornholt2018,Viser}), live
evaluation is a form of \emph{concrete} evaluation, adapting ideas from
\emph{contextual modal type theory}~\cite{CMTT}.
\citet[\S\! 5]{HazelnutLive} detail the relationship to related work on
\emph{partial evaluation}.
\hazelnutLive{} does not offer any form of synthesis; their ``fill-and-resume'' feature
refers to ordinary program edits by the user.

We note some technical differences in our formulation.
We choose a natural semantics presentation~\cite{Kahn:1987} for \coreSnsMyth{}
rather than one based on substitution.
Whereas their fill-and-resume mechanism is defined using contextual substitution,
our formulation instead defines evaluation resumption.
\hazelnutLive{} also includes hole types to support gradual
typing~\citep{Siek:2006,SiekSNAPL},
a language feature orthogonal to the (expression) synthesis motivations for our
work.
Finally, \citet{HazelnutLive} present a bidirectional type
system~\citep{PierceTurner,Chlipala:2005da} that, given type-annotated
functions, computes hole environments $\varHoleEnv$;
the same approach can be employed in our setting without complication.

\begin{comment}
Several proposals define \emph{unevaluators}, or \emph{backward evaluators},
that allow changes to the output value of an expression (without holes) to affect changes to the
expression.
%
\citet{Perera2012} propose an unevaluator that, given an output modified with
\emph{value holes}, slices away program expressions that do not contribute to
the parts of the output that remain, which is useful in an interactive debugging
session, for example.
%
\citet{Matsuda:2018} propose a bidirectional evaluator---which forms a
\emph{lens}~\citep{lenses}---for manipulating first-order values in a language
of \emph{residual expressions}, containing no function applications in
elimination positions.
%
\citet{sns-oopsla} generalize this approach to arbitrary programs and values in
a higher-order functional language, effectively mapping output value changes to
program repairs.
\end{comment}

\parahead{Bidirectional Evaluation (\sns{})}
Several proposals define \emph{unevaluators}, or \emph{backward evaluators},
that allow changes to the output value of an expression (without holes) to affect changes to the
expression~\citep{Perera2012,Matsuda:2018,sns-oopsla}.
Though related by analogy and terminology, our novel live unevaluation mechanism
shares essentially no technical overlap with the above techniques.
The prior backward evaluators essentially only modify constant literals of
base type---which can be thought of as ``non-empty'' holes that are subject to
replacement---at the leaves of an existing program, whereas our live unevaluator
propagates example constraints to holes of arbitrary type and in arbitrary
position.

An environment-style semantics is purposely chosen for each of the above
unevaluators, because value environments provide a sufficient mechanism for
tracing value provenance during evaluation.
In contrast, our unevaluator could just as easily be formulated with
substitution; in either style, hole expressions are labeled with unique
identifiers, which provide the necessary information to generate example
constraints.

\subsection{Program Synthesis}

We conclude with a broader discussion of the evaluator- and logic-based
synthesis techniques that we introduced in \autoref{sec:intro}.
We use the term ``functional programming''---in contrast to
``domain-specific''---to describe languages in which users (and synthesizers)
write unrestricted programs in a richly-typed functional language (\ie{}~with
directly recursive functions on algebraic datatypes).

\subsubsection{Evaluator-Based Synthesis Techniques}
We chose this term in \autoref{sec:intro} to describe synthesis algorithms in
which the core search strategy uses \emph{concrete} evaluation to ``check''
candidate terms,
typically against input-output example specifications.

\parahead{Programming-by-Example (PBE) for Domain-Specific Languages}

Programming-by-example techniques have been developed for numerous
domain-specific applications, including
string transformations~\citep{GulwaniPOPL2011}
(including bidirectional ones~\citep{Miltner2019}),
shell scripting~\citep{Gulwani2015},
web scraping~\citep{RousillonUIST2018},
parallel data processing~\citep{Smith2016},
and
generating vector graphics~\citep{sns-uist-2019}.
See \citet{DBLP:journals/ftpl/GulwaniPS17} for a recent survey of developments.
These approaches generally synthesize entire programs.
To allow experts to provide partial implementations, it should be possible to
formulate notions of live bidirectional evaluation of these domain-specific
techniques.

$\lambdaSquared$~\citep{Feser2015} synthesizes functions in a (first-order)
functional programming language (with higher-order components).
$\lambdaSquared$ enumerates \emph{open hypotheses} (\ie{}~sketches) involving
calls to a fixed set of primitive \inlinecode{List} and \inlinecode{Tree}
combinators~(\eg{}~\verb+filter+ and \verb+map+), and relies on axioms for
\emph{deductive reasoning} to convert examples for a goal into examples for the
subgoals.
This process is akin to refinement in \myth{}, and also helps prune
unsatisfiable example constraints (\eg{}~if a \verb+map+ hypothesis
requires input and output lists of different lengths).

However, function examples are not used used to ``refine'' the search; their
deduction rule for general recursion essentially falls back on raw term
enumeration, and their checking routine operates only on \emph{closed hypotheses}
(without holes).
In other words, examples need not be trace-complete because they are not used to
help synthesize recursive function literals.
Although the language supported by $\lambdaSquared$ nominally includes direct
recursive function literals~\citep[\S\!~3]{Feser2015},
in practice, their implementation synthesizes solutions \emph{only} by
composing the primitive data structure
combinators~\citep{FeserMSThesis,JohnFeser}, and furthermore does not introduce
non-trivial matches on inductive data~\citep{JohnFeser}.
$\lambdaSquared$ can synthesize a variety of functional programming tasks,
similar to the \myth{} and \smyth{} benchmarks, including with
randomly-generated examples~(\cf{}~\autoref{sec:experiments}) and with
significantly larger contexts than used in the \myth{} and \smyth{} experiments.
But because $\lambdaSquared$ does not search for directly recursive functions, it
is fundamentally a more domain-specific technique than \myth{} and \smyth{}.

For the domain of table transformations, \morpheus{} extends the approach of
$\lambdaSquared{}$ with (i) SMT-based reasoning to perform more powerful
deduction and (ii) partial evaluation of sketches.
\viser{}~\citep{Viser} further improves upon the techniques in \morpheus{}, by
providing \emph{backward} reasoning about program sketches using symbolic
reasoning over logical and subset constraints.
(\viser{} also integrates a domain-specific language for visualization,
resulting in a visualization-by-example tool.)
Sketches in both \morpheus{} and \viser{} are drawn from a first-order, domain-specific
language of table transformations. In contrast, \smyth{} performs bidirectional
reasoning about program sketches (a) in a \emph{general-purpose} richly-typed
functional programming language (as opposed to domain-specific table
transformation languages), (b) using techniques based on concrete evaluation
(rather than SMT solving and other symbolic reasoning techniques).

\parahead{PBE for Functional Programming}

Two prior evaluator-based systems synthesize recursive functions.
\escher{}~\citep{Escher} does so for an untyped, first-order functional language
(with base types rather than inductive datatypes),
relying on run-time type errors to help rule out candidate terms.
\myth{}~\citep{Myth} pioneered the idea to synthesize recursive functions over
algebraic datatypes using search techniques inspired by \emph{bidirectional
typing}~\citep{PierceTurner} and \emph{relevant proof
search}~\citep{ByrnesThesis,ProofRelevance}.
Both \escher{} and \myth{} require trace-complete examples.
As discussed next, the bidirectional typing approach of \myth{} has influenced
several logic-based approaches to synthesis.

\subsubsection{Logic-Based Synthesis Techniques}

We chose this term in \autoref{sec:intro} to describe synthesis algorithms that
use \emph{symbolic}, rather than concrete, evaluation to enumerate terms,
and which operate on more fine-grained, precise logical specifications than
examples.

\parahead{PBE for Functional Programming via Refinement Types}

\citet{Frankle2016} reformulate \myth{} by recasting concrete examples in a type
language of intersection and singleton types.
Rather than employing concrete evaluation, they perform (symbolic) proof search
within their rich type language.
Their formal development includes union and negation types, which allows more
than just examples (with concrete input and output values) to be specified.
Their implementation further supports type polymorphism, with symbolic values as
examples.
The combination of negation and polymorphism admit what \citet{Polikarpova2016}
dub ``generalized examples,'' which facilitate smaller specifications for
several \myth{} benchmarks.
(Generalized examples resemble the \emph{symbolic input-output examples}
supported by \leon{} for program repair~\citep{LeonRepair}.)
This reformulation of (generalized) examples suffers the same Limitations A and
B as \escher{} and \myth{}.
It would be valuable to extend \smyth{} in future work with similar typing
constructs.

\parahead{Program Sketching}

\sketch{}~\citep{SketchingPLDI2005,SketchingASPLOS2006,SketchingThesis,SketchingAPLAS2009}
is an imperative, C-like language that pioneered the approach of program
synthesis by sketching.
\rosette{}~\citep{Rosette2013,Rosette2014} further develops this approach within
the untyped functional language Racket.
Holes in \sketch{} and \rosette{} range only over integers and booleans, but
these can be used to define richer types of expressions.
The mechanisms for such \emph{syntax-guided synthesis}~\cite{SyGuS} are
particularly powerful in \rosette{}, which leverages the metaprogramming
facilities in Racket.
As \citet[\S\! 6]{Inala2017} suggest, one could embed the syntax and semantics of a
richly-typed, general-purpose functional programming language in \rosette{}.
There is no obvious reason to expect recursive functions over user-defined
algebraic datatypes embedded in this way to be readily synthesized, but this
approach would be an interesting experiment.

\parahead{Solver-Based Techniques for Functional Programming}

\synquid{}~\cite{Polikarpova2016} and \leon{}~\cite{Kneuss2013} directly support
sketching in richly-typed functional languages using solver-based techniques
driven by logical specifications.
\synquid{} employs bidirectional typing (like \myth{}) in a setting with
SMT-based refinement types~\citep{LiquidTypes,Vazou2013}.
\synquid{} furthermore introduces \emph{round-trip type checking}, which
propagates goal types ``through'' elimination forms, allowing errors to be
localized (\ie{}~found sooner) during type checking.
In a synthesis context, failing sooner means avoiding costly search~paths.

Example-based and logic-based specifications are complementary.
Combining support for such specifications is another interesting direction for
future work.
It would be interesting to consider whether live bidirectional evaluation could
help eliminate the inductive (\ie{}~trace-complete) requirement of partial
specifications in \synquid{}, so that its powerful logic-based reasoning could
better operate when given examples as partial specifications.

\begin{acks}
The authors would like to thank
Ian Voysey for guidance regarding proof strategies;
Nadia Polikarpova, Brian Hempel, Michael Adams, Youyou Cong, and anonymous reviewers for many helpful suggestions;
Aws Albarghouthi, John Feser, Viktor Kun\v{c}ak, and Nadia Polikarpova
for answering questions about
$\escher$, $\lambdaSquared{}$, $\leon{}$, and $\synquid{}$; and
Robert Rand---who coined the name \myth{}---for suggesting the name \snsMyth{},
thus further entangling our work with its predecessor. %% Get it?
This work was supported by
\grantsponsor{GS100000001}{NSF}{http://dx.doi.org/10.13039/100000001} grants
\textit{Semantic Foundations for Hole-Driven Development}
(CCF-\grantnum{GS100000001}{1814900} and CCF-\grantnum{GS100000001}{1817145}) and
\textit{Direct Manipulation Programming Systems}
(CCF-\grantnum{GS100000001}{1651794}).
\end{acks}

%% \clearpage

%% Bibliography
\bibliography{references}

%% Appendix
\appendix

\clearpage
\section{Additional Definitions and Proofs}
\label{sec:appendix}
\label{sec:appendix-dynamics}
\label{sec:appendix-synthesis}

%% example:
%%   \beginTheorem{ResDet}{Determinism of Resumption}{Res. Det.}{prop:res-det}
%%
%% generates commands:
%%   \autorefResDet        ==>   Theorem A.99
%%   \autorefResDetLong    ==>   Theorem A.99 (Determinism of Resumption)
%%   \autorefResDetShort   ==>   Theorem A.99 (Res. Det.)
%%
\newcommand{\beginTheorem}[4]
  {%
   \expandafter\newcommand\csname autoref#1\endcsname{\autoref{#4}}%
   \expandafter\newcommand\csname autoref#1Long\endcsname{\autoref{#4} (#2)}%
   \expandafter\newcommand\csname autoref#1Short\endcsname{\autoref{#4} (#3)}%
   \begin{theorem}[#2]\label{#4}%
  }

\newcommand{\beginLemma}[4]
  {%
   \expandafter\newcommand\csname autoref#1\endcsname{\autoref{#4}}%
   \expandafter\newcommand\csname autoref#1Long\endcsname{\autoref{#4} (#2)}%
   \expandafter\newcommand\csname autoref#1Short\endcsname{\autoref{#4} (#3)}%
   \begin{lemma}[#2]\label{#4}%
  }

This section provides additional definitions for \autoref{sec:live-bi-eval} and
\autoref{sec:synthesis}, as well as theorems and proofs.

\subsection{Syntax}
\label{sec:appendix-syntax}

\parahead{Datatypes}

Rather than supporting arbitrary-arity constructors (as in the technical
formulation of \citet{Myth}) we choose single-arity constructors and
products (following the formulation by \citet{MythFrankleThesis}) to lighten
the presentation of synthesis in \autoref{sec:synthesis}.

\parahead{Results}

\autoref{fig:final} defines result classification.
\\

\begin{figure}[h]

\judgementHeadNameOnly
  %% {Final Results / Final Environments}
  {\maybeUnderline{Final} Results and Environments}
  {$\JudgementBox{\isFinal{\varResult}}$ \hspace{0.00in}
   $\JudgementBox{\isFinal{\varEnv}}$}

\vsepRule

$
\inferrule* %%[lab=\ruleNameFig{}]
  {\isDet{\varResult}}
  {\isFinal{\varResult}}
$
\hsepRule
$
\inferrule* %%[lab=\ruleNameFig{}]
  {\isIndet{\varResult}}
  {\isFinal{\varResult}}
$
\hsepRule
%% \hspace{0.50in}
%
$
\inferrule* %%[lab=\ruleNameFig{}]
  {
  }
  {\isFinal{\emptyEnv}}
$
\hsepRule
$
\inferrule* %%[lab=\ruleNameFig{}]
  {
   \isFinal{\varEnv}
   %% \\\\
   \sepPremise
   \isFinal{\varResult}
  }
  {\isFinal{\envCat{\varEnv}{\envBind{\varVar}{\varResult}}}}
$

\vsepRule

\judgementHead
  %% {Values}
  %% {\maybeUnderline{De}s\maybeUnderline{t}ructible Results}
  {\maybeUnderline{Det}erminate Results}
  {\isDet{\varResult}}

\vsepRule

$
\inferrule* %%[lab=\ruleNameFig{}]
  {
  }
  {\isDet{\expUnit}}
$
\hsepRule
$
\inferrule* %%[lab=\ruleNameFig{}]
  {\multiPremise{\isFinal{\varResult_i}}{\sequenceSyntaxTwo{i}}}
  {\isDet{\pair{\varResult_1}{\varResult_2}}}
  %% {\isFinal{\triple{\varResult_1}{\ldots}{\varResult_n}}}
$
\hsepRule
$
\inferrule* %%[lab=\ruleNameFig{}]
  {\isFinal{\varResult}}
  {\isDet{\expApp{\varDataCon}{\varResult}}}
$
\hsepRule
%% \vsepRule
%
$
\inferrule* %%[lab=\ruleNameFig{}]
  {\isFinal{\varEnv}}
  {\isDet{\closure{\varEnv}{\valFixFun{\varVarF}{\varVar}{\varExp}}}}
$

\vsepRule

\judgementHead
  %% {Paused Results}
  %% {Undetermined Results}
  %% {\maybeUnderline{Inde}s\maybeUnderline{t}ructible Results}
  {\maybeUnderline{Indet}erminate Results}
  {\isIndet{\varResult}}

\vsepRule

$
\inferrule* %%[lab=\ruleNameFig{}]
  {\isFinal{\varEnv}}
  {\isIndet{\highlightHoleClosure{\holeClosure{\varEnv}{\varHoleName}}}}
$
\hsepRule
%% \hspace{0.13in} %% TODO
%
$
\inferrule* %%[lab=\ruleNameFig{}]
  {\isIndet{\highlightIndet{$\varResult_1$}}
   %% \\\\
   \sepPremise
   \isFinal{\varResult_2}
  }
  {\isIndet{\expAppIndet{\varResult_1}{\varResult_2}}}
$
\hsepRule
%% \hspace{0.13in} %% TODO
%
$
\inferrule* %%[lab=\ruleNameFig{}]
  {\isIndet{\highlightIndet{$\varResult$}}}
  {\isIndet{\expProjIndet{i}{\varResult}}}
$
%
%% \hsepRule
%% \hspace{0.13in} %% TODO
%

\vsepRule

$
\inferrule* %%[lab=\ruleNameFig{}]
  {\isFinal{\varEnv}
   %% \sepPremise
   \isIndet{\highlightIndet{$\varResult$}}
  }
  %% {\isIndet{\closure{\varEnv}{\expMatch{\varResult}{\varDataCon_i}{\varVar_i}{\varExp_i}}}}
  {\isIndet{\caseClosureIndet{\varEnv}{\varResult}{\varDataCon_i}{\varVar_i}{\varExp_i}}}
$

%% \vsepRule
%% 
%% \judgementHead
%%   {(Simple) Examples}
%%   {\isExample{\varResult}}
%% 
%% \vsepRule
%% 
%% $
%% \inferrule* %%[lab=\ruleNameFig{}]
%%   {
%%   }
%%   {\isExample{\expUnit}}
%% $
%% %
%% \hsepRule
%% %
%% $
%% \inferrule* %%[lab=\ruleNameFig{}]
%%   {\isExample{\varResult_i}}
%%   {\isExample{\pair{\varResult_1}{\varResult_2}}}
%% $
%% %
%% \hsepRule
%% %
%% $
%% \inferrule* %%[lab=\ruleNameFig{}]
%%   {\isExample{\varResult}}
%%   {\isExample{\expApp{\varDataCon}{\varResult}}}
%% $

\vsepBeforeCaption

%% \caption{Predicates regarding evaluation.}
\caption{Result Classification. Final results are determinate or indeterminate.}
\label{fig:final}
\end{figure}

\vsepRule

\vsepRule

\parahead{Examples}

We define three simple functions below.
The coercion $\liftSimpleEx{\varSimpleVal}$ ``upcasts'' a simple value to a
result.
The coercion $\coerceUndet{\varResult}{\varSimpleVal}$
``downcasts'' a result to a simple value.
The $\filterWorlds{\varWorlds}$ function removes top example constraints.

$$
\inferrule*
  {
  }
  {\coerceUndet
    {\expUnit}
    {\expUnit}}
\hsepRule
\inferrule*
  %% {\coerceUndet{\varResult_i}{\varSimpleVal_i}
  {\coerceUndet{\varResult_1}{\varSimpleVal_1}
   \sepPremise
   \coerceUndet{\varResult_2}{\varSimpleVal_2}
  }
  {\coerceUndet
    {\pair{\varResult_1}{\varResult_2}}
    {\pair{\varSimpleVal_1}{\varSimpleVal_2}}}
\hsepRule
\inferrule*
  {\coerceUndet{\varResult}{\varSimpleVal}
  }
  {\coerceUndet
    {\expApp{\varDataCon}{\varResult}}
    {\expApp{\varDataCon}{\varSimpleVal}}}
$$

$$
\filterWorlds{\varWorlds} =
  \setComp{\world{\varEnv}{\varEx} \in \varWorlds}
          {\varEx \neq \exHole}
$$

\vsepRule

%% \clearpage

\subsection{Type Checking}
\label{sec:appendix-type-checking}

\autoref{fig:result-typing} defines type checking for expressions, results, and
examples.
The result type checking $\typeCheckResult{\varResult}{\varType}$ and example
type checking $\typeCheckExample{\varEx}{\varType}$ judgements
do not require a type context $\varTypeEnv$ because results and expressions do
not contain free variables.
Result typing refers to expression typing because function closures and case
closures contain expressions and evaluation environments.
\autoref{fig:other-typing} defines type checking for constraints, solutions,
programs, and assertions.
\\

%% \begin{figure}[t]
\begin{figure}[h]

\input{fig-typing}

\vsepRule

\judgementHead
  {\maybeUnderline{R}esult \maybeUnderline{T}yping}
  {\typeCheckResult{\varResult}{\varType}}

\vsepRule

$
\inferrule*[lab=\ruleNameFig{RT-Fix}]
  {
   \typeCheckEnv{\varTypeEnv}{\varEnv}
   \sepPremise
   {\typeCheck
     {\varTypeEnv}
     {\expFixFun{\varVarF}{BLAH}{\varVar}{\varExp}}
     {\varType}}
  }
  {\typeCheckResult
    {\closure{\varEnv}{\expFixFun{\varVarF}{BLAH}{\varVar}{\varExp}}}
    {\varType}}
$
\hsepRule
$
\inferrule*[lab=\ruleNameFig{RT-Hole}]
  {
   \varHoleEnv(\expHole{\varHoleName}) = \tHole{\varTypeEnv}{\varType}
   \sepPremise
   \typeCheckEnv{\varTypeEnv}{\varEnv}
  }
  {\typeCheckResult
    {\closure{\varEnv}{\expHole{\varHoleName}}}
    {\varType}}
$

\vsepRule

$
\inferrule*[lab=\ruleNameFig{RT-Unit}]
  {
  }
  {\typeCheckResult
    {\expUnit}
    {\tUnit}}
$
\hsepRule
$
%% \inferrule*[lab=\ruleNameFig{RT-Tuple}]
\inferrule*[lab=\ruleNameFig{RT-Pair}]
  {
   \multiPremise{
   \typeCheckResult{\varResult_i}{\varType_i}
   }{\pairIndex{i}}
  }
  {\typeCheckResult
    {\pair{\varResult_1}{\varResult_2}}
    {\pair{\varType_1}{\varType_2}}}
    %% {\triple{\varResult_1}{\ldots}{\varResult_n}}
    %% {\triple{\varType_1}{\ldots}{\varType_n}}}
$
\hsepRule
$
\inferrule*[lab=\ruleNameFig{RT-Ctor}]
  {
   \lookupDataConArrowType{\varDataCon}{\varType}{\varTypeCon}
   \sepPremise
   \typeCheckResult{\varResult}{\varType}
  }
  {\typeCheckResult
    {\expApp{\varDataCon}{\varResult}}
    {\varTypeCon}}
$

\vsepRule

$
\inferrule*[lab=\ruleNameFig{RT-App}]
  {
   \typeCheckResult{\varResult_1}{\tArrow{\varType_2}{\varType}}
   \\\\ % \sepPremise
   \typeCheckResult{\varResult_2}{\varType_2}
  }
  {\typeCheckResult
    {\expApp{\varResult_1}{\varResult_2}}
    {\varType}}
$
\hsepRule
$
\inferrule*[lab=\ruleNameFig{RT-Prj}]
  {
   %% \typeCheckResult{\varResult}{\triple{\varType_1}{\ldots}{\varType_n}}
   \typeCheckResult{\varResult}{\tPair{\varType_1}{\varType_2}}
  }
  {\typeCheckResult
    {\expProj{i}{\varResult}}
    {\varType_i}}
$
\hsepRule
$
\inferrule*[lab=\ruleNameFig{RT-Case}]
  {
   \typeCheckResult{\varResult}{\varTypeCon}
   \sepPremise
   \lookupTypeConstructors{\varTypeCon}{\varDataCon_i}{\varType_i}
   \\\\
   \typeCheckEnv{\varTypeEnv}{\varEnv}
   \sepPremise
   \multiPremise{
   \typeCheck
     {\envCat{\varTypeEnv}{\envBindType{\varVar_i}{\varType_i}}}
     {\varExp_i}
     {\varType}
   }{\sequenceSyntax}
  }
  {\typeCheckResult
    {\closure{\varEnv}{\expMatch{\varResult}{\varDataCon_i}{\varVar_i}{\varExp_i}}}
    {\varType}}
$

\vsepRule

%% TODO
%% \rkc{TODO rule for unwrap ctor}

%% \dimColor{
%% $
%% \inferrule*[lab=\ruleNameFig{RT-Fun}]
%%   {
%%    \typeCheckEnv{\varTypeEnv}{\varEnv}
%%    \sepPremise
%%    \typeCheck
%%      {\envCat{\varTypeEnv}{\envBindType{\varVar}{\varType_1}}}
%%      {\varExp}
%%      {\varType_2}
%%   }
%%   {\typeCheckResult
%%     {\closure{\varEnv}{\expFun{\varVar}{\varExp}}}
%%     {\tArrow{\varType_1}{\varType_2}}}
%% $
%% }
%% 
%% \vsepRule

\judgementHead
  {Environment Typing}
  {\typeCheckEnv{\varTypeEnv}{\varEnv}}

$
\inferrule* %%[lab=\ruleNameFig{}]
  {
  }
  {\typeCheckEnv
    {\emptyEnv}
    {\emptyEnv}}
$
\hsepRule
$
\inferrule* %%[lab=\ruleNameFig{}]
  {
   \typeCheckEnv{\varTypeEnv}{\varEnv}
   \sepPremise
   \typeCheckResult{\varResult}{\varType}
  }
  {\typeCheckEnv
    {(\envCat{\varTypeEnv}{\envBindType{\varVar}{\varType}})}
    {(\envCat{\varEnv}{\envBind{\varVar}{\varResult}})}}
$

\vsepRule

\input{fig-ex-typing}

%% \vsepRule

\vsepBeforeCaption

%% \caption{Result Type Checking.}
\caption{Expression, Result, and Example Type Checking.}
\label{fig:typing}
\label{fig:result-typing}
\end{figure}

\begin{figure}[h]

\judgementHeadNameOnly
  %% {Example Constraints, Unsolved Constraints, and Solution Typing}
  {Example, Unsolved Con., and Solution Typing}
  {\JudgementBox{\typeCheckWorlds{\varHoleEnv}{\varWorlds}{\varTypeEnv}{\varType}}
   \JudgementBox{\typeCheckSolution{\varHoleEnv}{\varUnsolvedConstraints}}
   \JudgementBox{\typeCheckSolution{\varHoleEnv}{\varSolution}}}

\vsepRule

$
\inferrule*
  {
   \multiPremise{
   \typeCheckEnv{\varTypeEnv}{\varEnv_i}
   \sepPremise
   \typeCheckExample{\varEx_i}{\varType}
   }{\sequenceSyntax}
  }
  {\typeCheckWorlds
     {\varHoleEnv}
     {\envCatThree{\world{\varEnv_1}{\varEx_1}}{\ldots}{\world{\varEnv_n}{\varEx_n}}}
     {\varTypeEnv}
     {\varType}}
$
%
%% \hsepRule

\vsepRule
$
\inferrule*
  {
   \multiPremise{
   \varHoleEnv(\expHole{\varHoleName_i}) = \tHole{\varTypeEnv_i}{\varType_i}
   \sepPremise
   \typeCheckWorlds{\varHoleEnv}{\varWorlds_i}{\varTypeEnv_i}{\varType_i}
   }{\sequenceSyntax}
  }
  {\typeCheckUnsolved
     {\varHoleEnv}
      {(\envCatThree
         {\unsolvedConstraint{\varHoleName_1}{\varWorlds_1}}
         {\ldots}
         {\unsolvedConstraint{\varHoleName_n}{\varWorlds_n}})}}
$

\vsepRule

$
\inferrule*
  {
   \multiPremise{
   \varHoleEnv(\expHole{\varHoleName_i}) = \tHole{\varTypeEnv_i}{\varType_i}
   \sepPremise
   \typeCheck{\varTypeEnv_i}{\varExp_i}{\varType_i}
   }{\sequenceSyntax}
  }
  {\typeCheckSolution
     {\varHoleEnv}
      {(\envCatThree
         {\holeFilling{\varHoleName_1}{\varExp_1}}
         {\ldots}
         {\holeFilling{\varHoleName_n}{\varExp_n}})}}
$

\vsepRule

\judgementHeadNameOnly
  {Program and Assertion Typing}
  {\JudgementBox{\typeCheckProgramWrapper{\varProgram}{\varType}}
   \JudgementBox{\typeCheckProgram{\varProgram}{\varType}{\varType'}}
   \JudgementBox{\typeCheckAssertions{\varHoleEnv}{\varAssertions}}}

\vsepRule

$
\inferrule*
  {\typeCheckProgram{\expProgram{\varExp}{\varExp_1}{\varExp_2}}{\varType}{\varType'}}
  {\typeCheckProgramWrapper{\expProgram{\varExp}{\varExp_1}{\varExp_2}}{\varType}}
$

\vsepRule

$
\inferrule*
  {
   \typeCheck{\emptyEnv}{\varExp}{\varType}
   \sepPremise
   \multiPremise{\typeCheck{(\envBindType{\texttt{main}}{\varType})}{\varExp_i}{\varType'}}{\pairIndex{i}}
  }
  {\typeCheckProgram{\expProgram{\varExp}{\varExp_1}{\varExp_2}}{\varType}{\varType'}}
$

\vsepRule
$
\inferrule*
  {
   \multiPremise{
     \exists ~\varType
     \sepPremise
     \typeCheckResult{\varResult_i}{\varType}
     \sepPremise
     \typeCheckResult{\varSimpleVal_i}{\varType}
     }{\sequenceSyntax}
  }
  {\typeCheckAssertions{\varHoleEnv}{\multiPremise{\assertion{\varResult_i}{\varSimpleVal_i}}{\sequenceSyntax}}}
$

\vsepBeforeCaption

%% \caption{Result Type Checking.}
\caption{Constraint, Solution, Program, and Assertion Type Checking.}
\label{fig:other-typing}
\end{figure}

\clearpage

\subsection{Type Soundness}
\label{sec:appendix-type-soundness}

The progress property is complicated by the fact that, in a big-step semantics,
non-terminating computations are not necessarily distinguished from stuck
ones~\citep{Leroy:2009}.
Using a technique similar to that described by \citet{Ancona:2014}, we augment
evaluation with a natural $\varFuel$ that limits the beta-reduction depth of an
evaluation derivation.
The augmented evaluation judgment
$\reducesToN{\varEnv}{\varExp}{\varResult}{\varFuel}$
(\autoref{fig:eval-n}) asserts that evaluation produced a particular result or
that it reached the specified depth before doing so.

\autoref{fig:eval-n} shows how the evaluation judgment can be augmented
to add \emph{fuel} that limits the depth of beta reductions that can occur during
evaluation. Note that for simplicity, the fuel is only depleted in recursive
invocations that extend the environment. Also note that this relation is
exactly the same as the ordinary evaluation relation, except for the
beta-depth-limit $\varFuel$. As such, a progress theorem proven over this
relation reflects the properties of the original evaluation relation.
\\

%% \begin{figure}[t]
%% \begin{figure}[b]
\begin{figure}[h]

\judgementHead
  %% {Beta-Depth-\maybeUnderline{L}imited \maybeUnderline{E}valuation}
  {\maybeUnderline{A}ugmented \maybeUnderline{E}valuation}
  {\reducesToN{\varEnv}{\varExp}{\varResult}{\varFuel}}

\vsepRule

$
\inferrule*[lab=\ruleNameFig{E-Hole}]
  {
  }
  {\reducesToN
    {\varEnv}
    {\expHole{\varHoleName}}
    {\closure{\varEnv}{\expHole{\varHoleName}}}
    {{\varFuel}}}
$
\hsepRule
$
\inferrule*[lab=\ruleNameFig{E-Limit}]
  {
  }
  {\reducesToN{\varEnv}{\varExp}{\varResult}{0}}
$

\vsepRule

$
\inferrule*[lab=\ruleNameFig{E-Fix}]
  {
   \varExp = {\valFixFun{\varVarF}{\varVar}{\varExp}}
  }
  {\reducesToN
    {\varEnv}
    %% {\expFixFun{\varVarF}{\tArrow{\varType_1}{\varType_2}}{\varVar}{\varExp}}
    %% {\closure{\varEnv}{\valFixFun{\varVarF}{\varVar}{\varExp}}}
    {\varExp}
    {\closure{\varEnv}{\varExp}}
    {{\varFuel}}}
$
\hsepRule
$
\inferrule*[lab=\ruleNameFig{E-Var}]
  {
   \envBind{\varVar}{\varResult} \in \varEnv
   %% \envBindBoth{\varVar}{\varResult}{\varType} \in \varEnv
  }
  {\reducesToN{\varEnv}{\varVar}
    {\varResult}
    {{\varFuel}}}
$
\hsepRule
$
\inferrule*[lab=\ruleNameFig{E-Unit}]
  {
  }
  {\reducesToN
    {\varEnv}
    {\expUnit}
    {\tUnit}
    {{\varFuel}}}
$
%
%% \hsepRule

\vsepRule

$
%% \inferrule*[lab=\ruleNameFig{E-Tuple}]
\inferrule*[lab=\ruleNameFig{E-Pair}]
  {
   \multiPremise{\reducesToN{\varEnv}{\varExp_i}{\varResult_i}{{\varFuel}}}
                {\sequenceSyntaxTwo{i}}
  }
  {\reducesToN
    {\varEnv}
    {\pair{\varExp_1}{\varExp_2}}
    {\pair{\varResult_1}{\varResult_2}}
    {{\varFuel}}}
    %% {\triple{\varExp_1}{\ldots}{\varExp_n}}
    %% {\triple{\varResult_1}{\ldots}{\varResult_n}}}
$
\hsepRule
$
\inferrule*[lab=\ruleNameFig{E-Ctor}]
  {
   \reducesToN{\varEnv}{\varExp}{\varResult}{{\varFuel}}
  }
  {\reducesToN{\varEnv}{\expApp{\varDataCon}{\varExp}}{\expApp{\varDataCon}{\varResult}}{{\varFuel}}}
$

\vsepRule

$
\inferrule*[lab=\ruleNameFig{E-App}]
  {
   %% \reducesTo{\varEnv}{\varExp_1}{\closure{\varEnv_f}{\valFixFun{\varVarF}{\varVar}{\varExp_f}}}
   %% \reducesTo{\varEnv}{\varExp_1}{\varVal_1}
   \reducesToN{\varEnv}{\varExp_1}{\varResult_1}{{\varFuel}}
   \sepPremise
   %% \\\\
   \reducesToN{\varEnv}{\varExp_2}{\varResult_2}{{\varFuel}}
   %% \varVal_1 = \closure{\varEnv_f}{\valFixFun{\varVarF}{\varVar}{\varExp_f}}
   \\\\
   \varResult_1 = \closure{\varEnv_f}{\valFixFun{\varVarF}{\varVar}{\varExp_f}}
   \sepPremise
   %% \\\\
   \reducesToN{\envCatThree{\varEnv_f}
                          %% {\envBind{\varVarF}{\closure{\varEnv_f}{\valFixFun{\varVarF}{\varVar}{\varExp_f}}}}
                          %% {\envBind{\varVarF}{\varVal_1}}
                          {\envBind{\varVarF}{\varResult_1}}
                          {\envBind{\varVar}{\varResult_2}}}
             {\varExp_f}
             {\varResult}
             {{\varFuel - 1}}
  }
  {\reducesToN{\varEnv}{\expApp{\varExp_1}{\varExp_2}}{\varResult}{{\varFuel}}}
$
\hsepRule
$
\inferrule*[lab=\ruleNameFig{E-App-Indet}]
  {
   \reducesToN{\varEnv}{\varExp_1}{\varResult_1}{{\varFuel}}
   \sepPremise
   %% \\\\
   \reducesToN{\varEnv}{\varExp_2}{\varResult_2}{{\varFuel}}
   \\\\
   \varResult_1 \neq \closure{\varEnv_f}{\valFixFun{\varVarF}{\varVar}{\varExp_f}}
   %% \dimColor{\varResult_1 \neq \closure{\varEnv_f}{\expFun{\varVar}{\varExp_f}}}
  }
  {\reducesToN{\varEnv}{\expApp{\varExp_1}{\varExp_2}}{\expApp{\varResult_1}{\varResult_2}}{{\varFuel}}}
$

\vsepRule

$
%% \inferrule*[lab=\ruleNameFig{E-Get}]
\inferrule*[lab=\ruleNameFig{E-Prj}]
  {
   %% \reducesTo{\varEnv}{\varExp}{\triple{\varResult_1}{\ldots}{\varResult_n}}
   \reducesToN{\varEnv}{\varExp}{\pair{\varResult_1}{\varResult_2}}{{\varFuel}}
  }
  {\reducesToN{\varEnv}{\expProj{i}{\varExp}}{\varResult_i}{{\varFuel}}}
$
\hsepRule
$
%% \inferrule*[lab=\ruleNameFig{E-Get-Indet}]
\inferrule*[lab=\ruleNameFig{E-Prj-Indet}]
  {
   \reducesToN{\varEnv}{\varExp}{\varResult}{{\varFuel}}
   \sepPremise
   %% \varResult \neq \triple{\varResult_1}{\ldots}{\varResult_n}
   \varResult \neq \pair{\varResult_1}{\varResult_2}
  }
  {\reducesToN{\varEnv}{\expProj{i}{\varExp}}{\expProj{i}{\varResult}}{{\varFuel}}}
$

\vsepRule

$
\inferrule*[lab=\ruleNameFig{E-Case}]
  {
   %% \exists j \in [1,n]
   j \in [1,n]
   \sepPremise
   %% \\\\
   \reducesToN{\varEnv}{\varExp}{\expApp{\varDataCon_j}{\varResult}}{{\varFuel}}
   %% \sepPremise
   \\\\
   \reducesToN{\envCat{\varEnv}{\envBind{\varVar_j}{\varResult}}}{\varExp_j}{\varResult_j}{{\varFuel - 1}}
  }
  {\reducesToN
    {\varEnv}
    {\expMatch{\varExp}{\varDataCon_i}{\varVar_i}{\varExp_i}}
    {\varResult_j}
    {{\varFuel}}}
$
\hsepRule
%% \hfill %% TODO
%
$
\inferrule*[lab=\ruleNameFig{E-Case-Indet}]
  {
   \reducesToN{\varEnv}{\varExp}{\varResult}{{\varFuel}}
   \sepPremise
   %% \\\\
   %% \varResult \neq \expApp{\varDataCon_i}{\varResult}
   \not\exists j \in [1,n], \varResult_j \ s.t.\ 
   \varResult = \expApp{\varDataCon_j}{\varResult_j}
   \\\\
   \varResult' = 
     \closure{\varEnv}{\expMatch{\varResult}{\varDataCon_i}{\varVar_i}{\varExp_i}}
  }
  {\reducesToN
    {\varEnv}
    {\expMatch{\varExp}{\varDataCon_i}{\varVar_i}{\varExp_i}}
    {\varResult'}
    {{\varFuel}}}
    %% {\closure{\varEnv}{\expMatch{\varResult}{\varDataCon_i}{\varVar_i}{\varExp_i}}}}
$

\vsepBeforeCaption

%% \vsepRule
%% \caption{Big-step, environment-style evaluation.}
%% \caption{Evaluation augmented with a beta-depth-limit}

%% TODO \ruleName in caption...

\caption{Augmented Evaluation with beta-depth limit.
Only {E-App} and {E-Case} decrease the depth parameter.
}
\label{fig:eval-n}
\end{figure}

%\input{fig-fails}

%% \vsepRule

\clearpage

\begin{theorem}[Determinism of Evaluation]
\label{prop:eval-det}

\breakAndIndent
If $\reducesTo{\varEnv}{\varExp}{\varResult}$
and $\reducesTo{\varEnv}{\varExp}{\varResult'}$,
then $\varResult=\varResult'$.

\end{theorem}

\begin{theorem}[Finality of Evaluation]
\label{prop:eval-fin}

\breakAndIndent
If $\isFinal{\varEnv}$
and $\reducesTo{\varEnv}{\varExp}{\varResult}$,
then $\isFinal{\varResult}$.

\end{theorem}

Type checking and evaluation are related by the following properties.

\begin{theorem}[Type Preservation]
\label{prop:pres}

\breakAndIndent
If $\typeCheck{\varTypeEnv}{\varExp}{\varType}$
and $\typeCheckEnv{\varTypeEnv}{\varEnv}$
and $\reducesTo{\varEnv}{\varExp}{\varResult}$,
then $\typeCheckResult{\varResult}{\varType}$.

\end{theorem}

\beginTheorem{Prog}{Progress}{Progress}{prop:prog}

\breakAndIndent
For all $\varFuel$,
if $\typeCheck{\varTypeEnv}{\varExp}{\varType}$
and $\typeCheckEnv{\varTypeEnv}{\varEnv}$,
there exists $\varResult$ s.t.
$\reducesToN{\varEnv}{\varExp}{\varResult}{\varFuel}$
and $\typeCheckResult{\varResult}{\varType}$.
\end{theorem}

\subsection*{Proofs}

\vspace{0.5em}

\jtheorem{
  \jtref{\autoref{prop:eval-det}},
  \jtref{\autoref{prop:eval-fin}}, and
  \jtref{\autoref{prop:pres}}
}{
  Straightforward induction.
}

\jtheorem{
  \jtref{\autorefProgLong{}}
}{
  When $\varFuel = 0$, \ruleName{E-Limit} will go through for any result.  From
  the premise that $\varExp$ is well-typed
  ($\typeCheck{\varTypeEnv}{\varExp}{\varType}$), it is straightforward to
  derive a result of the same type.  When $\varFuel > 0$, the remaining cases go
  through by straightforward induction, thanks to the natural semantics.
}

\clearpage

\subsection{Resumption}
\label{sec:appendix-resumption}

\autoref{fig:resumption} defines how to resume partially evaluated expressions.
Resumption does not require an evaluation environment $\varEnv$ because results
do not contain free variables.

The definitions of \ruleName{R-Hole-Resume} and \ruleName{R-Hole-Indet}
below are slightly more complicated than the versions discussed in
\autoref{sec:live-eval}:
to account for the \ruleName{Defer} hole synthesis rule defined in
\autoref{sec:constraint-solving}, the rules below check whether
$\varSolution(\varHoleName)$ equals $\expHole{\varHoleName}$.
\\

%% \begin{figure}[t]
%% \begin{figure}[p]
\begin{figure}[h]

%% TURN OFF ASSERTS
\renewcommand{\reducesToUK}[4]
  {\ensuremath{\program{#1}{#2}\hspace{0.01in}\Rightarrow{#3}}}
\renewcommand{\resumesToUK}[4]
  {\ensuremath{\program{#1}{#2}\hspace{0.01in}\Rightarrow{#3}}}

\judgementHead
  {\maybeUnderline{R}esumption}
  {\resumesToUK{\varSolution}{\varResult}{\varResult'}{\varAssertions}}

\vsepRule

$
\inferrule*[lab=\ruleNameFig{R-Hole-Resume}]
  {
   \varSolution(h) = \varExp_h
   \sepPremise
   \varExp_h \not= \expHole{h}
   %% \sepPremise
   %% \rkc{\resumesTo{\varSolution}{\varEnv}{\varEnv'} before eval?}
   \sepPremise
   \reducesToUK{\varEnv}{\varExp_h}{\varResult}{\varAssertions}
   \sepPremise
   \resumesToUK{\varSolution}{\varResult}{\varResult'}{\varAssertions'}
  }
  {\resumesToUK{\varSolution}
               {\closure{\varEnv}{\expHole{h}}}
               {\varResult'}
               {\setUnion{\varAssertions}{\varAssertions'}}}
$
\hfill
$
\inferrule*[lab=\ruleNameFig{R-Hole-Indet}]
  {
   h \notin dom(\varSolution) \lor \varSolution(h) = \expHole{h}
   \sepPremise
   \resumesToUK{\varSolution}{\varEnv}{\varEnv'}{\varAssertions}
  }
  {\resumesToUK
    {\varSolution}
    {\closure{\varEnv}{\expHole{h}}}
    {\closure{\varEnv'}{\expHole{h}}}
    {\varAssertions}}
$

\vsepRule

$
\inferrule*[lab=\ruleNameFig{R-Fix}]
  {
   %% \varResult =
   %%   \closure{\varEnv}{\valFixFun{\varVarF}{\varVar}{\varExp}}
   %% \\\\
   \resumesToUK{\varSolution}{\varEnv}{\varEnv'}{\varAssertions}
   %% \\\\
   %% \varResult' =
   %%   \closure{\varEnv'}{\valFixFun{\varVarF}{\varVar}{\varExp}}
  }
  {\resumesToUK
    {\varSolution}
    %% {\varResult}
    %% {\varResult'}
    {\closure{\varEnv}{\valFixFun{\varVarF}{\varVar}{\varExp}}}
    {\closure{\varEnv'}{\valFixFun{\varVarF}{\varVar}{\varExp}}}
    {\varAssertions}
  }
$
\hsepRule
$
\inferrule*[lab=\ruleNameFig{R-Unit}]
  {
  }
  {\resumesToUK
    {\varSolution}
    {\expUnit}
    {\tUnit}
    {\emptySet}
  }
$
%
%% \hsepRule

\vsepRule
$
%% \inferrule*[lab=\ruleNameFig{R-Tuple}]
\inferrule*[lab=\ruleNameFig{R-Pair}]
  {
   %% \resumesTo{\varSolution}{\varResult_i}{\varResult'_i}
   \resumesToUK{\varSolution}{\varResult_1}{\varResult'_1}{\varAssertions_1}
   \sepPremise
   \resumesToUK{\varSolution}{\varResult_2}{\varResult'_2}{\varAssertions_2}
  }
  {\resumesToUK
    {\varSolution}
    {\pair{\varResult_1}{\varResult_2}}
    {\pair{\varResult'_1}{\varResult'_2}}
    {\setUnion{\varAssertions_1}{\varAssertions_2}}}
    %% {\triple{\varResult_1}{\ldots}{\varResult_n}}
    %% {\triple{\varResult'_1}{\ldots}{\varResult'_n}}}
$
\hsepRule
$
\inferrule*[lab=\ruleNameFig{R-Ctor}]
  {
   \resumesToUK{\varSolution}{\varResult}{\varResult'}{\varAssertions}
  }
  {\resumesToUK{\varSolution}{\expApp{\varDataCon}{\varResult}}{\expApp{\varDataCon}{\varResult'}}{\varAssertions}}
$

\vsepRule

$
\inferrule*[lab=\ruleNameFig{R-App}]
  {
   %% \resumesTo{\varSolution}{\varExp_1}{\varVal_1}
   \resumesToUK{\varSolution}{\varResult_1}{\varResult_1'}{\varAssertions_1}
   \sepPremise
   \resumesToUK{\varSolution}{\varResult_2}{\varResult_2'}{\varAssertions_2}
   \sepPremise
   \varResult_1' = \closure{\varEnv_f}{\valFixFun{\varVarF}{\varVar}{\varExp_f}}
   \\\\
   %% \varVal_1 = \closure{\varEnv_f}{\valFixFun{\varVarF}{\varVar}{\varExp_f}}
   \reducesToUK{\envCatThree{\varEnv_f}
                          %% {\envBind{\varVarF}{\varVal_1}}
                          {\envBind{\varVarF}{\varResult_1'}}
                          {\envBind{\varVar}{\varResult_2'}}
               }
             {\varExp_f}{\varResult}{\varAssertions_f}
   \sepPremise
   \resumesToUK{\varSolution}{\varResult}{\varResult'}{\varAssertions'}
  }
  {\resumesToUK
      {\varSolution}
      {\expApp{\varResult_1}{\varResult_2}}
      {\varResult'}
      {\setUnionFour{\varAssertions_1}{\varAssertions_2}{\varAssertions_f}{\varAssertions'}}
  }
$
\hsepRule
$
\inferrule*[lab=\ruleNameFig{R-App-Indet}]
  {
   \resumesToUK{\varSolution}{\varResult_1}{\varResult'_1}{\varAssertions_1}
   \sepPremise
   \resumesToUK{\varSolution}{\varResult_2}{\varResult'_2}{\varAssertions_2}
   \\\\
   \varResult'_1 \neq \closure{\varEnv_f}{\valFixFun{\varVarF}{\varVar}{\varExp_f}}
   %% \dimColor{\varResult'_1 \neq \closure{\varEnv_f}{\expFun{\varVar}{\varExp_f}}}
  }
  {\resumesToUK
     {\varSolution}
     {\expApp{\varResult_1}{\varResult_2}}
     {\expApp{\varResult'_1}{\varResult'_2}}
     {\setUnion{\varAssertions_1}{\varAssertions_2}}
  }
$

\vsepRule

$
%% \inferrule*[lab=\ruleNameFig{R-Get}]
\inferrule*[lab=\ruleNameFig{R-Prj}]
  {
   %% \resumesTo{\varSolution}{\varResult}{\triple{\varResult_1}{\ldots}{\varResult_n}}
   \resumesToUK{\varSolution}{\varResult}{\pair{\varResult_1}{\varResult_2}}{\varAssertions}
  }
  {\resumesToUK{\varSolution}{\expProj{i}{\varResult}}{\varResult_i}{\varAssertions}}
$
\hsepRule
$
%% \inferrule*[lab=\ruleNameFig{R-Get-Indet}]
\inferrule*[lab=\ruleNameFig{R-Prj-Indet}]
  {
   \resumesToUK{\varSolution}{\varResult}{\varResult'}{\varAssertions}
   \sepPremise
   %% \varResult' \neq \triple{\varResult_1}{\ldots}{\varResult_n}
   \varResult' \neq \pair{\varResult_1}{\varResult_2}
  }
  {\resumesToUK{\varSolution}{\expProj{i}{\varResult}}{\expProj{i}{\varResult'}}{\varAssertions}}
$

\vsepRule

%% $
%% \dimColor{
%% \inferrule*[lab=\ruleNameFig{R-Match}]
%%   {
%%    \exists j \in [1,n]
%%    \sepPremise
%%    \resumesTo{\varSolution}{\varResult}{\expApp{\varDataCon_j}{\varResult'}}
%%    \\\\
%%    %% \reducesTo{\envCat{\varEnv}{\envBind{\varVar_j}{\varResult'}}}{\varExp_j}{\varResult_j}
%%    \reducesTo{\envCat{\varEnv}{\envBind{\varVar_j}{\expUnwrap{\varDataCon_j}{\varResult}}}}{\varExp_j}{\varResult_j}
%%    \sepPremise
%%    \resumesTo{\varSolution}{\varResult_j}{\varResult_j'}
%%   }
%%   {\resumesTo
%%     {\varSolution}
%%     {\closure{\varEnv}{\expMatch{\varResult}{\varDataCon_i}{\varVar_i}{\varExp_i}}}
%%     {\varResult_j'}}
%% }
%% $
%

$
\inferrule*[lab=\ruleNameFig{R-Case}]
  {
   \exists j \in [1,n]
   \sepPremise
   \resumesToUK{\varSolution}{\varResult}{\expApp{\varDataCon_j}{\varResult'}}{\varAssertions}
   \\\\
   \resumesToUK
     {\varSolution}
     {\expApp{(\closure{\varEnv}{\expFun{\varVar_j}{\varExp_j}})}{\varResult'}}
     {\varResult_j}
     {\varAssertions'}
  }
  {\resumesToUK
    {\varSolution}
    {\closure{\varEnv}{\expMatch{\varResult}{\varDataCon_i}{\varVar_i}{\varExp_i}}}
    {\varResult_j}
    {\setUnion{\varAssertions}{\varAssertions'}}
  }
$
%% \hsepRule
\hfill %% TODO
$
\inferrule*[lab=\ruleNameFig{R-Case-Indet}]
  {
   \resumesToUK{\varSolution}{\varResult}{\varResult'}{\varAssertions}
   \sepPremise
   \not\exists j \in [1,n], \varResult_j \ s.t.\ 
   \varResult' = \expApp{\varDataCon_j}{\varResult_j}
   \\\\
   \resumesToUK{\varSolution}{\varEnv}{\varEnv'}{\varAssertions'}
   \sepPremise
   \varResult'' =
     \closure{\varEnv'}{\expMatch{\varResult'}{\varDataCon_i}{\varVar_i}{\varExp_i}}
  }
  {\resumesToUK
    {\varSolution}
    {\closure{\varEnv}{\expMatch{\varResult}{\varDataCon_i}{\varVar_i}{\varExp_i}}}
    {\varResult''}
    {\setUnion{\varAssertions}{\varAssertions'}}
  }
    %% {\closure{\varEnv}{\expMatch{\varResult'}{\varDataCon_i}{\varVar_i}{\varExp_i}}}}
$

%% \vsepRule
%% 
%% \dimColor{
%% $
%% \inferrule*[lab=\ruleNameFig{R-Fun}]
%%   {
%%   }
%%   {\resumesTo
%%     {\varSolution}
%%     {\closure{\varEnv}{\expFun{\varVar}{\varExp}}}
%%     {\closure{\varEnv}{\expFun{\varVar}{\varExp}}}}
%% $
%% }
%% %
%% \hsepRule
%% %
%% \dimColor{
%% $
%% \inferrule*[lab=\ruleNameFig{R-App}]
%%   {
%%    \resumesTo{\varSolution}{\varResult_1}{\closure{\varEnv_f}{\expFun{\varVar}{\varExp_f}}}
%%    \sepPremise
%%    \resumesTo{\varSolution}{\varResult_2}{\varResult'_2}
%%    \\\\
%%    \reducesTo{\envCat{\varEnv_f}{\envBind{\varVar}{\varResult'_2}}}{\varExp_f}{\varResult}
%%   }
%%   {\resumesTo{\varSolution}{\expApp{\varResult_1}{\varResult_2}}{\varResult}}
%% $
%% }

\vsepRule

$
\inferrule*[lab=\ruleNameFig{R-Unwrap-Ctor}]
  {
   \resumesToUK{\varSolution}{\varResult}{\expApp{\varDataCon_j}{\varResult_j}}{\varAssertions}
  }
  {\resumesToUK
    {\varSolution}
    {\expUnwrap{\varDataCon_j}{\varResult}}
    {\varResult_j}
    {\varAssertions}}
$
\hsepRule
$
\inferrule*[lab=\ruleNameFig{R-Unwrap-Ctor-Indet}]
  {
   \resumesToUK{\varSolution}{\varResult}{\varResult'}{\varAssertions}
   \sepPremise
   %% \isIndet{\varResult'}
   \varResult' \neq \expApp{\varDataCon_i}{\varResult_i} \textrm{ (for any $i$)}
  }
  {\resumesToUK
    {\varSolution}
    {\expUnwrap{\varDataCon}{\varResult}}
    {\expUnwrap{\varDataCon}{\varResult'}}
    {\varAssertions}}
$

\vsepRule

\judgementHead
  {\maybeUnderline{E}nvironment \maybeUnderline{R}esumption}
  {\resumesToUK{\varSolution}{\varEnv}{\varEnv'}{\varAssertions}}

\vsepRuleNoNeed

$
\inferrule* %% [lab=\ruleNameFig{Blah}]
  {
  }
  {\resumesToUK
    {\varSolution}
    {\emptyEnv}
    {\emptyEnv}
    {\emptySet}}
$
\hsepRule
$
\inferrule* %% [lab=\ruleNameFig{Blah}]
  {
   \resumesToUK{\varSolution}{\varEnv}{\varEnv'}{\varAssertions}
   \sepPremise
   \resumesToUK{\varSolution}{\varResult}{\varResult'}{\varAssertions'}
  }
  {\resumesToUK
    {\varSolution}
    {\envCat{\varEnv}{\envBind{\varVar}{\varResult}}}
    {\envCat{\varEnv'}{\envBind{\varVar}{\varResult'}}}
    {\setUnion{\varAssertions}{\varAssertions'}}
  }
$

%% \vsepRule

%% \judgementHeadThree
%%   {\maybeUnderline{R}esumption}
%%   {(extended)}
%%   {\resumesToUK{\varSolution}{\varResult}{\varResult'}{\varAssertions}}
%% 
%% \rkc{TODO}

%% \vsepRule
%% 
%% $
%% \inferrule*[lab=\ruleNameFig{R-Constraints}]
%%   {
%%    \resumesToUK{\varSolution}{\varResult_1}{\varResult_1'}{\varConstraints_1}
%%    \sepPremise
%%    \resumesToUK{\varSolution}{\varResult_2}{\varResult_2'}{\varConstraints_2}
%%    \sepPremise
%%    \resultConsistent{\varResult_1'}{\varResult_2'}{\varConstraints_3}
%%   }
%%   {\resumesToUK
%%     {\varSolution}
%%     {\expPbeConstraints{\varResult_1}{\varResult_2}}
%%     {\expUnit}
%%     {\setUnionThree{\varConstraints_1}{\varConstraints_2}{\varConstraints_3}}}
%% $

\vsepBeforeCaption

%% {\small
%% Changes to rules above are straightforward:
%% constraints are propagated from premises to conclusion.
%% }

%% \caption{Big-step, environment-style evaluation and resumption.}
%% \caption{Big-step, environment-style resumption.}
\caption{Resumption.}
\label{fig:resumption}
\end{figure}

%% \vsepRule

\clearpage

\beginTheorem{ResDet}{Determinism of Resumption}{Res. Det.}{prop:res-det}

\breakAndIndent
If $\resumesTo{\varSolution}{\varResult}{\varResult}$
and $\resumesTo{\varSolution}{\varResult}{\varResult'}$,
then $\varResult = \varResult'$.

\end{theorem}

\begin{theorem}[Finality of Resumption]
\label{prop:res-fin}

\breakAndIndent
If $\resumesTo{\varSolution}{\varResult}{\varResult'}$,
then $\isFinal{\varResult'}$.

\end{theorem}

\begin{theorem}[Type Preservation of Resumption]
\label{prop:pres-res}

\breakAndIndent
If $\typeCheckSolution{\varHoleEnv}{\varSolution}$
and $\typeCheckResult{\varResult}{\varType}$
and $\resumesTo{\varSolution}{\varResult}{\varResult'}$,
then $\typeCheckResult{\varResult'}{\varType}$.

\end{theorem}

\beginLemma{ResIdemp}{Idempotency of Resumption}{Res. Idemp.}{prop:res-idemp}

\breakAndIndent
If $\resumesTo{\varSolution}{\varResult_0}{\varResult}$,
then $\resumesTo{\varSolution}{\varResult}{\varResult}$.

\end{lemma}

\begin{lemma}[Simple Value Resumption]
\label{prop:v-res}

\breakAndIndent
If $\coerceUndet{\varResult}{\varSimpleVal}$,
then $\resumesTo{\varSolution}{\varResult}{\varResult}$.

\end{lemma}

\beginLemma{ResApp}{Resumption of App Operator}{{Res. App. Op.}}{prop:res-app}

\breakAndIndent
If $\resumesTo{\varSolution}{\varResult_1}{\varResult_1'}$
and $\resumesTo{\varSolution}{\expApp{\varResult_1'}{\varResult_2}}{\varResult}$,
then $\resumesTo{\varSolution}{\expApp{\varResult_1}{\varResult_2}}{\varResult}$.

\end{lemma}

\beginLemma{ResComp}{Resumption Composition}{Res. Comp.}{prop:res-comp}

\breakAndIndent
If $\resumesTo{\varSolution_1}{\varResult}{\varResult_1}$
and $\resumesTo{\addSolutions{\varSolution_1}{\varSolution_2}}{\varResult_1}{\varResult_2}$,
then $\resumesTo{\addSolutions{\varSolution_1}{\varSolution_2}}{\varResult}{\varResult_2}$.

\end{lemma}

\beginLemma{EvalRes}{Evaluation Respects Environment Resumption}{Eval. Respects Env. Res.}{prop:eval-res}

\breakAndIndent
If $\resumesTo{\varSolution_1}{\varEnv}{\varEnv'}$
and $\reducesTo{\varEnv}{\varExp}{\varResult_1}$
and $\reducesTo{\varEnv'}{\varExp}{\varResult_2}$
and $\resumesTo{\addSolutions{\varSolution_1}{\varSolution_2}}{\varResult_1}{\varResult_1'}$
and $\resumesTo{\addSolutions{\varSolution_1}{\varSolution_2}}{\varResult_2}{\varResult_2'}$,

\justIndent
then $\varResult_1' = \varResult_2'$.

\end{lemma}

\subsection*{Proofs}
{
\label{sec:appendix-res-proofs}

In the proofs below, we assume that evaluation and resumption are total. A
priori, this assumption is unfounded; however, there are simple modifications we
can make to \smyth{} to ensure that this property holds.
One approach described in \cite{OseraThesis} is to annotate type contexts with
tags that guarantee that all recursion is structurally decreasing (and thus
terminating). This is the approach we used in our implementation of \smyth{}.
Moreover, the premise $ \varExp_h \not= \expHole{h} $ of
\ruleName{R-Hole-Resume} ensures that resumption is total, even in the presence
of the \ruleName{Defer} synthesis rule.

This totality assumption is needed because otherwise the \ruleName{Refine-Fix}
rule could synthesize non-terminating functions which could then prevent
evaluation (or resumption) from going through cleanly in the proof terms.

\vspace{0.5em}

\jtheorem{
  \jtref{\autoref{prop:res-det}},
  \jtref{\autoref{prop:res-fin}},
  \jtref{\autoref{prop:pres-res}},
  \jtref{\autoref{prop:res-idemp}}, and
  \jtref{\autoref{prop:v-res}}
}{
  Straightforward induction.
}

%% \parahead{\autoref{prop:res-app} (Resumption of App Operator)}
\jtheorem{\jtref{\autorefResAppLong{}}}{

\jgivengoal{
  \caseFact{1} $\resumesTo{\varSolution}{\varResult_1}{\varResult_1'}$

  \caseFact{2} $\resumesTo{\varSolution}{\expApp{\varResult_1'}{\varResult_2}}{\varResult}$
}{
  $\resumesTo{\varSolution}{\expApp{\varResult_1}{\varResult_2}}{\varResult}$
}
By inversion of resumption on (2), we get two cases.
\pagebreak

\jcase{1}{\ruleName{R-App}}{
  \caseFact{3} $\resumesTo{\varSolution}{\varResult_1'}{\varResult_1''}$

  \caseFact{4} $\resumesTo{\varSolution}{\varResult_2}{\varResult_2'}$

  \caseFact{5} $\varResult_1'' = \closure{\varEnv_f}{\valFixFun{\varVarF}{\varVar}{\varExp_f}}$

  \caseFact{6} $\reducesTo
                        {\envCatThree{\varEnv_f}{\envBind{\varVarF}{\varResult_1''}}{\envBind{\varVar}{\varResult_2'}}}
                        {\varExp_f}{\varResult^*}$

  \caseFact{7} $\resumesTo{\varSolution}{\varResult^*}{\varResult}$
  %% \caseText{By \autoref{prop:res-idemp} on (1)}

  \caseText{By \autoref{prop:res-idemp} on (1)}

  \caseFact{8} $\resumesTo{\varSolution}{\varResult_1'}{\varResult_1'}$
  %% \caseText{By \autorefResDet{} on (3) and (8)}

  \caseText{By \autorefResDet{} on (3) and (8)}

  \caseFact{9} $\varResult_1' = \varResult_1''$
  %% \caseText{Goal is given by \ruleName{R-App} on (1) (observing (9)), (4), (5), (6), and (7).}

  \caseText{Goal is given by \ruleName{R-App} on (1) (observing (9)), (4), (5),
  (6), and (7).}
}

\jcase{2}{\ruleName{R-App-Indet}}{
  \caseFact{3} $\resumesTo{\varSolution}{\varResult_1'}{\varResult_1''}$

  \caseFact{4} $\resumesTo{\varSolution}{\varResult_2}{\varResult_2'}$

  \caseFact{5} $\varResult_1'' \ne \closure{\varEnv}{\valFixFun{\varVarF}{\varVar}{\varExp_f}}$

  \caseText{By \autoref{prop:res-idemp} on (1)}

  \caseFact{6} $\resumesTo{\varSolution}{\varResult_1'}{\varResult_1'}$

  \caseText{By \autorefResDetShort{} on (3) and (6)}

  \caseFact{7} $\varResult_1' = \varResult_1''$

  \caseText{Goal is given by \ruleName{R-App-Indet} on (1) (observing (7)), (4),
  and (5)}
}

}

%% \parahead{\autoref{prop:res-comp} (Resumption composes)}
\jtheorem{\jtref{\autorefResCompLong{}}}{
Most cases are trivial or go through by straightforward induction,
along with the evaluation and resumption assumptions.
The non-trivial cases are considered in detail here.

\jcase{1}{First premise through \ruleName{R-Hole-Resume}}{

\jgivengoal{
  \caseFact{1} $\resumesTo{\varSolution}{\holeClosure{\varEnv}{\varHoleName}}{\varResult'}$

  \caseFact{2} $\resumesTo{\addSolutions{\varSolution}{\varSolution'}}{\varResult'}{\varResult''}$
}{
  $\resumesTo{\addSolutions{\varSolution}{\varSolution'}}{\holeClosure{\varEnv}{\varHoleName}}{\varResult''}$
}
\caseText{Because this is the case where (1) goes through \ruleName{R-Hole-Resume}, we can, by inversion,
establish the premises of \ruleName{R-Hole-Resume}}

\caseFact{3}{$\varSolution(\varHoleName) = \varExp$}

\caseFact{4}{$\varExp \not= \expHole{h}$}

\caseFact{5}{$\reducesTo{\varEnv}{\varExp}{\varResult}$}

\caseFact{6}{$\resumesTo{\varSolution}{\varResult}{\varResult'}$}

\caseText{By the definition of $\oplus$, and (3)}

\caseFact{8}{$(\addSolutions{\varSolution}{\varSolution'})(\varHoleName) = \varExp$}

\caseText{By the induction hypothesis on (6) and (2)}

\caseFact{9}{$\resumesTo{\addSolutions{\varSolution}{\varSolution'}}{\varResult}{\varResult''}$}

\caseText{Goal is given by \ruleName{R-Hole-Resume} on (8), (5), and
(9)}
}

\jcase{2}{First premise through \ruleName{R-Hole-Indet}, second
premise through \ruleName{R-Hole-Resume}}{
\jgivengoal{
  \caseFact{1}{$\resumesTo{\varSolution}{\holeClosure{\varEnv}{\varHoleName}}{\varResult'}$}

  \caseFact{2}{$\resumesTo{\addSolutions{\varSolution}{\varSolution'}}{\varResult'}{\varResult''}$}
}{
  $\resumesTo{\addSolutions{\varSolution}{\varSolution'}}{\holeClosure{\varEnv}{\varHoleName}}{\varResult''}$
}
\caseText{Because this is a case where (1) goes through \ruleName{R-Hole-Indet}, we can, by inversion,
establish the premises of \ruleName{R-Hole-Indet}}

\caseFact{3}{$\varHoleName \notin \dom \varSolution \lor
\varSolution(\varHoleName) = \expHole{h} $}

\caseFact{4}{$\resumesTo{\varSolution}{\varEnv}{\varEnv'}$}

\caseFact{5}{$\varResult' = \holeClosure{\varEnv'}{\varHoleName}$}

\caseText{Likewise, (2) goes through \ruleName{R-Hole-Resume} (noting (5))}

\caseFact{6}{$(\addSolutions{\varSolution}{\varSolution'})(\varHoleName) = \varExp$}

\caseFact{7}{$\reducesTo{\varEnv'}{\varExp}{\varResult}$}

\caseFact{8}{$\resumesTo{\addSolutions{\varSolution}{\varSolution'}}{\varResult}{\varResult''}$}

\caseText{By the evaluation assumption}

\caseFact{9}{$\reducesTo{\varEnv}{\varExp}{\varResult^*}$}

\caseText{By the resumption assumption}

\caseFact{10}{$\resumesTo{\addSolutions{\varSolution}{\varSolution'}}{\varResult^*}{\varResult^+}$}

\caseText{By \autorefEvalResShort{} on (4), (9), (7), (10), and (8)}

\caseFact{11}{$\varResult'' = \varResult^+$}

\caseText{Goal is given by \ruleName{R-Hole-Resume} on (6), (9), and (10),
observing (11)}
}

\jcase{3}{First premise through \ruleName{R-App}}{
\jgivengoal{
  \caseFact{1}{$\resumesTo{\varSolution}{\expApp{\varResult_1}{\varResult_2}}{\varResult'}$}

  \caseFact{2}{$\resumesTo{\addSolutions{\varSolution}{\varSolution'}}{\varResult'}{\varResult''}$}
}{
  $\resumesTo{\addSolutions{\varSolution}{\varSolution'}}{\expApp{\varResult_1}{\varResult_2}}{\varResult''}$
}
\caseText{Because this is the case where the first premise goes through
\ruleName{R-App}, we can, by inversion, establish the premises of
\ruleName{R-App}}

\caseFact{3}{$\resumesTo{\varSolution}{\varResult_1}{\varResult_1'}$}

\caseFact{4}{$\resumesTo{\varSolution}{\varResult_2}{\varResult_2'}$}

\caseFact{5}{$\varResult_1' = \closure{\varEnv_f'}{\valFixFun{\varVarF}{\varVar}{\varExp_f}}$}

\caseFact{6}{$\reducesTo
                      {\envCatThree{\varEnv_f'}{\envBind{\varVarF}{\varResult_1'}}{\envBind{\varVar}{\varResult_2'}}}
                      {\varExp_f}{\varResult^*}$}
                      j
\caseFact{7}{$\resumesTo{\varSolution}{\varResult^*}{\varResult'}$}

\caseText{By the resumption assumption}

\caseFact{8}{$\resumesTo{\addSolutions{\varSolution}{\varSolution'}}{\varEnv_f'}{\varEnv_f'^+}$}

\caseFact{9}{$\resumesTo{\addSolutions{\varSolution}{\varSolution'}}{\varResult_2'}{\varResult_2'^+}$}

\caseText{By \ruleName{R-Fix} (observing (8))}

\caseFact{10}{$\resumesTo
                      {\addSolutions{\varSolution}{\varSolution'}}
                      {\closure{\varEnv_f'}{\valFixFun{\varVarF}{\varVar}{\varExp_f}}}
                      {\closure{\varEnv_f'^+}{\valFixFun{\varVarF}{\varVar}{\varExp_f}}}$}

\caseText{By the definition of environment resumption, (8), (9), and (10)}

\caseFact{11}{$\resumesToMultiLine
                      {\addSolutions{\varSolution}{\varSolution'}}
                      {(\envCatThree
                         {\varEnv_f'}
                         {\envBind{\varVarF}{\closure{\varEnv_f'}{\valFixFun{\varVarF}{\varVar}{\varExp_f}}}}
                         {\envBind{\varVar}{\varResult_2'}})}
                      {(\envCatThree
                         {\varEnv_f'^+}
                         {\envBind{\varVarF}{\closure{\varEnv_f'^+}{\valFixFun{\varVarF}{\varVar}{\varExp_f}}}}
                         {\envBind{\varVar}{\varResult_2'^+}})}$}

\caseText{By the evaluation assumption}

\caseFact{12}{$\reducesTo
                       {(\envCatThree
                         {\varEnv_f'^+}
                         {\envBind{\varVarF}{\closure{\varEnv_f'^+}{\valFixFun{\varVarF}{\varVar}{\varExp_f}}}}
                         {\envBind{\varVar}{\varResult_2'^+}})}
                       {\varExp_f}
                       {\varResult^{*+}}$}

\caseText{By the induction hypothesis on (7) and (2)}

\caseFact{13}{$\resumesTo{\addSolutions{\varSolution}{\varSolution'}}{\varResult^*}{\varResult''}$}

\caseText{By the resumption assumption}

\caseFact{14}{$\resumesTo{\addSolutions{\varSolution}{\varSolution'}}{\varResult^{*+}}{\varResult^{*++}}$}

\caseText{By \autorefEvalResShort{} on (11), (6), (12), (13), and (14)}

\caseFact{15}{$\varResult'' = \varResult^{*++}$}

\caseText{By the induction hypothesis on (3) and (10) (observing (5))}

\caseFact{16}{$\resumesTo
                      {\addSolutions{\varSolution}{\varSolution'}}
                      {\varResult_1}
                      {\closure{\varEnv_f'^+}{\valFixFun{\varVarF}{\varVar}{\varExp_f}}}$}

\caseText{By the induction hypothesis on (4) and (9)}

\caseFact{17}{$\resumesTo{\addSolutions{\varSolution}{\varSolution'}}{\varResult_2}{\varResult_2'^+}$}

\caseText{By \ruleName{R-App} on (16), (17), (trivial), (12), and (14)}

\caseFact{18}{$\resumesTo{\addSolutions{\varSolution}{\varSolution'}}{\expApp{\varResult_1}{\varResult_2}}{\varResult^{*++}}$}

\caseText{The goal is given by combining (15) and (18)}
}

\jcase{4}{First premise goes through \ruleName{R-Case}}{
\jgivengoal{
  \caseFact{1}{$\resumesTo{\varSolution}{\caseClosureIndet{\varEnv}{\varResult}{\varDataCon_i}{\varVar_i}{\varExp_i}}{\varResult'}$}

  \caseFact{2}{$\resumesTo{\addSolutions{\varSolution}{\varSolution'}}{\varResult'}{\varResult''}$}
}{
  $\resumesTo{\addSolutions{\varSolution}{\varSolution'}}{\caseClosureIndet{\varEnv}{\varResult}{\varDataCon_i}{\varVar_i}{\varExp_i}}{\varResult''}$
}
\caseText{Because this is the case where the first premise goes through
\ruleName{R-Case}, we can, by inversion, establish the premises of
\ruleName{R-Case}}

\caseFact{3}{$\resumesTo{\varSolution}{\varResult}{\expApp{\varDataCon_j}{\varResult_j'}}$}

\caseFact{4}{$\resumesTo{\varSolution}{\expApp{(\closure{\varEnv}{\expFun{\varVar_j}{\varExp_j}})}{\varResult_j'}}{\varResult'}$}

\caseText
{By inversion of resumption on (4), we find that (4) can only go through
\ruleName{R-App} since the first argument is a syntactic \inlinecode{fix}
(which by \ruleName{R-Fix} will resume to a syntactic \inlinecode{fix}); so,
by inversion, we can establish the premises of \ruleName{R-App}
(to establish premises (5) and (6), we use inversion again)}

\caseFact{5}{$\resumesTo{\varSolution}{\varEnv}{\varEnv'}$}

\caseFact{6}{$\resumesTo{\varSolution}{\closure{\varEnv}{\expFun{\varVar_j}{\varExp_j}}}{\closure{\varEnv'}{\expFun{\varVar_j}{\varExp_j}}}$}

\caseFact{7}{$\resumesTo{\varSolution}{\varResult_j'}{\varResult_2'}$}

\caseFact{8}{$\reducesTo{(\envCat{\varEnv'}{\envBind{\varVar_j}{\varResult_2'}})}{\varExp_j}{\varResult^*}$}

\caseFact{9}{$\resumesTo{\varSolution}{\varResult^*}{\varResult'}$}

\caseText{By \ruleName{R-Ctor} on (7)}

\caseFact{10}{$\resumesTo{\varSolution}{\expApp{\varDataCon_j}{\varResult_j'}}{\expApp{\varDataCon_j}{\varResult_2'}}$}

\caseText{By \autoref{prop:res-idemp} on (3)}

\caseFact{11}{$\resumesTo{\varSolution}{\expApp{\varDataCon_j}{\varResult_j'}}{\expApp{\varDataCon_j}{\varResult_j'}}$}

\caseText{By \autorefResDetShort{} on (10) and (11)}

\caseFact{12}{$\varResult_j' = \varResult_2'$}

\caseText{By the resumption assumption}

\caseFact{13}{$\resumesTo{\addSolutions{\varSolution}{\varSolution'}}{\varResult_j'}{\varResult_j'^+}$}

\caseFact{14}{$\resumesTo{\addSolutions{\varSolution}{\varSolution'}}{\varEnv'}{\varEnv'^+}$}

\caseText{By \ruleName{R-Ctor} on (13)}

\caseFact{15}{$\resumesTo{\addSolutions{\varSolution}{\varSolution'}}{\expApp{\varDataCon_j}{\varResult_j'}}{\expApp{\varDataCon_j}{\varResult_j'^+}}$}

\caseText{By the induction hypothesis on (3) and (15)}

\caseFact{16}{$\resumesTo{\addSolutions{\varSolution}{\varSolution'}}{\varResult}{\expApp{\varDataCon_j}{\varResult_j'^+}}$}

\caseText{By \ruleName{R-Fix} on (14)}

\caseFact{17}{$\resumesTo{\addSolutions{\varSolution}{\varSolution'}}{\closure{\varEnv'}{\expFun{\varVar_j}{\varExp_j}}}{\closure{\varEnv'^+}{\expFun{\varVar_j}{\varExp_j}}}$}

\caseText{By the induction hypothesis on (6) and (17)}

\caseFact{18}{$\resumesTo{\addSolutions{\varSolution}{\varSolution'}}{\closure{\varEnv}{\expFun{\varVar_j}{\varExp_j}}}{\closure{\varEnv'^+}{\expFun{\varVar_j}{\varExp_j}}}$}

\caseText{By \autoref{prop:res-idemp} on (13)}

\caseFact{19}{$\resumesTo{\addSolutions{\varSolution}{\varSolution'}}{\varResult_j'^+}{\varResult_j'^+}$}

\caseText{By the evaluation assumption}

\caseFact{20}{$\reducesTo{(\envCat{\varEnv'^+}{\envBind{\varVar_j}{\varResult_j'^+}})}{\varExp_j}{\varResult^{*+}}$}

\caseText{By the resumption assumption}

\caseFact{21}{$\resumesTo{\addSolutions{\varSolution}{\varSolution'}}{\varResult^*}{\varResult^{*++}}$}

\caseFact{22}{$\resumesTo{\addSolutions{\varSolution}{\varSolution'}}{\varResult^{*+}}{\varResult^{*\prime+}}$}

\caseText{By the definition of environment resumption, (14), and (13) (observing (12))}

\caseFact{23}{$\resumesTo{\addSolutions{\varSolution}{\varSolution'}}{(\envCat{\varEnv'}{\envBind{\varVar_j}{\varResult_2'}})}{(\envCat{\varEnv'^+}{\envBind{\varVar_j}{\varResult_j'^+}})}$}

\caseText{By \autorefEvalResShort{} on (23), (8), (20), (21), and (22)}

\caseFact{24}{$\varResult^{*++} = \varResult^{*\prime+}$}

\caseText{By \ruleName{R-App} on (18), (19), (trivial), (20), and (22) (observing (24))}

\caseFact{25}{$\resumesTo{\addSolutions{\varSolution}{\varSolution'}}{\expApp{(\closure{\varEnv}{\expFun{\varVar_j}{\varExp_j}})}{\varResult_j'^+}}{\varResult^{*++}}$}

\caseText{By the induction hypothesis on (9) and (2)}

\caseFact{26}{$\resumesTo{\addSolutions{\varSolution}{\varSolution'}}{\varResult^*}{\varResult''}$}

\caseText{By \autorefResDetShort{} on (21) and (26)}

\caseFact{27}{$\varResult'' = \varResult^{*++}$}

\caseText{Goal is given by \ruleName{R-Case} on (16) and (25) (observing (27))}
}

\jcase{5}{First premise through \ruleName{R-Case-Indet}, second premise through
\ruleName{R-Case}}{
\caseText{We use inversion to establish the premises of these rules as
givens.}
\jgivengoal{
  \caseFact{1}{$\resumesTo{\varSolution}{\varResult}{\varResult'}$}

  \caseFact{2}{$\varResult' \ne {\expApp{\varDataCon_j}{\varResult_j}}$}

  \caseFact{3}{$\resumesTo{\varSolution}{\varEnv}{\varEnv'}$}

  \caseFact{4}{$\resumesTo{\varSolution}{\caseClosureIndet{\varEnv}{\varResult}{\varDataCon_i}{\varVar_i}{\varExp_i}}{\caseClosureIndet{\varEnv'}{\varResult'}{\varDataCon_i}{\varVar_i}{\varExp_i}}$}

  \caseFact{5}{$\resumesTo{\addSolutions{\varSolution}{\varSolution'}}{\varResult'}{\expApp{\varDataCon_j}{\varResult_j'^+}}$}

  \caseFact{6}{$\resumesTo{\addSolutions{\varSolution}{\varSolution'}}{\expApp{(\closure{\varEnv'}{\expFun{\varVar_j}{\varExp_j}})}{\varResult_j'^+}}{\varResult''}$}
}{
  $\resumesTo{\addSolutions{\varSolution}{\varSolution'}}{\caseClosureIndet{\varEnv}{\varResult}{\varDataCon_i}{\varVar_i}{\varExp_i}}{\varResult''}$
}
\caseText{By the resumption assumption}

\caseFact{7}{$\resumesTo{\addSolutions{\varSolution}{\varSolution'}}{\varEnv'}{\varEnv'^+}$}

\caseFact{8}{$\resumesTo{\addSolutions{\varSolution}{\varSolution'}}{\expApp{(\closure{\varEnv'^+}{\expFun{\varVar_j}{\varExp_j}})}{\varResult_j'^+}}{\varResult''^+}$}

\caseText{By \ruleName{R-Fix} on (7)}

\caseFact{9}{$\resumesTo
                     {\addSolutions{\varSolution}{\varSolution'}}
                     {\closure{\varEnv'}{\expFun{\varVar_j}{\varExp_j}}}
                     {\closure{\varEnv'^+}{\expFun{\varVar_j}{\varExp_j}}}$}

\caseText{By \autoref{prop:res-app} (Res. of App Op.) on (9) and (8)}

\caseFact{10}{$\resumesTo{\addSolutions{\varSolution}{\varSolution'}}{\expApp{(\closure{\varEnv'}{\expFun{\varVar_j}{\varExp_j}})}{\varResult_j'^+}}{\varResult''^+}$}

\caseText{By \autorefResDetShort{} on (6) and (10)}

\caseFact{11}{$\varResult'' = \varResult''^+$}

\caseText{By the induction hypothesis on (3) and (7)}

\caseFact{12}{$\resumesTo{\addSolutions{\varSolution}{\varSolution'}}{\varEnv}{\varEnv'^+}$}

\caseText{By \ruleName{R-Fix} on (12)}

\caseFact{13}{$\resumesTo
                     {\varSolution}
                     {\closure{\varEnv}{\expFun{\varVar_j}{\varExp_j}}}
                     {\closure{\varEnv'^+}{\expFun{\varVar_j}{\varExp_j}}}$}

\caseText{By \autoref{prop:res-app} (Res. of App Op.) on (13) and (8) (observing
(11)}

\caseFact{14}{$\resumesTo{\addSolutions{\varSolution}{\varSolution'}}{\expApp{(\closure{\varEnv}{\expFun{\varVar_j}{\varExp_j}})}{\varResult_j'^+}}{\varResult''}$}

\caseText{By the induction hypothesis on (1) and (5)}

\caseFact{15}{$\resumesTo{\addSolutions{\varSolution}{\varSolution'}}{\varResult}{\expApp{\varDataCon_j}{\varResult_j'^+}}$}

\caseText{The goal is given by \ruleName{R-Case} on (15) and (14)}
}

}

\vspace{-0.5em}
%% \parahead{\autoref{prop:eval-res} (Evaluation respects environment resumption)}
\jtheorem{\jtref{\autorefEvalResLong{}}}{
The \ruleName{E-Unit} case is trivial.
The cases for \ruleName{E-Ctor} and \ruleName{E-Pair} go through by
straightforward induction, and likewise for \ruleName{E-Hole} if the hole is
filled.
The unfilled case for \ruleName{E-Hole} is analagous to the proof for the
\ruleName{E-Fix} case below.
The remaining cases are considered in detail here.

\jcase{1}{\ruleName{E-Fix}}{
\jgivengoal{
  \caseFact{1}{$\resumesTo{\varSolution}{\varEnv_1}{\varEnv_2}$}

  \caseFact{2}{$\reducesTo{\varEnv_1}{\valFixFun{\varVarF}{\varVar}{\varExp}}{\varResult_1}$}

  \caseFact{3}{$\reducesTo{\varEnv_2}{\valFixFun{\varVarF}{\varVar}{\varExp}}{\varResult_2}$}

  \caseFact{4}{$\resumesTo{\mergeSolutions{\varSolution}{\varSolution'}}{\varResult_1}{\varResult_1'}$}

  \caseFact{5}{$\resumesTo{\mergeSolutions{\varSolution}{\varSolution'}}{\varResult_2}{\varResult_2'}$}
}{
  $\varResult_1' = \varResult_2'$
}
\caseText{By inversion of evaluation on (2)}

\caseFact{6}{$\varResult_1 = \closure{\varEnv_1}{\valFixFun{\varVarF}{\varVar}{\varExp}}$}

\caseText{By inversion of evaluation on (3)}

\caseFact{7}{$\varResult_2 = \closure{\varEnv_2}{\valFixFun{\varVarF}{\varVar}{\varExp}}$}

\caseText{By inversion of resumption on (6)}

\caseFact{8}{$\resumesTo{\mergeSolutions{\varSolution}{\varSolution'}}{\varEnv_1}{\varEnv_1'}$}

\caseFact{9}{$\varResult_1' = \closure{\varEnv_1'}{\valFixFun{\varVarF}{\varVar}{\varExp}}$}

\caseText{By inversion of resumption on (7)}

\caseFact{10}{$\resumesTo{\mergeSolutions{\varSolution}{\varSolution'}}{\varEnv_2}{\varEnv_2'}$}

\caseFact{11}{$\varResult_2 = \closure{\varEnv_2'}{\valFixFun{\varVarF}{\varVar}{\varExp}}$}

\caseText{By \autorefResCompShort{} (across all bindings in environment) on
(1) and (10)}

\caseFact{12}{$\resumesTo{\mergeSolutions{\varSolution}{\varSolution'}}{\varEnv_1}{\varEnv_2'}$}

\caseText{By \autorefResDetShort{} (across all bindings in environment) on
(8) and (12)}

\caseFact{13}{$\varEnv_1' = \varEnv_2'$}

\caseText{Observing (13), (9) and (11) equate to form the goal}
}

\pagebreak

\jcase{2}{\ruleName{E-Var}}{
\jgivengoal{
  \caseFact{1}{$\resumesTo{\varSolution}{\varEnv_1}{\varEnv_2}$}

  \caseFact{2}{$\reducesTo{\varEnv_1}{\varVar}{\varResult_1}$}

  \caseFact{3}{$\reducesTo{\varEnv_2}{\varVar}{\varResult_2}$}

  \caseFact{4}{$\resumesTo{\mergeSolutions{\varSolution}{\varSolution'}}{\varResult_1}{\varResult_1'}$}

  \caseFact{5}{$\resumesTo{\mergeSolutions{\varSolution}{\varSolution'}}{\varResult_2}{\varResult_2'}$}
}{
  $\varResult_1' = \varResult_2'$
}
\caseText{By inversion of evaluation}

\caseFact{6}{$\varEnv_1(\varVar)=\varResult_1$}

\caseFact{7}{$\varEnv_2(\varVar)=\varResult_2$}

\caseText{By (1), (6), and (7)}

\caseFact{8}{$\resumesTo{\varSolution}{\varResult_1}{\varResult_2}$}

\caseText{By \autorefResCompShort{} on (8) and (5)}

\caseFact{9}{$\resumesTo{\mergeSolutions{\varSolution}{\varSolution'}}{\varResult_1}{\varResult_2'}$}

\caseText{By \autorefResDetShort{} on (4) and (9)}

\caseFact{Goal}{$\varResult_1' = \varResult_2'$}
}
For some expressions, evaluation can go through different rules,
so the names of these cases will be given by the expression type
rather than by the evaluation rule they go through.

\newcommand{\efun}{\varExp_{\textit{fun}}}
\newcommand{\earg}{\varExp_{\textit{arg}}}
\newcommand{\rfun}[1]{\varResult_{\textit{fun}#1}}
\newcommand{\rarg}[1]{\varResult_{\textit{arg}#1}}

\jcase{3}{Applications}{
\jgivengoal{
  \caseFact{1}{$\resumesTo{\varSolution}{\varEnv_1}{\varEnv_2}$}

  \caseFact{2}{$\reducesTo{\varEnv_1}{\expApp{\efun}{\earg}}{\varResult_1}$}

  \caseFact{3}{$\reducesTo{\varEnv_2}{\expApp{\efun}{\earg}}{\varResult_2}$}

  \caseFact{4}{$\resumesTo{\mergeSolutions{\varSolution}{\varSolution'}}{\varResult_1}{\varResult_1'}$}

  \caseFact{5}{$\resumesTo{\mergeSolutions{\varSolution}{\varSolution'}}{\varResult_2}{\varResult_2'}$}
}{
  $\varResult_1' = \varResult_2'$
}
\caseText{By inversion of eval on (2) and (3), we get four subcases:}

\jsubcase{1}{\ruleName{E-App}, \ruleName{E-App}}{
\caseFact{6}{$\reducesTo{\varEnv_1}{\efun}{\rfun{1}}$}

\caseFact{7}{$\rfun{1} = \closure{\varEnv_{f1}}{\valFixFun{\varVarF_1}{\varVar_1}{\varExp_{f1}}}$}

\caseFact{8}{$\reducesTo{\varEnv_1}{\earg}{\rarg{1}}$}

\caseFact{9}{$\reducesTo{(\envCatThree{\varEnv_{f1}}{\envBind{\varVarF_1}{\rfun{1}}}{\envBind{\varVar_1}{\rarg{1}}})}{\varExp_{f1}}{\varResult_1}$}

\caseFact{10}{$\reducesTo{\varEnv_2}{\efun}{\rfun{2}}$}

\caseFact{11}{$\rfun{2} = \closure{\varEnv_{f2}}{\valFixFun{\varVarF_2}{\varVar_2}{\varExp_{f2}}}$}

\caseFact{12}{$\reducesTo{\varEnv_2}{\earg}{\rarg{2}}$}

\caseFact{13}{$\reducesTo{(\envCatThree{\varEnv_{f2}}{\envBind{\varVarF_2}{\rfun{2}}}{\envBind{\varVar_2}{\rarg{2}}})}{\varExp_{f2}}{\varResult_2}$}

\caseText{By \ruleName{R-Fix} (observing (7) and (11))}

\caseFact{14}{$\resumesTo{\mergeSolutions{\varSolution}{\varSolution'}}{\varEnv_{f1}}{\varEnv_{f1}'}$}

\caseFact{15}{$\resumesTo{\mergeSolutions{\varSolution}{\varSolution'}}{\rfun{1}}{\closure{\varEnv_{f1}'}{\valFixFun{\varVarF_1}{\varVar_1}{\varExp_{f1}}}}$}

\caseFact{16}{$\resumesTo{\mergeSolutions{\varSolution}{\varSolution'}}{\varEnv_{f2}}{\varEnv_{f2}'}$}

\caseFact{17}{$\resumesTo{\mergeSolutions{\varSolution}{\varSolution'}}{\rfun{2}}{\closure{\varEnv_{f2}'}{\valFixFun{\varVarF_2}{\varVar_2}{\varExp_{f2}}}}$}

\caseText{By the induction hypothesis on (1), (6), (10), (15), and (17)}

\caseFact{18}{$\closure{\varEnv_{f1}'}{\valFixFun{\varVarF_1}{\varVar_1}{\varExp_{f1}}} = \closure{\varEnv_{f2}'}{\valFixFun{\varVarF_2}{\varVar_2}{\varExp_{f2}}}$}

\caseText{By the resumption assumption}

\caseFact{19}{$\resumesTo{\mergeSolutions{\varSolution}{\varSolution'}}{\rarg{1}}{\rarg{1}'}$}

\caseFact{20}{$\resumesTo{\mergeSolutions{\varSolution}{\varSolution'}}{\rarg{2}}{\rarg{2}'}$}

\caseText{By the induction hypothesis on (1), (8), (12), (19), and (20)}

\caseFact{21}{$\rarg{1}' = \rarg{2}'$}

\caseText{By the definition of environment resumption, (16), (17), and (20)}

\caseFact{22}{$\resumesTo
                      {\mergeSolutions{\varSolution}{\varSolution'}}
                      {(\envCatThree{\varEnv_{f2}}{\envBind{\varVarF_2}{\rfun{2}}}{\envBind{\varVar_2}{\rarg{2}}})}
                      {(\envCatThree{\varEnv_{f2}'}{\envBind{\varVarF_2}{\rfun{2}}}{\envBind{\varVar_2}{\rarg{2}'}})}$}

\caseText{By the evaluation assumption}

\caseFact{23}{$\reducesTo{(\envCatThree{\varEnv_{f2}'}
                                            {\envBind{\varVarF_2}{\closure{\varEnv_{f2}'}{\valFixFun{\varVarF_2}{\varVar_2}{\varExp_{f2}}}}}
                                            {\envBind{\varVar_2}{\rarg{2}'}})}
                              {\varExp_{f2}}
                              {\varResult_2^*}$}

\caseText{By the resumption assumption}

\caseFact{24}{$\resumesTo{\mergeSolutions{\varSolution}{\varSolution'}}{\varResult_2^*}{\varResult_2^{*\prime}}$}

\caseText{By the induction hypothesis on (22), (13), (23), (5), and (24)}

\caseFact{25}{$\varResult_2^{*\prime} = \varResult_2'$}

\caseText{By the definition of environment resumption, (14), (18), (15), (19),
and (21)}

\caseFact{26}{$
  \resumesToMultiLine
    {\mergeSolutions{\varSolution}{\varSolution'}}
    {(\envCatThree{\varEnv_{f1}}{\envBind{\varVarF_1}{\rfun{1}}}{\envBind{\varVar_1}{\rarg{1}}})}
    {(\envCatThree{\varEnv_{f2}'}{\envBind{\varVarF_2}{\closure{\varEnv_{f2}'}{\valFixFun{\varVarF_2}{\varVar_2}{\varExp_{f2}}}}}{\envBind{\varVar_2}{\rarg{2}'}})}
  $}

\caseText{By the induction hypothesis on (26), (9) (noting (18)), (23), (4), and
(24)}

\caseFact{27}{$\varResult_2^{*\prime} = \varResult_1'$}

\caseText{Combining (25) and (27) gives the goal}
}

\jsubcase{2}{\ruleName{E-App}, \ruleName{E-App-Indet}}{
\caseFact{6}{$\reducesTo{\varEnv_1}{\efun}{\rfun{1}}$}

\caseFact{7}{$\rfun{1} = \closure{\varEnv_{f1}}{\valFixFun{\varVarF_1}{\varVar_1}{\varExp_{f1}}}$}

\caseFact{8}{$\reducesTo{\varEnv_1}{\earg}{\rarg{1}}$}

\caseFact{9}{$\reducesTo{(\envCatThree{\varEnv_{f1}}{\envBind{\varVarF_1}{\rfun{1}}}{\envBind{\varVar_1}{\rarg{1}}})}{\varExp_{f1}}{\varResult_1}$}

\caseFact{10}{$\varResult_2 = \expApp{\rfun{2}}{\rarg{2}}$}

\caseFact{11}{$\reducesTo{\varEnv_2}{\efun}{\rfun{2}}$}

\caseFact{12}{$\rfun{2} \neq \closure{\varEnv_{f2}}{\valFixFun{\varVarF_2}{\varVar_2}{\varExp_{f2}}}$}

\caseFact{13}{$\reducesTo{\varEnv_2}{\earg}{\rarg{2}}$}

\caseText{By \ruleName{R-Fix} (observing (7))}

\caseFact{14}{$\resumesTo{\mergeSolutions{\varSolution}{\varSolution'}}{\varEnv_{f1}}{\varEnv_{f1}'}$}

\caseFact{15}{$\resumesTo{\mergeSolutions{\varSolution}{\varSolution'}}{\rfun{1}}{\closure{\varEnv_{f1}'}{\valFixFun{\varVarF_1}{\varVar_1}{\varExp_{f1}}}}$}

\caseText{By the resumption assumption}

\caseFact{16}{$\resumesTo{\mergeSolutions{\varSolution}{\varSolution'}}{\rfun{2}}{\rfun{2}'}$}

\caseFact{17}{$\resumesTo{\mergeSolutions{\varSolution}{\varSolution'}}{\rarg{1}}{\rarg{1}'}$}

\caseFact{18}{$\resumesTo{\mergeSolutions{\varSolution}{\varSolution'}}{\rarg{2}}{\rarg{2}'}$}

\caseText{By the induction hypothesis on (1), (8), (13), (17), and (18)}

\caseFact{19}{$\rarg{1}' = \rarg{2}'$}

\caseText{By the induction hypothesis on (1), (6), (11), (15), and (16)}

\caseFact{20}{$\closure{\varEnv_{f1}'}{\valFixFun{\varVar_{f1}}{\varVar_1}{\varExp_{f1}}} = \rfun{2}'$}

\caseText{By the evaluation assumption}

\caseFact{21}{$
  \reducesTo{(\envCatThree{\varEnv_{f1}'}
                          {\envBind{\varVarF}{\closure{\varEnv_{f1}'}{\valFixFun{\varVarF_1}{\varVar_1}{\varExp_{f1}}}}}
                          {\envBind{\varVar}{\rarg{1}'}})}
            {\varExp_{f1}}
            {\varResult^*}$}

\caseText{By the resumption assumption}

\caseFact{22}{$\resumesTo{\mergeSolutions{\varSolution}{\varSolution'}}{\varResult^*}{\varResult^{*\prime}}$}

\caseText{By \ruleName{R-App} on (16), (18), (20), (21) (observing (19)), and
(22)}

\caseFact{23}{$\resumesTo{\mergeSolutions{\varSolution}{\varSolution'}}{\expApp{\rfun{2}}{\rarg{2}}}{\varResult^{*\prime}}$}

\caseText{By \autorefResDetLong{} on (5) and (23)}

\caseFact{24}{$\varResult^{*\prime} = \varResult_2'$}

\caseText{By the definition of environment resumption, (14), (15), and (17)}

\caseFact{25}{$
  \resumesToMultiLine
    {\mergeSolutions{\varSolution}{\varSolution'}}
    {(\envCatThree{\varEnv_{f1}}{\envBind{\varVarF_1}{\rfun{1}}}{\envBind{\varVar_1}{\rarg{1}}})}
    {(\envCatThree{\varEnv_{f1}'}{\envBind{\varVarF_1}{\closure{\varEnv_{f1}'}{\valFixFun{\varVarF_1}{\varVar_1}{\varExp_{f1}}}}}{\envBind{\varVar_1}{\rarg{1}'}})}
  $}

\caseText{By the induction hypothesis on (25), (9), (21), (4), and (22)}

\caseFact{26}{$\varResult^{*\prime} = \varResult_1'$}

\caseText{Combining (24) and (26) gives the goal}
}

\jsubcase{3}{\ruleName{E-App-Indet}, \ruleName{E-App}}{
  \caseText{Analagous to previous case.}
}

\jsubcase{4}{\ruleName{E-App-Indet}, \ruleName{E-App-Indet}}{
\caseFact{7}{$\reducesTo{\varEnv_1}{\efun}{\rfun{1}}$}

\caseFact{8}{$\rfun{1} \neq \closure{\varEnv_{f1}}{\valFixFun{\varVarF_1}{\varVar_1}{\varExp_{f1}}}$}

\caseFact{9}{$\reducesTo{\varEnv}{\earg}{\rarg{1}}$}

\caseFact{10}{$\varResult_2 = \expApp{\rfun{2}}{\rarg{2}}$}

\caseFact{11}{$\reducesTo{\varEnv_2}{\efun}{\rfun{2}}$}

\caseFact{12}{$\rfun{2} \neq \closure{\varEnv_{f2}}{\valFixFun{\varVarF_2}{\varVar_2}{\varExp_{f2}}}$}

\caseFact{13}{$\reducesTo{\varEnv_2}{\earg}{\rarg{2}}$}

\caseText{By the resumption assumption}

\caseFact{14}{$\resumesTo{\mergeSolutions{\varSolution}{\varSolution'}}{\rfun{1}}{\rfun{1}'}$}

\caseFact{15}{$\resumesTo{\mergeSolutions{\varSolution}{\varSolution'}}{\rarg{1}}{\rarg{1}'}$}

\caseFact{16}{$\resumesTo{\mergeSolutions{\varSolution}{\varSolution'}}{\rfun{2}}{\rfun{2}'}$}

\caseFact{17}{$\resumesTo{\mergeSolutions{\varSolution}{\varSolution'}}{\rarg{2}}{\rarg{2}'}$}

\caseText{By the induction hypothesis on (1), (7), (11), (14), and (16)}

\caseFact{18}{$\rfun{1}' = \rfun{2}'$}

\caseText{By the induction hypothesis on (1), (9), (13), (15), and (17)}

\caseFact{19}{$\rarg{1}' = \rarg{2}'$}

\caseText{Resumption of $\varResult_1$ and $\varResult_2$ could go through
\ruleName{R-App} or \ruleName{R-Indet}, but in either case, the premises and
conclusion are entirely determined by the resumptions of $\rfun{1}$ and
$\rfun{2}$, equated by (18), and $\rarg{1}$ and $\rarg{2}$, equated by (19). As
such, we can conclude that $\varResult_1' = \varResult_2'$.}
}

}

\vspace{1em}

\jcase{4}{Projections}{
  \caseText{Without loss of generality, we will only detail the $\expFst{\varExp}$
  case}
\jgivengoal{
  \caseFact{1}{$\resumesTo{\varSolution}{\varEnv_1}{\varEnv_2}$}

  \caseFact{2}{$\reducesTo{\varEnv_1}{\expFst{\varExp}}{\varResult_1}$}

  \caseFact{3}{$\reducesTo{\varEnv_2}{\expFst{\varExp}}{\varResult_2}$}

  \caseFact{4}{$\resumesTo{\mergeSolutions{\varSolution}{\varSolution'}}{\varResult_1}{\varResult_1'}$}

  \caseFact{5}{$\resumesTo{\mergeSolutions{\varSolution}{\varSolution'}}{\varResult_2}{\varResult_2'}$}
}{
  $\varResult_1' = \varResult_2'$
}
\caseText{By inversion of evaluation on (2) and (3), we get four subcases:}

\jsubcase{1}{\ruleName{E-Prj}, \ruleName{E-Prj}}{
\caseFact{6}{$\reducesTo{\varEnv_1}{\varExp}{\pair{\varResult_1}{\varResult_1^*}}$}

\caseFact{7}{$\reducesTo{\varEnv_2}{\varExp}{\pair{\varResult_2}{\varResult_2^*}}$}

\caseText{By the resumption assumption}

\caseFact{8}{$\resumesTo{\mergeSolutions{\varSolution}{\varSolution'}}{\varResult_1^*}{\varResult_1^{*\prime}}$}

\caseFact{9}{$\resumesTo{\mergeSolutions{\varSolution}{\varSolution'}}{\varResult_2^*}{\varResult_2^{*\prime}}$}

\caseText{By \ruleName{R-Pair} on (4) and (8)}

\caseFact{10}{$\resumesTo{\mergeSolutions{\varSolution}{\varSolution'}}
                              {\pair{\varResult_1}{\varResult_1^*}}
                              {\pair{\varResult_1'}{\varResult_1^{*\prime}}}
                   $}

\caseText{By \ruleName{R-Pair} on (5) and (9)}

\caseFact{11}{$\resumesTo{\mergeSolutions{\varSolution}{\varSolution'}}
                              {\pair{\varResult_2}{\varResult_2^*}}
                              {\pair{\varResult_2'}{\varResult_2^{*\prime}}}
                   $}

\caseText{By the induction hypothesis on (1), (6), (7), (10), and (11)}

\caseFact{12}{$\pair{\varResult_1'}{\varResult_1^{*\prime}} =
                    \pair{\varResult_2'}{\varResult_2^{*\prime}}$}

\caseText{The goal follows from (12)}
}

\pagebreak

\jsubcase{2}{\ruleName{E-Prj}, \ruleName{E-Prj-Indet}}{
\caseFact{6}{$\reducesTo{\varEnv_1}{\varExp}{\pair{\varResult_1}{\varResult_1^*}}$}

\caseFact{7}{$\varResult_2 = \expFst{\varResult_2^*}$}

\caseFact{8}{$\reducesTo{\varEnv_2}{\varExp}{\varResult_2^*}$} 

\caseFact{9}{$\varResult_2^* \neq \pair{\varResult_a}{\varResult_b}$}

\caseText{By the resumption assumption}

\caseFact{10}{$\resumesTo{\mergeSolutions{\varSolution}{\varSolution'}}{\varResult_1^*}{\varResult_1^{*\prime}}$}

\caseFact{11}{$\resumesTo{\mergeSolutions{\varSolution}{\varSolution'}}{\varResult_2^*}{\varResult_2^{*\prime}}$}

\caseText{By \ruleName{R-Pair} on (4) and (10)}

\caseFact{12}{$\resumesTo{\mergeSolutions{\varSolution}{\varSolution'}}
                              {\pair{\varResult_1}{\varResult_1^*}}
                              {\pair{\varResult_1'}{\varResult_1^{*\prime}}}
                   $}

\caseText{By the induction hypothesis on (1), (6), (8), (12), and (11)}

\caseFact{13}{$\pair{\varResult_1'}{\varResult_1^{*\prime}} = \varResult_2^{*\prime}$}

\caseText{By \ruleName{R-Prj} on (11) (observing (13) and (7))}

\caseFact{14}{$\resumesTo{\mergeSolutions{\varSolution}{\varSolution'}}{\varResult_2}{\varResult_1'}$}

\caseText{\autorefResDetLong{} combined with (5) and (14) yield the goal}
}

\jsubcase{3}{\ruleName{E-Prj-Indet}, \ruleName{E-Prj}}{
  Analagous to previous case.
}

\jsubcase{4}{\ruleName{E-Prj-Indet}, \ruleName{E-Prj-Indet}}{
\caseFact{6}{$\varResult_1 = \expFst{\varResult_1^*}$}

\caseFact{7}{$\reducesTo{\varEnv_1}{\varExp}{\varResult_1^*}$}

\caseFact{8}{$\varResult_1^* \neq \pair{\varResult_a}{\varResult_b}$}

\caseFact{9}{$\varResult_2 = \expFst{\varResult_2^*}$}

\caseFact{10}{$\reducesTo{\varEnv_2}{\varExp}{\varResult_2^*}$}

\caseFact{11}{$\varResult_2^* \neq \pair{\varResult_c}{\varResult_d}$}

\caseText{By the resumption assumption}

\caseFact{12}{$\resumesTo{\mergeSolutions{\varSolution}{\varSolution'}}{\varResult_1^*}{\varResult_1^{*\prime}}$}

\caseFact{13}{$\resumesTo{\mergeSolutions{\varSolution}{\varSolution'}}{\varResult_2^*}{\varResult_2^{*\prime}}$}

\caseText{By the induction hypothesis on (1), (7), (10), (12), and (13)}

\caseFact{14}{$\varResult_1^{*\prime} = \varResult_2^{*\prime}$}

\caseText{Resumption of $\varResult_1$ and $\varResult_2$ could go through
\ruleName{R-Prj} or \ruleName{R-Prj-Indet}, but in either case, the premises
and conclusion are entirely determined by the resumptions of
$\varResult_1^*$ and $\varResult_2^*$, equated by (14).  As such, we can
conclude that $\varResult_1' = \varResult_2'$}
}
}

\jcase{5}{Case/Match}{
\jgivengoal{
  \caseFact{1}{$\resumesTo{\varSolution}{\varEnv_1}{\varEnv_2}$}

  \caseFact{2}{$\reducesTo{\varEnv_1}{\expMatch{\varExp}{\varDataCon_i}{\varVar_i}{\varExp_i}}{\varResult_1}$}

  \caseFact{3}{$\reducesTo{\varEnv_1}{\expMatch{\varExp}{\varDataCon_i}{\varVar_i}{\varExp_i}}{\varResult_2}$}

  \caseFact{4}{$\resumesTo{\mergeSolutions{\varSolution}{\varSolution'}}{\varResult_1}{\varResult_1'}$}

  \caseFact{5}{$\resumesTo{\mergeSolutions{\varSolution}{\varSolution'}}{\varResult_2}{\varResult_2'}$}
}{
  $\varResult_1' = \varResult_2'$
}
\caseText{By inversion of evaluation on (2) and (3), we get four subcases:}

\jsubcase{1}{\ruleName{E-Case}, \ruleName{E-Case}}{
\caseFact{6}{$\reducesTo{\varEnv_1}{\varExp}{\expApp{\varDataCon_{j_1}}{\varResult_1^*}}$ (for some $j_1 < n$)}

\caseFact{7}{$\reducesTo{(\envCat{\varEnv_1}{\envBind{\varVar_{j_1}}{\varResult_1^*}})}{\varExp_{j_1}}{\varResult_1}$}

\caseFact{8}{$\reducesTo{\varEnv_2}{\varExp}{\expApp{\varDataCon_{j_2}}{\varResult_2^*}}$ (for some $j_2 < n$)}

\caseFact{9}{$\reducesTo{(\envCat{\varEnv_2}{\envBind{\varVar_{j_2}}{\varResult_2^*}})}{\varExp_{j_2}}{\varResult_2}$}

\caseText{By the resumption assumption}

\caseFact{10}{$\resumesTo{\mergeSolutions{\varSolution}{\varSolution'}}{\varResult_1^*}{\varResult_1^{*\prime}}$}

\caseFact{11}{$\resumesTo{\mergeSolutions{\varSolution}{\varSolution'}}{\varResult_2^*}{\varResult_2^{*\prime}}$}

\caseText{By \ruleName{R-Ctor} on (10)}

\caseFact{12}{$\reducesTo{\mergeSolutions{\varSolution}{\varSolution'}}
                              {\expApp{\varDataCon_{j_1}}{\varResult_1^*}}
                              {\expApp{\varDataCon_{j_1}}{\varResult_1^{*\prime}}}
                  $}

\caseText{By \ruleName{R-Ctor} on (11)}

\caseFact{13}{$\reducesTo{\mergeSolutions{\varSolution}{\varSolution'}}
                              {\expApp{\varDataCon_{j_2}}{\varResult_2^*}}
                              {\expApp{\varDataCon_{j_2}}{\varResult_2^{*\prime}}}
                  $}

\caseText{By the induction hypothesis on (1), (6), (8), (12), and (13)}

\caseFact{14}{$\expApp{\varDataCon_{j_1}}{\varResult_1^{*\prime}} = \expApp{\varDataCon_{j_2}}{\varResult_2^{*\prime}}$}

\caseText{By (14), $\varDataCon_{j_1} = \varDataCon_{j_2}$.}

\caseText{Each constructor of a given type is unique, so $j_1 = j_2$, and thus

$\varVar_{j_1} = \varVar_{j_2}$ and $\varExp_{j_1} = \varExp_{j_2}$.}

\caseText{By the resumption assumption}

\caseFact{15}{$\resumesTo{\mergeSolutions{\varSolution}{\varSolution'}}{\varEnv_2}{\varEnv_2'}$}

\caseText{By \autorefResCompShort{} (across all bindings in environment) on
(1) and (15)}

\caseFact{16}{$\resumesTo{\mergeSolutions{\varSolution}{\varSolution'}}{\varEnv_1}{\varEnv_2'}$}

\caseText{By the evaluation assumption}

\caseFact{17}{$\reducesTo{(\envCat{\varEnv_2'}{\envBind{\varVar_{j_2}}{\varResult_2^{*\prime}}})}{\varExp_{j_2}}{\varResult^*}$} 

\caseText{By the resumption assumption}

\caseFact{18}{$\resumesTo{\mergeSolutions{\varSolution}{\varSolution'}}{\varResult^*}{\varResult^{*\prime}}$}

\caseText{By the definition of environment resumption, (16), (10), and (14)}

\caseFact{19}{$\resumesTo{\mergeSolutions{\varSolution}{\varSolution'}}
                              {(\envCat{\varEnv_1}{\envBind{\varVar_{j_1}}{\varResult_1^*}})}
                              {(\envCat{\varEnv_2'}{\envBind{\varVar_{j_2}}{\varResult_2^{*\prime}}})}
                   $}

\caseText{By the induction hypothesis on (19), (7), (17), (4), and (18)}

\caseFact{20}{$\varResult^{*\prime} = \varResult_1'$}

\caseText{By the definition of environment resumption, (15), and (11)}

\caseFact{21}{$\resumesTo{\mergeSolutions{\varSolution}{\varSolution'}}
                              {(\envCat{\varEnv_2}{\envBind{\varVar_{j_2}}{\varResult_2^*}})}
                              {(\envCat{\varEnv_2'}{\envBind{\varVar_{j_2}}{\varResult_2^{*\prime}}})}
                   $}

\caseText{By the induction hypothesis on (21), (9), (17), (5), and (18)}

\caseFact{22}{$\varResult^{*\prime} = \varResult_2'$}

\caseText{Combining (20) and (22) gives the goal}
}

\jsubcase{2}{\ruleName{E-Case}, \ruleName{E-Case-Indet}}{
\caseFact{6}{$\reducesTo{\varEnv_1}{\varExp}{\expApp{\varDataCon_{j_1}}{\varResult_1^*}}$ (for some $j_1 < n$)}

\caseFact{7}{$\reducesTo{(\envCat{\varEnv_1}{\envBind{\varVar_{j_1}}{\varResult_1^*}})}{\varExp_{j_1}}{\varResult_1}$}

\caseFact{8}{$\reducesTo{\varEnv_2}{\varExp}{\varResult_2^*}$}

\caseFact{9}{$\varResult_2^* \neq \expApp{\varDataCon_j}{\varResult_j}$}

\caseFact{10}{$\varResult_2 = \caseClosure{\varEnv_2}{\varResult_2^*}{\varDataCon_i}{\varVar_i}{\varExp_i}$}

\caseText{By the resumption assumption}

\caseFact{11}{$\resumesTo{\mergeSolutions{\varSolution}{\varSolution'}}{\varResult_1^*}{\varResult_1^{*\prime}}$}

\caseFact{12}{$\resumesTo{\mergeSolutions{\varSolution}{\varSolution'}}{\varResult_2^*}{\varResult_2^{*\prime}}$}

\caseFact{13}{$\resumesTo{\mergeSolutions{\varSolution}{\varSolution'}}{\varEnv_2}{\varEnv_2'}$}

\caseText{By \ruleName{R-Ctor} on (11)}

\caseFact{14}{$\resumesTo{\mergeSolutions{\varSolution}{\varSolution'}}
                              {\expApp{\varDataCon_{j_1}}{\varResult_1^*}}
                              {\expApp{\varDataCon_{j_1}}{\varResult_1^{*\prime}}}
                   $}

\caseText{By the induction hypothesis on (1), (6), (8), (14), and (12)}

\caseFact{15}{$\expApp{\varDataCon_{j_1}}{\varResult_1^{*\prime}} = \varResult_2^{*\prime}$}

\caseText{By \ruleName{R-Fix} on (13)}

\caseFact{16}{$\resumesTo{\mergeSolutions{\varSolution}{\varSolution'}}
                              {\closure{\varEnv_2}{\valFixFun{\varVarF_{j_1}}{\varVar_{j_1}}{\varExp_{j_1}}}}
                              {\closure{\varEnv_2'}{\valFixFun{\varVarF_{j_1}}{\varVar_{j_1}}{\varExp_{j_1}}}}
                   $}

\caseText{By \autorefResIdempLong{} on (11)}

\caseFact{17}{$\resumesTo{\mergeSolutions{\varSolution}{\varSolution'}}{\varResult_1^{*\prime}}{\varResult_1^{*\prime}}$}

\caseText{By the evaluation assumption}

\caseFact{18}{$\reducesTo{(\envCat{\varEnv_2'}{\envBind{\varVar_{j_1}}{\varResult_1^{*\prime}}})}{\varExp_{j_1}}{\varResult^*}$}

\caseText{By the resumption assumption}

\caseFact{19}{$\resumesTo{\mergeSolutions{\varSolution}{\varSolution'}}{\varResult^*}{\varResult^{*\prime}}$}

\caseText{By \ruleName{R-App} on (16), (17), (trivial), (18), and (19)}

\caseFact{20}{$\resumesTo{\mergeSolutions{\varSolution}{\varSolution'}}
                              {\expApp{(\closure{\varEnv_2}{\valFixFun{\varVarF_{j_1}}{\varVar_{j_1}}{\varExp_{j_1}}})}{\varResult_1^{*\prime}}}
                              {\varResult^{*\prime}}
                   $}

\caseText{By \ruleName{R-Case} on (12) (noting (15)) and (20)}

\caseFact{21}{$\resumesTo{\mergeSolutions{\varSolution}{\varSolution'}}{\varResult_2}{\varResult^{*\prime}}$}

\caseText{By \autorefResDetLong{} on (5) and (21)}

\caseFact{22}{$\varResult^{*\prime} = \varResult_2'$}

\caseText{By \autorefResCompShort{} (across all bindings in environment) on (1)
and (13)}

\caseFact{23}{$\resumesTo{\mergeSolutions{\varSolution}{\varSolution'}}{\varEnv_1}{\varEnv_2'}$}

\caseText{By the definition of environment resumption, (23), and (11)}

\caseFact{24}{$\resumesTo{\mergeSolutions{\varSolution}{\varSolution'}}
                              {(\envCat{\varEnv_1}{\envBind{\varVar_{j_1}}{\varResult_1^*}})}
                              {(\envCat{\varEnv_2'}{\envBind{\varVar_{j_1}}{\varResult_1^*}})}
                   $}

\caseText{By the induction hypothesis on (24), (7), (18), (4), and (19)}

\caseFact{25}{$\varResult^{*\prime} = \varResult_1'$}

\caseText{Combining (22) and (25) gives the goal}
}

\jsubcase{3}{\ruleName{E-Case-Indet}, \ruleName{E-Case}}{
  Analagous to previous case.
}

\jsubcase{4}{\ruleName{E-Case-Indet}, \ruleName{E-Case-Indet}}{
\caseFact{6}{$\reducesTo{\varEnv_1}{\varExp}{\varResult_1^*}$}

\caseFact{7}{$\varResult_1^* \neq \expApp{\varDataCon_j}{\varResult_j}$}

\caseFact{8}{$\varResult_1 = \caseClosure{\varEnv_1}{\varResult_1^*}{\varDataCon_i}{\varVar_i}{\varExp_i}$}

\caseFact{9}{$\reducesTo{\varEnv_2}{\varExp}{\varResult_2^*}$}

\caseFact{10}{$\varResult_2^* \neq \expApp{\varDataCon_j}{\varResult_j}$}

\caseFact{11}{$\varResult_2 = \caseClosure{\varEnv_2}{\varResult_2^*}{\varDataCon_i}{\varVar_i}{\varExp_i}$}

\caseText{By the resumption assumption}

\caseFact{12}{$\resumesTo{\mergeSolutions{\varSolution}{\varSolution'}}{\varResult_1^*}{\varResult_1^{*\prime}}$}

\caseFact{13}{$\resumesTo{\mergeSolutions{\varSolution}{\varSolution'}}{\varResult_2^*}{\varResult_2^{*\prime}}$}

\caseText{By the induction hypothesis on (1), (6), (8), (12), and (13)}

\caseFact{14}{$\varResult_1^{*\prime} = \varResult_2^{*\prime}$}

\caseText{By the resumption assumption}

\caseFact{15}{$\resumesTo{\mergeSolutions{\varSolution}{\varSolution'}}{\varEnv_2}{\varEnv_2'}$}

\caseText{By \autorefResCompShort{} (across all bindings in environment) on
(1) and (15)}

\caseFact{16}{$\resumesTo{\mergeSolutions{\varSolution}{\varSolution'}}{\varEnv_1}{\varEnv_2'}$}

\caseText{Resumption of $\varResult_1$ and $\varResult_2$ could go through either \ruleName{R-Case} or \ruleName{R-Case-Indet}.
    For \ruleName{R-Case}, all aspects of the premises and conclusion depend only on the
    resumptions of $\varResult_1^*$ and $\varResult_2^*$---which are equated by (14)---except for the
    $\varEnv$ in the premise
    $\resumesTo{\varSolution}{\expApp{(\closure{\varEnv}{\expFun{\varVar_j}{\varExp_j}})}{\varResult'}}{\varResult_j'}$.
    Noting R-Fix, this premise
    must go through R-App, whose premises and conclusions are entirely dependent
    on the resumptions of the operands $\varResult_1$ $\varResult_2$. By \ruleName{R-Fix}, (15), and (16),
    the resumptions of
    $\closure{\varEnv_1}{\valFixFun{\varVarF_j}{\varVar_j}{\varExp_j}}$ and
    $\closure{\varEnv_2}{\valFixFun{\varVarF_j}{\varVar_j}{\varExp_j}}$ are equal.
    The situation for \ruleName{R-Case-Indet} is similar, though simpler, since its premises
    and conclusion depend only on the resumptions of $\varResult_1^*$ and $\varResult_2^*$ and the resumptions
    of $\varEnv_1$ and $\varEnv_2$, which are equated by (15) and (16).
    Since the resumptions of $\varResult_1$ and $\varResult_2$ (i.e. $\varResult_1'$ and $\varResult_2'$) depend entirely on values
    which are equated, they must themselves be equated.}
}
}
}
% \autoref{v-res}, \autoref{res-idemp}, and \autoref{empty-res} are proven by straightforward induction.

}
\clearpage

\subsection{Unevaluation Constraint Merging}
\label{sec:appendix-constraint-merge}

\autoref{fig:constraint-merge} defines the merge operations for constraints.
\\

%% \begin{figure}[t]
\begin{figure}[h]

\judgementHeadNameOnly
  {(Syntactic) Constraint Merging}
  {
   $\JudgementBox{\mergeConstraintsEquals{\varSolution_1}{\varSolution_2}{\varSolution}}$
   \hspace{0.00in}
   $\JudgementBox{\mergeConstraintsEquals{\varUnsolvedConstraints_1}{\varUnsolvedConstraints_2}{\varUnsolvedConstraints}}$
   \hspace{0.00in}
   $\JudgementBox{\mergeConstraintsEquals{\varConstraints_1}{\varConstraints_2}{\varConstraints}}$
  }

\vsepRule

$
\inferrule*
  {
   \forall
     \expHole{\varHoleName} \in
       \dom{\varSolution_1} \cap \dom{\varSolution_2}.\ 
         \varSolution_1(\expHole{\varHoleName}) =
         \varSolution_2(\expHole{\varHoleName})
  }
  {\mergeConstraintsEquals
    {\varSolution_1}
    {\varSolution_2}
    {\setUnion{\varSolution_1}{\varSolution_2}}
  }
$

\vsepRule

$
\inferrule*
  {
   \varUnsolvedConstraints_1' =
     \setMinus{\varUnsolvedConstraints_1}{\dom{\varUnsolvedConstraints_2}}
   \sepPremise
   \varUnsolvedConstraints_2' =
     \setMinus{\varUnsolvedConstraints_2}{\dom{\varUnsolvedConstraints_1}}
   \\\\
   \varUnsolvedConstraints_{12} =
     \setComp
       {\unsolvedConstraint
         {\varHoleName}
         {\setUnion
           {\varUnsolvedConstraints_1(\expHole{\varHoleName})}
           {\varUnsolvedConstraints_2(\expHole{\varHoleName})}}
       }
       {\expHole{\varHoleName} \in
          \dom{\varSolution_1} \cap \dom{\varSolution_2}
       }
       %% {\unsolvedConstraint{\varHoleName}{\setUnion{\varWorlds_1}{\varWorlds_2}}}
       %% {
       %%  \unsolvedConstraint{\varHoleName}{\varWorlds_1} \in \varUnsolvedConstraints_2
       %%  \wedge
       %%  \unsolvedConstraint{\varHoleName}{\varWorlds_2} \in \varUnsolvedConstraints_2
       %% }
  }
  {\mergeConstraintsEquals
    {\varUnsolvedConstraints_1}
    {\varUnsolvedConstraints_2}
    {\setUnionThree
      {\varUnsolvedConstraints_1'}
      {\varUnsolvedConstraints_{12}}
      {\varUnsolvedConstraints_2'}
    }
  }
$

\vsepRule

$
\inferrule*
  {
   \mergeConstraintsEquals
     {\varSolution_1}
     {\varSolution_2}
     {\varSolution'}
   \sepPremise
   \mergeConstraintsEquals
     {\varUnsolvedConstraints_1}
     {\varUnsolvedConstraints_2}
     {\varUnsolvedConstraints'}
  }
  {\mergeConstraintsEquals
    {\pairConstraints{\varSolution_1}{\varUnsolvedConstraints_1}}
    {\pairConstraints{\varSolution_2}{\varUnsolvedConstraints_2}}
    {\pairConstraints{\varSolution'}{\varUnsolvedConstraints'}}
  }
$

\vsepRule

%% \judgementHeadNameOnly{Constraint Simplification}
\judgementHeadNameOnly
  {(Semantic) Constraint Merging}
  $\JudgementBox{\simplifyConstraintsEquals
    {\varHoleEnv}
    {\varDatatypeEnv}
    {(\varConstraints)}
    {\varConstraints'}}$

\vsepRule

$
\inferrule*
  { \varSolution(\expHole{\varHoleName}) = \varExp
    \sepPremise
    \worldConsistentFull
      {\varHoleEnv}{\varDatatypeEnv}
      {\varSolution}{\varExp}{\varWorlds}{\varConstraints}
  }
  { \resolveEquals
      {\varHoleEnv}
      {\varDatatypeEnv}
      {\varSolution}
      {\unsolvedConstraint{\varHoleName}{\varWorlds}}
      {\varConstraints}
  }
$
\hsepRule
$
\inferrule*
  { \expHole{\varHoleName} \not \in \varSolution
  }
  { \resolveEquals
      {\varHoleEnv}
      {\varDatatypeEnv}
      {\varSolution}
      {\unsolvedConstraint{\varHoleName}{\varWorlds}}
      { \pairConstraints
          {\emptySet}
          {\unsolvedConstraint{\varHoleName}{\varWorlds}}
      }
  }
$

\vsepRule

$
\inferrule*
  { \multiPremise{
      \resolveEquals
        {\varHoleEnv}
        {\varDatatypeEnv}
        {\varSolution}
        {\unsolvedConstraint{\varHoleName_i}{\varWorlds}}
        {\varConstraints'_i}
    }{\sequenceSyntax}
  }
  { \stepSimplifyConstraintsEquals
      {\varHoleEnv}
      {\varDatatypeEnv}
      {\pairConstraints
        {\varSolution}
        {\envCatThree
          {\unsolvedConstraint{\varHoleName_1}{\varWorlds_1}}
          {\ldots}
          {\unsolvedConstraint{\varHoleName_n}{\varWorlds_n}}}}
      {\mergeConstraints
          {\pairConstraints{\varSolution}{\emptySet}}
          {\mergeConstraintsThree{\varConstraints'_1}{\cdots}{\varConstraints'_n}}}
  }
$

\vsepRule

$
\inferrule*
  { \stepSimplifyConstraintsEquals
      {\varHoleEnv}
      {\varDatatypeEnv}
      {(\varConstraints)}
      {\varConstraints'}
      \sepPremise
      \varConstraints \not= \varConstraints'
      \\\\
      \simplifyConstraintsEquals
        {\varHoleEnv}
        {\varDatatypeEnv}
        {(\varConstraints')}
        {\varConstraints''}
  }
  { \simplifyConstraintsEquals
      {\varHoleEnv}
      {\varDatatypeEnv}
      {(\varConstraints)}
      {\varConstraints''}
  }
$
\hsepRule
$
\inferrule*
  { \stepSimplifyConstraintsEquals
      {\varHoleEnv}
      {\varDatatypeEnv}
      {(\varConstraints)}
      {\varConstraints'}
      \sepPremise
      \varConstraints = \varConstraints'
  }
  { \simplifyConstraintsEquals
      {\varHoleEnv}
      {\varDatatypeEnv}
      {(\varConstraints)}
      {\varConstraints'}
  }
$

\vsepBeforeCaption

\caption{Constraint Merging.}
\label{fig:constraint-merge}
\end{figure}

\clearpage

\subsection{Type-Directed Guessing}
\label{sec:appendix-guessing}

\autoref{fig:synthesis-guess} defines type-directed guessing rules analogous to
expression type rules (\autoref{fig:typing}).
\\

%% \begin{figure}[t]
\begin{figure}[h]

%% \judgementHead
%%   {``\maybeUnderline{EGuess}ing'' \rkc{drop this in favor of version below?}}
%%   {\guessFull{\varHoleEnv}{\varDatatypeEnv}{\varTypeEnv}{\varType}{\varExp}{\varHoleEnv'}}
%% 
%% \vsepRule
%% 
%% $
%% \inferrule*[lab=\ruleNameFig{EGuess-Var}]
%%   {
%%    \varTypeEnv(x) = \varType
%%   }
%%   {\guess{\varTypeEnv}{\varType}{\varVar}{\emptyEnv}}
%% $
%% %
%% \hsepRule
%% %
%% $
%% \inferrule*[lab=\ruleNameFig{EGuess-App}]
%%   {
%%    \guess{\varTypeEnv}{\tArrow{\varType'}{\varType}}{\varExp}{\varHoleEnv'}
%%    \\\\
%%    \varHoleName_1 \textrm { fresh}
%%    \\\\
%%    \varHoleEnv_1 =
%%      \envBindHole{\varHoleName}{\tHole{\varTypeEnv}{\varType'}}
%%   }
%%   {\guess{\varTypeEnv}{\varType}{\expApp{\varExp}{\expHole{\varHoleName}}}{\setUnion{\varHoleEnv'}{\varHoleEnv_1}}}
%% $
%% %
%% \hsepRule
%% %
%% $
%% \inferrule*[lab=\ruleNameFig{EGuess-Get}]
%%   {
%%    \guess{\varTypeEnv}{\tPair{\varType_1}{\varType_2}}{\varExp}{\varHoleEnv'}
%%   }
%%   {\guess{\varTypeEnv}{\varType_i}{\expProj{i}{\varExp}}{\varHoleEnv'}}
%% $
%% 
%% \vsepRule

\tightJudgementHead
  %% {``\maybeUnderline{EGuess}ing''}
  %% {Type-Directed \maybeUnderline{Guess}ing}
  {\maybeUnderline{Guess}ing}
  {\guessFull{\varDatatypeEnv}{\varTypeEnv}{\varType}{\varExp}}

\vsepRule

%% $
%% \inferrule*[lab=\ruleNameFig{EGuess-Var}]
%%   {
%%    \varTypeEnv(x) = \varType
%%   }
%%   {\guess{\varTypeEnv}{\varType}{\varVar}}
%% $
%% %
%% \hsepRule
%% %
%% $
%% \inferrule*[lab=\ruleNameFig{EGuess-App}]
%%   {
%%    \termGen{\varTypeEnv}{\tArrow{\varType_2}{\varType}}{\varExp_1}
%%    \\\\
%%    \termGen{\varTypeEnv}{\varType_2}{\varExp_2}
%%   }
%%   {\guess{\varTypeEnv}{\varType}{\expApp{\varExp_1}{\varExp_2}}}
%% $
%% %
%% \hsepRule
%% %
%% $
%% \inferrule*[lab=\ruleNameFig{EGuess-Get}]
%%   {
%%    \termGen{\varTypeEnv}{\tPair{\varType_1}{\varType_2}}{\varExp}
%%   }
%%   {\guess{\varTypeEnv}{\varType_i}{\expProj{i}{\varExp}}}
%% $
%% 
%% \vsepRule
%% 
%% \judgementHead
%%   {\maybeUnderline{Term} \maybeUnderline{Gen}eration}
%%   {\termGenFull{\varDatatypeEnv}{\varTypeEnv}{\varType}{\varExp}}
%% 
%% \vsepRule

%% $
%% \inferrule*[lab=\ruleNameFig{TermGen-EGuess}]
%%   {
%%    \guess{\varTypeEnv}{\varType}{\varExp}
%%   }
%%   {\termGen{\varTypeEnv}{\varType}{\varExp}}
%% $
%% %
%% \hsepRule
%% %
$
\inferrule*[lab=\ruleNameFig{Guess-Unit}]
  {
  }
  {\guess{\varTypeEnv}{\tUnit}{\expUnit}}
$
\hsepRule
$
\inferrule*[lab=\ruleNameFig{Guess-Pair}]
  {
   \multiPremise{
   \guess{\varTypeEnv}{\varType_i}{\varExp_i}
   }{\pairIndex{i}}
  }
  {\guess{\varTypeEnv}{\tPair{\varType_1}{\varType_2}}{\pair{\varExp_1}{\varExp_2}}}
$

\vsepRule

$
\inferrule*[lab=\ruleNameFig{Guess-Ctor}]
  {
   \lookupDataConArrowType{\varDataCon}{\varType}{\varTypeCon}
   \sepPremise
   \guess{\varTypeEnv}{\varType}{\varExp}
  }
  {\guess{\varTypeEnv}{\varTypeCon}{\expApp{\varDataCon}{\varExp}}}
$
\hsepRule
$
\inferrule*[lab=\ruleNameFig{Guess-Fix}]
  {
   \guess{\envCatThree{\varTypeEnv}
                        {\envBindType{\varVarF}{\tArrow{\varType_1}{\varType_2}}}
                        {\envBindType{\varVar}{\varType_1}}
           }{\varType_2}{\varExp}
  }
  {\guess{\varTypeEnv}{\tArrow{\varType_1}{\varType_2}}{\expFixFun{\varVarF}{}{\varVar}{\varExp}}}
$

\vsepRule

$
\inferrule*[lab=\ruleNameFig{Guess-Case}]
  {
   \lookupTypeConstructors{\varTypeCon}{\varDataCon_i}{\varType_i}
   \sepPremise
   \guess{\varTypeEnv}{\varTypeCon}{\varExp}
   \sepPremise
   \multiPremise{
   \guess{\envCat{\varTypeEnv}{\envBindType{\varVar_i}{\varType_i}}}
           {\varType}
           {\varExp_i}
   }{\sequenceSyntax}
  }
  {\guess
    {\varTypeEnv}
    {\varType}
    {\expMatch{\varExp}{\varDataCon_i}{\varVar_i}{\varExp_i}}
  }
$

\vsepRule

$
\inferrule*[lab=\ruleNameFig{Guess-Var}]
  {
   \varTypeEnv(x) = \varType
  }
  {\guess{\varTypeEnv}{\varType}{\varVar}}
$
%
%% \hsepRule
\hspace{0.12in} %% TODO
$
\inferrule*[lab=\ruleNameFig{Guess-App}]
  {
   \guess{\varTypeEnv}{\tArrow{\varType_2}{\varType}}{\varExp_1}
   \\\\
   \guess{\varTypeEnv}{\varType_2}{\varExp_2}
  }
  {\guess{\varTypeEnv}{\varType}{\expApp{\varExp_1}{\varExp_2}}}
$
%
%% \hsepRule
\hspace{0.12in} %% TODO
$
\inferrule*[lab=\ruleNameFig{Guess-Prj}]
  {
   \guess{\varTypeEnv}{\tPair{\varType_1}{\varType_2}}{\varExp}
  }
  {\guess{\varTypeEnv}{\varType_i}{\expProj{i}{\varExp}}}
$

\vsepBeforeCaption

\caption{Type-Directed Guessing.}
\label{fig:synthesis-guess}
\end{figure}

\vsepRule

\vsepRule

\parahead{Guessing Recursive Sketches}

Guessing does not generate hole expressions.
Guessing is, furthermore, limited to small terms and elimination forms in
practice.
However, if guessing were to generate recursive function sketches,
the \ruleName{Guess-and-Check} rule provides an additional antidote for
trace-completeness:
when guessing an expression $\expFixFun{\varVarF}{}{\varVar}{\varExp}$ to fill
$\expHole{\varHoleName}$, the extended hole-filling
$\envCat{\varSolution}{\holeFilling{\varHoleName}{\expFixFun{\varVarF}{}{\varVar}{\varExp}}}$
``ties the recursive knot'' before checking example consistency.

\clearpage

\subsection{Synthesis Soundness}
\label{sec:appendix-synthesis-soundness}

\begin{theorem}[Type Soundness of Unevaluation]
\label{prop:sound-type-uneval}

\breakAndIndent
If $\typeCheckSolution{\varHoleEnv}{\varSolution'}$
and $\typeCheckResult{\varResult}{\varType}$
and $\typeCheckExample{\varEx}{\varType}$
and $\uneval{\varSolution'}{\varResult}{\varEx}{(\varUnsolvedConstraints, \varSolution)}$,

\justIndent
then $\typeCheckUnsolved{\varHoleEnv}{\varUnsolvedConstraints}$
and $\typeCheckSolution{\varHoleEnv}{\varSolution}$.

\end{theorem}

\begin{theorem}[Type Soundness of Checking]
\label{prop:sound-type-check}

\breakAndIndent
If $\typeCheckSolution{\varHoleEnv}{\varSolution'}$
and $\typeCheckWorlds{\varHoleEnv}{\varWorlds}{\varTypeEnv}{\varType}$
and $\typeCheck{\varTypeEnv}{\varExp}{\varType}$
and $\worldConsistent{\varSolution'}{\varExp}{\varWorlds}{(\varUnsolvedConstraints, \varSolution)}$,

\justIndent
then $\typeCheckUnsolved{\varHoleEnv}{\varUnsolvedConstraints}$
and $\typeCheckSolution{\varHoleEnv}{\varSolution}$.

\end{theorem}

\begin{theorem}[Type Soundness of Guess]
\label{prop:sound-type-guess}

\breakAndIndent
If $\guessFull{\varDatatypeEnv}{\varTypeEnv}{\varType}{\varExp}$,
then $\typeCheck{\varTypeEnv}{\varExp}{\varType}$.

\end{theorem}

\begin{theorem}[Type Soundness of Refine/Branch]
\label{prop:sound-type-refine-branch}

\breakAndIndent
If $\typeCheckWorlds{\varHoleEnv}{\varWorlds}{\varTypeEnv}{\varType}$
and $\typeCheckSolution{\varHoleEnv}{\varSolution'}$

\justIndent
and ${\refineOrBranch
       {\varSolution'}{\varTypeEnv}{\varWorlds}{\varType}{\varExp}
       {
         \multiPremise{
         \problemNameTypeWorlds{\varHoleName_i}{\varTypeEnv_i}{\varType_i}{\varWorlds_i}
         }{\sequenceSyntax}
       }
       {(\varUnsolvedConstraints, \varSolution)}
     }$,

\justIndent
then $\typeCheckIn
        {\setUnion{\varHoleEnv}{\multiPremise{\envBindHole{\varHoleName_i}{\tHole{\varTypeEnv_i}{\varType_i}}}{\sequenceSyntax}}}
        {\varTypeEnv}{\varExp}{\varType}$
and $\typeCheckWorlds
        {\varHoleEnv}{\varWorlds_i}{\varTypeEnv_i}{\varType_i}$.

\justIndent
and $\typeCheckUnsolved{\varHoleEnv}{\varUnsolvedConstraints}$
and $\typeCheckSolution{\varHoleEnv}{\varSolution}$

\end{theorem}

\begin{theorem}[Type Soundness of Fill]
\label{prop:sound-type-fill}

\breakAndIndent
If $\typeCheckSolution{\varHoleEnv}{\varSolution'}$
and $\typeCheckWorlds{\varHoleEnv}{\varWorlds}{\varTypeEnv}{\varType}$
and $\fillHoleFull{\varHoleEnv}{\varDatatypeEnv}{\varHoleName}
                  {\varSolution'}{\varTypeEnv}{\varWorlds}{\varType}
                  {\pairConstraints{\varSolution}{\varUnsolvedConstraints}}
                  {\varHoleEnv'}$

\justIndent
%%
%then $\varHoleName \in F$
%
then $\typeCheckSolution
       {\setUnionThree{\varHoleEnv}{\varHoleEnv'}{\left(\envBindHole{\varHoleName}{\tHole{\varTypeEnv}{\varType}}\right)}}
       {\left(\varUnsolvedConstraints, \varSolution\right)}$.%{\left(\holeFilling{\varHoleName}{\varSolution\left(\expHole{\varHoleName}\right)}\right)}$.

\end{theorem}

\begin{theorem}[Type Soundness of Result Consistency]
\label{prop:sound-type-consistency}

\breakAndIndent
If $\typeCheckResult{\varResult}{\varType}$
and $\typeCheckResult{\varResult'}{\varType}$
and $\resultConsistent{\varResult}{\varResult'}{\varAssertions}$
then $\typeCheckAssertions{\varHoleEnv}{\varAssertions}$.

\end{theorem}

\begin{theorem}[Type Soundness of Simplify]
\label{prop:sound-type-simplify}

\breakAndIndent
If $\typeCheckAssertions{\varHoleEnv}{\varAssertions}$
and $\simplifyEquals{\varAssertions}{(\varUnsolvedConstraints, \varSolution)}$,
then $\typeCheckUnsolved{\varHoleEnv}{\varUnsolvedConstraints}$
and $\typeCheckSolution{\varHoleEnv}{\varSolution}$.

\end{theorem}

\begin{theorem}[Type Soundness of Program Evaluation]
\label{prop:sound-type-prog}

\breakAndIndent
If $\typeCheckProgram{\varProgram}{\varType}{\varType'}$
and $\programReducesTo{\varProgram}{\varResult}{\varAssertions}$,
then $\typeCheckResult{\varResult}{\varType}$
and $\typeCheckAssertions{\varHoleEnv}{\varAssertions}$.

\end{theorem}

\beginTheorem{ExUneval}{Soundness of Example Unevaluation}{Ex. Uneval.}{prop:sound-uneval}

\breakAndIndent
If $\constraintSat{\addSolutions{\varSolution}{\varSolution'}}{\varConstraints}$
and $\isFinal{\varResult}$
and $\uneval{\varSolution}{\varResult}{\varEx}{\varConstraints}$
and $\resumesTo{\addSolutions{\varSolution}{\varSolution'}}{\varResult}{\varResult'}$
%
%% \breakAndIndent
then $\exSat{\addSolutions{\varSolution}{\varSolution'}}{\varResult'}{\varEx}$.

\end{theorem}

\beginTheorem{LBECheck}{Soundness of Live Bidirectional Example Checking}{Ex. Check.}{prop:sound-check}

\breakAndIndent
If $\constraintSat{\addSolutions{\varSolution}{\varSolution'}}{\varConstraints}$
and $\worldConsistent{\varSolution}{\varExp}{\varWorlds}{\varConstraints}$,
then $\worldSat{\addSolutions{\varSolution}{\varSolution'}}{\varExp}{\varWorlds}$.

\end{theorem}

\begin{theorem}[Example Soundness of Refine]
\label{prop:sound-ex-refine}

\breakAndIndent
If ${\refineFull
      {\varHoleEnv}{\varDatatypeEnv}
      {\varTypeEnv}{\varWorlds}{\varType}{\varExp}
      {
        \multiPremise{
        \problemNameTypeWorlds{\varHoleName_i}{\varTypeEnv_i}{\varType_i}{\varWorlds_i}
        }{\sequenceSyntax}
      }
    }$
and $\multiPremise
       {\worldSat{\varSolution}{\expHole{\varHoleName_i}}{\varWorlds_i}}
       {\sequenceSyntax}$,

\justIndent
then $\worldSat{\varSolution}{\varExp}{\varWorlds}$.

\end{theorem}

\begin{theorem}[Example Soundness of Branch]
\label{prop:sound-ex-branch}

\breakAndIndent
If ${\branchFull
      {\varHoleEnv}{\varDatatypeEnv}
      {\varTypeEnv}{\varWorlds}{\varType}{\varExp}
      {
        \multiPremise{
        \problemNameTypeWorlds{\varHoleName_i}{\varTypeEnv_i}{\varType_i}{\varWorlds_i}
        }{\sequenceSyntax}
      }
      {\varConstraints}
      {\varSolution}
    }$

\justIndent
and $\constraintSat{\addSolutions{\varSolution}{\varSolution'}}{\varConstraints}$
and $\multiPremise
       {\worldSat{\addSolutions{\varSolution}{\varSolution'}}{\expHole{\varHoleName_i}}{\varWorlds_i}}
       {\sequenceSyntax}$,

\justIndent
then $\worldSat{\addSolutions{\varSolution}{\varSolution'}}{\varExp}{\varWorlds}$.

\end{theorem}

\begin{theorem}[Example Soundness of Fill]
\label{prop:sound-ex-fill}

\breakAndIndent
If $\fillHoleFull{\varHoleEnv}{\varDatatypeEnv}{\varHoleName}
                 {\varSolution}{\varTypeEnv}{\varWorlds}{\varType}
                 {\varConstraints}
                 {\varHoleEnv'}$
and $\constraintSat{\addSolutions{\varSolution}{\varSolution'}}{\varConstraints}$,

\justIndent
then $\left(\addSolutions{\varSolution}{\varSolution'}\right) \left(\expHole{\varHoleName}\right) = \varExp$
and $\worldSat{\addSolutions{\varSolution}{\varSolution'}}{\varExp}{\varWorlds}$.

\end{theorem}

\begin{theorem}[Type Soundness of Semantic Merge]
\label{prop:sound-type-merge}

\breakAndIndent
If $\typeCheckSolution{\varHoleEnv}{\varConstraints}$
and $\simplifyConstraintsEquals{\varHoleEnv}{\varDatatypeEnv}{(\varConstraints)}{\varConstraints'}$,
then $\typeCheckSolution{\varHoleEnv}{\varConstraints'}$.

\end{theorem}

\begin{theorem}[Example Soundness of Semantic Merge]
\label{prop:sound-merge}

\breakAndIndent
If $\constraintSat{\varSolution}{\varConstraints'}$
and $\simplifyConstraintsEquals{\varHoleEnv}{\varDatatypeEnv}{(\varConstraints)}{\varConstraints'}$,
then $\constraintSat{\varSolution}{\varConstraints}$.

\end{theorem}

\begin{theorem}[Soundness of Solve]
\label{prop:sound-solve}

\breakAndIndent
If $\typeCheckUnsolved{\varHoleEnv}{\varUnsolvedConstraints}$
and $\typeCheckUnsolved{\varHoleEnv}{\varSolution}$
and $\iterSolveEqualsFull{\varHoleEnv}{\varDatatypeEnv}{(\varUnsolvedConstraints, \varSolution)}{\varSolution'}{\varHoleEnv'}$,
%
%% \justIndent
%
then $\typeCheckSolution{\varHoleEnv'}{\varSolution'}$
and $\constraintSat{\varSolution'}{(\varUnsolvedConstraints, \varSolution)}$.

\end{theorem}

\begin{theorem}[Soundness of Assertion Simplification]
\label{prop:sound-asrt-simpl}

\breakAndIndent
If $\simplifyEquals{\varAssertions}{\varConstraints}$
and $\constraintSat{\varSolution}{\varConstraints}$,
then $\constraintSat{\varSolution}{\varAssertions}$.

\end{theorem}

\begin{theorem}[Soundness of Synthesis]
\label{prop:sound-synth}

\breakAndIndent
If $\typeCheckProgram{\varProgram}{\varType}{\varType'}$
and $\programReducesTo{\varProgram}{\varResult}{\varAssertions}$
and $\simplifyEquals{\varAssertions}{\varConstraints}$
and $\iterSolveEqualsFull{\varHoleEnv}{\varDatatypeEnv}{(\varConstraints)}{\varSolution}{\varHoleEnv'}$,

\justIndent
then $\typeCheckSolution{\varHoleEnv'}{\varSolution}$
and $\assertionSat{\varSolution}{\varAssertions}$.

\end{theorem}

\begin{lemma}[Example Satisfaction of Simple Value]
\label{prop:ex-sat-val}

\breakAndIndent
If $\coerceUndet{\varEx}{\varSimpleVal}$
and $\exSat{\varSolution}{\varResult}{\varEx}$,
then $\coerceUndet{\varResult}{\varSimpleVal}$.

\end{lemma}

\begin{lemma}[Constraint Satisfaction Implies Complete Resumption]
\label{prop:sat->res}

\breakAndIndent
If $\coerceUndet{\varEx}{\varSimpleVal}$
and $\isFinal{\varResult}$
and $\uneval{\emptyEnv}{\varResult}{\varEx}{\varConstraints}$
and $\constraintSat{\varSolution}{\varConstraints}$,
%
%% \justIndent
%
then $\resumesTo{\varSolution}{\varResult}{\varResult'}$
and $\coerceUndet{\varResult'}{\varSimpleVal}$.

\end{lemma}

\subsection*{Proofs}
{
\label{sec:appendix-synthesis-proofs}

\jtheorem{\jtref{\autoref{prop:sound-type-uneval} (Uneval)} and \jtref{\autoref{prop:sound-type-check} (Check)}}{
Straightforward mutual induction.
}

\jtheorem{\jtref{\autoref{prop:sound-type-guess} (Guess)}}{
Straightforward induction.
}

\jtheorem{\jtref{\autoref{prop:sound-type-refine-branch} (Refine/Branch)}}{
Straightforward by way of \autoref{prop:sound-type-guess}, \autoref{prop:pres}, and \autoref{prop:sound-type-check}.
}

\jtheorem{\jtref{\autoref{prop:sound-type-fill} (Fill)}}{
The \ruleName{Defer} case is trivial.
The \ruleName{Refine,Branch} case is straightforward by way of \autoref{prop:sound-type-refine-branch}.
The \ruleName{Guess-and-Check} case goes through by \autoref{prop:sound-type-check}.
}

\jtheorem{\jtref{\autoref{prop:sound-type-consistency} (Result consistency)}}{
Straightforward induction.
}

\jtheorem{\jtref{\autoref{prop:sound-type-simplify} (Simplify)}}{
Straightforward by way of \autoref{prop:sound-type-uneval}.
}

\jtheorem{\jtref{\autoref{prop:sound-type-prog} (Program evaluation)}}{
Straightforward by way of \autoref{prop:pres} and \autoref{prop:sound-type-consistency}.
}

\jtheorem{\jtref{\autorefExUnevalLong{}}}{
The cases \ruleName{U-Top}, \ruleName{U-Unit}, \ruleName{U-Pair} and \ruleName{U-Ctor} are straightforward applications of their respective \textsc{XS} rules and induction.
\ruleName{U-Hole} goes through because the premise $\constraintSat{\addSolutions{\varSolution}{\varSolution'}}{\varConstraints}$
proves example satisfaction for the single generated constraint.
The remaining cases are considered in detail here.

\jcase{1}{\ruleName{U-Fix}}{
\jgivengoal{
  \caseFact{1} $\constraintSat{\addSolutions{\varSolution}{\varSolution'}}{\varConstraints}$

  \caseFact{2} $\isFinal{\closure{\varEnv}{\valFixFun{\varVarF}{\varVar}{\varExp}}}$

  \caseFact{3} $\uneval{\varSolution}{\closure{\varEnv}{\valFixFun{\varVarF}{\varVar}{\varExp}}}{\ioExample{\varSimpleVal}{\varEx}}{\varConstraints}$

  \caseFact{4} $\resumesTo{\addSolutions{\varSolution}{\varSolution'}}{\closure{\varEnv}{\valFixFun{\varVarF}{\varVar}{\varExp}}}{\varResult'}$
}{
  $\exSat{\addSolutions{\varSolution}{\varSolution'}}{\varResult'}{\ioExample{\varSimpleVal}{\varEx}}$
}
\caseText{By inversion of unevaluation on (3)}

\caseFact{5} $\worldConsistent
                {\varSolution}
                {\varExp}
                {\world
                   {(\envCatThree
                          {\varEnv}
                          {\envBind{\varVarF}{\closure{\varEnv}{\valFixFun{\varVarF}{\varVar}{\varExp}}}}
                          {\envBind{\varVar}{\varSimpleVal}})}
                   {\varEx}}
                {\varConstraints}$

\caseText{By inversion of the check judgment on (5)}

\caseFact{6} $\reducesTo{(\envCatThree
                          {\varEnv}
                          {\envBind{\varVarF}{\closure{\varEnv}{\valFixFun{\varVarF}{\varVar}{\varExp}}}}
                          {\envBind{\varVar}{\varSimpleVal}})}{\varExp}{\varResult^*}$

\caseFact{7} $\resumesTo{\varSolution}{\varResult^*}{\varResult^{**}}$

\caseFact{8} $\uneval{\varSolution}{\varResult^{**}}{\varEx}{\varConstraints}$

\caseText{By the resumption assumption}

\caseFact{9} $\resumesTo{\addSolutions{\varSolution}{\varSolution'}}{\varResult^{**}}{\varResult^{**+}}$

\caseText{By \autorefResCompShort{} on (7) and (9)}

\caseFact{10} $\resumesTo{\addSolutions{\varSolution}{\varSolution'}}{\varResult^{*}}{\varResult^{**+}}$

\caseText{By the induction hypothesis on (1), \autoref{prop:res-fin}, (8), and (9)}

\caseFact{11} $\exSat{\addSolutions{\varSolution}{\varSolution'}}{\varResult^{**+}}{\varEx}$

\caseText{By inversion of resumption on (4)}

\caseFact{12} $\resumesTo{\addSolutions{\varSolution}{\varSolution'}}{\varEnv}{\varEnv'}$

\caseFact{13} $\varResult' = {\closure{\varEnv'}{\valFixFun{\varVarF}{\varVar}{\varExp}}}$

\caseText{By the evaluation assumption}

\caseFact{14} $\reducesTo{(\envCatThree
                          {\varEnv'}
                          {\envBind{\varVarF}{\closure{\varEnv'}{\valFixFun{\varVarF}{\varVar}{\varExp}}}}
                          {\envBind{\varVar}{\varSimpleVal}})}{\varExp}{\varResult^{*'}}$

\caseText{By \autoref{prop:v-res} (Simple Value Resumption)}

\caseFact{15} $\resumesTo{\addSolutions{\varSolution}{\varSolution'}}{\varSimpleVal}{\varSimpleVal}$

\caseText{By the resumption assumption}

\caseFact{16} $\resumesTo{\addSolutions{\varSolution}{\varSolution'}}{\varResult^{*'}}{\varResult^{*''}}$

\caseText{By \autorefResIdempLong{} on (4)}

\caseFact{17} $\resumesTo{\addSolutions{\varSolution}{\varSolution'}}{\closure{\varEnv'}{\valFixFun{\varVarF}{\varVar}{\varExp}}}{\closure{\varEnv'}{\valFixFun{\varVarF}{\varVar}{\varExp}}}$

\caseText{By \ruleName{R-App} on (17), (15), (13), (14), and (16)}

\caseFact{18} $\resumesTo{\addSolutions{\varSolution}{\varSolution'}}{\expApp{(\closure{\varEnv'}{\valFixFun{\varVarF}{\varVar}{\varExp}})}{\varSimpleVal}}{\varResult^{*''}}$

\caseText{By the definition of environment resumption, (12), (4), and (15)}

\caseFact{19} $\resumesTo
                 {\addSolutions{\varSolution}{\varSolution'}}
                 {(\envCatThree
                          {\varEnv}
                          {\envBind{\varVarF}{\closure{\varEnv}{\valFixFun{\varVarF}{\varVar}{\varExp}}}}
                          {\envBind{\varVar}{\varSimpleVal}})}
                 {(\envCatThree
                          {\varEnv'}
                          {\envBind{\varVarF}{\closure{\varEnv'}{\valFixFun{\varVarF}{\varVar}{\varExp}}}}
                          {\envBind{\varVar}{\varSimpleVal}})}$

\caseText{By \autorefEvalResShort{} on (19), (6), (14), (10), and (16)}

\caseFact{20} $\varResult^{**+} = \varResult^{*''}$

\caseText{By \ruleName{XS-Input-Output} on (18), (20), and (11)}

\caseFact{21} $\exSat{\addSolutions{\varSolution}{\varSolution'}}{\closure{\varEnv'}{\valFixFun{\varVarF}{\varVar}{\varExp}}}{\ioExample{\varSimpleVal}{\varEx}}$

\caseText{Observing (13), (21) is the goal}
}

\jcase{2}{\ruleName{U-App}}{
\jgivengoal{
  \caseFact{1} $\constraintSat{\addSolutions{\varSolution}{\varSolution'}}{\varConstraints}$

  \caseFact{2} $\isFinal{(\expApp{\varResult_1}{\varResult_2})}$

  \caseFact{3} $\uneval{\varSolution}{\expApp{\varResult_1}{\varResult_2}}{\varEx}{\varConstraints}$

  \caseFact{4} $\resumesTo{\addSolutions{\varSolution}{\varSolution'}}{\expApp{\varResult_1}{\varResult_2}}{\varResult'}$
}{
  $\exSat{\addSolutions{\varSolution}{\varSolution'}}{\varResult'}{\varEx}$
}
\caseText{By inversion of unevaluation on (3)}

\caseFact{5} $\coerceUndet{\varResult_2}{\varSimpleVal_2}$

\caseFact{6} $\uneval{\varSolution}{\varResult_1}{\ioExample{\varSimpleVal_2}{\varEx}}{\varConstraints}$

\caseText{By the resumption assumption}

\caseFact{7} $\resumesTo{\addSolutions{\varSolution}{\varSolution'}}{\varResult_1}{\varResult_1'}$

\caseText{By the induction hypothesis on (1), inversion of final on (2), (6) and (7)}

\caseFact{8} $\exSat{\addSolutions{\varSolution}{\varSolution'}}{\varResult_1'}{\ioExample{\varSimpleVal_2}{\varEx}}$

\caseText{By inversion of example satisfaction on (8)}

\caseFact{9} $\resumesTo{\addSolutions{\varSolution}{\varSolution'}}{\expApp{\varResult_1'}{\varResult_2}}{\varResult}$

\caseFact{10} $\exSat{\addSolutions{\varSolution}{\varSolution'}}{\varResult}{\varEx}$

\caseText{By \autorefResAppShort{} on (7) and (9)}

\caseFact{11} $\resumesTo{\addSolutions{\varSolution}{\varSolution'}}{\expApp{\varResult_1}{\varResult_2}}{\varResult}$

\caseText{By \autorefResDetLong{} on (4) and (11)}

\caseFact{12} $\varResult = \varResult'$

\caseText{Goal is given by (10), observing (12)}
}

\jcase{3}{Projections}{
\caseText{Without loss of generality, we will only detail the $\ruleName{U-Prj-1}$ case}

\jgivengoal{
\caseFact{1} $\constraintSat{\addSolutions{\varSolution}{\varSolution'}}{\varConstraints}$

\caseFact{2} $\isFinal{(\expFst{\varResult})}$

\caseFact{3} $\uneval{\varSolution}{\expFst{\varResult}}{\varEx}{\varConstraints}$

\caseFact{4} $\resumesTo{\addSolutions{\varSolution}{\varSolution'}}{\expFst{\varResult}}{\varResult'}$
}{
  $\exSat{\addSolutions{\varSolution}{\varSolution'}}{\varResult'}{\varEx}$
}
\caseText{By inversion of unevaluation on (3)}

\caseFact{5} $\uneval{\varSolution}{\varResult}{\pair{\varEx}{\exHole}}{\varConstraints}$

\caseText{By the resumption assumption}

\caseFact{6} $\resumesTo{\addSolutions{\varSolution}{\varSolution'}}{\varResult}{\varResult^+}$

\caseText{By the induction hypothesis on (1), inversion of final on (2), (5), and (6)}

\caseFact{7} $\exSat{\addSolutions{\varSolution}{\varSolution'}}{\varResult^+}{\pair{\varEx}{\exHole}}$

\caseText{By inversion of example satisfaction on (7)}

\caseFact{8} $\varResult^+ = \pair{\varResult_1^+}{\varResult_2^+}$

\caseFact{9} $\exSat{\addSolutions{\varSolution}{\varSolution'}}{\varResult_1^+}{\varEx}$

\caseText{By \ruleName{R-Prj} on (6) (observing (8))}

\caseFact{10} $\resumesTo{\addSolutions{\varSolution}{\varSolution'}}{\expFst{\varResult}}{\varResult_1^+}$

\caseText{By \autorefResDetLong{} on (4) and (10)}

\caseFact{11} $\varResult' = \varResult_1^+$

\caseText{Goal is given by (9), observing (11)}
}

\jcase{4}{\ruleName{U-Case}}{
\jgivengoal{
  \caseFact{1} $\constraintSat{\addSolutions{\varSolution}{\varSolution'}}{\mergeConstraints{\varConstraints_1}{\varConstraints_2}}$

  \caseFact{2} $\isFinal{(\closure{\varEnv}{\expMatch{\varResult}{\varDataCon_i}{\varVar_i}{\varExp_i}})}$

  \caseFact{3} $\uneval{\varSolution}{\closure{\varEnv}{\expMatch{\varResult}{\varDataCon_i}{\varVar_i}{\varExp_i}}}{\varEx}{\mergeConstraints{\varConstraints_1}{\varConstraints_2}}$

  \caseFact{4} $\resumesTo{\addSolutions{\varSolution}{\varSolution'}}{\closure{\varEnv}{\expMatch{\varResult}{\varDataCon_i}{\varVar_i}{\varExp_i}}}{\varResult'}$
}{
  $\exSat{\addSolutions{\varSolution}{\varSolution'}}{\varResult'}{\varEx}$
}
\caseText{By inversion of unevaluation on (3), going through \ruleName{U-Case}}

\caseFact{5} $\uneval{\varSolution}{\varResult}{\expApp{\varDataCon_j}{\exHole}}{\varConstraints_1}$

\caseFact{6} $\worldConsistent
                {\varSolution}
                {\varExp_j}
                {\world
                   {(\envCat{\varEnv}{\envBind{\varVar_j}{\expUnwrap{\varDataCon_j}{\varResult}}})}
                   {\varEx}}
                {\varConstraints_2}$

\caseText{By inversion of checking on (6)}

\caseFact{7} $\reducesTo{(\envCat{\varEnv}{\envBind{\varVar_j}{\expUnwrap{\varDataCon_j}{\varResult}}})}{\varExp_j}{\varResult_0}$

\caseFact{8} $\resumesTo{\varSolution}{\varResult_0}{\varResult_0'}$

\caseFact{9} $\uneval{\varSolution}{\varResult_0'}{\varEx}{\varConstraints_2}$

\caseText{By \autoref{prop:res-fin} on (8)}

\caseFact{10} $\isFinal{\varResult_0'}$

\caseText{By the resumption assumption}

\caseFact{11} $\resumesTo{\addSolutions{\varSolution}{\varSolution'}}{\varResult_0'}{\varResult_0^{+'}}$

\caseFact{12} $\resumesTo{\addSolutions{\varSolution}{\varSolution'}}{\varResult}{\varResult^+}$

\caseText{By the induction hypothesis on (1), (10), (9), and (11)}

\caseFact{13} $\exSat{\addSolutions{\varSolution}{\varSolution'}}{\varResult_0^{+'}}{\varEx}$

\caseText{By the induction hypothesis on (1), inversion of final on (2), (5), and (12)}

\caseFact{14} $\exSat{\addSolutions{\varSolution}{\varSolution'}}{\varResult^+}{\expApp{\varDataCon_j}{\exHole}}$

\caseText{By inversion of example satisfaction on (14)}

\caseFact{15} $\varResult^+ = \expApp{\varDataCon_j}{\varResult^{+'}}$

\caseText{By inversion of resumption on (4), noting that on account of (15), (12) is the first premise of \ruleName{R-Case} and precludes \ruleName{R-Case-Indet}}

\caseFact{16} $\resumesTo{\addSolutions{\varSolution}{\varSolution'}}{\expApp{(\closure{\varEnv}{\valFixFun{\varVar_j}{\varVar_j}{\varExp_j}})}{\varResult^{+'}}}{\varResult'}$

\caseText{By inversion of resumption on (16)}

\caseFact{17} $\resumesTo{\addSolutions{\varSolution}{\varSolution'}}{\closure{\varEnv}{\valFixFun{\varVar_j}{\varVar_j}{\varExp_j}}}{\closure{\varEnv'}{\valFixFun{\varVar_j}{\varVar_j}{\varExp_j}}}$

\caseFact{18} $\resumesTo{\addSolutions{\varSolution}{\varSolution'}}{\varResult^{+'}}{\varResult^{+'}}$

\caseFact{19} $\reducesTo{(\envCat{\varEnv'}{\envBind{\varVar_j}{\varResult^{+'}}})}{\varExp_j}{\varResult^*}$

\caseFact{20} $\resumesTo{\addSolutions{\varSolution}{\varSolution'}}{\varResult^{*}}{\varResult'}$

\caseText{By \autorefResCompShort{} on (8) and (11)}

\caseFact{21} $\resumesTo{\addSolutions{\varSolution}{\varSolution'}}{\varResult_0}{\varResult_0^{+'}}$

\caseText{By \ruleName{R-Unwrap-Ctor} on (12), observing (15)}

\caseFact{22} $\resumesTo{\addSolutions{\varSolution}{\varSolution'}}{\expUnwrap{\varDataCon_j}{\varResult}}{\varResult^{+'}}$

\caseText{By \autorefEvalResShort{} on (17/22), (7), (19), (21), and (20)}

\caseFact{23} $\varResult' = \varResult_0^{+'}$

\caseText{Goal is given by (13), noting (23)}
}

\jcase{5}{\ruleName{U-Inverse-Ctor}}{
\jgivengoal{
  \caseFact{1} $\constraintSat{\addSolutions{\varSolution}{\varSolution'}}{\varConstraints}$

  \caseFact{2} $\isFinal{\expUnwrap{\varDataCon}{\varResult}}$

  \caseFact{3} $\uneval{\varSolution}{\expUnwrap{\varDataCon}{\varResult}}{\varEx}{\varConstraints}$

  \caseFact{4} $\resumesTo{\addSolutions{\varSolution}{\varSolution'}}{\expUnwrap{\varDataCon}{\varResult}}{\varResult'}$
}{
  $\exSat{\addSolutions{\varSolution}{\varSolution'}}{\varResult'}{\varEx}$
}
\caseText{By inversion of unevaluation on (3)}

\caseFact{5} $\uneval{\varSolution}{\varResult}{\expApp{\varDataCon}{\varEx}}{\varConstraints}$

\caseText{By the resumption assumption}

\caseFact{6} $\resumesTo{\addSolutions{\varSolution}{\varSolution'}}{\varResult}{\varResult^+}$

\caseText{By the induction hypothesis on (1), inversion of final on (2), (5), and (6)}

\caseFact{7} $\exSat{\addSolutions{\varSolution}{\varSolution'}}{\varResult^+}{\expApp{\varDataCon}{\varEx}}$

\caseText{By inversion of example satisfaction on (7)}

\caseFact{8} $\varResult^+ = {\expApp{\varDataCon}{\varResult^{+'}}}$

\caseFact{9} $\exSat{\addSolutions{\varSolution}{\varSolution'}}{\varResult^{+'}}{\varEx}$

\caseText{By \ruleName{R-Unwrap-Ctor} on (6), observing (8)}

\caseFact{10} $\resumesTo{\addSolutions{\varSolution}{\varSolution'}}{\expUnwrap{\varDataCon}{\varResult}}{\varResult^{+'}}$

\caseText{By \autorefResDetLong{} on (4) and (10)}

\caseFact{11} $\varResult' = \varResult^{+'}$

\caseText{Goal is given by (9), observing (11)}
}

\jcase{6}{\ruleName{U-Case-Guess}}{
\jgivengoal{
  \caseFact{1} $\constraintSat{\addSolutions{\varSolution}{\varSolution'}}{\mergeConstraints{\pairConstraints{\varSolution_g}{\emptyEnv}}{\varConstraints}}$

  \caseFact{2} $\isFinal{(\closure{\varEnv}{\expMatch{\varResult}{\varDataCon_i}{\varVar_i}{\varExp_i}})}$

  \caseFact{3} $\uneval{\varSolution}{\closure{\varEnv}{\expMatch{\varResult}{\varDataCon_i}{\varVar_i}{\varExp_i}}}{\varEx}{\mergeConstraints{\pairConstraints{\varSolution_g}{\emptyEnv}}{\varConstraints}}$

  \caseFact{4} $\resumesTo{\addSolutions{\varSolution}{\varSolution'}}{\closure{\varEnv}{\expMatch{\varResult}{\varDataCon_i}{\varVar_i}{\varExp_i}}}{\varResult'}$
}{
  $\exSat{\addSolutions{\varSolution}{\varSolution'}}{\varResult'}{\varEx}$
}
\caseText{By inversion of unevaluation on (3)}

\caseFact{5} $\varSolution_g = \guessesForMatch{\varHoleEnv}{\varDatatypeEnv}{\varResult}$

\caseFact{6} $\resumesTo{\addSolutions{\varSolution}{\varSolution_g}}{\varResult}{\expApp{\varDataCon_j}{\varResult_j}}$

\caseFact{7} $\worldConsistent
                {\addSolutions{\varSolution}{\varSolution_g}}
                {\varExp_j}
                {\world
                   {(\envCat{\varEnv}{\envBind{\varVar_j}{\varResult_j}})}
                   {\varEx}}
                {\varConstraints}$

\caseText{By inversion of the check judgment on (7)}

\caseFact{8} $\reducesTo{(\envCat{\varEnv}{\envBind{\varVar_j}{\varResult_j}})}{\varExp_j}{\varResult_0}$

\caseFact{9} $\resumesTo{\addSolutions{\varSolution}{\varSolution_g}}{\varResult_0}{\varResult_0'}$

\caseFact{10} $\uneval{\addSolutions{\varSolution}{\varSolution_g}}{\varResult_0'}{\varEx}{\varConstraints}$

\caseText{By (1) and the definition of $\oplus$, $\varSolution_g$ and the second component of $\varConstraints$ must be consistent.
     Likewise, by (6) and others, $\varSolution$ and $\varSolution_g$ are consistent. By the definition of constraint satisfaction,
     $\addSolutions{\varSolution}{\varSolution'}$ must be a supermapping of the second component of
     ${\mergeConstraints{\pairConstraints{\varSolution_g}{\emptyEnv}}{\varConstraints}}$,
     which, noting the previous observations, means $\varSolution'$ is a supermapping of $\varSolution_g$}

\caseFact{11} $\addSolutions{\varSolution}{\varSolution'} = \mergeSolutionsThree{\varSolution}{\varSolution_g}{(\setMinus{\varSolution'}{\varSolution_g})}$

\caseFact{12} $\constraintSat{\addSolutions{\varSolution}{\varSolution'}}{\varConstraints}$

\caseText{By the resumption assumption}

\caseFact{13} $\resumesTo{\addSolutions{\varSolution}{\varSolution'}}{\varResult_0'}{\varResult_0^+}$

\caseText{By the induction hypothesis on (12) (observing (11)), \autoref{prop:res-fin}, (10), and (13)}

\caseFact{14} $\exSat{\addSolutions{\varSolution}{\varSolution'}}{\varResult_0^+}{\varEx}$

\caseText{By Resumption Composition on (9) and (13) (observing (11))}

\caseFact{15} $\resumesTo{\addSolutions{\varSolution}{\varSolution'}}{\varResult_0}{\varResult_0^+}$

\caseText{By the resumption assumption}

\caseFact{16} $\resumesTo{\addSolutions{\varSolution}{\varSolution'}}{\varResult_j}{\varResult_j^+}$

\caseText{By \ruleName{R-Ctor} on (16)}

\caseFact{17} $\resumesTo{\addSolutions{\varSolution}{\varSolution'}}{\expApp{\varDataCon_j}{\varResult_j}}{\expApp{\varDataCon_j}{\varResult_j^+}}$

\caseText{By \autorefResCompShort{} on (6) and (17) (observing (11))}

\caseFact{18} $\resumesTo{\addSolutions{\varSolution}{\varSolution'}}{\varResult}{\expApp{\varDataCon_j}{\varResult_j^+}}$

\caseText{By inversion of resumption on (4), noting that (18) is the first premise of \ruleName{R-Case} and precludes \ruleName{R-Case-Indet}}

\caseFact{19} $\resumesTo{\addSolutions{\varSolution}{\varSolution'}}{\expApp{(\closure{\varEnv}{\valFixFun{\varVar_j}{\varVar_j}{\varExp_j}})}{\varResult_j^+}}{\varResult'}$

\caseText{By the resumption assumption}

\caseFact{20} $\resumesTo{\addSolutions{\varSolution}{\varSolution'}}{\varEnv}{\varEnv^+}$

\caseText{By \ruleName{R-Fix} on (20)}

\caseFact{22} $\resumesTo{\addSolutions{\varSolution}{\varSolution'}}{\closure{\varEnv}{\valFixFun{\varVar_j}{\varVar_j}{\varExp_j}}}{\closure{\varEnv^+}{\valFixFun{\varVar_j}{\varVar_j}{\varExp_j}}}$

\caseText{By inversion of resumption on (19), noting that (22) is the first premise of \ruleName{R-App} and precludes \ruleName{R-App-Indet}}

\caseFact{23} $\resumesTo{\addSolutions{\varSolution}{\varSolution'}}{\varResult_j^+}{\varResult_j^+}$ (noting \autorefResIdempShort{} on (16))

\caseFact{24} $\reducesTo{(\envCat{\varEnv^+}{\envBind{\varVar_j}{\varResult_j^+}})}{\varExp_j}{\varResult^*}$

\caseFact{25} $\resumesTo{\addSolutions{\varSolution}{\varSolution'}}{\varResult^*}{\varResult'}$

\caseText{By the definition of environment resumption, (20), and (16)}

\caseFact{26} $\resumesTo
                 {\addSolutions{\varSolution}{\varSolution'}}
                 {(\envCat{\varEnv}{\envBind{\varVar_j}{\varResult_j}})}
                 {(\envCat{\varEnv^+}{\envBind{\varVar_j}{\varResult_j^+}})}$

\caseText{By \autorefEvalResShort{} on (26), (8), (24), (15), and (25)}

\caseFact{27} $\varResult' = \varResult_0^+$

\caseText{Goal is given by (14), observing (27)}
}
}

\jtheorem{\jtref{\autoref{prop:sound-check} (Check)}}{
\jgivengoal{
\caseFact{1} $\constraintSat{\addSolutions{\varSolution}{\varSolution'}}{\varConstraints}$

\caseFact{2} $\worldConsistent
                {\varSolution}
                {\varExp}
                {\multiPremise{
                 \world{\varEnv_i}{\varEx_i}
                 }{\sequenceSyntax}}
                {\varConstraints}$
}{
  $\worldSat{\addSolutions{\varSolution}{\varSolution'}}{\varExp}{\multiPremise{\world{\varEnv_i}{\varEx_i}}{\sequenceSyntax}}$
}

\vspace{0.1em}

\jnocase{
\caseText{By inversion of checking on (2)}

\caseFact{3} $\reducesTo{\varEnv_i}{\varExp}{\varResult_i}$

\caseFact{4} $\resumesTo{\varSolution}{\varResult_i}{\varResult_i'}$

\caseFact{5} $\uneval{\varSolution}{\varResult_i'}{\varEx_i}{\varConstraints_i}$

\caseFact{6} $\varConstraints = \mergeConstraintsThree{\varConstraints_1}{\ldots}{\varConstraints_n}$

\caseText{By \autoref{prop:res-fin} on (4)}

\caseFact{7} $\isFinal{\varResult_i'}$

\caseText{By the resumption assumption}

\caseFact{8} $\resumesTo{\addSolutions{\varSolution}{\varSolution'}}{\varResult_i'}{\varResult_i''}$

\caseText{By \autorefExUnevalShort{} on (1) (observing (6)), (7), (5), and (8)}

\caseFact{9} $\exSat{\addSolutions{\varSolution}{\varSolution'}}{\varResult_i''}{\varEx_i}$

\caseText{By \autorefResCompLong{} on (4) and (8)}

\caseFact{10} $\resumesTo{\addSolutions{\varSolution}{\varSolution'}}{\varResult_i}{\varResult_i''}$

\caseText{Goal is given by \ruleName{Sat} on (3), (10), and (9)}
}
}

\jtheorem{\jtref{\autoref{prop:sound-ex-refine} (Refine)}}{
We consider only the most complicated case, \ruleName{Refine-Fix}, in detail.
The other cases are straightforward by similar reasoning.

\jgivengoal{
\caseFact{1}
    ${\refineFull
      {\varHoleEnv}{\varDatatypeEnv}
      {\varTypeEnv}{\varWorlds}{\varType}{\varExp}
      {
        \multiPremise{
        \problemNameTypeWorlds{\varHoleName_i'}{\varTypeEnv_i}{\varType_i'}{\varWorlds_i}
        }{\sequenceSyntax}
      }
    }$

\caseFact{2}
        $\multiPremise{
        \worldSat{\varSolution}{\expHole{\varHoleName_i'}}{\varWorlds_i}
        }{\sequenceSyntax}$
}{
  $\worldSat{\varSolution}{\varExp}{\varWorlds}$
}
\vspace{0.1em}

\jnocase{
\caseText{By inversion of Refine on (1), assuming we go through \ruleName{Refine-Fix}}

\caseFact{3} $\filterWorlds{\varWorlds} = \multiPremise{\world{\varEnv_j}{\ioExample{\varSimpleVal_j}{\varEx_j}}}{\generalSequenceSyntax{j}{m}}$

\caseFact{4} $\varHoleName_1 \textrm{ fresh}$

\caseFact{5} $\varExp = \valFixFun{\varVarF}{\varVar}{\expHole{\varHoleName_1}}$

\caseFact{6} $\problemNameTypeWorlds{\varHoleName_1'}{\varTypeEnv_1}{\varType_1'}{\varWorlds_1} =
              \problemNameTypeWorlds
                {\varHoleName_1}
                {(\envCatThree
                  {\varTypeEnv}
                  {\envBind{\varVarF}{(\tArrow{\varType_1}{\varType_2})}}
                  {\envBind{\varVar}{\varType_1}})}
                {\varType_2}
                {\varWorlds_1}$

\caseFact{7} $\varWorlds_1 = \multiPremise{
                \world
                  {(\envCatThree
                    {\varEnv_j}
                    {\envBind{\varVarF}{\closure{\varEnv_j}{\valFixFun{\varVarF}{\varVar}{\expHole{\varHoleName_1}}}}}
                    {\envBind{\varVar}{\varSimpleVal_j}})}
                  {\varEx_j}}{\generalSequenceSyntax{j}{m}}$

\caseText{By inversion of Sat on (2), observing (6) and (7)}

\caseFact{8} $\resumesTo
                {(\envCatThree
                    {\varEnv_j}
                    {\envBind{\varVarF}{\closure{\varEnv_j}{\valFixFun{\varVarF}{\varVar}{\expHole{\varHoleName_1}}}}}
                    {\envBind{\varVar}{\varSimpleVal_j}})}
                {\expHole{\varHoleName_1}}
                {\varResult_j^*}$

\caseFact{9} $\resumesTo{\varSolution}{\varResult_j^*}{\varResult_j^{*'}}$

\caseFact{10} $\exSat{\varSolution}{\varResult_j^{*'}}{\varEx_j}$

\caseText{By \ruleName{E-Fix}, observing (5)}

\caseFact{11} $\reducesTo{\varEnv_j}{\varExp}{\closure{\varEnv_j}{\valFixFun{\varVarF}{\varVar}{\expHole{\varHoleName_1}}}}$

\caseText{By the resumption assumption}

\caseFact{12} $\resumesTo{\varSolution}{\varEnv_j}{\varEnv_j'}$

\caseText{By \ruleName{R-Fix} on (12)}

\caseFact{13} $\resumesTo{\varSolution}{\closure{\varEnv_j}{\valFixFun{\varVarF}{\varVar}{\expHole{\varHoleName_1}}}}{\closure{\varEnv_j'}{\valFixFun{\varVarF}{\varVar}{\expHole{\varHoleName_1}}}}$

\caseText{By \autoref{prop:v-res} (Simple Value Resumption)}

\caseFact{14} $\resumesTo{\varSolution}{\varSimpleVal_j}{\varSimpleVal_j}$

\caseText{By \autorefResIdempLong{} on (13)}

\caseFact{15} $\resumesTo{\varSolution}{\closure{\varEnv_j'}{\valFixFun{\varVarF}{\varVar}{\expHole{\varHoleName_1}}}}{\closure{\varEnv_j'}{\valFixFun{\varVarF}{\varVar}{\expHole{\varHoleName_1}}}}$

\caseText{By the evaluation assumption}

\caseFact{16} $\reducesTo
                  {(\envCatThree
                    {\varEnv_j'}
                    {\envBind{\varVarF}{\closure{\varEnv_j'}{\valFixFun{\varVarF}{\varVar}{\expHole{\varHoleName_1}}}}}
                    {\envBind{\varVar}{\varSimpleVal_j}})}
                  {\expHole{\varHoleName_1}}
                  {\varResult_j^{**}}$

\caseText{By the resumption assumption}

\caseFact{17} $\resumesTo{\varSolution}{\varResult_j^{**}}{\varResult_j^{**'}}$

\caseText{By the definition of environment resumption, (12), (13), and (14)}

\caseFact{18} $\resumesTo
                 {\varSolution}
                 {(\envCatThree
                   {\varEnv_j}
                   {\envBind{\varVarF}{\closure{\varEnv_j}{\valFixFun{\varVarF}{\varVar}{\expHole{\varHoleName_1}}}}}
                   {\envBind{\varVar}{\varSimpleVal_j}})}
                 {(\envCatThree
                   {\varEnv_j'}
                   {\envBind{\varVarF}{\closure{\varEnv_j'}{\valFixFun{\varVarF}{\varVar}{\expHole{\varHoleName_1}}}}}
                   {\envBind{\varVar}{\varSimpleVal_j}})}$

\caseText{By \autorefEvalResShort{} on (18), (8), (16), (9), and (17)}

\caseFact{19} $\varResult_j^{**'} = \varResult_j^{*'}$

\caseText{By \ruleName{R-App} on (15), (14), (trivial), (16), and (17), observing (19)}

\caseFact{20} $\resumesTo{\varSolution}{\expApp{(\closure{\varEnv_j'}{\valFixFun{\varVarF}{\varVar}{\expHole{\varHoleName_1}}})}{\varSimpleVal_j}}{\varResult_j^{*'}}$

\caseText{By \ruleName{XS-Input-Output} on (20) and (10)}

\caseFact{21} $\exSat{\varSolution}{\closure{\varEnv_j'}{\valFixFun{\varVarF}{\varVar}{\expHole{\varHoleName_1}}}}{\ioExample{\varSimpleVal_j}{\varEx_j}}$

\caseText{Goal is given by \ruleName{Sat} on (11), (13), and (21),
observing (3) and the fact that the filtered-out example constraints
are trivially satisfied.}
}
}

%% \hfill \\
%% \indent

\jtheorem{\jtref{\autoref{prop:sound-ex-branch} (Branch)}}{
\jgivengoal{
\caseFact{1}
   ${\branchFull
      {\varHoleEnv}{\varDatatypeEnv}
      {\varTypeEnv}{\varWorlds}{\varType}{\varExp'}
      {
        \multiPremise{
        \problemNameTypeWorlds{\varHoleName_i}{\varTypeEnv_i}{\varType_i'}{\varWorlds_i}
        }{\sequenceSyntax}
      }
      {\varConstraints}
      {\varSolution}
    }$

\caseFact{2} $\constraintSat{\addSolutions{\varSolution}{\varSolution'}}{\varConstraints}$

\caseFact{3}
    $\multiPremise{
    \worldSat{\addSolutions{\varSolution}{\varSolution'}}{\expHole{\varHoleName_i}}{\varWorlds_i}
    }{\sequenceSyntax}$
}{
  $\worldSat{\addSolutions{\varSolution}{\varSolution'}}{\varExp'}{\varWorlds}$
}

\vspace{0.1em}

\jnocase{
\caseText{By inversion of branch on (1)}

\caseFact{4} $\lookupTypeConstructors{\varTypeCon}{\varDataCon_i}{\varType_i}$

\caseFact{5} $\guess{\varTypeEnv}{\varTypeCon}{\varExp}$

\caseFact{6} $\reducesTo{\varEnv_j}{\varExp}{\varResult_j}$

\caseFact{7} $\worldConsistent
                {\varSolution}
                {\varExp}
                {\world{\varEnv_j}{\expApp{\varDataConChoice{j}}{\exHole}}}
                {\varConstraints_j}$

\caseFact{8} $\varHoleName_i \textrm{ fresh}$

\caseFact{9} $\problemNameTypeWorlds{\varHoleName_i}{\varTypeEnv_i}{\varType_i'}{\varWorlds_i} =
              \problemNameTypeWorlds
                {\varHoleName_i}
                {(\envCat{\varTypeEnv}{\envBind{\varVar_i}{\varType_i}})}
                {\varType}
                {\varWorlds_i}$

\caseFact{10} $\varWorlds_i =
              \multiPremise
                {\world{(\envCat{\varEnv_j}{\envBind{\varVar_i}{\collapseUnwrapWrap{\expUnwrap{\varDataCon_i}{\varResult_j}}}})}{\varEx_j}}
                {\generalSequenceSyntax{j}{m} \land \varDataConChoice{j} = \varDataCon_i}$

\caseFact{11} $\filterWorlds{\varWorlds} = \multiPremise{\world{\varEnv_j}{\varEx_j}}{\generalSequenceSyntax{j}{m}}$

\caseFact{12} $\varExp' = \expMatch{\varExp}{\varDataCon_i}{\varVar_i}{\expHole{\varHoleName_i}}$

\caseFact{13} $\varConstraints = \mergeConstraintsThree{\varConstraints_1}{\ldots}{\varConstraints_m}$

\caseText{Now, for each $j$, there are two cases, depending on whether or not $\varExp$ evaluates to a
constructor form ($\expApp{\varDataCon_i}{\varResult_j^*}$ for $i = \alpha_j$).}
}

\jcase{A}{$e$ evaluates to $\expApp{\varDataCon_i}{\varResult_j^*}$ for $i = \alpha_j$}{
\caseFactPl{A1} $\alpha_j = i$

\caseFactPl{A2} $\varResult_j = \expApp{\varDataCon_i}{\varResult_j^*}$

\caseText{By inversion of Sat on (3), observing (10), (A2), and \ruleName{E-Hole}}

\caseFactPl{A3} $\reducesTo
                  {(\envCat{\varEnv_j}{\envBind{\varVar_i}{\varResult_j^*}})}
                  {\expHole{\varHoleName_i}}
                  {\closure{\envCat{\varEnv_j}{\envBind{\varVar_i}{\varResult_j^*}}}{\expHole{\varHoleName_i}}}$

\caseFactPl{A4} $\resumesTo{\addSolutions{\varSolution}{\varSolution'}}{\closure{\envCat{\varEnv_j}{\envBind{\varVar_i}{\varResult_j^*}}}{\expHole{\varHoleName_i}}}{\varResult_j^{*+}}$

\caseFactPl{A5} $\exSat{\addSolutions{\varSolution}{\varSolution'}}{\varResult_j^{*+}}{\varEx_j}$

\caseText{By \ruleName{E-Case} on (6) (observing (A2)) and (A3), observing (12)}

\caseFactPl{A6} $\reducesTo
                  {\varEnv_j}
                  {\varExp'}
                  {\closure{\envCat{\varEnv_j}{\envBind{\varVar_i}{\varResult_j^*}}}{\expHole{\varHoleName_i}}}$

\caseText {Goal is given by \ruleName{Sat} on (A6), (A4), and (A5),
          observing (11) and the fact that the filtered-out example constraints
          are trivially satisfied.}
}

\jcase{B}{$e$ does not evaluate to a constructor form}{
\caseFactPl{B1} $\varResult_j \ne \expApp{\varDataCon_i}{\varResult_j^*}$

\caseText{By the resumption assumption}

\caseFactPl{B2} $\resumesTo{\addSolutions{\varSolution}{\varSolution'}}{\varEnv_j}{\varEnv_j'}$

\caseText{By \autorefLBECheckShort{} on (2) (observing (13)), and (7)}

\caseFactPl{B3} $\worldSat{\addSolutions{\varSolution}{\varSolution'}}{\varExp}{\world{\varEnv_j}{\expApp{\varDataConChoice{j}}{\exHole}}}$

\caseText{By inversion of Sat on (B3), observing (6)}

\caseFactPl{B4} $\resumesTo{\addSolutions{\varSolution}{\varSolution'}}{\varResult_j}{\varResult_j^+}$

\caseFactPl{B5} $\exSat{\addSolutions{\varSolution}{\varSolution'}}{\varResult_j^+}{\expApp{\varDataConChoice{j}}{\exHole}}$

\caseText{By inversion of example satisfaction on (B5)}

\caseFactPl{B6} $\varResult_j^+ = \expApp{\varDataConChoice{j}}{\varResult_j^{+*}}$

\caseText{By \ruleName{R-Unwrap-Ctor} on (B4), observing (B6)}

\caseFactPl{B7} $\resumesTo{\addSolutions{\varSolution}{\varSolution'}}{\expUnwrap{\varDataConChoice{j}}{\varResult_j}}{\varResult_j^{+*}}$

\caseText{By \autorefResIdempLong{} on (B4), observing (B6)}

\caseFactPl{B8} $\resumesTo{\addSolutions{\varSolution}{\varSolution'}}{\expApp{\varDataConChoice{j}}{\varResult_j^{+*}}}{\expApp{\varDataConChoice{j}}{\varResult_j^{+*}}}$

\caseText{By inversion of resumption on (B8)}

\caseFactPl{B9} $\resumesTo{\addSolutions{\varSolution}{\varSolution'}}{\varResult_j^{+*}}{\varResult_j^{+*}}$

\caseText{Below, unless otherwise noted, $i = \alpha_j$}

\caseText{By the definition of environment resumption, (B2), and (B9)}

\caseFactPl{B10} $\resumesTo{\addSolutions{\varSolution}{\varSolution'}}{(\envCat{\varEnv_j}{\envBind{\varVar_i}{\varResult_j^{+*}}})}{(\envCat{\varEnv_j'}{\envBind{\varVar_i}{\varResult_j^{+*}}})}$

\caseText{By \ruleName{E-Hole}}

\caseFactPl{B11} $\reducesTo
                  {(\envCat{\varEnv_j'}{\envBind{\varVar_i}{\varResult_j^{+*}}})}
                  {\expHole{\varHoleName_i}}
                  {\closure{\envCat{\varEnv_j'}{\envBind{\varVar_i}{\varResult_j^{+*}}}}{\expHole{\varHoleName_i}}}$

\caseText{By the resumption assumption}

\caseFactPl{B12} $\resumesTo{\addSolutions{\varSolution}{\varSolution'}}{\closure{\envCat{\varEnv_j'}{\envBind{\varVar_i}{\varResult_j^{+*}}}}{\expHole{\varHoleName_i}}}{\varResult_j^{++'}}$

\caseText{By inversion of Sat on (3), observing (10), (B1), and \ruleName{E-Hole}}

\caseFactPl{B13} $\reducesTo
                  {(\envCat{\varEnv_j}{\envBind{\varVar_i}{\expUnwrap{\varDataCon_i}{\varResult_j}}})}
                  {\expHole{\varHoleName_i}}
                  {\closure{\envCat{\varEnv_j}{\envBind{\varVar_i}{\expUnwrap{\varDataCon_i}{\varResult_j}}}}{\expHole{\varHoleName_i}}}$

\caseFactPl{B14} $\resumesTo
                  {\addSolutions{\varSolution}{\varSolution'}}
                  {\closure{\envCat{\varEnv_j}{\envBind{\varVar_i}{\expUnwrap{\varDataCon_i}{\varResult_j}}}}{\expHole{\varHoleName_i}}}
                  {\varResult_j^{*+}}$

\caseFactPl{B15} $\exSat{\addSolutions{\varSolution}{\varSolution'}}{\varResult_j^{*+}}{\varEx_j}$

\caseText{By the definition of environment resumption, (B2), and (B7)}

\caseFactPl{B16} $\resumesTo
                  {\addSolutions{\varSolution}{\varSolution'}}
                  {(\envCat{\varEnv_j}{\envBind{\varVar_i}{\expUnwrap{\varDataCon_i}{\varResult_j}}})}
                  {(\envCat{\varEnv_j'}{\envBind{\varVar_i}{\varResult_j^{+*}}})}$

\caseText{By \autorefEvalResShort{} on (B16), (B13), (B11), (B14), and (B12)}

\caseFactPl{B17} $\varResult_j^{*+} = \varResult_j^{++'}$

\caseText{By \ruleName{E-Case-Indet} on (6), and (B1), observing (12)}

\caseFactPl{B18} $\reducesTo{\varEnv_j}{\varExp'}{\closure{\varEnv_j}{\expMatch{\varResult_j}{\varDataCon_i}{\varVar_i}{\expHole{\varHoleName_i}}}}$

\caseText{By \ruleName{R-Fix} on (B2)}

\caseFactPl{B19} $\resumesTo
                    {\addSolutions{\varSolution}{\varSolution'}}
                    {\closure{\varEnv_j}{\valFixFun{\varVar_i}{\varVar_i}{\expHole{\varHoleName_i}}}}
                    {\closure{\varEnv_j'}{\valFixFun{\varVar_i}{\varVar_i}{\expHole{\varHoleName_i}}}}$

\caseText{By \ruleName{R-App} on (B19), (B9), (trivial), (B11), and (B12)}

\caseFactPl{B20} $\resumesTo
                    {\addSolutions{\varSolution}{\varSolution'}}
                    {\expApp{(\closure{\varEnv_j}{\valFixFun{\varVar_i}{\varVar_i}{\expHole{\varHoleName_i}}})}{\varResult_j^{+*}}}
                    {\varResult_j^{++'}}$

\caseText{By \ruleName{R-Case} on (B4) (observing (B6)) and (B20)}

\caseFactPl{B21} $\resumesTo
                    {\addSolutions{\varSolution}{\varSolution'}}
                    {\closure{\varEnv_j}{\expMatch{\varResult_j}{\varDataCon_i}{\varVar_i}{\expHole{\varHoleName_i}}}}
                    {\varResult_j^{++'}}$

\caseText {Goal is given by \ruleName{Sat} on (B18), (B21), and (B15) (observing (B17)),
          observing (11) and the fact that the filtered-out example constraints
          are trivially satisfied.}
}
}

\vspace{-0.15em}
\jtheorem{\jtref{\autoref{prop:sound-ex-fill} (Fill)}}{
The \ruleName{Defer} case is trivial.
The \ruleName{Refine, Branch} case is straightforward by way of
\autoref{prop:sound-ex-refine} (Refine) and \autoref{prop:sound-ex-branch} (Branch).
The \ruleName{Guess-And-Check}
case is straightforward by way of \autoref{prop:sound-check} (Check).
}

\vspace{-0.15em}
\jtheorem{\jtref{\autoref{prop:sound-type-merge} (Type soundness of merge)}}{
Straightforward by way of induction and (eventually) \autoref{prop:sound-type-check}.
Technically, we must establish similar lemmas applying to \ruleName{Step} and \ruleName{Resolve},
but the definitions and proofs of these lemmas are straightforward.
}

\vspace{-0.15em}
\jtheorem{\jtref{\autoref{prop:sound-merge} (Example soundness of merge)}}{
Straightforward by way of induction and (eventually) \autoref{prop:sound-check} (Check).
Technically, we must establish similar lemmas applying to \ruleName{Step} and \ruleName{Resolve},
but the definitions and proofs of these lemmas are straightforward.
}

\vspace{-0.15em}
\jtheorem{\jtref{\autoref{prop:sound-solve} (Soundness of solve)}}{
\jgivengoalTwo{
  \caseFact{1} $\typeCheckUnsolved{\varHoleEnv}{\varUnsolvedConstraints}$

  \caseFact{2} $\typeCheckSolution{\varHoleEnv}{\varSolution}$

  \caseFact{3} $\iterSolveEqualsFull{\varHoleEnv}{\varDatatypeEnv}{(\varUnsolvedConstraints, \varSolution)}{\varSolution'}{\varHoleEnv''}$
}{
  $\typeCheckSolution{\varHoleEnv''}{\varSolution'}$
}{
  $\constraintSat{\varSolution'}{\pairConstraints{\varSolution}{\varUnsolvedConstraints}}$
}

\vspace{0.1em}

\jnocase{
\caseText{By inversion of Solve on (3), going through \ruleName{Solve-One} rule since \ruleName{Solve-Done} is trivial}

\caseFact{4} $\varHoleName \in \dom{\varUnsolvedConstraints}$

\caseFact{5} $\varHoleEnv({\varHoleName}) = \tHole{\varTypeEnv}{\varType}$

\caseFact{6} $\varUnsolvedConstraints({\varHoleName}) = \varWorlds$

\caseFact{7} $\fillHole
                {\varHoleName}
                {\varSolution}
                {\varTypeEnv}
                {\varWorlds}
                {\varType}
                {\varConstraints}
                {\varHoleEnv'}$

\caseFact{8}
   \simplifyConstraintsEquals
     {\setUnion{\varHoleEnv}{\varHoleEnv'}}
     {\varDatatypeEnv}
     {(\mergeConstraints
       {\pairConstraints{\varSolution}{\setMinus{\varUnsolvedConstraints}{\varHoleName}}}
       {\varConstraints})}
     {\varConstraints'}

\caseFact{9}
   \iterSolveEqualsFull
     {\setUnion{\varHoleEnv}{\varHoleEnv'}}
     {\varDatatypeEnv}
     {(\varConstraints')}
     {\varSolution'}
     {\varHoleEnv''}

\caseText{By definition of constraints typing on (1), (5), and (6)}

\caseFact{10} $\typeCheckWorlds{\varHoleEnv}{\varWorlds}{\varTypeEnv}{\varType}$

\caseText{By \autoref{prop:sound-type-fill} (Type Soundness of Fill) on (2), (10), and (7)}

\caseFact{11} $\typeCheckSolution{\setUnionThree{\varHoleEnv}{\varHoleEnv'}{(\envBindHole{\varHoleName}{\tHole{\varTypeEnv}{\varType}})}}{\varConstraints}$

\caseText{By (11), observing (4) and (5)}

\caseFact{12} $\typeCheckSolution{\setUnion{\varHoleEnv}{\varHoleEnv'}}{\varConstraints}$

\caseText{By observing that freshness premises ensure that $\varHoleEnv'$ is disjoint from $\varHoleEnv$}

\caseFact{13} $\typeCheckUnsolved{\setUnion{\varHoleEnv}{\varHoleEnv'}}{\varUnsolvedConstraints}$

\caseFact{14} $\typeCheckSolution{\setUnion{\varHoleEnv}{\varHoleEnv'}}{\varSolution}$

\caseText{By \autoref{prop:sound-type-merge} (Type Soundness of Sem. Merge) on (12+13+14) and (8)}

\caseFact{15} $\typeCheckUnsolved{\setUnion{\varHoleEnv}{\varHoleEnv'}}{\varConstraints'}$

\caseText{By the induction hypothesis on (15), (15), and (9)}

\caseFact{16} $\typeCheckSolution{\varHoleEnv''}{\varSolution'}$

\caseFact{17} $\constraintSat{\varSolution'}{\varConstraints'}$

\caseText{By \autoref{prop:sound-merge} (Ex. Soundness of Sem. Merge) on (17) and (8)}

\caseFact{18}
   $\constraintSat{\varSolution'}{\mergeConstraints{\pairConstraints{\varSolution}{\setMinus{\varUnsolvedConstraints}{\varHoleName}}}{\varConstraints}}$

\caseText{By the definition of constraint satisfaction and (18)}

\caseFact{19} $\constraintSat{\varSolution'}{\pairConstraints{\varSolution}{\setMinus{\varUnsolvedConstraints}{\varHoleName}}}$

\caseFact{20} $\constraintSat{\varSolution'}{\varConstraints}$

\caseText{By \autoref{prop:sound-ex-fill} (Ex. Soundness of Fill) on (7) and (20) (observing (19))}

\caseFact{21} $\varSolution'(\varHoleName) = \varExp$

\caseFact{22} $\worldSat{\varSolution'}{\varExp}{\varWorlds}$

\caseText{By straightforward reasoning on (21) and (22)}

\caseFact{23} $\worldSat{\varSolution'}{\expHole{\varHoleName}}{\varWorlds}$

\caseText{Goal A is given by (16)}

\caseText{Goal B is given by combining (19), (6), and (23)}
}
}

\jtheorem{\jtref{\autoref{prop:sound-asrt-simpl} (Soundness of assertion simplification)}}{
Straightforward by way of \autoref{prop:sat->res}.
}

\jtheorem{\jtref{\autoref{prop:sound-synth} (Soundness of synthesis)}}{
Straightforward by way of
\autoref{prop:sound-type-prog}, \autoref{prop:sound-type-simplify},
\autoref{prop:sound-solve} (solve) and \autoref{prop:sound-asrt-simpl}.
}

\jtheorem{\jtref{\autoref{prop:ex-sat-val}}}{
Straightforward induction.
}

\jtheorem{\jtref{\autoref{prop:sat->res}}}{
\jgivengoal{
  \caseFact{1} $\varSimpleVal \textrm{ simple value}$

  \caseFact{2} $\isFinal{\varResult}$

  \caseFact{3} $\uneval{\emptyEnv}{\varResult}{\varSimpleVal}{\varConstraints}$

  \caseFact{4} $\constraintSat{\varSolution}{\varConstraints}$
}{
  $\resumesTo{\varSolution}{\varResult}{\varSimpleVal}$
}

\vspace{0.1em}

\jnocase{

\caseText{By the resumption assumption}

\caseFact{5} $\resumesTo{\varSolution}{\varResult}{\varResult'}$

\caseText{By \autorefExUnevalLong{} on (4), (2), (3), and (5)}

\caseFact{6} $\exSat{\varSolution}{\varResult'}{\varSimpleVal}$

\caseText{By \autoref{prop:ex-sat-val} (Example Satisfaction of Simple Value) on (1) and (6)}

\caseFact{7} $\varResult' = \varSimpleVal$

\caseText{Goal is given by (5), observing (7)}
}
}

}
\clearpage

\clearpage
\section{Additional Experimental Results}
\label{sec:appendix-random-graphs}

\newcommand{\randomGraphWidth}{0.333\textwidth}

\newcommand{\randomGraph}[2]{
  %% arxiv-version
  %% for f in `ls ../random-graphs/2b/`; do cp ../random-graphs/2b/$f 2b_$f; done
  %% for f in `ls ../random-graphs/3b/`; do cp ../random-graphs/3b/$f 3b_$f; done
  %% \includegraphics[width=\randomGraphWidth{}]{random-graphs/#1/#2}
  \includegraphics[width=\randomGraphWidth{}]{#1_#2}
}

\newcommand{\randomHistogramWidth}{0.5\textwidth}

\newcommand{\randomHistogram}[2]{
  %% arxiv-version
  %% \includegraphics[width=\randomHistogramWidth{}]{random-graphs/#1/#2}
  \includegraphics[width=\randomHistogramWidth{}]{#1_#2}
}

%% \setcounter{subsection}{1} %% start with B.2

%% \subsection{Experiment 2b (No Sketch)}
\subsection{Experiment 2b: No Sketch + Random Examples}

\randomGraph{2b}{bool_band}
\randomGraph{2b}{bool_bor}
\randomGraph{2b}{bool_impl}
\randomGraph{2b}{bool_neg}
\randomGraph{2b}{bool_xor}
\randomGraph{2b}{list_append}
\randomGraph{2b}{list_concat}
\randomGraph{2b}{list_drop}
\randomGraph{2b}{list_even_parity}
\randomGraph{2b}{list_hd}
\randomGraph{2b}{list_inc}
\randomGraph{2b}{list_last}
\randomGraph{2b}{list_length}
\randomGraph{2b}{list_nth}
\randomGraph{2b}{list_rev_append}
\randomGraph{2b}{list_rev_fold}
\randomGraph{2b}{list_rev_snoc}
\randomGraph{2b}{list_rev_tailcall}
\randomGraph{2b}{list_snoc}
\randomGraph{2b}{list_sort_sorted_insert}
\randomGraph{2b}{list_stutter}
\randomGraph{2b}{list_sum}
\randomGraph{2b}{list_take}
\randomGraph{2b}{list_tl}
\randomGraph{2b}{nat_add}
\randomGraph{2b}{nat_iseven}
\randomGraph{2b}{nat_max}
\randomGraph{2b}{nat_pred}
\randomGraph{2b}{tree_collect_leaves}
\randomGraph{2b}{tree_count_nodes}
\randomGraph{2b}{tree_inorder}
\randomGraph{2b}{tree_preorder}

\vspace{0.20in}

\randomHistogram{2b}{k50}
\randomHistogram{2b}{k90}

% Do not manually edit this data; use the 'histogram-statistics' tool in the
% Smyth repo

{
\newcommand{\kfMedian}{0}
\newcommand{\kfMax}{2}
\newcommand{\kfAboveOne}{2 (7\%)}
\newcommand{\kfAboveTwo}{0 (0\%)}
\newcommand{\knMedian}{1}
\newcommand{\knMax}{9}
\newcommand{\knAboveOne}{8 (27\%)}
\newcommand{\knAboveTwo}{5 (17\%)}

\begin{figure}[h]
  \footnotesize
  \begin{tabular}{c|cccc}
    & median $k'_p$
    & max $k'_p$
    & \# $ > $ median $ k'_p + 1 $
    & \# $ > $ median $ k'_p + 2 $ \\
    \hline
    $ p = 50\% $ & \kfMedian{} & \kfMax{} & \kfAboveOne{} & \kfAboveTwo{} \\
    $ p = 90\% $ & \knMedian{} & \knMax{} & \knAboveOne{} & \knAboveTwo{}
  \end{tabular}
  \caption{A summary of the distribution of $k'_p$ for Experiment 2b.}
  \label{fig:kprime2}
\end{figure}
}

\clearpage

%% \subsection{Experiment 3b (Base Case Sketch)}
\subsection{Experiment 3b: Base Case Sketch + Random Examples}

\randomGraph{3b}{list_append}
\randomGraph{3b}{list_concat}
\randomGraph{3b}{list_drop}
\randomGraph{3b}{list_even_parity}
\randomGraph{3b}{list_last}
\randomGraph{3b}{list_length}
\randomGraph{3b}{list_nth}
\randomGraph{3b}{list_rev_append}
\randomGraph{3b}{list_rev_snoc}
\randomGraph{3b}{list_rev_tailcall}
\randomGraph{3b}{list_snoc}
\randomGraph{3b}{list_sort_sorted_insert}
\randomGraph{3b}{list_stutter}
\randomGraph{3b}{list_take}
\randomGraph{3b}{nat_add}
\randomGraph{3b}{nat_iseven}
\randomGraph{3b}{nat_max}
\randomGraph{3b}{tree_collect_leaves}
\randomGraph{3b}{tree_count_nodes}
\randomGraph{3b}{tree_inorder}
\randomGraph{3b}{tree_preorder}

\vspace{0.20in}

\randomHistogram{3b}{k50}
\randomHistogram{3b}{k90}

% Do not manually edit this data; use the 'histogram-statistics' tool in the
% Smyth repo

{
\newcommand{\kfMedian}{2}
\newcommand{\kfMax}{6}
\newcommand{\kfAboveOne}{4 (22\%)}
\newcommand{\kfAboveTwo}{3 (17\%)}
\newcommand{\knMedian}{4}
\newcommand{\knMax}{14}
\newcommand{\knAboveOne}{4 (22\%)}
\newcommand{\knAboveTwo}{4 (22\%)}

\begin{figure}[h]
  \footnotesize
  \begin{tabular}{c|cccc}
    & median $k'_p$
    & max $k'_p$
    & \# $ > $ median $ k'_p + 1 $
    & \# $ > $ median $ k'_p + 2 $ \\
    \hline
    $ p = 50\% $ & \kfMedian{} & \kfMax{} & \kfAboveOne{} & \kfAboveTwo{} \\
    $ p = 90\% $ & \knMedian{} & \knMax{} & \knAboveOne{} & \knAboveTwo{}
  \end{tabular}
  \captionsetup{justification=centering}
  \caption{A summary of the distribution of $k'_p$ for Experiment 3b. \\ (Does not
  include \texttt{list\_concat} due to failure in Experiment 3a.)}
  \label{fig:kprime3}
\end{figure}
}

\subsection{Experimental Setup (circa February 2020)}
\vspace{2em}
\begin{center}
\includegraphics[scale=0.30]{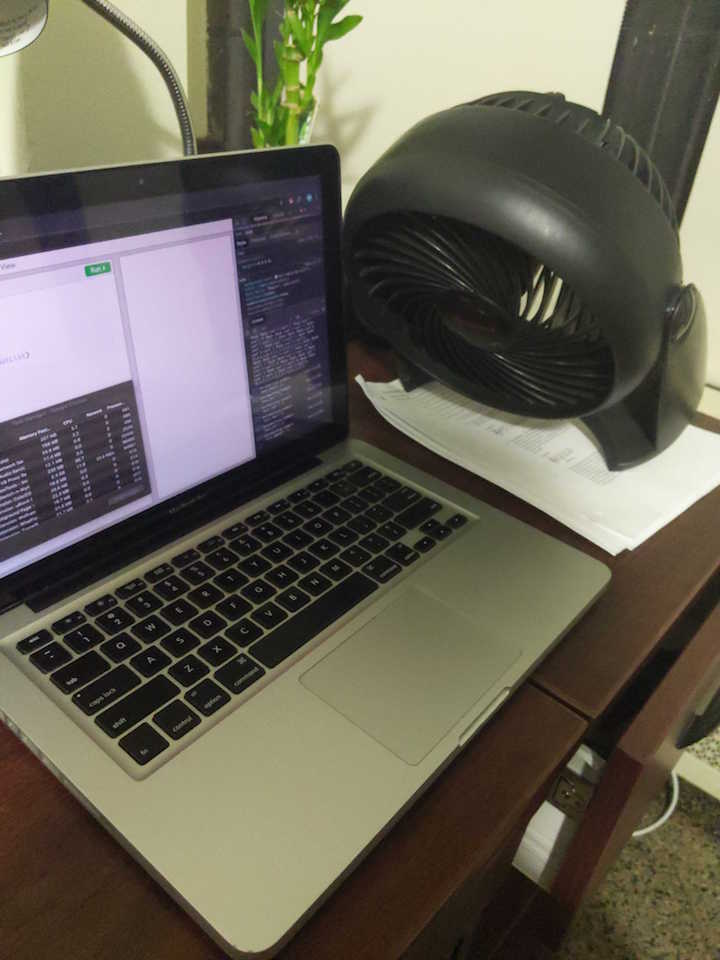}
\end{center}

\clearpage
\section{Polymorphism}
\label{sec:appendix-polymorphism}

\subsection{Implementation}

The \smyth{} implementation supports System~F universal polymorphism, as well as user-defined
polymorphic type operators such as the following:
\begin{blockcode}
  type List a = Nil | Cons a (List a)
\end{blockcode}
The \myth{} thesis \citep[Ch. 9]{OseraThesis} details how to extend \myth{} to include System F
universal polymorphism (and the details generalize as expected in \smyth{}), but
does not include a description of how to support polymorphic type operators.

\parahead{Polymorphic Type Operators}
The changes to the \smyth{} codebase to support polymorphic type operators were
largely straightforward except for in one place: synthesis of \inlinecode{case}
scrutinees. When synthesizing a \inlinecode{case} scrutinee, \myth{} and
\smyth{} attempt synthesis at every datatype in scope. But with the inclusion
of a single polymorphic type operator (and a base type), there are an infinite
number of datatypes in scope---for example, \inlinecode{List Nat},
\inlinecode{List (List Nat)}, \inlinecode{List (List (List Nat))}, etc. All
of these are valid types for the scrutinee of a \inlinecode{case} expression.
Raw term enumeration occurs at a single type, but in this instance there is an
infinite family of types that serve as the goal to term enumeration.

To capture this notion of an infinite class of types, we introduced the simple
notion of a \textit{type wildcard} ($ * $) to \smyth{} and straightforwardly
extended the standard syntactic notion of type equality ($ = $) to \textit{type
matching} ($ \equiv_* $):
\begin{equation*}
  \inferrule*[lab=\ruleNameFig{Match-Equality}]
    { \tau \text{ Type} \\
      \sigma \text{ Type} \\
      \tau = \sigma
    }
    { \tau \equiv_* \sigma
    }
  \hsepRule
  \inferrule*[lab=\ruleNameFig{Match-Left}]
    { \tau \text{ Type} \\
    }
    { * \equiv_* \tau
    }
  \hsepRule
  \inferrule*[lab=\ruleNameFig{Match-Right}]
    { \tau \text{ Type} \\
    }
    { \tau \equiv_* *
    }
\end{equation*}
Scrutinee synthesis then occurs as before (once per datatype), but with
polymorphic datatypes instantiated with the wildcard type. Raw term enumeration
then substitutes equality for type matching wherever necessary to compensate.

\parahead{Examples for Polymorphic Types}

For the purpose of specifying examples for polymorphic functions,
the \myth{} thesis \citep[Ch. 9]{OseraThesis} introduces
``polymorphic constants'' (called ``abstract refinements'' by \citet{Frankle2016}).
Later in the chapter, ``boxed'' concrete examples are presented
as an equally-expressive alternative to polymorphic constants.

Neither of these apparatuses is necessary in \smyth{}; examples can be specified
by normal function application and type argument application, and live
unevaluation will transform the examples to hole constraints, albeit with a
polymorphic type so that concrete refinements of these examples cannot be
performed. (This is the crux of why fewer examples are needed to correctly
synthesize polymorphic functions in \smyth{}.) For example, consider the
following synthesis task:
%%   type Nat = Z | S Nat
%% 
%%   type List a = Nil | Cons a (List a)
%% 
\begin{blockcode}
  stutter : forall a . List a -> List a
  stutter <a> xs = `\color{CadetBlue}\texttt{??}`

  spec (stutter <Nat>)
    [ ([], [])
    , ([1, 0], [1, 1, 0, 0])
    ]
\end{blockcode}
\smyth{} correctly synthesis a polymorphic version of the \inlinecode{stutter}
function when given this sketch. Notice that \inlinecode{spec} is called with
the argument \inlinecode{stutter <Nat>} (a type argument aplication), so the
examples can be provided monomorphically (the implementation requires a
few additional annotations to simplify typechecking).
The assertions could alternatively be specified as follows:
\begin{blockcode}
  spec2 stutter
    [ (<Nat>, [], [])
    , (<Nat>, [1, 0], [1, 1, 0, 0])
    ]
\end{blockcode}
demonstrating that no special machinery is needed to handle examples for
polymorphic functions other than the live unevaluation rules for type argument
application.

\subsection{Experiments 5 and 6}

Of the \numBenchmarks{} tasks that succeeded in Experiment~1, 23 can be
specified with a polymorphic type signature rather than a monomorphic one.
\autoref{fig:poly-experiments} summarizes the results of re-running
Experiments~2~and~3 on these 23 tasks given polymorphic type signatures;
Experiment~5 is the polymorphic version of Experiment~2, and
Experiment~6 is the polymorphic version of Experiment~3.
The process for correctness checking, expert
example selection, and random example generation are the same as in earlier
experiments.

In summary, polymorphic examples offer a modest reduction in the number of
examples needed for synthesis. More qualitatively, they ensure that example
providers need not worry about specifically crafting examples that do not
``overlap'' in the sense that they happen to share incidental refinements that
do not generalize to the correct solution.

\vspace{0.08in}

%% NOTE: Don't change this file.
%% Make the changes in the smyth repo and copy over via make ours.
\begin{figure}[h]

\experimentTableSize

\begin{tabular}{l|cc|cc}
& \multicolumn{4}{c}{\textbf{\snsMyth{}}}
\\\hline
\multicolumn{1}{r|}{\textbf{Experiment}} &
\textbf{5a} & \textbf{5b} & \textbf{6a} & \textbf{6b}
\\\hline
\multicolumn{1}{r|}{{Sketch / Objective}} &
\multicolumn{2}{c|}{\textit{None / Top-1}} &
\multicolumn{2}{c}{\textit{Base Case / Top-1-R}}
\\\hline
\multicolumn{1}{r|}{Type Specification{}} &
\multicolumn{2}{c|}{\textit{Polymorphic}} &
\multicolumn{2}{c}{\textit{Polymorphic}}
\\\hline
\textbf{Name} &
\textbf{Expert} & \textbf{Random} &
\textbf{Expert} & \textbf{Random}
\\
&
& {(50\%, 90\%)} &
& {(50\%, 90\%)}
\\
\input{figure-20-data}
\end{tabular}

\vsepBeforeCaption
  \captionsetup{justification=centering}
  \caption{
    Experiments with Polymorphic Types.
    %% Tasks without polymorphic specifications are marked ``\labelColorSkipped{---}''.
    \\
    \textbf{5a}:
      Percentages w.r.t. to number of examples in Experiment 2a.
    \\
    \textbf{6a}:
      Percentages w.r.t. to total specification size in Experiment 3a.
  }
\label{fig:poly-experiments}
\end{figure}

\end{document}